# Gene and Gene-Set Analysis for Genome-Wide Association Studies

Inti Pedroso

National Institute of Health Research Specialist Biomedical Research
Centre for Mental Health
&
Medical Research Council Social, Genetic and Developmental Psychiatry
Research Centre

King's College London

This dissertation is submitted for the degree of
Doctor of Philosophy
March 2011

# Abstract

Genome-wide association studies (GWAS) have identified hundreds of loci at very stringent levels of statistical significance across many different human traits. However, it is now clear that very large samples ($n\sim10^4$-$10^5$) are needed to find the majority of genetic variants underlying risk for most human diseases. Therefore, the field has engaged itself in a race to increase study sample sizes with some studies yielding very successful results but also studies which provide little or no new insights. This project started early on in this new wave of studies and I decided to use an alternative approach that uses prior biological knowledge to improve both interpretation and power of GWAS. The project aimed to a) implement and develop new gene-based methods to derive gene-level statistics to use GWAS in well established system biology tools; b) use of these gene-level statistics in networks and gene-set analyses of GWAS data; c) mine GWAS of neuropsychiatric disorders using gene, gene-sets and integrative biology analyses with gene-expression studies; and d) explore the ability of these methods to improve the analysis GWAS on disease sub-phenotypes which usually suffer of very small sample sizes.

In this project, we focused on the analysis of GWAS on bipolar disorder, however, GWAS from other complex disorders have also been analysed and are used throughout to compare the performance of the methods developed across disorders that, very likely, have different genetic architecture. From the analysis of these datasets, it was possible to draw that conclusion.

I found that from a computational perspective the application of gene and gene-set analyses to GWAS is feasible, even if calculation are performed on a common desktop computer. As a genetic mapping tool, gene-based association of GWAS provides a valuable complement to single SNP analysis methods. It highlights true disease loci and generates an accurate association statistic for a gene. These gene-level statistics proved of merit for meta-analyses of GWAS and integrative analysis with protein-protein interaction networks or gene-expression studies. In this thesis, I combined results of GWAS and gene-expression studies and I was able to provide evidence of association with bipolar disorder for a protein-interaction network. This demonstrates the possibility of using GWAS to extract information of biological systems from GWAS of neuropsychiatric disorders. This information is not readily available from single SNP analyses, which also suffer from genetic heterogeneity. Therefore, these methods help to improve the interpretation and provide an alternative to the simple but costly method of improving statistical power through increasing sample size. This ability proved to be important in our analyses of bipolar disorder sub-phenotypes, which had significantly smaller sample sizes. Analysis of disease sub-phenotypes showed that some genetic associations are shared with the main diagnosis but that others seemed largely specific of the sub-phenotypes. This evidence allowed me to suggest that genetic association with sub-phenotypes should be analysed as both a follow up of association in a main diagnosis and an independently if the available evidence is suggesting that the genetic associations observed are largely independent of those with the main diagnosis (e.g., as we found for age of onset of bipolar disorder).

Despite the interesting new insights and potential relevance, our gene and gene-set associations with bipolar disorder require replication in independent samples. This is especially important for the association found with the bipolar disorder sub-phenotypes because their smaller samples size also makes them more prone to be false positives.

This project has also generated the software suite FORGE which allows users to perform fast and robust analysis of GWAS, allowing researchers with a lack of experience in bioinformatics and statistical genetics, e.g. wet-laboratory based biologists, to perform both gene-based and pathway analyses to generate a systems biology interpretation of their phenotype of interest.

# Acknowledgements

I am most grateful to everyone that contributed to the work of this thesis, for giving advice and encouragement.

First of all, I would like to thank my supervisor Gerome Breen. His imagination and scientific twist has always prompted me to look at scientific problems from a different perspective. I would also like to thank Mike Barnes, who never spared words for strong criticism but also to give support and prompt me to be critic, ruthless and grasp beyond the obvious. Finally, Tom Price, who joined my PhD supervision in the last turn of the race but provided timely and sharp advice and time for enjoyable scientific conversation. Thanks to you since you have helped the most.

Of all others I shall first thank my wife Katherinne for cooking and gardening our love. I must thank my parents for giving me space to be or not to be and for giving me a whole life of love. I must also thanks my grand mother and my uncle Gonzalo for bringing passion into this endless route of discovery and for making me believe in people and what they can do. Thanks to you because you made me what I am.

I cannot forget my friends, both in Chile and UK, who have helped me from the beginning and continue we help me now. In particular I would like to thank Anbarasu L, James R, Sara Campos, Sarah Cohen, Sarah Jugurnauth, Margarita, Shaza, Katherine Tansey, Ursula, Cerisse, Jo G, Carol Shum, Sietske, Evangelos Vassos, Bahare Azadi, Chloe Wong and many other people has made my years as student in the SGDP a pleasure. Because of my sweet tooth I must also express my gratitude to the Fellowship of the Cake, for baking every month and sharing with me the sweetest of the friendships.

Thanks to David Collier, Cathryn Lewis, Stuart Newman, Ammar Al-Chalabi, all member of the Depression Genetic Consortium led by Prof Peter McGuffin and many others for small and big gestures of help or advice.

I am immensely grateful for the funding provided by the NIHR BRC for Mental Health and the Overseas Research Studentship Award Scheme.

Finally, I must thank the patients and researchers who gathered the data I analysed. Without their hard work this thesis would not be possible.

# Do we need to do all of this?

As a first year PhD student I used to ask myself, Do we need all these genetic studies? Are they relevant at all? Are we just teasing ourselves while pleasing our big brain's curiosity? Is the answer relevant just because we want to know it or can know it? Unfortunately, I still have no answer.

I have learnt that beyond pleasing ourselves genetic research can be tremendously valuable when a real breakthrough occurs and people out there benefit from it. That is the reason I focused my efforts in help transforming GWAS into something useful to biologists, with the hope that it would help produce solutions for real people. I do think more genetic research should be funded and more genetic findings should end up in systems biology efforts, where causality and complexity can be studied systematically. At the end of the day complexity is at the heart of why we are in love with biology and why is so challenging. However, I have also learnt that we may be missing an important bit in here. Many diseases do have a strong genetic liability but also have an important environmental contribution. Many have risen in frequency beyond what genetics alone can explain. We inherited our genes and not our diseases. Changes in lifestyle may be an important aspect of decreasing disease morbidity, probably in some cases beyond what genetic findings can do.

I think human genetics will soon face great scientific challenges. The massive amounts of incoming data will enable us to test some of our wildest hypothesis with an accuracy and depth not seen before. New sequencing technologies are opening the door to systematic and almost complete sampling of genetic material. Protein measurements may probably follow soon as will quantifications and sequencing of lipids and sugars. Biology will move aways from a reductionistic science into a data driven technology.

Quantitative models will supersede speculation and medicine will become a science of forecasting, in the same way that weather forecasting changed in the last 80 years. Nevertheless, I think you still will have to go to the gym to lose weight and get fit for the summer.

# Contents

## III   GENE-BASED GENOME-WIDE ASSOCIATION REVEALS TRUE DISEASE LOCI NOT FOUND BY SINGLE SNP

# List of Tables and Figures

## Figures

**Figure 3.3. In silico validation of gene-based associations.** The ability of gene-based analyses to identify true disease loci was benchmarked by: i) assessing how many of their significant findings are replicated in other studies; and ii) how many of these significant genes were not identified by using single SNP analyses alone. Venn diagram and workflow boxes' colours represent gene counts and steps, respectively, of the different analyses: light green (left) = gene-based association, light blue (centre) = single SNP analyses and purple (right) = benchmark set from NHGRI GWAS Catalogue. Results were analysed regarding the proportion genes of the benchmark set identified only by either single SNP analyses or gene-based association (see yellow boxes on "In silico benchmark"). In addition we evaluated the use of gene-based

**Figure 4.1. Data analysis strategy.** Gene-wide p-values for association with BD were calculated on four GWAS and 2 gene-expression studies. Gene-wide results were mapped to the PPIN and a greedy search strategy (Ideker et al., 2002) was used to identify subnetworks of interacting proteins enriched with BD association signals. A meta-analysis of the GWAS gene-wide association p-values and PPIN subnetworks results was carried out. GWAS PPIN results were combined with gene-expression subnetwork results and significant subnetworks were characterised with regard to their biological functions and locus heterogeneity. Colour codes are: Grey = statistical/ bioinformatic analyses, yellow = final results, blue = GWAS data and green = gene-expression data. DLPFC: doroslateral prefrontal cortex and OFC = orbitofrontal cortex tissue. .....................................................................................................................112

**Figure 4.2. Significant network from PPIN-wide search and CGNet analysis.** Blue nodes had a p-value < 0.05 in at least one study of the GWAS. Grey node had p-values > 0.05 in all studies. A) PPIN-1576 sub-network found by PPIN-wide search and B) CACNA1C sub-networks merged. CACNA1C is highlighted in yellow.................117

**Figure 4.3. Gene-wide p-values for PPIN-1576.** Plotted are -log10 gene-wide p-values on the y-axis for each gene within PPIN-1576. Different studies have been coloured as indicated in the legend and their bar plots were stacked to avoid over-plotting and allow comparison. The main figure presents data for genes with p-values < 0.05 in at least one study GWAS or the gene-expression DLPFC. Insert (top right)

# Tables

# Chapter I

# Introduction

# 1.1. Genetic Mapping

Genetic mapping can be defined as the identification of correlations between genetic variants and phenotypes (Altshuler et al., 2008). Its simplest form was introduced by Sturtevant for fruit flies in 1913. However, it was not until the 1980s that its application to human diseases became common and not until the last 6 years that it became widespread. We are currently at the middle of extraordinary growth in the data describing human genetic variation and its correlation with multiple phenotypes.

Until recently, genetic mapping of human diseases, especially for complex traits, was a fierce battle of ideas and theoretical models often with little or no empirical data, simply because the technology to collect the data was not available. This changed with the development of enabling technologies, such as high density oligonucleotide arrays and high throughput sequencing, allowing an international consortium of geneticists to develop a foundational resource called the HapMap (described in section 1.1.3). The HapMap project enabled the development of a new generation of genetic studies. After the resulting first wave of large scale studies, new questions have emerged while old ones have been reformulated and the theoretical models underlying much of our analysis have been challenged and greatly enriched.

### 1.1.1.    Sweet and sour: From Mendelian to complex diseases

Initially, genetic mapping of human diseases was impeded by human's relatively small family size, their mating (which cannot be experimentally controlled) and the few classical genetic markers available to trace inheritance patterns (Lander and Schork, 1994). Development of linkage maps in the 1980s and 1990s provided the genetic markers to systematically study families and identify genomic regions associated with human diseases (Altshuler et al., 2008). These regions were usually followed up by cloning and sequencing to identify the causal mutations and the affected gene in a process termed *Reverse Genetics*, the most well known example of which is the localisation of the CFTR gene as a cause of cystic fibrosis (Kerem et al., 1989; Collins, 1995). The approach led to a rapid increase in the number of human Mendelian disorders mapped to specific loci, from ~100 by the late 80s to >2000 today (www.ncbi.nlm.nih.gov/Omim).

Soon researchers aimed to disentangle the genetic basis of complex traits. The most common and chronic human diseases are complex traits, many of which present significant familial clustering and heritability and yet have no clearly discernible Mendelian inheritance pattern (Lander and Schork, 1994). Linkage mapping in complex disorders resulted in mixed results. Success was found with Mendelian-subtypes of complex traits, e.g., in breast cancer (Welcsh and King, 2001), hypertension (Lifton, 2004), and diabetes (Bell and Polonsky, 2001), but these variants were mostly rare and explained only a small proportion of the population risk [1] . In common forms of the diseases, linkage results provided poorly replicated results and it gradually emerged that complex traits were not driven by one or several genes of major effect but instead were likely to driven by polygenic risk resulting from the cumulative contribution of many risk loci throughout the genome (Altshuler et al., 2008).

---

[1] A commonly used method to calculate the additive genetic variance attributable to a genetic variant is described in Risch (2000). Briefly, for continuously measurable traits, we consider a single locus L with two variant alleles A and a with population frequencies p and q = 1 − p, respectively. We first defined two parameters. Displacement (t) is the number of standard deviations difference between the mean values of the two homozygotes AA and aa (assuming equal variance within each genotype). The second parameter, d, represents the mean value of heterozygotes Aa relative to the two homozygotes: d = 1 corresponds to equal means for genotypes AA and Aa (that is, A is dominant), d = 0 corresponds to equal means for genotypes Aa and aa (that is, A is recessive) and d = 0.5 corresponds to the additive situation. Then the population variance attributable to segregation of the alleles is given by $V_G(L) = V_A(L) + V_D(L)$, where $V_A(L) = 2pqt^2(p(1 − d)+qd)^2$ is the 'additive' genetic variance and $V_D(L) = p^2q^2t^2(d − 0.5)^2$ is the 'dominance' genetic variance. The proportion of total variance attributable to locus A, which we denote as $h^2(L)$, is then given by $V_G(L)/(1 + V_G(L))$, assuming the variance within genotype to be 1.0. The formulas for the dichotomous case are analogous but we replace t by the genotypic risk ratio (GRR).

### 1.1.2. The common disease-common variant model

In 1996, Risch and Merikangas proposed that association studies, using a case-control or sib-pair design, should supersede linkage studies in the search for risk alleles of complex traits. The argument was relatively straightforward: linkage and association have good statistical power to detect different kinds of diseases-susceptibility alleles. Linkage provides good power to pick loci harbouring one or multiple common or rare alleles with large effects but is not well suited to the identification of alleles with low effect-sizes. Association studies are more suited to the identification of risk alleles that are of intermediate frequency. However, there was an additional proposal in this model: "*Despite the small effect size of common alleles the magnitude of their attributable risk (the proportion of people affected due to them) may be large because they are quite frequent in the population, making them of public health significance...*" (Risch and Merikangas, 1996). This argument forms the basis of the so called common disease-common variant (CDCV) hypothesis, also proposed by others (Lander, 1996; Cargill et al., 1999; Chakravarti, 1999). The opposite model proposes that common diseases can be genetically underpinned by multiple rare variants each explaining a small fraction of the population risk, known as the common disease-rare variant (CDRV) hypothesis. In this model, it is possible that initial linkage studies did not find more loci simply because they had a small effect size but it argues that these could still be rare (<1% frequency) and would be picked up by larger linkage but not association studies. The genetic architecture of most traits likely lies in between the two extreme models and variants across the frequency and effect size range will be associated with human complex traits (Altshuler et al., 2008).

The CDCV strategy fits better in a simple model conceiving a single risk variant per loci. If there are 2 or more independent susceptibility variants within the same locus (allelic heterogeneity) or genetic risk is spread across multiple locus (locus heterogeneity), association studies have less power (Terwilliger and Weiss, 1998; Pritchard, 2001). There is empirical and theoretical evidence supporting the presence of allelic and locus heterogeneity for alleles associated with human diseases at low frequency and intermediate to large effect sizes (Lander and Schork, 1994; Terwilliger and Weiss, 1998; Pritchard, 2001; Eyre-Walker, 2010).

After Risch and Merikangas (1996), it would take almost another decade until the CDCV hypothesis could be tested extensively because it demanded the generation of an unprecedented volume of data and new technological advances. The advent of this technology has also allowed further exploration of the CDRV hypothesis but not as thoroughly. Both models will be tested more comprehensively in the coming years.

### 1.1.3.    Construction of public resources to analyse human genetic variation

A major limitation for large association studies was the lack of a comprehensive map of genetic variants across the human genome. The SNP Consortium was launched at the end of the 1990s and generated an initial map of 1.4 million single nucleotide polymorphisms (SNPs)[2] by 2001 (Sachidanandam et al., 2001). The International HapMap Project was launched in 2002, with the goal of characterising SNP frequencies and linkage disequilibrium (LD)[3] patterns across the human genome in 270 samples from Europe, Asia, and West Africa. The project genotyped ~1 million SNPs by 2005 (International HapMap Consortium, 2005) and more than 3 million by 2007 (Frazer et al., 2007). In a subsequent phase more than 1 million polymorphic sites, including many rare and copy number variants, were genotyped in a larger panel of 1184 individuals (Altshuler et al., 2010). It was also found that, association studies would need to test every variant in the genome, but because genetic variants are correlated due to recombination patterns, a sub-set of markers could suffice to capture most of the genetic information of common variants (Gabriel et al., 2002; International HapMap Consortium, 2005). Thus it became feasible to perform genome-wide association testing using an informative sub-set of common variants to detect association with complex traits. This model has been severely criticised because correlations between genetic variants may not always provide as much information on disease susceptibility as originally proposed, simply because the assumptions built into these models do not always hold, e.g. multiplicativity of correlations (Terwilliger and Hiekkalinna, 2006). Since the initial HapMap report, the evidence showed that most common SNPs have high correlation with nearby genetic variants and suggested that about half a million SNPs would provide coverage for more than 90% of common SNP variation in non-African populations and that a million SNPs would provide the same level of coverage for African populations (International HapMap Consortium, 2005).

_________________

[2] Single nucleotide polymorphisms constitute the most common genetic variation across the human genome and represent position. They are classically defined as genomic positions where more than one allele is present in the population. SNPs are usually bi-allelic but tri-allelic are also found (Huebner et al., 2007). Although deletions or insertions of a single nucleotide may also be considered SNPs, they are usually not.

This fundamental data set became the basis for the development of a range of competing high-throughput microarray genotyping technologies, which made dense genome-wide analysis of large sample sets possible. SNP genotyping was initially performed one SNP at a time, at a cost of ~$1 per measurement. Multiplex genotyping of hundreds of SNPs on DNA microarrays was demonstrated in the late 1990s (Wang et al., 1998), and capacity per array grew from 10,000 to 100,000 SNPs in 2002 and from 500,000 to 1 million SNPs by 2007. In parallel, cost fell to $0.001 per genotype, or less than $1000 per sample for a whole-genome analysis. By 2006, multiple technological platforms could genotype hundreds of thousands of SNPs at high levels of completeness and accuracy (for details see Altshuler et al. (2008) and references therein).

---

[3] Linkage disequilibrium is a measure of the correlation between alleles at two loci. Its calculation is not equal but numerically it is equivalent to the Pearson's correlation between the alleles counts (Devlin and Risch, 1995).

In 2007, the 1000 genomes project (www.1000genomes.org) was launched to discover and describe haplotype information for all types of DNA polymorphisms in multiple human populations. Specifically, its goal was to characterise over 95% of variants that are in genomic regions accessible to second-generation sequencing technologies and have an allele frequency of 1% or higher in five human population groups (Durbin et al., 2010). By combining microarray-based genotyping with imputation technologies (Marchini and Howie, 2010), geneticists are now able to access millions of common genetic variants and many thousands of rare variants for association tests (Altshuler et al., 2010). Ongoing projects, such as the UK10K (www.uk10k.org), will provide whole genome sequences for thousands of individuals, enabling for the first time a systematic analysis of variants across the frequency spectrum. However, some variants at the low frequency extreme (~1/10,000) may need larger cohorts and probably additional pedigree or population information to be called or imputed accurately; for examples see (Kong et al., 2008; Stefansson et al., 2008).

### 1.1.4.    Genome-wide association studies

Genome-wide association studies systematically analyse genetic variation across the genome to study its effects on phenotypic change. Their success can be determined by factors like sample size, genotyping quality and coverage of genetic variants (Wang et al., 2005). Initial studies had modest sample sizes of several thousands of individuals and generally had sufficient power only for allelic odd-ratios > 1.5 and incomplete coverage of common genetic variants (Figure 1.1) (Wang et al., 2005).

The early landmark study using these technologies was the Wellcome Trust Case Control Consortium (WTCCC), which reported genetic mapping results for over 500,000 SNPs in 7 disease sample sets of ~2000 individuals each and ~3000 control individuals (WTCCC, 2007). Several of the diseases studied by the WTCCC had notable results. For example, known disease susceptibility loci were supported, e.g., SNPs within NOD2 in Crohn's disease and the human leukocyte antigen (HLA) loci at chromosome six in immune-system related disease (Figure 1.2), and new loci were highlighted at high significance levels (WTCCC, 2007). The WTCCC established its significance threshold for multiple testing correction at 7.5 x $10^{-7}$ by assuming the number of risk variants at ~10 and a statistical power of ~50% (WTCCC, 2007). Currently, genome-wide significance level is widely regarded as 5 x $10^{-8}$ as proposed by Dudbridge and Gusnanto (2008). Thus, some of the WTCCC results would currently only be considered as having suggestive significance (approximate p-value = 0.5/(# tests)).

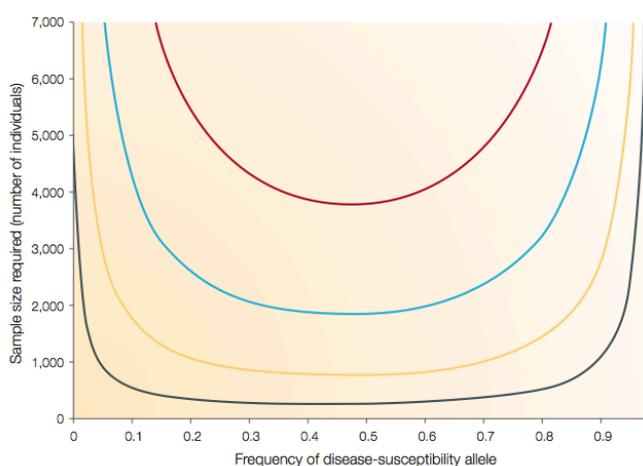

**Figure 1.1. Effects of allele frequency and effect-size on sample-size requirements.** The numbers of cases and controls that are required in an association study to detect disease variants with allelic odds ratios of 1.2 (red), 1.3 (blue), 1.5 (yellow) and 2 (black) are shown. Numbers shown are for a statistical power of 80% at a significance level of p <10–6, assuming a multiplicative model for the effects of alleles and perfect correlative linkage disequilibrium between alleles of test markers and disease variants. Figure reproduced from (Wang et al., 2005).

An important result of the WTCCC

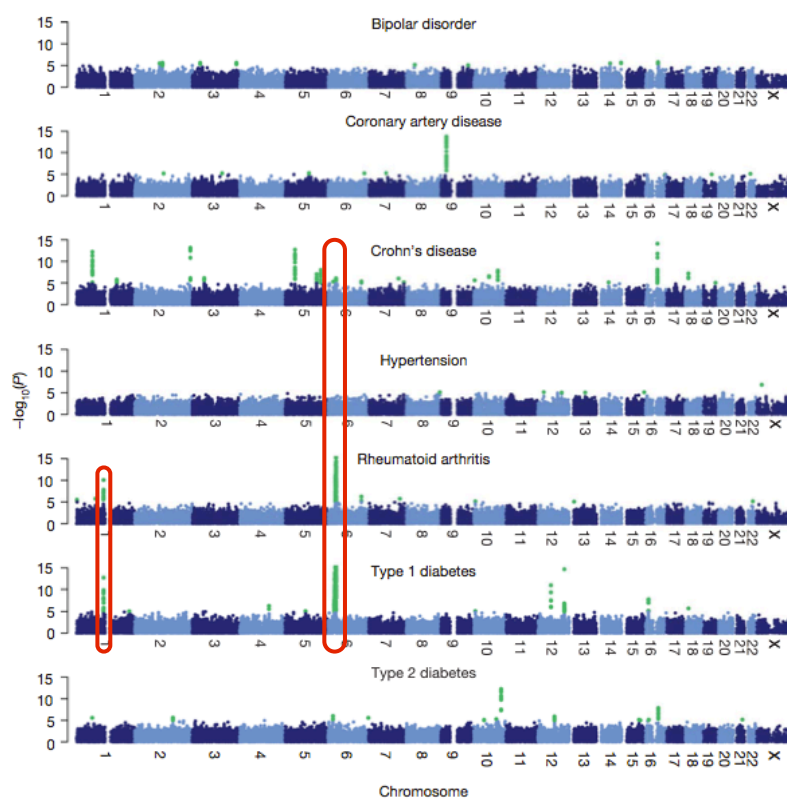

**Figure 1.2. Summary of Wellcome Trust Case Control Consortium GWAS results.** Single SNP association significance is plotted on y-axis as -log10 of p-values and genomic position along the x-axis. Chromosome position names are shown on the x-axis. We have highlighted immune-system regions showing consistent association across the immune-system related diseases. Figure reproduced from (WTCCC, 2007).

study was to show that common risk variants with large effect-sizes (>2) were unlikely to exist for complex diseases, in agreement with linkage studies. Therefore, larger sample sizes would be needed to establish compelling evidence of association with alleles of low effect size. For example, it took another 2 years to pool together sample sizes in the order of tens of thousands of individuals to robustly identify hypertension risk alleles with odd-ratios around 1.1 to 1.2 (Levy et al., 2009; Padmanabhan et al., 2010). Few (if any) single research centres can undertake such a study on their own and international collaborations are usually set up to coordinate these large meta-analyses. These efforts confirmed that replication in several independent studies is crucial to identify alleles of low effect size and overcome potential false positives, e.g. due to biases of individual studies or the winner's curse effect (Frayling et al., 2007; Yu et al., 2007; Zollner and Pritchard, 2007; Kraft et al., 2009; Lehrer, 2010). Increasing sample size has recently led to GWAS of very large size, e.g., with more than 50,000 for Crohn's disease (Franke et al., 2010), 100,000 for blood lipid levels (Teslovich et al., 2010) and 200,000 samples for body mass index (Speliotes et al., 2010). These efforts have identified many genetic variants associated with diverse human traits, robustly implicating common genetic variants in the aetiology of common human diseases. The reader is referred to the NIHGRI Catalog of Published GWAS for a compendium of published studies (www.genome.gov/gwastudies). Interestingly, as happened with low frequency variants discovered by linkage and positional cloning, researchers were puzzled by the fact that common variants reaching significance in GWAS explained only a small proportion of the estimated additive genetic contribution of a trait variation (Manolio et al., 2009). However, significant variants from the latest meta-analysis in Crohn's disease explained up to 23% (Franke et al., 2010) of the phenotypic variance attributable to genetic variation, for BMI between 6-11% (Speliotes et al., 2010), for lipid levels between 25–30% (Speliotes et al., 2010) and, for height, all SNPs simultaneously explained up to 45% (Yang et al., 2010a).

It is likely that by augmenting the sample size, the contribution of common variants alone will be better realised. Recent development in statistical technologies have shown that a large proportion of the phenotypic variance can be explained by combining additively information on hundred of thousands of genetic variants (first shown by Purcell et al., 2009 and more completely analysed and explored in Yang et al., 2010a; Lee et al., 2011; Yang et al., 2011). These studies show clearly how many and how much genetic variants with statistical associations under the genome-wide significance threshold contribute to a trait variation and disease risk. The statistical methods used in these studies only provide an overall measure of the heritability explained by all these genetic variants together and additional methods, e.g., those discussed in this thesis, are needed to identify their genomic location and biological meaning.

One significant issue is that most reported significant GWAS results to date are from studies in a few traits that have reached large sample sizes, such as Crohn's disease, height, type 2 diabetes and blood lipid levels, and might likely benefit from additional GWAS of common variants, but what does the future hold for studies of phenotypes with unsuccessful GWAS?

Until now, GWAS have only provided statistical power to detect high frequency disease susceptibility variants (Figure 1.1 and Wang et al. (2005)). An exception is some disorders that have been analysed for low frequency structural variants, e.g., in schizophrenia some variants identified have odds ratios of 3-10 but their frequency tends to be <0.1% in the cases and <0.01% in controls (Stefansson et al., 2008). This means that diseases with a genetic architecture with a contribution from predominantly low frequency alleles (< 1%) will consistently have poor results in GWAS. For example, for psychiatric traits, it has been proposed that rare variants play a predominant role because mental illness usually has an onset early in reproductive age and is associated with a substantial reproductive disadvantage. Therefore, susceptibility variants should be under strong negative selection (Uher, 2009). Although there is evidence to sustain this argument (Stefansson et al., 2008; Uher, 2009), the picture is likely more complicated. GWAS in psychiatric genetics may offer poor results because an important contribution to the genetic aetiology comes from low frequency variants but, on the other hand, a genome-wide search for rare variants requires substantial sample sizes to achieve statistical power. Family linkage studies are aimed at identifying these loci but have not been notably successful at identifying rare variants in fields such as neuropsychiatry. This is probably due to issues such as phenotypic heterogeneity, incomplete penetrance and disease diagnostic heterogeneity. Furthermore, recent studies have shown that there is a huge amount of rare variation in the human genome and that normal individuals carry about 250 to 300 loss-of-function variants in annotated genes and 50 to 100 variants previously implicated in inherited disorders (Durbin et al., 2010). So, even if we can afford the large samples sizes that provide statistical power for a rare variant search, how are we going to distinguish normal variation from pathogenic or pathogenic variation with low penetrance?

We think that a key problem here is our poor understanding of how genetic variants lead to phenotypic change, how they relate, add and interact in a non-linear fashion and how to build experimental models leading to the study of these phenomena (Weiss and Fullerton, 2000; Weiss and Terwilliger, 2000; Weiss, 2008). Some studies have been able to combine risk variants identified by genetic studies in humans with experiments aiming to understand their effect on cellular and physiological phenotypes relevant for the disease (Badano et al., 2005; Zaghloul and Katsanis, 2010; Davis et al., 2011). Others have, for example, resorted to system genetic analyses to infer the causal effect of risk alleles on cellular and physiological pathways that are later the focus of detailed research (Chen et al., 2008; Schadt et al., 2009a; Schadt et al., 2009b; Schadt, 2009; Webster et al., 2009; Yang et al., 2009; Quigley et al., 2011). The first strategy requires good animal models to dissect in detail the link between genotype, biological mechanisms and disease risk. Such models are not available for many diseases but when available, they have proven an important complement to genetic mapping in humans (Zaghloul and Katsanis, 2010). The ciliopathies are one such case, which have focused efforts to assess the effect of risk variants on the function of the cilia (Badano et al., 2006). Interestingly, the number of proteins in the cilia is ~1000 (Gerdes et al., 2009), which is similar to that detected in the neuronal post-synaptic density in human and mouse (~1500 proteins (Bayés and Grant, 2009)), suggesting that an experimental system to study the effect of risk variants of mental illness on neurone function may be feasible and would be fruitful. A systems genetic approach requires enough statistical power to perform causal modelling and to test for interaction effects. These types of datasets are not abundant yet, but are likely to be become more common and be a powerful complement to other genomic technologies, e.g., Webster et al. (2009) and Choi et al. (2011) are examples of the application of systems genetic to neuropsychiatric disorders.

Undoubtedly, psychiatric genetics would greatly benefit from a strategy incorporating both large genetic studies of common and rare variants and adequate experimental models targeting discrete biological systems (e.g. post-synaptic density) to understand the effect and properties of risk conferred by genetic variants. However, a major challenge for the psychiatric genetic field will be to define biological systems or functions that are relevant to understand disease aetiology or treatment and that merit detailed scientific study. Access to good quality post-mortem brains from psychiatric patients is not easy and thus genomic technologies cannot be powerfully leveraged to provide biological clues as in other diseases with easier tissue access (e.g. cancers). We think that genetics and GWAS have great potential to take on this challenge.

## 1.2. Challenges of current GWAS

GWAS of neuropsychiatric diseases have provided few robust genetic associations, e.g., Ferreira et al. (2008) and Stefansson et al. (2009). Large odds-ratios for common variants are not common and may not exist in many disorders. In addition, there is mounting evidence to suggest that genetic and phenotypic heterogeneity are important concerns (for example in bipolar disorder, see (Alda, 1999; Alda, 2004; Schulze et al., 2009)). Phenotypic heterogeneity may be particularly problematic in psychiatric genetics, where tests for physical symptoms or biomarkers are not available or are poorly defined, and environment is thought to play a key role (Caspi et al., 2003). Genetic heterogeneity is to be expected based on empirical evidence and theoretical models (Lander and Schork, 1994; Terwilliger and Weiss, 1998; Pritchard, 2001).

There has been strong criticism of GWAS due to the low predictive power of common risk alleles (e.g., Goldstein (2009)). A notable exception in psychiatric genetics is the APOE ε-4 allele (Pericak-Vance et al., 1991), which may account for more than half the genetic variance in risk for Alzheimer's disease (AD). Although it was discovered by positional cloning of a linkage region in the pre-GWAS era, it is one of the few replicated loci for AD in GWAS. Furthermore, its discovery has not so far led directly to new treatments or genetic testing for AD, although it is showing potential to influence clinical practice and treatment on the basis of sub-classification of the disease into early and late onset forms (Roses et al., 2007).

It is important to make a distinction between acquiring predictive power and increasing biological knowledge. As noted by Altshuler et al. (2008), a good example is Brown and Goldstein's studies of hypercholesterolemia (Brown and Goldstein, 1974a; Brown and Goldstein, 1974b), which affects ~0.2% of the population and accounts for a tiny fraction of the heritability of low-density lipoprotein (LDL) and myocardial infarction. Studies of hypercholesterolemia led to the discovery of the LDL receptor and supported the development of 3-hydroxy-3-methyl glutaryl–coenzyme A reductase (HMGCR) inhibitors (also known as statins) for lowering LDL, the use of which is not limited to hypercholesterolemia risk variants carriers. Although SNPs in HMGCR have only a small effect (~5%) on LDL levels (Kathiresan et al., 2008; Willer et al., 2008), drugs targeting the encoded protein decrease LDL levels by a much greater extent (~30%). This is because the effect of an inherited variant is limited by natural selection and pleiotropy, whereas the effect of a drug treatment is not. Thus, the primary objective of GWAS may not be to produce powerful predictors but to reveal clues about the biological underpinnings of a disorder which can be gathered to improve treatment and diagnosis.

I propose that some of the most important challenges of GWAS concern data interpretation and that these will transcend the analysis of both common and rare variants. I have formulated my concerns as follows:

i) How to extract biologically meaningful results from GWAS. After applying the stringent statistical thresholds necessary for multiple testing correction, there are very few significant variants (often none!) presenting little information about the underlying biology of phenotypic variation.

ii) How to identify the many real disease susceptibility loci that do not reach genome-wide significance in individual studies, e.g. PPARG in type 2 diabetes (Altshuler et al., 2000). Although combining data sets by meta-analysis has been a successful approach, it faces problems in the presence of genetic heterogeneity (and data availability).

iii) How to tackle genetic heterogeneity present as both allelic and locus heterogeneity. It is known that genetic heterogeneity will decrease the power of GWAS and methods robust to it are needed to extract information from GWAS. An additional challenge is to quantify this genetic heterogeneity.

The many thousands of marginally significant results could hold the "higher hanging fruit" of real associations that fail to reach genome-wide significance. This may be due to a lack of power caused by either low allele frequency, effect size or phenotypic heterogeneity. It may prove useful to interpret these genetic associations in the context of the genes and biological processes they may influence, the aim being to identify biological processes acting as (subtle) drivers of the cellular and physiological changes that confer disease susceptibility. Therefore, improving interpretation of GWAS and helping to increase its power to detect small effect size alleles by clustering them in common biological processes may in turn prove a fruitful target to develop new treatments (for an example see Fleming et al. (2011)).

## 1.3. Gene-set and network analyses

GWAS analyses are most commonly performed by testing the strength of association between individual genetic variants and phenotypes. This univariate strategy has parallels with analysis of gene-expression data where the association between each probe and the phenotype is tested. We will exploit this similarity to present gene-set analyses (GSA) and network analyses for GWAS.

In contrast to gene-centric approaches, such as assessing expression differences of individual genes or gene products, GSA and network analysis aim to capture the collective activity of sets of molecules by testing the group instead of each component individually. For example, large expression differences in one molecule provide supporting evidence for its perturbed state but little information to answer relevant questions like *What are the biological processes that explain this large differential expression? How do the molecules showing large differences relate to each other? or Are there any small expression changes likely to be biologically significant?* These are some of the questions that GSA aims to address.

There are hundreds of variants of GSA (reviewed in (Nam and Kim, 2008; Ackermann and Strimmer, 2009)). All aim to find association with gene-sets rather than individual genes. Gene-sets are simply groups of gene products that have been put together using some criteria, e.g., participating in the same biological process or having the same sub-cellular localisation. Thus, GSA centres the focus of attention on changes in biological processes. This strategy is supported by the fact that cellular and physiological changes are driven by groups of gene products acting coordinately, e.g., to execute cellular functions or process signalling information. Finding an association with a biological process can, potentially, help to infer something about that biological process on the condition of interest. For example using evidence from a gene-expression study one can suggest that genes of a signalling pathway are up-regulated by 20% in cases compared with controls.

Ackermann and Strimmer (2009) analysed 261 different GSA methods and proposed a general framework to describe them. An interesting result from their report is that most methods, although differing in the statistical methods used, provide very similar results. Major differences in sensitivity were related to differences in relatively few methodological steps, such as summarising the evidence at the gene-set level or the choice of the null hypothesis. For example, calculating a simple statistics like the mean gene-expression change within a gene-set had more statistical power than complicated methods, such as the popular gene-set enrichment analysis (GSEA) (Subramanian et al., 2005) or Hotelling-type statistics (Kong et al., 2006; Dinu et al., 2007).

Gene-sets (see Table 1.1 for some sources) can be constructed via literature review or compilation of results from high throughput experiments and allow the incorporation of previous knowledge, such as tissue specific expression profiles or manually curated gene-gene interaction information. They allow us to test specific biological hypothesis across datasets and methodological techniques (Lehner and Lee, 2008). A drawback of GSA is its strong dependence on the availability and quality of gene-sets. Better annotation is expected for well studied processes, for example cancer related pathways, compared with biological processes that are difficult to study or the focus of limited research. Manual curation of bionformatic resources is greatly valued and of commercial value, for example see www.genego.com or www.ingenuity.com. Unfortunately, manual curation has not scaled up to the speed of data generation by newer genomic technologies. Generation of gene-sets from high-throughput experiments is currently a challenge (Davies et al., 2010) and how to incorporate the uncertainty of each gene-set and experimental source into the GSA is currently very much unexplored.

| Pathway definition resources | |
|---|---|
| BioCarta | www.biocarta.com |
| KEGG | www.genome.jp/kegg |
| Gene Ontology | www.geneontology.org |
| MSigDB | www.broad.mit.edu/gsea/msigdb |
| **Gene Set Analysis Tools for GWAS** | |
| FORGE | https://github.com/inti/FORGE/wiki |
| GSEA | www.broad.mit.edu/gsea |
| FatiGO | http://fatigo.bioinfo.cnio.es |
| GenGen Package | www.openbioinformatics.org/gengen |
| GSEA-SNP | www.nr.no/pages/samba/area_emr_smbi_gseasnp |
| ALIGATOR | http://x004.psycm.uwcm.ac.uk/~peter |
| i-GSEA4GWAS | http://gsea4gwas.psych.ac.cn |
| GESBAP | http://bioinfo.cipf.es/gesbap |
| GRASS | http://linchen.fhcrc.org/grass.html |
| GSA-SNP | http://gsa.muldas.org |
| PLINK set-test | http://pngu.mgh.harvard.edu/~purcell/plink |

**Table 1.1.** Tools to explore GWAS using GSA. We present a list of resources to obtain gene-set definitions and software to perform GSA on GWAS.

Much of the information used as input in constructing gene-sets relates to the interaction between biological molecules, e.g. phosphorylation or co-expression, which can also be represented as an interaction network. Network analysis makes direct use of the interaction information[4] and aims to identify groups of interacting gene products associated with a phenotype (Ideker and Sharan, 2008). They are somewhat more flexible than GSA because the groups of genes being tested, equivalent to gene-sets, can be defined in different ways (for examples see Ideker et al. (2002), Chuang et al. (2007) and Dittrich et al. (2008)). Some of the limitations of network analysis are: a) finding significant sub-networks is an NP-problem, i.e. execution time scales exponentially and usually computationally intensive algorithms are used, and b) the exponential nature of the interaction information makes its acquisition slow and biased against interactions or molecules difficult to study, such as transmembrane proteins or interactions involving sugars or RNA molecules (see review by Schwartz et al. (2009)). It is possible to overcome some of these biases by constructing networks using functional genomics data (Lage et al., 2007; Webster et al., 2009; Lage et al., 2010). This strategy makes used of information from different sources, such as gene-expression, proteomics, text mining and protein-protein interaction assays. Each of these technologies has its own biases. Therefore, it is possible to combine interactions gathered with different technological platforms to construct an integrated network with better coverage and more information than any produced with a single experimental source (for examples see Lee et al. (2008), Lehner and Lee (2008) and Zhu et al. (2008)).

---

[4] A gene-set does not provide any information on how the molecules interact with each other whereas in a network the pairwise interactions are explicit. Often, additional information is available for these interactions. For example, the network can be undirected providing only information about the pairwise connections or it can be a directed network, in which case the interactions have a directionality. Furthermore, each interaction can be characterised by its strength or nature (e.g. phosphorylation in the case or kinases or DNA-binding for transcription factors). These information can be explicitly integrated into the statistical analyses, e.g., giving more importance to molecules with more or stronger connections. We refer the interested reader to an introductory book by Newman (2010) and a review article by Ideker and Sharan (2008) for additional details.

Both GSA and network analysis have been successfully used to characterise, prioritise and shed new light on the molecular processes underlying phenotypes in human and model organisms (Fuller et al., 2007; Chen et al., 2008; Emilsson et al., 2008; Ideker and Sharan, 2008; Presson et al., 2008; Zhu et al., 2008). Thus their development as a tool to interrogate GWAS is promising. For example, replication of results across studies and experimental platforms is often easier at the gene-set level than using individual genes (Chuang et al., 2007; Chen et al., 2008; Emilsson et al., 2008). GSA and network analyses can help to reduce multiple testing and facilitate follow up studies because often the biological role of a gene-set is better understood that that of single genes (Schadt et al., 2009b).

## 1.4. GSA and network analyses in GWAS

Before commenting on GSA and network analyses in GWAS, I need to point out some caveats. GSA was originally developed for and has been extensively applied to gene-expression data. In gene-expression data is common to observe high correlation of the gene's expression. This high autocorrelation makes the analysis of gene-sets easier because gene expression changes occur in a concerted manner. Furthermore, gene-expression analyses capture changes that can be the cause or consequence of the phenotype (Schadt, 2009) and such changes are often of several orders of magnitude. On the other hand, genotype-phenotype correlations of common variants with complex traits discovered with GWAS have subtle effect sizes of much smaller magnitude than those of gene-expression studies and are not correlated with each other beyond what is expected due to local LD. These differences have important practical consequences. Whereas a strong GSA finding on a gene-expression study is usually supported by several genes, providing evidence that a significant portion of a biological process has changed. On the other hand, in a GWAS one significant association in the same gene-set may not be statistically significant in GSA but its functional consequences may indeed perturb the entire biological process. Although this does not invalidate the application of GSA methods, it suggests that biological processes affected by a single SNP will not be found by GSA of GWAS. This is an important aspect to consider when interpreting the results of this thesis. On the other hand, significant association in a GSA or network analysis of GWAS provides strong evidence for the biological processes contributing to phenotype variation because each of the signals is independent from each other (in as far as the genetic variants are not in LD), i.e. each represents additional evidence and they are not a consequence of the phenotype.

There is a theoretical background as well as empirical evidence supporting the use of gene-set methods in GWAS (Gibson, 2009; Wang et al., 2010). The main arguments being: a) the underlying cause of phenotypic variability are changes in biological processes at the cellular and organismal level; b) perturbations can arise from any element acting on these process; and c) many alleles with small affect size will explain phenotypic variation of complex traits. Wang et al. (2007) introduced GSA to GWAS with an implementation of the GSEA algorithm (Subramanian et al., 2005). Since then, there has been an increasing number of studies applying or proposing new GSA and network methods for GWAS, witnessing the interest of the field for their development. Published articles range from the application of gene-expression analysis techniques to new approaches exploiting additional information for genetic mapping, such as genetic interactions or integration with functional data like gene-expression. Table 1.2 provides a representative list of publications using GSA and network analyses in GWAS. Figure 1.3 presents the pipeline for the application of GSA to GWAS proposed by Pedroso and Breen (2009).

| Reference | Phenotype | Method | Software |
|-----------|-----------|--------|----------|
| Wang et al., 2007 | PD, AMD | GSEA | GenGen Package |
| Inada et al., 2008 | TD | Fisher exact test | Ingenuity Pathway Analysis |
| Lesnick et al., 2007 | PD | Regression methods | |
| Iossifov et al., 2008 | BD, SZC, AU | Gene mixture generative model | |
| Walsh et al., 2008 | SZC | Fisher exact test | Ingenuity Pathway Analysis |
| Craddock et al., 2010 | SABP | mean chi-square | Manually curated |
| Breuer et al., 2010 | SABP | mean chi-square | Manually curated |
| Holden et al., 2008 | | GSEA-SNP | GSEA-SNP |
| Torkamani et al., 2008 | BD, CD, CAD, T2D, T1D, RA, HT | Hypergeometric test | MetaCore |
| Srinivasan et al., 2009 | PD | Boosted decision trees | |
| Wang et al., 2009 | CD | GSEA | GenGen Package |
| Perry et al., 2009 | T2D | GSEA | GenGen Package |
| Yu et al., 2009 | CSB | Adaptive Combination of P-Values | |
| Askland et al., 2009 | BD | Exploratory visual analysis | Exploratory Visual Analysis (EVA) |
| Chen et al., 2009 | BD, CD, CAD, T2D, HT | Pathway total risk estimation | |
| Elbers et al., 2009 | T2D | mixed | Webgestalt, GATHER, DAVID, PANTHER, BioCarta |
| Baranzini et al., 2009 | MS | Network analysis | jActive Modules |
| Peng et al., 2010 | BD, CD, CAD, T2D, HT, PD, AREDS, ALS | multivariate method | |
| Luo et al., 2010 | RA | multivariate method | |

**Table 1.2: Reports of GSA of GWAS.** BD = bipolar disorder, CD = crohn's disease, CAD = coronary artery disease, T2D = type 2 diabetes, HT = hypertension, T1D = type 1 diabetes, CSB= cigarette smoking behaviours, PD = Parkinson's disease, RA = rheumatoid arthritis, SZC = schizophrenia, AU = autism, TD = treatment-resistant tarditive dyskinesia, AMD = age-related macular degeneration, SABP = schizoaffective bipolar disorder sub type, AREDS = age-related eye disease, ALS = amyotrophic lateral sclerosis.

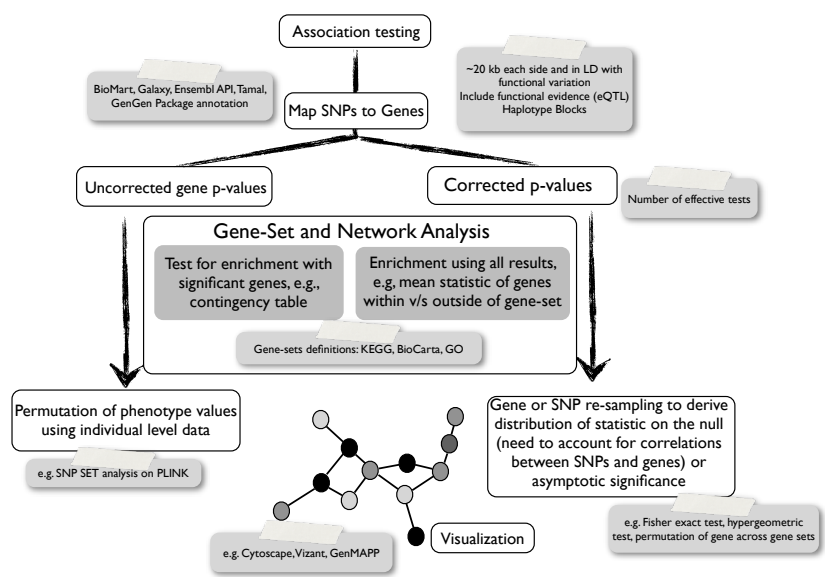

**Figure 1.3. Proposed pipeline to apply Gene-Set Analyses to GWAS.** SNP summary statistics are mapped to genes and correction for gene-size biases is applied by either phenotype permutations or a statistical method using a measure of the number of effective tests. Genes-wide statistics are later mapped to networks and biological pathways to test for over-representation of association signals. We have highlighted databases and software publicly available to perform different steps of the analysis. For additional details see review by Pedroso and Breen (2009).

Probably the most relevant methodological finding to date has been the ability of GSA to provide successful replication at the pathway level in the presence of allelic and locus heterogeneity (Baranzini et al., 2009; Wang et al., 2009; Luo et al., 2010; Peng et al., 2010). These examples open the door for the application of GSA in large scale studies, which in the long term will offer the evidence to judge the relative contribution of GSA to genetic mapping studies.

In the psychiatric genetics field, several reports have found significant association with biological processes using GWAS. Analysis of the WTCCC bipolar disorder GWAS and its meta-analysis with another GWAS provided evidence of association within biological processes involved in the modulation of transcription and cellular activity, including that of hormone action and adherens junctions (Baranzini et al., 2009; Holmans et al., 2009; O'Dushlaine et al., 2011). Craddock et al. (2010) observed a strong enrichment for associations within the GABRB1 genes in patients diagnosed with the subtype schizoaffective bipolar disorder type (SABP). Associations with SNPs within the genes GABRA4, GABRB3, GABRA5 and GABRR1, all subunits of the GABRA A receptor, providing a system wide association p-value = 7 x $10^{-5}$. Recently, Breuer et al. (2010) successfully replicated the association between GABRA A receptor genes and SABP (p-value = 0.003). However, as we will see in Chapter V, these two associations with the GABRA A receptor genes may not be as clear as they currently look. O'Dushlaine et al. (2011) found replicated association with five gene-sets in two large schizophrenia studies. One of these gene-sets, cell adhesion molecules (hsa04514), was also associated with bipolar disorder. Baranzini et al. (2009) overlaid results of eleven GWAS of 9 diseases in a protein-protein interaction network to find network modules associated with one or more disorders. They found that in general, each disorder is represented by a distinctive group of modules but some modules are shared across disorders, such as modules associated with HLA gene products across autoimmune disorders or modules shared between central nervous system disorders like bipolar disorder, Alzheimer's disease and multiple sclerosis. This suggests that common genetic variation can be used not only to reveal associations with network modules but also to explore the molecular basis of similar, and often comorbid, disorders.

However, caution is required when interpreting reports of GSA of GWAS. For example, some authors do not control for the number of variants tested within a gene-set and a large number of false positives may be observed in GSA simply because some gene-sets have more or larger genes (Elbers et al., 2009; Jia et al., 2010). During my literature review, I noted several of these cases, suggesting that some results may be the product of the gene-set size bias; examples are  Torkamani et al. (2008), Walsh et al. (2008) and Askland et al. (2009).

# 1.5. Objectives of this thesis

GSA and network analysis are proving to be a good complement to conventional approaches for the analysis of GWAS. There are several areas in which methods can be developed in order to extract the most from their results. For example, most studies have used the minimum p-value in the gene to represent association. However, recent evidence suggests that allelic heterogeneity in studies of genetic association with common alleles may be more common than expected, for example (Schulze et al., 2009; Lango Allen et al., 2010). A sensible strategy might therefore be to combine the effect of multiple variants at the gene and pathway level. In addition, by combining information from multiple SNPs it is possible to improve the information capture of poorly tagged variants (Chapman and Whittaker, 2008). To this end it may be crucial to integrate information on the predicted functionality of genetic variants in order to find biologically relevant signals. As more functional information for the human genome is accumulated, this avenue may be useful to prioritise and understand the effect of the identified variations (Curtis et al., 2007; Eskin, 2008).

Methods to mine GWAS in a systems biology approach may potentially uncover findings of potential relevance to understand the genetic nature of phenotypic variation. Integration with functional genomic data will be crucial to identify key biological molecules to monitor and manipulate biological processes to improve drug efficacy and disease diagnosis and treatment. Although there have been significant advances in this direction (Schadt et al., 2009b; Yang et al., 2009), this is still an open field. Additional development of data mining and statistical methods is required to integrate and interrogate a combination of GWAS and functional genomic data sets to identify and prioritise biological systems.

This PhD project started soon after the WTCCC publication (WTCCC, 2007), during the first wave of GWAS, when most efforts were focused on increasing statistical power, e.g., by increasing the sample size or imputation to increase genomic coverage (Marchini et al., 2007). I decided to focus on improving interpretation of GWAS by incorporating independent biological knowledge, such as information of gene function and biological processes. The aims of the thesis were to:

i)     Develop, implement and provide the wider scientific community with software tools to perform gene-set and network analyses of GWAS. Chapter II;

ii)    Assess the ability of these methods to find and replicate genetic association. Chapter III and IV;

iii)   Evaluate the feasibility of GWAS for data integration with other genomic datasets in a systems biology framework. Chapter IV and V; and

iv)    Provide association evidence for genes and biological processes to help the wider scientific community to define biological systems which merit additional scientific study, e.g. detailed experimental study of risk variants. Chapters IV and V.

# Chapter II

# A software suite to perform gene-based and gene-set analyses of Genome-Wide Association Studies

## 2.1. Introduction

Genome-Wide Association Studies (GWAS) have led to the discovery of many replicated variants associated with diverse human phenotypes (Hindorff et al., 2009). However, most GWAS have low statistical power to detect true effects due to low effect-sizes and the need for stringent genome-wide significance thresholds to correct for multiple testing, e.g., 5 x $10^{-8}$ (Dudbridge and Gusnanto, 2008). Multivariate analyses considering a group of genetic markers across a genomic region can potentially increase statistical power by a) improving tagging of a single causal variant by combining information from several SNPs and enabling the capture of more information than any single SNP alone, b) combining information from unlinked alleles potentially associated with the phenotype, c) generating fewer tests genome-wide and d) generating gene p-values for system biology and integrative analyses (Neale and Sham, 2004; Chapman and Whittaker, 2008; Pedroso and Breen, 2009; Wang et al., 2010). Currently a major limitation to their application is the lack of software implementing available gene-wide association methods to make them more accessible to the wider community, e.g., methods proposed by Galwey et al. (2009) and Luo et al. (2010).

FORGE is a software suite implementing different methods to perform region-wide, gene-wide and gene-set association with GWAS summary results. The software offers the possibility to calculate significance with asymptotic approximations or by simulation using a multivariate normal distribution approximation. In Section 2 (below) the association methods implemented in FORGE are described, example data are then used to highlight some of the features of the software suite.

## 2.2. Gene-based analyses of GWAS

There is increasing interest in the application of gene-based association methods to GWAS. This strategy can potentially allow for genetic heterogeneity and provide gene level statistics that can be used for meta-analysis or systems biology approaches (Pedroso, 2010). Gene-wide analyses of GWAS have used methods that generate a corrected minimum p-value per gene or methods that combine association statistics across genetic variants within a gene. The minimum p-value per gene has been used with correction for the number of SNPs or, more accurately, the number of effective or independent tests per gene by application of Sidak's correction (Sidak, 1971) or Simers' correction (Simers, 1986). To combine association statistics from several loci across a genomic region authors have used statistical methods such as rank tests (Yu et al., 2009), truncated products (Moskvina et al., 2008) and multivariate tests (Wessel and Schork, 2006; Chapman and Whittaker, 2008; Galwey, 2009; Lin and Schaid, 2009). Each strategy has its merits. For example, correcting the minimum p-value is likely to be a better choice when there is one well tagged causal variant within a gene but it may underestimate evidence in the presence of multiple causal variants or a single causal variant weakly tagged by several markers (Chapman and Whittaker, 2008).

Linkage Disequilibrium (LD) among genetic variants introduces correlation between their association statistics and makes the distribution of the gene-wide statistic under the null hypothesis of no association with the phenotype difficult to estimate asymptotically. Therefore, calculation of the significance of gene-wide statistics requires either estimation of the number of effective tests (which is usually less than the number of polymorphisms tested) within the gene, permutations to estimate significance empirically or simulations strategies, e.g., the method of Liu et al. (2010) described later in this chapter. Empirical estimation of significance, via permutations, is the "gold-standard" approach but for tens of thousands of genes and increasingly large GWAS datasets imposes a large computational burden, and are practically unfeasible for most researchers. Permutation also requires access to raw rather than summary data. There is a need for fast gene-wide association methods that provide both accuracy similar to empirical estimates and which can be computed from SNP level summary statistics or raw data. Several authors have addressed this, proposing statistics to calculate the number of effective tests for groups of genetic variants (Nyholt, 2004; Li and Ji, 2005; Galwey, 2009), allowing fast calculation of gene-wide p-values. The method proposed by Galwey (2009) was and improvement on that of (Nyholt, 2004; Li and Ji, 2005) but, as I will show in Chapter III, gene p-values calculated with this method do not correlate well with empirical estimates. Recently, Liu et al. (2010) introduced a sampling-based estimation of significance for gene-wide statistics. The method models the individual SNP statistics within a gene using a multivariate normal distribution (MND) with a covariance matrix defined by the correlation between these SNPs. This strategy is slower than asymptotic estimates but it was shown to provide statistics with very good agreement with those calculated by phenotype permutations.

## 2.3. Gene-set analyses

In section 1.3, I provided an introduction to GSA. Here I continue the discussion of this topic. To recapitulate, GSA aims to find association with gene-sets rather than individual genes. Gene-sets are simply groups of gene's products that have been put together using some criteria, e.g., participating in the same biological process or having the same sub-cellular localisation. Thus, GSA focuses attention on changes in biological processes. GSA analyses can be characterised by two features: the null hypothesis tested and the statistics used in the analysis (Goeman and Buhlmann, 2007; Liu et al., 2007; Wang et al., 2010).

### 2.3.1.    Different null hypotheses

Goeman and Buhlmann (2007) proposed a characterisation of the null hypothesis of GSA: a self-contained strategy tests if the association in the gene-set is greater than expected by chance and a competitive if the association is greater than expected compared with the rest of the genome. The self-contained strategy only needs information on the statistics of the genes in the gene-set of interest. On the other hand, competitive methods compare the statistics of genes in the gene-set with that of genes outside the gene-set. In order to perform this comparison it is necessary to account for the correlation structure of genes in and outside the gene-set, which remains a challenge in GSA for GWAS (Wang et al., 2010).

### 2.3.2.    Using SNP versus gene-based statistics

If the analysis is based on SNP statistics, gene-set statistics may be calculated based on all SNPs mapped to the genes within the gene-set but an alternative strategy is calculation of gene p-values as an intermediate step. Gene p-values are then aggregated to obtain a gene-set level association statistics. This also provides useful information at the gene level and allows the use of network and other methods in addition to GSA.

### 2.3.3.    Methods implemented in FORGE

In FORGE self-contained statistics are implemented to avoid bias due to the correlation between SNPs. GSA methods have been implemented with and without need of calculation of gene p-values as an intermediate step.

# 2.4. Methods

### 2.4.1.    Region or gene-wide statistics

Data consists of $m$ genetic markers, e.g., within a gene or biological pathway, and their p-values obtained from a two-sided test of significance. A single statistic is calculated for the group of genetic markers.

#### 2.4.1.1.    Sidak's correction of the minimum p-value

Sidak's correction of the best SNP p-value, $p_{corrected} = 1 - (1 - p_{raw})^k$ where $p_{raw}$ is the gene's minimum p-value and $k$ is the number of effective tests (Gao et al., 2008).

#### 2.4.1.2.    Modified Fisher's method to combine correlated p-values

In order to obtain a combined test (T) taking the correlation among the genetic markers into account, I use the method originally derived by Brown (1975) and modified by Makambi (2003) and Kost and McDermott (2002) which leads to the approximately distributed chi-square variable with $\upsilon$ degrees of freedom $T = 0.5 \cdot \upsilon \cdot M_{F,m}$, where $M_{F,m} = -2 \cdot \sum_{i=1}^{m} w_i \cdot \log(p_i)$ is the weighted version of the Fisher's method, $p_i$ is the p-value of $i$th marker and $w_i$ are weights greater than zero that sum to one. The possible use of weights will be discussed later. The degrees of freedom are $\upsilon = 8/\mathrm{var}(M_{F,m})$ with $\mathrm{var}(M_{F,m}) = \sum_{i=1}^{m} \sum_{j=1}^{m} w_i w_j (3.263 |\rho_{ij}| + 0.71 |\rho_{ij}|^2 + 0.027 |\rho_{ij}|^3)$, where $\rho_{ij}$ is the correlation between the $\log(p_i)$ and $\log(p_j)$. How to determine $\rho_{ij}$ will be discussed later.

#### 2.4.1.3.    Fixed-effect z-score statistic

I also calculate a fixed-effects estimate of association by $Z_{fix} = \left( \sum_{i=1}^{m} z_i w_i / \sum_{i=1}^{m} w_i \right) \cdot V_{fix}^{-0.5}$ where $z_i$ are the p-values transformed to z-scores using the standard normal distribution inverse cumulative distribution function (c.d.f.) and

$V_{fix} = \sum_{i=1}^{m} \sum_{j=1}^{m} w_i w_j \rho_{ij}$ is the variance of the test (Huedo-Medina et al., 2006). This

method is equivalent to that presented by Luo et al. (2010).

### 2.4.1.4.    Random-effect z-score statistic

A random-effects estimate is given by $Z_{random} = \left( \sum_{i=1}^{m} z_i w_i^* / \sum_{i=1}^{m} w_i^* \right) \cdot V_{random}^{-0.5}$, with variance $V_{random} = \sum_{i=1}^{m} \sum_{j=1}^{m} w_i^* w_i^* \rho_{ij}$ and weights equal to $w_i^* = \left( \tau^2 + w_i^{-1} \right)^{-1}$

which are adjusted with the heterogeneity measure

$\tau^2 = \max \left[ 0, (Q' - (m-1)) / \left( \sum w_i - \sum w_i^2 / \sum w_i \right) \right]$. In calculating $\tau^2$ Cochran's heterogeneity

$$Q = \sum_{i=1}^{m} w_i \cdot \left[ z_i - \left( \frac{\sum_{i=1}^{m} z_i w_i}{\sum_{i=1}^{m} w_i} \right) \right]^2$$

statistics                                                                is normally used. This is an approximately

distributed chi-square variable with $m$-1 degrees of freedom when the combined

statistics are uncorrelated (Huedo-Medina et al., 2006). However, in my application, the

z-scores are correlated and the distribution of Q does not follow a chi-square

distribution with $m$-1 degrees of freedom. Since Q is obtained by re-scaling and adding

the original z-scores, we approached the correlation problem using the modified

Fisher's method for correlated statistics described in section 2.4.1.2 and calculate $\tau^2$

using Q' as we describe now. In order to obtain Q', I first calculate the tail probability of

Q ($p_Q$) using the modified Fisher's method (describe in section 2.4.1.2) and then calculate Q' as the chi-square value from a chi-square distribution with $m$-1 degrees of freedom for the probability $p_Q$.

### 2.4.1.5.    Significance of gene-wide statistics by sampling from a MND

To estimate the significance of the gene-wide statistics using the multivariate normal distribution (MND) approximation proposed by Liu et al. (2010), the algorithm depicted in Figure 2.1 and described below is used:

I.    Specify a set of $m$ SNP loci in the gene under consideration. For each locus, obtain the p-value representing the association between genotype and phenotype. From this set of $m$ p-values, obtain a gene-wide statistic, $o$, using one of the methods described in Section 2.4.1. Steps II to VII of this algorithm produce a set of values of the same statistic, designated $o^*$, simulated on the null hypothesis (i.e. no association at any locus), to provide a distribution with which this observed value can be compared.

II.    Obtain the $m$ x $m$ correlation matrix among the genotypes, $\mathbf{C}_{snp}$. For this purpose, a heterozygote is a given a score intermediate between the corresponding homozygotes: for example, the genotypes AA, AG, GG may be coded as 2,3,4.

III.    Obtain the Choleski decomposition matrix of $\mathbf{C}_{snp}$, designated $\mathbf{A}_{snp}$.

IV.    Generate a vector, designated $\mathbf{x_m}$, comprising $m$ random numbers from a MND with variance-covariance structure $\mathbf{C}_{snp}$. This is done by the following steps:

 a.    Obtain a vector $\mathbf{l}$, comprising $m$ independent random values from a standard Normal distribution. That is, $\mathbf{l}$ is an observation of $\mathbf{L}$, where $\mathbf{L} \sim \mathbf{N(0, I)}$, with $\mathbf{0}$ equal to a vector of zeroes of length $m$ and $\mathbf{I}$ = the m x m identity matrix.

 b.    Obtain $\mathbf{x_m} = \boldsymbol{\mu} + \mathbf{A_{snp}} \bullet \mathbf{l}$, where $\boldsymbol{\mu}$ is a vector of length $m$ of the means of the multivariate normal distribution. In my application $\boldsymbol{\mu}$ is a vector of zeroes.

V.    Transform $\mathbf{x_m}$ to a set of p-values, designated $\mathbf{p^*}$, a vector of length $m$. The sign of the association between each locus and the phenotype is not meaningful, and therefore this is done by comparing the square of value in $\mathbf{x_m}$ with the chi-square distribution (two-tailed test), rather than by comparing the untransformed value with the standard Normal distribution (one-tail test). Thus $p_i = P(Q > z_i^2)$ where $p_i$ = $i$th value in $\mathbf{p^*}$, $z_i^2$ = $i$th value in $\mathbf{x_m}$ and $Q \sim \chi_1^2$.

VI.   Obtain the gene-wide statistic $o^*$ and $\mathbf{p^*}$, using the same method as in Step I.

VII.  Repeat Steps IV-VI $n$ times stopping when #($o^* > o$) = 10, i.e. when 10 values of $o^*$ that surpass the observed values $o$ have been obtained.

VIII. Obtain the estimated gene p-value = $(1 + \#(o^* > o))/(1 + n)$.

It is notable that while the observed p-values and the gene-wide statistic derived from them ($o$) are obtained using both the genotype and the phenotype data, the corresponding values $\mathbf{p^*}$ and $o^*$ simulated on the null hypothesis are obtained from the genotype data only. This method for obtaining an empirical distribution for $o$ on the null hypothesis is much less computationally intensive than the standard approach, for two reasons, namely:

• it does not require random permutation of a large set of phenotype values

• the test statistic is samples from multivariate normal distribution, rather than being derived by calculations performed in a large data set.

## 2.5.    Gene-set Analyses

### 2.5.1.    SNP to gene-sets strategy

In this case we treat gene-sets as large genes, i.e. map to it all SNPs of its constituent genes and apply the statistics described in Section 2.4.1, also allowing for correlations between genes, such as those that exist in gene-clusters.

#### 2.5.1.1.    Gene-sets analysis using gene p-values

GSA is performed using the same basic algorithm described in Figure 2.1 and section 2.4.1.5. The main difference is that a correlation matrix between the gene-statistics, denoted $\mathbf{C_G}$, is needed and replaces $\mathbf{C_{snp}}$ in Figure 2.1. This correlation matrix is derived as described by Luo et al. (2010) using the genotypes of each gene in the gene-set.

Briefly, the correlation between the $i$th and $j$th gene-statistics of two genes with $m_i$ and

$m_j$ number of SNPs, respectively, is defined by

$$C_{G_{ij}} = \frac{\sum_{u=1}^{m_i} \sum_{v=1}^{m_j} corr(M_{i_u}, M_{j_v})}{\sqrt{V_{g_i} \cdot V_{g_j}}}$$ , where $u$ and $v$ are indexes running from 1

to $m_i$ and $m_j$, respectively; $M_{iu}$ and $M_{jv}$ correspond to the genotypes at the $u$th and $v$th

SNP in the $i$th and $j$th genes, respectively; and $V_{g_i} = \sum C_{snp,g_i}$ and

$V_{g_j} = \sum C_{snp,g_j}$ are the variances of the $g_i$ and $g_j$ gene statistics that equal the sum

over the gene's SNP p-value correlation matrix ($\mathbf{C_{snp,\ g_i}}$) as defined in section 2.4.1.3

(note that differently from section 2.4.1.5 I have added the $g_i$ and $g_j$ to the notation of

the correlation matrices to clarify which genes I am referring to). It is important to note

that this expression for the correlation matrix of gene statistics is valid for the Z FIX

method as was shown by Luo et al (2010). It is also valid for the Z RANDOM method

after including the weights and it is possibly a good approximation for the modified

Fisher's methods but I did not explore its use with the latter method. Similarly to the

gene-wide association the GSA can be performed asymptotically or using the MND

simulation method. The asymptotic approximation is performed by using the methods described in Section 2.4.1 and replacing SNP p-values for gene p-values and the correlation matrix of the SNP statistics for the $\mathbf{C_G}$ matrix describe above. The methods described in Section 2.4.1 lead to z-scores or chi-square values whose statistical significance can be estimated by comparing them to the standard Normal distribution or a chi-square distribution with degrees of freedom calculated with the modified Fisher's method, respectively.

### 2.5.1.2.    Reducing the number of simulations needed of p-values << 1.

The strategies introduced in section 2.4 allow calculation of a minimum p-value of 1/ (1+n), with n being the number of samplings from a MND performed. If very small p-values are present in the data-set this strategy may require a prohibitively large number of samplings to produce accurate p-values. Knijnenburg et al. (2009) proposed to use extreme value distribution theory to approximate the p-values by using a much smaller number of simulations. They showed it allows to estimate p-values below $10^{-9}$ accurately with as little as $10^6$ simulations. I have implemented their methods with the difference that parameter estimation is done with the method described by Zhang (2010), which has smaller error over a wider range of parameter values. This method has been implemented for completeness of the software but it was used in my analyses here. I refer the reader to the original reports for details, see (Knijnenburg et al., 2009; Zhang, 2010).

## 2.6. Implementation

I implemented these methods in a program called FORGE written in Perl (www.perl.com) using the PDL, PDL::Stats and PDL::LinearAlgebra libraries, all freely available at the Comprehensive Perl Archive Network (www.cpan.org) and PDL (Perl Data Language, http://pdl.perl.org). Perl provides single numerical precision natively but the PDL library allows to performed calculation in double numerical precision and threading in memory for matrix and vectorial operations.

The FORGE code is freely available at https://github.com/inti/FORGE/wiki with documentation and instructions to use the example files. A master branch is maintained with the official release version and several parallel development branches to produce new code and features. The code is developed under controlled version and is currently open for the wider community to contribute or use the software in independent projects. Currently the code contains over 30,000 words in more than 5000 lines of code and documentation. However, all major function have been implemented as separated modules or routines to allows an easy modification and improvement of the code.

### 2.6.1.  Software features

#### 2.6.1.1.  Input files

The program reads input and output files and genotype file formats of commonly used GWAS analysis software, e.g., genotype files in Linkage Pedigree, Pedigree Binary Format, genotype probability files in the BEAGLE (http:// faculty.washington.edu/browning/beagle/beagle.html) as well as IMPUTE (http:// www.stats.ox.ac.uk/%7Emarchini/software/gwas/file_format.html) formats. SNP association files with columns labelled in a header are accepted, e.g., files produced by PLINK (Purcell et al., 2007). Gene-set definitions may also be provided and the above described statistics are calculated from SNPs in all the genes of a given gene-set.

#### 2.6.1.2.  SNP-SNP correlations.

Our formulas use estimates of the correlation between the test statistics. The user can calculate them prior to using FORGE by, e.g., performing permutations and randomising the phenotype to derive statistics under then null distribution and then calculating the correlation between these SNP statistics. This can be done in PLINK by using it with the "--mperm 1000 --mperm-save-all" flags. The resulting output files with extension *.mperm.dump.all have permuted statistics that can be input to forge.pl to use the SNP-SNP correlations ($\rho_{ij}$) as the correlation among the test statistics.

Alternatively, $\rho_{ij}$ can be approximated as the correlation between the $M_i$ and $M_j$ markers, either Pearson's correlation or LD. Despite the latter being an approximation to the correlation between the test statistics, both strategies provide approximately equivalent results. Pearson's correlations are calculated with the methods proposed by Schafer and Strimmer (2005) that ensure the correlation matrix is positive definite, which is important for the Cholesky decomposition needed for the MND sampling.

### 2.6.1.3.    SNP-to-Gene annotations

Pre-computed files with genetic variants mapped to genes up to 500 kb from gene coordinates are available at the FORGE website. A Perl script to update the annotation using the Ensembl API (Rios et al., 2010) is distributed as a utility of FORGE.

### 2.6.1.4.    Additional features

User provided SNP weights, e.g. functional scores, can be used and will be re-scaled to sum to 1 within each gene. Genomic-control correction (Devlin and Roeder, 1999) can be automatically performed within the program. Analyses can be restricted to chromosomes, gene lists or gene types (e.g. protein coding or miRNA genes). Affymetrix SNP identifiers are accepted. The program also reports the $I^2$ heterogeneity statistics (Huedo-Medina et al., 2006). Example files, documentation and tutorials are available on the software's website. These methods have been implemented as a web-server where the user can run the analyses by simply uploading a file with SNP p-values. The web-sever was developed in collaboration with David To and Richard Dobson at the BRC Nucleus (IoP) and it is available at https://compbio.brc.iop.kcl.ac.uk:8443/forgeweb.

### 2.6.2.    Running time

Analysis of an example file which involved the calculation of a p-value using the asymptotic approximation for a region of 19 SNPs required approximately 10 seconds and in my experience a whole genome analysis involving ~40,000 genes and ~450,000 SNPs requires ~3-4 hours on a single 2.2 GHz 32 bit computer with 4 gigabytes of RAM. Using the multivariate normal distribution approximation with $10^6$ sampling for the same region of 19 SNPs took ~6 minutes.

# 2.7. Results and conclusion

To demonstrate the use of FORGE to calculate gene p-values I applied it to the GWAS of total blood haemoglobin levels reported by Ferreira et al. (Ferreira et al., 2009). Identifying alleles regulating total blood haemoglobin level is of potential interest to understand haemoglobin level abnormalities present in some human diseases. This GWAS was made publicly available by the authors together with GWAS of another twelve haematological traits. I choose to analyse the GWAS of total blood haemoglobin level because it had a lambda for genomic control of 1.01 (reported in the Supplementary Material of Ferreira et al. (2009)) and had no genome-wide significant results at $p < 5 \times 10^{-8}$). Thus, this is a GWAS with no inflation of the SNP statistics (i.e. SNP associations are mostly under the null hypothesis) that offers us the possibility to analyse a real data set instead of a simulated one that might not present the complexities of a real GWAS. In addition, by not having genome-wide significant SNPs this data sets enable use to test if my methods can identify as significant some potentially interesting loci. I will explore further this argument in Chapter III of this thesis. I downloaded summary statistics of the GWAS of total blood haemoglobin level for 1,872,255 SNPs (file "HB.qimr.mdd.assoc") from the authors' website, used genotypes of the CEU samples of the HapMap Project Phase 2 (files hapmap-ceu.bed, hapmap-ceu.bim and

hapmap-ceu.fam available at the PLINK software website (Purcell et al., 2007)) to

calculate SNP-SNP correlation and used the SNP-to-gene mapping files distributed with

FORGE. HapMap Phase 2 genotyped were used because the authors of the original

study performed imputation using those genotypes as a reference panel and therefore

these genotypes cover most of the SNP association statistics available in the GWAS

results. To perform the analysis using the asymptotic approximation I used the

command line:

$perl forge.pl –bfile hapmap-ceu –assoc HB.qimr.mdd.assoc –snpmap

ensemblv59_SNP_2_GENE.txt –out HB.qimr.forge.txt

This took approximately 6 hours to run on a single CPU of computer with 8 GB of

RAM. To use the multivariate normal distribution sampling I used the command:

$perl forge.pl –bfile hapmap-ceu –assoc HB.qimr.mdd.assoc –snpmap

ensemblv59_SNP_2_GENE.txt –out HB.qimr.forge_mnd.txt -mnd -mnd_max 1000000

The -mnd command sets the program to use the MND sampling strategy and the -

mnd_max sets the maximum number of samplings. This command took ~ 60 hours to run on the same computer.

Comparison of gene p-values calculated with both strategies shows an overall good

agreement (Figure 2.1). In Figure 2.2 I present the quantile-quantile plots for the four

gene p-values methods calculated with both strategies. The p-values calculated with

sampling provide a better fit to the uniform distribution expected under the null

hypothesis than those calculated asymptotically. However, it is also important to note

that p-values will have a uniform distribution if they are representative of the null

hypothesis (i.e. in my case that is no association with the phenotype). When analysing

real datasets it is possible that deviation from the uniform distribution is due to real genetic associations rather than unreliability of the gene-wide association methods. Although this complicates the use of the quantile-quantile plots as technical diagnostic tool for the gene-wide p-values, it is important to remember that this GWAS has a lambda for genomic control of 1.01 indicating little or no inflation of the statistics, suggesting most genetic association signal are representative of the null hypothesis. Therefore, I would expect most of the gene-wide p-values generated from this GWAS to agree with the null hypothesis.

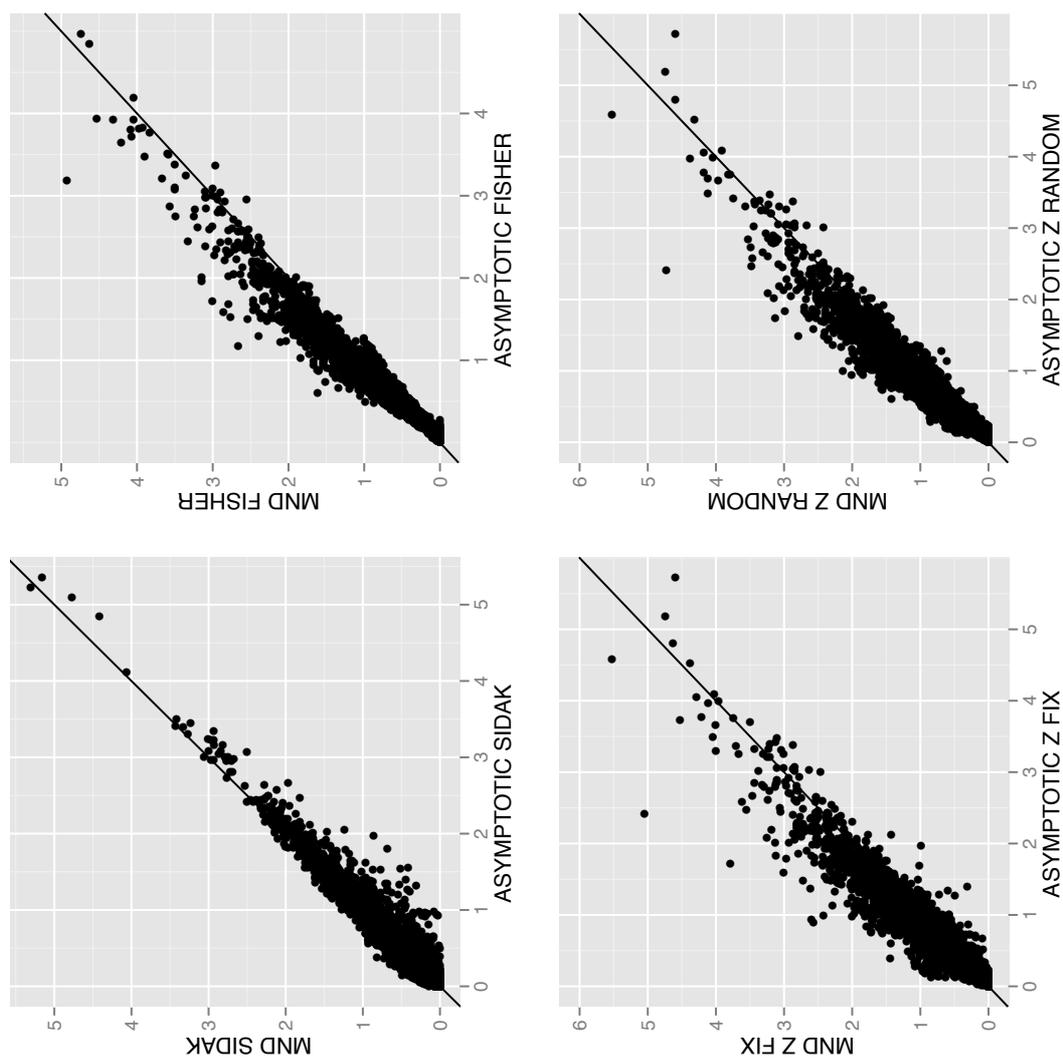

**Figure 2.1. Comparison of gene-wide p-values calculated with asymptotic and MND strategies.** Each panel presents ~30,000 points. Axis scales are -log10 of gene p-value. Filled line represents the 1-to-1 diagonal.

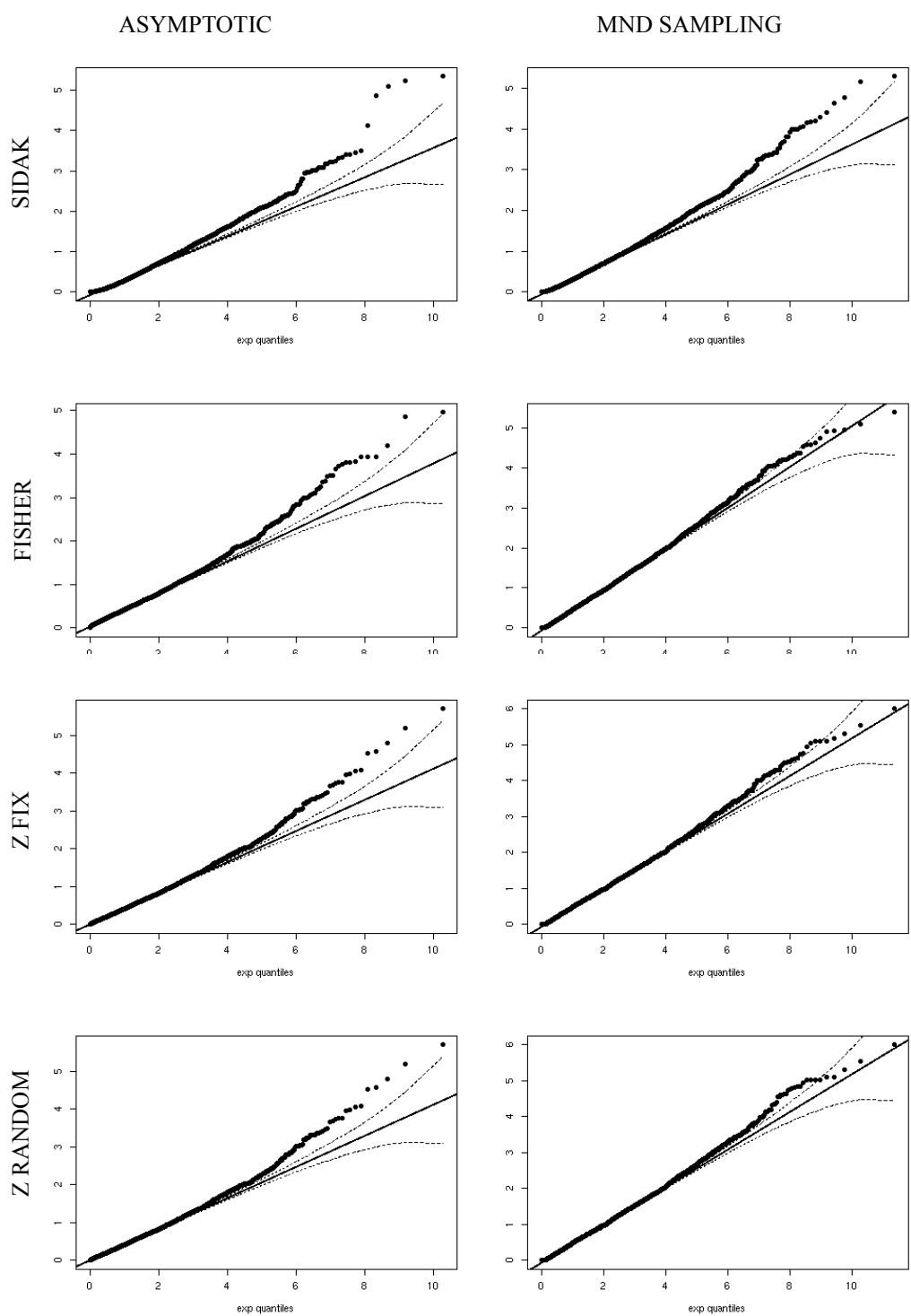

**Figure 2.2. Quantile-quantile plots for gene p-values.** Axis scales are -log10 of gene p-values. Y and x-axis present the observed and expected quantiles, respectively. Filled and dotted line represent the 1-to-1 diagonal and the 95% confidence interval calculated for an assumed a null exponential distribution.

| Gene symbol | Asymptotic | | | | MND Sampling | | | |
|---|---|---|---|---|---|---|---|---|
| | SIDAK | FISHER | Z FIX | Z RANDOM | SIDAK | FISHER | Z FIX | Z RANDOM |
| C14orf159 | 6E-02 | 1E-05 | 3E-06 | 3E-06 | 5E-02 | 6E-04 | 3E-05 | 3E-05 |
| TTLL9 | 1E-03 | 3E-05 | 2E-05 | 2E-05 | 6E-04 | 1E-04 | 6E-06 | 6E-06 |
| IL7 | 2E-04 | 2E-04 | 2E-05 | 2E-05 | 8E-05 | 9E-04 | 1E-04 | 1E-04 |
| PDRG1 | 4E-04 | 2E-05 | 3E-05 | 3E-05 | 3E-04 | 1E-05 | 2E-06 | 2E-06 |
| HFE | 2E-04 | 1E-05 | 3E-05 | 5E-05 | 1E-04 | 5E-05 | 2E-04 | 3E-04 |
| CIZ1 | 5E-04 | 9E-05 | 3E-05 | 3E-05 | 3E-04 | 2E-04 | 8E-05 | 8E-05 |
| LTA4H | 6E-04 | 5E-05 | 4E-05 | 5E-05 | 4E-04 | 1E-04 | 3E-05 | 3E-05 |
| KIRREL3 | 3E-02 | 1E-05 | 5E-05 | 5E-05 | 1E-02 | 3E-04 | 2E-04 | 2E-04 |
| CCDC54 | 4E-03 | 6E-05 | 6E-05 | 1E-04 | 2E-03 | 5E-04 | 2E-04 | 3E-04 |
| ENSG00000205754 | 2E-03 | 1E-04 | 8E-05 | 4E-05 | 1E-03 | 2E-04 | 1E-04 | 1E-04 |

**Table 2.1. Top 10 genes on haemoglobin GWAS.** Gene p-values calculated with asymptotic and MND sampling strategy for top 10 genes on the MND sampling Z FIX method. Rows have been sorted in ascending order by the Asymptotic Z FIX method.

As an example of FORGE's results, Table 2.1 presents the gene p-values for the top 10 genes for the Z FIX model using both the asymptotic and MND sampling methods. One of the top ten genes, the hemochromatosis gene, HFE is a known locus underlying haemoglobin levels based on evidence from rare variants (Beutler et al., 2003) and GWAS (Chambers et al., 2009; Ganesh et al., 2009). Table 2.2 presents the correlation between the gene p-values and the gene's number of SNPs for each method.

In summary, FORGE produces gene p-values unbiased by gene size (Table 2.2) and complements single SNP associations analysis (Table 2.1). It is flexible regarding input file formats and implements for the first time, to the best of my knowledge, a random-effects model and the use of SNP weights for gene-wide association. In addition, it implements both asymptotic and simulation based p-values providing users the ability to balance the trade off between speed and accuracy.

| Signficance strategy | Gene p-values | Spearman's correlation | Correlation p-value |
|---|---|---|---|
| Asymptotic | SIDAK | 0.011 | 0.021 |
| | FISHER | -0.011 | 0.024 |
| | Z FIX | 0.015 | 0.003 |
| | Z RANDOM | 0.014 | 0.003 |
| MND Sampling | SIDAK | 0.085 | $< 2.2 \times 10^{-16}$ |
| | FISHER | 0.004 | 4.1E-01 |
| | Z FIX | 0.014 | 4.2E-03 |
| | Z RANDOM | 0.015 | 2.2E-03 |

**Table 2.2. Correlation between gene p-values and number of SNPs within the gene.** Spearman's rank correlation between gene p-values and number of SNPs mapped to the gene (within 20 kb of genes coordinates). N = ~20,000.

The FORGE version described here and the code presented in Appendix A corresponds to the software version 1 as of Feb 2011. Results presented in the following chapters were performed with a previous version providing the same functionality except for the MND sampling strategy, which was reported after completion of the analyses.

# Chapter III

# Gene-based genome-wide association reveals true disease loci not found by single SNP analyses.

# 3.1. Introduction

As I review in Chapter I, genome-wide association studies (GWAS) can be a powerful tool to discover genetic associations and have been successful for many human phenotypes (see www.genome.gov/gwastudies for a catalogue of studies). It is not without its drawbacks, however. The stringent genome-wide significance thresholds needed to correct for the millions of univariate tests performed in each study (e.g., 5 x $10^{-8}$, Dudbridge and Gusnanto (2008)) mean that in many GWAS projects few, if any, genetic variants reach genome-wide significance. Indeed, 81% (199 out of 246) of the traits recorded in the NIHGRI GWAS Catalogue (as of 10 December 2010)(Hindorff et al., 2009) have no genome-wide significant associations. Clearly, for any trait, researchers will greatly benefit from gathering large sample sizes ($10^4$ or $10^5$ samples) (for examples see Franke et al., 2010; Speliotes et al., 2010; Teslovich et al., 2010). However, assembling sample sets of this magnitude requires extensive international collaborations and may prove very difficult for traits with low population prevalence such as Anorexia Nervosa (4.2 to 8.3 cases per 100,000) (Miller and Golden, 2010) or phenotypes requiring costly characterisation, such as neuroimaging. Thus, it is desirable to increase power through alternative strategies in parallel to efforts aimed at increasing sample size.

Multivariate analysis strategies, such as gene-wide testing, are an attractive complement to individual SNP analysis of GWAS (Neale and Sham, 2004). I and others have shown that this can allow for allelic heterogeneity, result in fewer genome-wide tests and provide p-values for each gene that can be used with pathway analysis methods or simply to find genes associated with a phenotype of interest (Furney et al., 2010; Pedroso, 2010; Peng et al., 2010; Wang et al., 2010). Shifting the focus towards genes and biological processes can aid interpretation of GWAS results and generate hypotheses for follow up studies (Pedroso, 2010; Peng et al., 2010; Wang et al., 2010). Gene-wide analyses can provide better replication than single SNP analyses and establish accurate estimates of the statistical evidence within a gene (Liu et al., 2010; Luo et al., 2010; Peng et al., 2010; Wang et al., 2010). In addition, some reports have demonstrated that gene-based association can find a wider range of loci than single SNP analyses, including loci with smaller effect-sizes (Furney et al., 2010; Liu et al., 2010; Luo et al., 2010; Peng et al., 2010). In this study, I re-analysed the seven GWAS of the Wellcome Trust Case Control Consortium phase 1 (WTCCC, 2007) using a set of gene-wide analysis methods and compared these results with a set of known susceptibility loci for each disorder, aiming to quantify the ability of gene-based association methods to discover true disease susceptibility genes.

## 3.2. Methods

### 3.2.1. Genotypic data analyses

SNP association summary statistics of the Wellcome Trust Case-Control Consortium study (WTCCC, 2007) were downloaded from the European Genotype Archive (EGA) with formal data access permission of the WTCCC Data Access Committee. (See (WTCCC, 2007) for details of quality control (QC) undertaken by the WTCCC.) I also obtained WTCCC bipolar disorder (BD) GWAS genotype and phenotype data from the EGA website, for analysis in conjunction with the association data on this disease. Routine QC was performed by excluding samples as indicated in the files provided by the WTCCC and SNPs with rate of missingness $\geq$ 1%, minor allele frequency (MAF) $\leq$ 1%, Hardy-Weinberg equilibrium (HWE) p-value $\leq 10^{-3}$ in controls and p-value $\leq 10^{-5}$ in cases as previously described (WTCCC, 2007). I calculated multidimensional scaling (MDS) values using PLINK (Purcell et al., 2007) and used the first five as covariates in a logistic regression model in the BD GWAS to correct for population structure. To investigate the effect of genotype calling quality, I analysed the same GWAS but with genotypes recalled with BEAGLECALL, which has been shown to provide a significant improvement in data quality (Browning and Yu, 2009). BEAGLECALL genotypes were obtained from the authors of Browning and Yu (2009) with authorisation of the WTCCC Data Access Committee. QC was performed by excluding samples as indicated in the files provided by the WTCCC and SNPs with a rate of missing genotypes $\geq 1\%$, MAF $\leq 1\%$, HWE p-value $\leq 10^{-3}$ and p-value $\leq 10^{-5}$ in controls and cases, respectively.

### 3.2.2. Gene-wide Association

Gene-wide statistics were calculated for approximately 19,550 protein-coding, long intergenic non-coding RNA and micro-RNA genes annotated in Ensembl version 59 and whose SNPs passed QC in the WTCCC studies. I mapped SNPs to genes if the SNP was within 20 kb upstream or downstream of the annotated coordinates aiming to include 95% of potential expression quantitative trait loci (eQTL) (Veyrieras et al., 2008). Gene-wide p-values were calculated with the Perl software FORGE, which is freely available at https://github.com/inti/FORGE/wiki and described in detail in Chapter II. Along with the four gene-wide p-values described in Chapter II, I implemented the estimate of effective number of tests proposed by Galwey (2009), the Eigen value ratio (EVR) method. Briefly, this method estimates the effective number of tests ($M_{eff}$) of a group of

$$M_{eff} = \frac{\left( \sum_{i=1}^{e} \sqrt{\lambda_i} \right)^2}{\sum_{i=1}^{e} \lambda_i}$$

genetic variants by , where $\lambda_i$ are the $e$ positive eigenvalues of the genotype correlation matrix. $M_{eff}$ is used to calculate gene-wide statistics (EVR gene-wide p-value) with an re-scaled Fisher's method to combine p-values (Galwey, 2009) by

$$\chi^2_{adjusted} = -2\frac{M_{eff}}{m}\sum_{i=1}^{m}\ln(p_i)$$, where $\chi^2_{adjusted}$ is a variable with an approximate chi-square distribution with $2*M_{eff}$ degrees of freedom under the null hypothesis.

I approximated the correlation between test statistics at each SNP locus as the Pearson's correlation between the allele counts. I used a local False Discovery Rate (FDR) < 0.1 (Efron et al., 2001; Efron, 2010) as implemented in the "fdrtools" package version 1.2.6 (Strimmer, 2008) for R (R Development Core Team, 2010) to select a group of interesting genes.

### 3.2.3.    Gene-Set Analysis

Gene-Set Analysis (GSA) was performed with the FORGE software suite using the gene-wide p-values calculated as describe above (see Chapter II for details). I compiled gene-sets ranging from 2 to 200 genes in size from three sources: a) 161 from the KEGG database (Kanehisa et al., 2008); b) 1,181 from Gene Ontology (GO) level 4 databases (Carbon et al., 2009); and c) 5,384 gene-sets derived from the Human Protein Reference Database protein-protein interaction network (PPIN)(Keshava Prasad et al., 2009). The PPIN gene-sets were constructed with the following algorithm. Sub-network searches started from each node in the PPIN, termed the seed node. A sub-network was defined by adding sequentially the direct neighbours of every node in the sub-network (initially only the seed node). I allowed searches to go to a maximum of five interactions from the seed node and generated sub-networks ranging from 2 to 200 nodes in size. Each of these sub-networks was used as a gene-set.

As with gene-based analyses I used a FDR threshold equal to 0.1 to select interesting sub-networks. I interpreted the biological function of the significant sub-networks using MetaCore (GeneGo, Inc.; www.genego.com). GeneGO generates biological networks using information curated by systematic literature review of genes and interactions of gene products. It also provides bioinformatic tools to analyse genomic data sets for enrichment in these biological networks. GeneGO's tools assess the significance of the overlap between the PPIN sub-networks and their manually curated biological processes using hypergeometric distribution statistics. The p-value from this statistical analysis essentially represents the probability that the overlap between genes of a significant network and a GeneGO's biological processes arise by chance (assuming a null hypergeometric distribution), considering the numbers of genes in the PPIN sub-network, the number of genes in the GeneGO biological processes included in the experiment and the total number of genes included in the experiment mapped to any GeneGO biological process. Additional details of the calculation of this p-value can be found at GeneGO's website (https://portal.genego.com/help/P-value_calculations.pdf) and the report by Ekins et al. (2006).

### 3.2.4.    Permutation study

To compare FORGE's asymptotic statistics with their equivalent obtained empirically I generated SNP association statistics under the null hypothesis (i.e. no association with the phenotype) by permuting the case-control labels. For each gene I generated sets of SNP statistics under the null hypothesis and recalculated the gene-wide statistic until the asymptotic gene-wide p-value was observed at least ten times (Doerge and Churchill, 1996). I used this strategy to calculate empirical p-values for all genes on chromosome 22 (905 genes) and the top 20 genes from the bipolar disorder genome-wide analysis. The empirical p-value was calculated as described by North et al. (2003) using $p_{empirical} = (R+1)/(N+1)$, where $R$ is the number of times the observed statistics was surpassed during simulations and $N$ is the total number of simulations.

### 3.2.5.    In silico validation of FORGE results

For each disease studied in the WTCCC dataset (WTCCC, 2007), I constructed a list of replication genes using the studies recorded in the NIHGRI GWAS Catalogue (as of 10 December, 2010)(Hindorff et al., 2009). I removed results derived directly from the WTCCC paper (WTCCC, 2007) and mapped the SNPs with p-values < $5 \times 10^{-8}$ to genes as explained in Section 3.2.2, generating a set of replication genes for each disease. I excluded SNPs entries of the GWAS Catalogue that did not pass QC in the WTCCC GWASs. I calculated two measures of enrichment for the WTCCC results among the genes of the replication sets. First, I selected genes highlighted by significant SNPs and gene-based association from the WTCCC studies and calculated their overlap with the GWAS Catalogue replication sets. The significance of this overlap was tested against an assumed null binomial distribution using R (R Development Core Team, 2010). This analysis was performed at commonly used significance thresholds, i.e., FDR of 0.1 for gene-based analyses and a p-value of $5 \times 10^{-8}$ for single SNP analyses. I calculated a second enrichment measure that allowed me to explore the enrichment in replicated genes at different significance thresholds. Receiver operating characteristic (ROC) curves were calculated for the seven GWAS datasets using the GWAS Catalogue genes as a benchmark set. The area under the curve (AUC) is related to a rank correlation statistics, e.g., rank correlation between SNP p-values and a binary variable characterising if the SNP was replicated or not. It can take values between 0 and 1 (or 0 and 100% depending on the notation) and in my application represents as a measure of how well the GWAS results prioritise or find disease genes. An AUC value of 0.5 represents the classification expected simply by chance and a value of 1 a perfect classification, i.e. maximum sensitivity and specificity. I calculated the AUC at different p-value thresholds for the results of single SNP analyses and different gene-wide association methods to provide a measure of their ability to prioritise genes of the GWAS Catalogue replication sets at different levels of significance. AUC values and their confidence intervals were calculated by bootstrapping with the package "pROC" version 1.3.2 (Robin et al., 2011) for R (R Development Core Team, 2010).

### 3.2.6.    Hierarchical clustering

Hierarchical clustering was performed in R (R Development Core Team, 2010) using Euclidean distance and the complete linkage method. As input for the clustering algorithm I used the z-scores for each gene-set obtained from the test of association with the phenotype. Heat-maps were drawn using R (R Development Core Team, 2010).

## 3.3. Results

### 3.3.1.    Performance of gene-wide association methods

Calculation of gene-wide p-values requires computing of the correlation between SNP genotypes.  I found that the analysis was approximately 50 times faster when using a correlation matrix derived from the genotypes of a reference population, e.g., HapMap project samples, compared with using the WTCCC BD GWAS data. There were Pearson's correlation values of $r > 0.99$ for all gene-wide p-values when calculated with either the WTCCC BD or the HapMap Phase 3 CEU population genotypes. All results presented hereafter were calculated using the HapMap Phase 3 CEU genotypes. Analysis of a GWAS took roughly 3-4 hours for approximately 20,000 genes annotated in Ensembl v59 on a single 2.2 GHz 32-bit dual core computer with 4 gigabytes of RAM.

There was a high correlation between the asymptotic gene-wide p-values and empirical estimates (minimum Pearson's correlation $r = 0.98$; Figure 3.1). There was appreciable over-dispersion in the gene-wide p-values derived from the EVR method (Figure 3.1). The results presented before and hereafter are based on the SIDAK, FISHER, Z FIX and Z RANDOM methods.

Multivariate methods might be more sensitive to confounders like population stratification and genotyping error than single SNP analyses (Moskvina et al., 2006) because there is potential to combine the noise in the same way they combine association signals. To address the effects of these confounders, the analysis was repeated in the BD GWAS with two modifications: a) I re-analysed the WTCCC genotypes and included covariates to correct for population structure and b) I used a set of better genotype calls, obtained with the software BEAGLECALL (Browning and Yu, 2009). In Table 3.1 I report the correlation of SNP and gene p-values calculated with and without correction for possible confounders. The correlation of the gene-wide associations calculated under different corrections is not expected to be higher that of the single SNP analysis itself because they depend directly on single SNP statistics and I expect gene p-values to change at least as much as single SNP statistics do after correcting for a confounder. Therefore, in Table 3.1 I also provide a correlation for the gene p-values re-scaled with that of the single SNP analyses. This ratio provides a better insight because it accounts for the change in gene p-values beyond that of single SNP p-values alone. There was marginal evidence for gene-wide association being more sensitive than single SNP analysis to potential confounders, such as population stratification or genotyping error (Table 3.1). I also inspected the agreement between the different gene-wide p-values. There was a high correlation between all gene-wide p-values (minimum correlation r > 0.7) and the correlations between multivariate methods were higher (minimum correlation r > 0.95) (Figure 3.2).

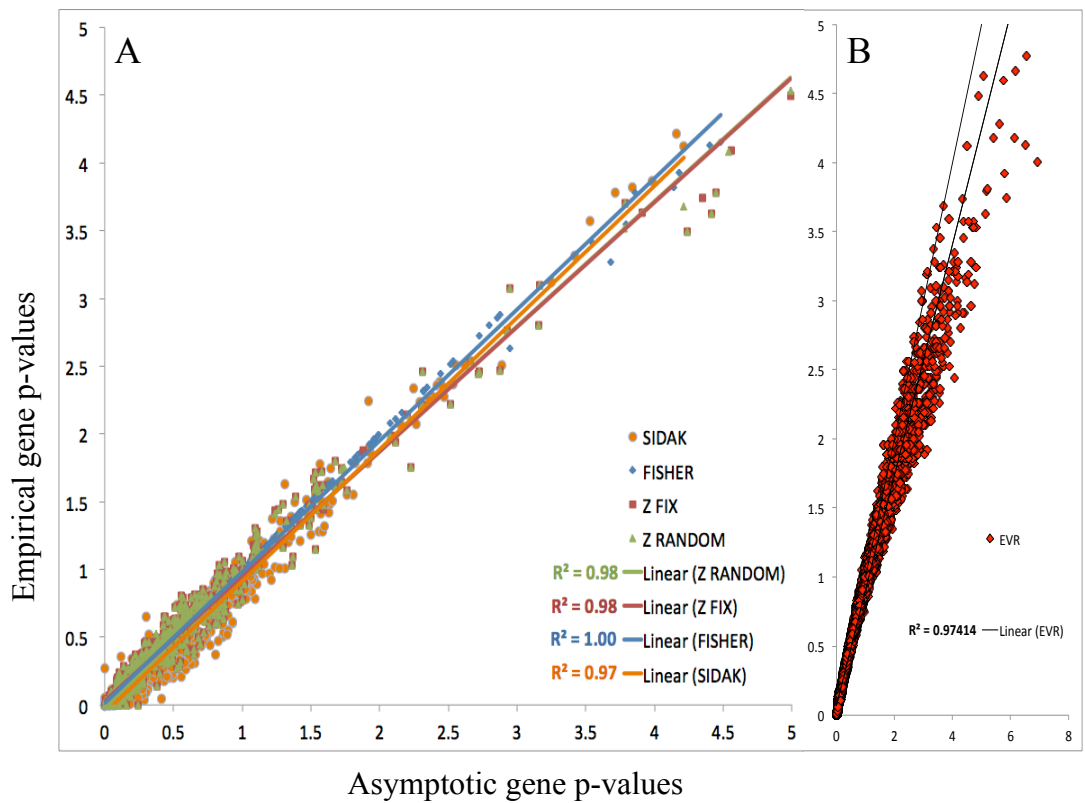

**Figure 3.1. Correlation between asymptotic and empirical p-values.** Trend lines were fitted using a linear regression model and the r2 is reported. Panel A; four methods implemented in FORGE. Panel B; Eigen value ratio (EVR) method. In panel A blue line is equivalent to the diagonal, in panel B we represented the 1-to-1 diagonal with the left most black line. Note that the scale of the x-axis differs in both panels.

|  | MDS | MDS re-scaled | BEAGLECALL | BEAGLECALL re-scaled |
|---|---|---|---|---|
| **Single SNP** | 0.959 | | 0.982 | |
| **SIDAK** | 0.955 | 0.996 | 0.978 | 0.996 |
| **FISHER** | 0.959 | 1.000 | 0.987 | 1.006 |
| **Z FIX** | 0.922 | 0.962 | 0.979 | 0.997 |
| **Z RANDOM** | 0.922 | 0.962 | 0.979 | 0.997 |

**Table 3.1. Effect of possible confounders on gene p-values**. We report Spearman's rank correlation between the gene p-values calculated from the summary statistics provided by the WTCCC and those using correction for population stratification (Multidimensional Scaling, MDS) and genotypes called with BEAGLECALL. Re-scaled correlations were constructed by dividing the correlation of each gene-wide association method by that of the single SNP association results. This ratio informs on the additional sensitivity to possible confounders of the gene-wide association methods.

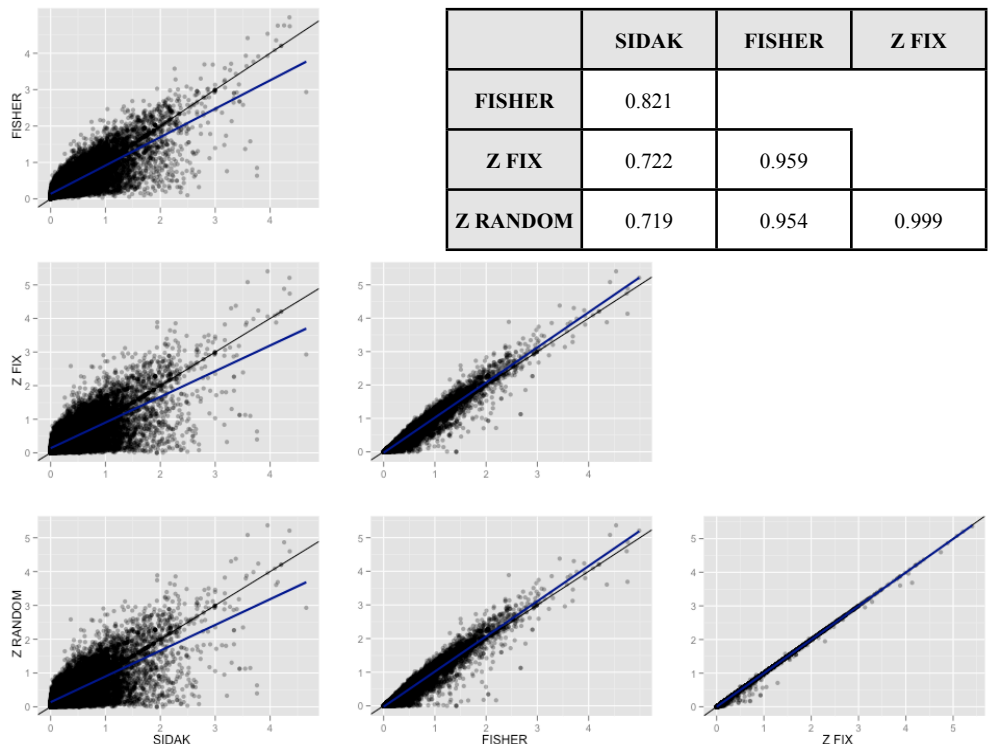

|  | SIDAK | FISHER | Z FIX |
|---|---|---|---|
| **FISHER** | 0.821 | | |
| **Z FIX** | 0.722 | 0.959 | |
| **Z RANDOM** | 0.719 | 0.954 | 0.999 |

**Figure 3.2. Correlation between different asymptotic p-values.** Trend lines (blue) were fitted using a linear model and the correlation is reported in the insert table.

### 3.3.2.    Gene-wide analysis of GWAS of seven common disorders

Figure 3.3 presents the analysis workflow performed on the WTCCC GWASs. Gene-wide association p-values were calculated in the GWAS of Crohn's disease (CD), type 1 diabetes (T1D), type 2 diabetes (T2D), rheumatoid arthritis (RA), bipolar disorder (BD), coronary artery disease (CAD) and hypertension (HT) (WTCCC, 2007) using the summary statistics provided by the WTCCC and HapMap Phase 3 CEU population genotypes.

For each disease, I calculated the overlap between genes identified by the gene-based methods (FDR < 0.1) and those highlighted by SNPs (p-value < 5 x $10^{-8}$) in studies reported in the NHGRI GWAS catalogue (Hindorff et al., 2009). I excluded SNPs reported solely by the initial WTCCC article (WTCCC, 2007). Comparison of FDR and SNP p-value thresholds showed that in general a FDR < 0.1 corresponded to SNP p-values < 5 x $10^{-8}$ (not shown).

SNPs reached significance in four out of the seven WTCCC GWAS. These SNPs showed significant enrichment in replicated loci (Table 3.2). All gene-based methods identified genes with FDR < 0.1 and these were also overrepresented within the replicated set. To take an example, the Venn diagram depicted in Figure 3.2 presents the results of the CD analysis. In this case, 40% (13 out of 32) of the replicated genes were only identified by the gene-based association, this is approximately twice the number of

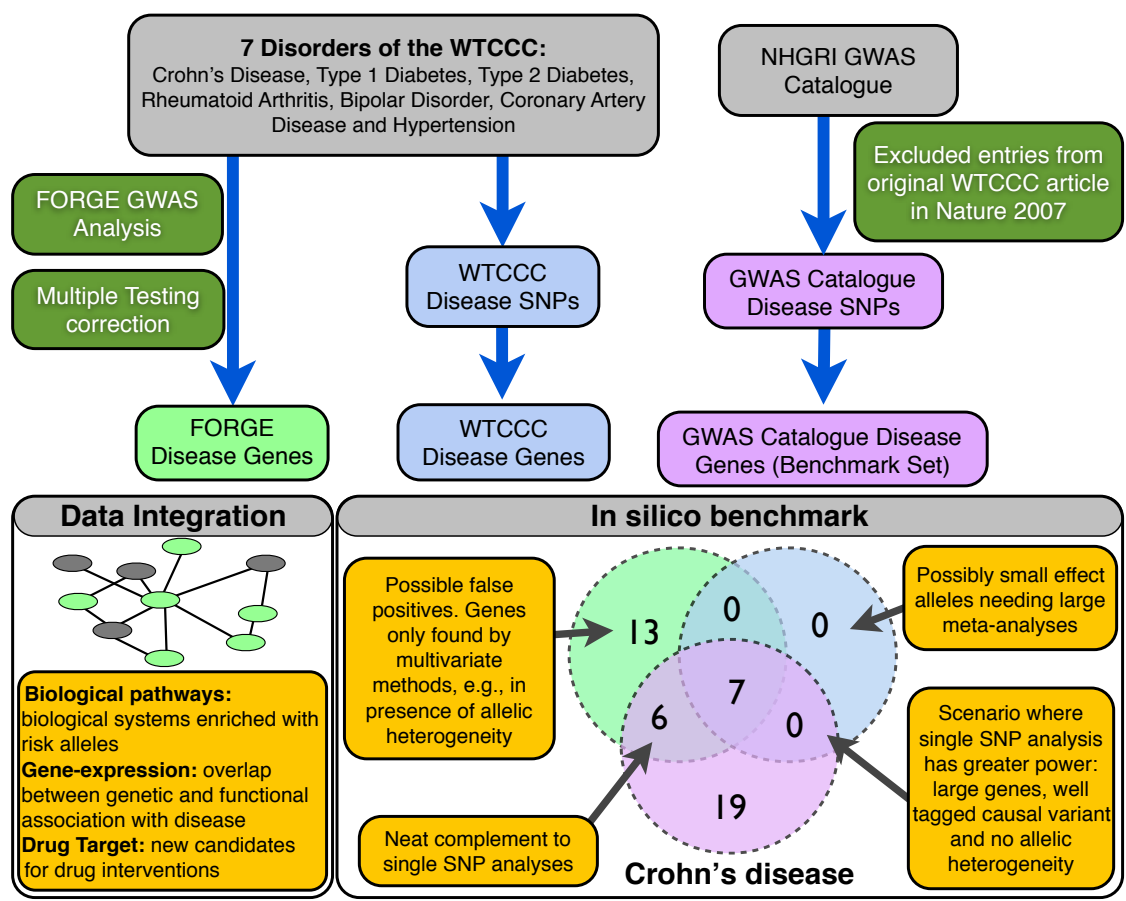

**Figure 3.3. In silico validation of gene-based associations.** The ability of gene-based analyses to identify true disease loci was benchmarked by: i) assessing how many of their significant findings are replicated in other studies; and ii) how many of these significant genes were not identified by using single SNP analyses alone. Venn diagram and workflow boxes' colours represent gene counts and steps, respectively, of the different analyses: light green (left) = gene-based association, light blue (centre) = single SNP analyses and purple (right) = benchmark set from NHGRI GWAS Catalogue. Results were analysed regarding the proportion genes of the benchmark set identified only by either single SNP analyses or gene-based association (see yellow boxes on "In silico benchmark"). In addition we evaluated the use of gene-based association in integrative analyses with the protein-protein interaction network and biological pathways (see main text for details). Data presented on Venn diagram are the results of the Z FIX method on the WTCCC Crohn's disease data set. (see Table 3.2, Figure 3.4 and main text for details).

| Data Set | Genes found (replicated) | | | | | Significance | | | | | GWAS Catalogue |
|---|---|---|---|---|---|---|---|---|---|---|---|
| | SNP | SIDAK | FISHER | Z FIX | Z RANDOM | SNP | SIDAK | FISHER | Z FIX | Z RANDOM | |
| CD | 7(7) | 9(9) | 84(19) | 26(13) | 25(12) | $6 \times 10^{-20}$ | $2 \times 10^{-25}$ | $2 \times 10^{-34}$ | $2 \times 10^{-29}$ | $6 \times 10^{-27}$ | 35 |
| T1D | 47(10) | 75(11) | 146(14) | 111(13) | 123(13) | $2 \times 10^{-16}$ | $6 \times 10^{-16}$ | $3 \times 10^{-17}$ | $3 \times 10^{-17}$ | $1 \times 10^{-16}$ | 58 |
| T2D | 2(2) | 1(1) | 39(4) | 24(2) | 32(2) | $2 \times 10^{-6}$ | $2 \times 10^{-3}$ | $5 \times 10^{-7}$ | $7 \times 10^{-4}$ | $1 \times 10^{-3}$ | 31 |
| RA | 9(6) | 9(6) | 37(7) | 26(5) | 31(6) | $4 \times 10^{-16}$ | $5 \times 10^{-16}$ | $7 \times 10^{-14}$ | $7 \times 10^{-10}$ | $4 \times 10^{-12}$ | 26 |
| BD | 0 | 0 | 26(0) | 15(0) | 21(0) | NA | NA | NA | NA | NA | 3 |
| CAD | 2(0) | 2(0) | 22(2) | 22(0) | 21(0) | NA | NA | $3 \times 10^{-5}$ | NA | NA | 7 |
| HT | 0 | 0 | 35(0) | 21(0) | 28(0) | NA | NA | NA | NA | NA | 0 |
| WTCCC Combined | 67(25) | 96(27) | 389(46) | 245(33) | 281(33) | $8 \times 10^{-56}$ | $3 \times 10^{-49}$ | $1 \times 10^{-75}$ | $1 \times 10^{-56}$ | $1 \times 10^{-54}$ | 160 |

**Table 3.2. Overlap recombination intervals with significant SNP or genes among replicated genes.** Gene counts correspond to the number of recombination intervals, as defined by the HapMap project phase 2, with genes highlighted by single SNP analysis or gene-based methods. Within parentheses is the number of genes indexed in the NIHGRI GWAS Catalogue for that disorder (Hindorff et al., 2009). Enrichment p-values are calculated with the binomial distribution. CD = Crohn's disease, T1D = type 1 diabetes, T2D = type 2 diabetes, RA = rheumatoid arthritis, BD = Bipolar Disorder, CAD = coronary artery disease and HT = hypertension.

| Data Set | Genes found (replicated) | | | | Significance | | | | GWAS Catalogue |
|---|---|---|---|---|---|---|---|---|---|
| | SIDAK | FISHER | Z FIX | Z RANDOM | SIDAK | FISHER | Z FIX | Z RANDOM | |
| **CD** | 2(2) | 77(12) | 20(7) | 20(7) | $3 \times 10^{-6}$ | $4 \times 10^{-20}$ | $5 \times 10^{-15}$ | $5 \times 10^{-15}$ | 35 |
| **T1D** | 28(1) | 100(4) | 68(3) | 80(3) | 0.08 | $3 \times 10^{-4}$ | 0.001 | 0.002 | 58 |
| **T2D** | 0 | 38(3) | 23(1) | 31(1) | NA | $3 \times 10^{-5}$ | 0.04 | 0.05 | 31 |
| **RA** | 0 | 28(1) | 19(0) | 23(1) | NA | 0.04 | 0.98 | 0.97 | 26 |
| **BD** | 0 | 26(0) | 15(0) | 21(0) | NA | NA | NA | NA | 3 |
| **CAD** | 2(0) | 20(2) | 20(0) | 19(0) | NA | $2 \times 10^{-5}$ | NA | NA | 7 |
| **HT** | 0 | 35(0) | 21(0) | 28(0) | NA | NA | NA | NA | 0 |
| **WTCCC Combined** | 30(3) | 324(22) | 186(11) | 222(11) | $6 \times 10^{-6}$ | $2 \times 10^{-31}$ | $8 \times 10^{-16}$ | $6 \times 10^{-15}$ | 160 |

**Table 3.3. Enrichment among replicated genes of loci only found by gene-wide association methods.** Enrichment p-values calculated with the binomial distribution. Gene counts correspond to the number of genes found by gene-based methods after removing those with SNPs with p-value < 5 x 10-8. Within parentheses is the number of genes indexed in the NIHGRI GWAS Catalogue for that disorder (Hindorff et al., 2009). CD = Crohn's disease, T1D = type 1 diabetes, T2D = type 2 diabetes, RA = rheumatoid arthritis, BD = Bipolar Disorder, CAD = coronary artery disease and HT = hypertension.

replicated genes identified by the single SNP analysis alone (Table 3.2). Interestingly,

gene-based association found most genes identified by the single SNP analysis (SIDAK

= 99%, FISHER = 94%, Z FIX = 88% and Z-RANDOM = 87%). After exclusion of genes found by single SNP analysis, the gene-wide association results were still significantly enriched with replicated genes (Table 3.3). A list of these replicated genes highlighted by gene-based but not single SNP analysis can be found in Table 3.4.

These results were derived with thresholds of FDR < 0.1 and an SNP to gene

mapping distance of 20 kb. These were my initial criteria and I performed post-hoc

| GWAS | Hugo Symbol | SNP | FDR | | | | References |
|------|-------------|-----|-------|--------|-------|-----------|-----------|
| | | | SIDAK | FISHER | Z FIX | Z RANDOM | |
| CD | CUL2 | $1\times10^{-4}$ | 0.83 | 0.03 | 0.19 | 0.18 | Barrett et al., 2008 |
| | KIF21B | $6\times10^{-5}$ | 0.84 | 0.03 | 0.14 | 0.14 | Barrett et al., 2008 |
| | ZNF300 | $1\times10^{-7}$ | 0.06 | $3\times10^{-7}$ | $1\times10^{-2}$ | 0.14 | Barrett et al., 2008 |
| | APEH | $2\times10^{-6}$ | 0.18 | $2\times10^{-4}$ | $2\times10^{-3}$ | $2\times10^{-3}$ | Parkes et al., 2007; Barrett et al., 2008 |
| | RNF123 | $4\times10^{-5}$ | 0.59 | $8\times10^{-3}$ | 0.09 | 0.09 | Barrett et al., 2008 |
| | STAT3 | $2\times10^{-5}$ | 0.67 | 0.03 | 0.14 | 0.13 | Barrett et al., 2008 |
| | MST1 | $1\times10^{-6}$ | 0.18 | $1\times10^{-4}$ | $1\times10^{-3}$ | $9\times10^{-4}$ | Parkes et al., 2007; Barrett et al., 2008 |
| | PTPN2 | $1\times10^{-7}$ | 0.1 | $5\times10^{-3}$ | 0.02 | 0.02 | Parkes et al., 2007; Barrett et al., 2008 |
| | TNFSF15 | $1\times10^{-4}$ | 0.83 | 0.06 | 0.19 | 0.18 | Barrett et al., 2008 |
| | C5orf56 | $1\times10^{-6}$ | 0.25 | $1\times10^{-5}$ | $3\times10^{-4}$ | $2\times10^{-4}$ | Barrett et al., 2008 |
| | AL390240.1 | $1\times10^{-4}$ | 0.83 | 0.08 | 0.3 | 0.3 | Barrett et al., 2008 |
| | IRGM | $1\times10^{-7}$ | 0.06 | $4\times10^{-7}$ | $1\times10^{-4}$ | $1\times10^{-4}$ | Parkes et al., 2007 |
| CAD | PSRC1 | $2\times10^{-5}$ | 1 | 0.04 | 0.25 | 0.24 | Samani et al., 2007 |
| | MYBPHL | $2\times10^{-5}$ | 1 | 0.04 | 0.3 | 0.29 | Samani et al., 2007 |
| RA | C5orf30 | $2\times10^{-4}$ | 1 | 0.05 | 0.28 | 0.28 | Stahl et al., 2010 |
| T1D | SMARCE1 | $2\times10^{-5}$ | 0.3 | $8\times10^{-3}$ | 0.04 | 0.04 | Barrett et al., 2009 |
| | SH2B3 | $4\times10^{-7}$ | 0.04 | $2\times10^{-4}$ | $1\times10^{-2}$ | 0.02 | Barrett et al., 2009 |
| | CTLA4 | $3\times10^{-5}$ | 0.37 | $9\times10^{-3}$ | 0.23 | 0.57 | Cooper et al., 2008; Barrett et al., 2009 |
| | ATXN2 | $3\times10^{-5}$ | 0.28 | $9\times10^{-3}$ | 0.18 | 0.27 | Barrett et al., 2009 |
| T2D | ZFAND6 | $1\times10^{-5}$ | 1 | 0.07 | 0.22 | 0.23 | Voight et al., 2010 |
| | TSPAN8 | $2\times10^{-6}$ | 1 | $3\times10^{-3}$ | 0.05 | 0.07 | Zeggini et al., 2008 |
| | RBMS1 | $3\times10^{-6}$ | 1 | 0.04 | 0.18 | 0.17 | Qi et al., 2010 |

**Table 3.4. Genes highlighted only by gene-based analyses that were replicated in other studies.** References as reported in the GWAS Catalogue (Hindorff et al., 2009). CD = Crohn's disease, T1D = type 1 diabetes and T2D = type 2 diabetes.

| Data Set | FDR < 0.01 | | FDR < 0.05 | | FDR < 0.1 | | FDR < 0.2 | | FDR < 0.5 | | GWAS Catalogue genes | Total Genes Study |
|---|---|---|---|---|---|---|---|---|---|---|---|---|
| | Replicated (Total) | OR | Replicated (Total) | OR | Replicated (Total) | OR | Replicated (Total) | OR | Replicated (Total) | OR | | |
| RA | 0(22) | -- | 0(37) | -- | 0(59) | -- | 0(82) | -- | 0(166) | -- | 16 | 21006 |
| CD | 5(9) | 932 (187-4204) | 9(36) | 285 (107-725) | 10(57) | 192 (77-446) | 8(97) | 74 (28-177) | 7(272) | 21(8-50) | 33 | 21002 |
| T1D | 1(30) | 10(0.3-68) | 3(30) | 36(7-123) | 5(32) | 63 (18-172) | 5(68) | 26(8-69) | 2(190) | 3(0.4-12) | 67 | 20999 |
| T2D | 1(1) | inf(34-inf) | 2(15) | 228 (23-1162) | 2(27) | 119 (12-572) | 3(50) | 102 (18-392) | 4(277) | 25(6-84) | 16 | 20976 |
| CAD | 0(0) | -- | 0(6) | -- | 0(10) | -- | 0(20) | -- | 0(96) | -- | 5 | 20993 |
| HT | 0(8) | -- | 0(27) | -- | 0(44) | -- | 0(82) | -- | 0(377) | -- | 2 | 20993 |
| BD | 0(0) | -- | 0(10) | -- | 0(27) | -- | 0(93) | -- | 1(347) | 60 (0.8-4469) | 2 | 20977 |

**Table 3.5. Effect of FDR threshold on enrichment of different FISHER method results in replicated loci.** FDR: local false discovery rate; OR: Odd ratio. The 95% confidence interval (CI) of the OR is show in brackets after the actual ratio. Total: FORGE positive genes excluding those found by single SNP analyses. The set of replicated genes are genes mapped to SNPs reported in the NHGRI GWAS Catalogue reaching p-value = $5 \times 10^{-8}$ (see Material and Method for details), excluding findings reported in (WTCCC, 2007). CAD = coronary artery disease, CD = Crohn's disease, HT = hypertension, RA = rheumatoid arthritis, T1D = type 1 diabetes, T2D = type 2 diabetes and BD = bipolar disorder.

| Data Set | 0 kb | | | 20 kb | | | 50 kb | | | 100 kb | | | 500 kb | | |
|---|---|---|---|---|---|---|---|---|---|---|---|---|---|---|---|
| | Replicated | Total | OR | Replicated | Total | OR | Replicated | Total | OR | Replicated | Total | OR | Replicated | Total | OR |
| RA | 0(4) | 17 (13607) | -- | 0(16) | 59 (21006) | -- | 0(28) | 66 (22026) | -- | 1(47) | 87 (222425) | 5 (0.2-33) | 1(177) | 47 (22880) | 3 (0.1-16) |
| CD | 4(13) | 26 (13594) | 271 (57-1024) | 10(33) | 57 (21002) | 192 (77-446) | 19(53) | 103 (22024) | 144 (75-272) | 31(92) | 192 (22423) | 70 (42-113) | 88(427) | 307 (22885) | 26 (20-34) |
| T1D | 0(18) | 22 (13591) | -- | 5(65) | 32 (20999) | 64 (19-177) | 4(77) | 98 (22022) | 13(3-35) | 1(121) | 51 (22423) | 4 (0.1-22) | 19(506) | 356 (22885) | 2(1-4) |
| T2D | 0(11) | 0 (13567) | -- | 2(15) | 27 (20976) | 128 (13-625) | 2(21) | 31 (22018) | 80 (9-357) | 2(32) | 53 (22418) | 29 (3-121) | 7(123) | 131 (22885) | 11(4-24) |
| CAD | 0(5) | 4 (13595) | -- | 0(5) | 10 (20993) | -- | 0(6) | 21 (22019) | -- | 0(8) | 34 (22422) | -- | 0(22) | 94 (22885) | -- |
| HT | 0(2) | 24 (13586) | -- | 0(2) | 44 (20993) | -- | 0(2) | 30 (22024) | -- | 0(3) | 41 (22422) | -- | 0(13) | 275 (22885) | -- |
| BD | 0(2) | 13 (13554) | -- | 0(2) | 27 (20977) | -- | 0(3) | 48 (22022) | -- | 0(3) | 136 (22423) | -- | 0(15) | 168 (22885) | -- |

**Table 3.6. Effect of mapping distance threshold on enrichment of different FISHER method results in replicated loci.** FDR: false discovery rate; OR: Odd ratio and 95% confidence intervals. Total: genes found by FORGE, excluding those found by single SNP analyses, and number of genes included. Replicated: replicated genes and total number of replicated genes. See Methods in main text for more details. CAD = coronary artery disease, CD = Crohn's disease, HT = hypertension, RA = rheumatoid arthritis, T1D = type 1 diabetes, T2D = type 2 diabetes and BD = bipolar disorder.

analyses to assess their effects on my results. First, using the FISHER method with different FDR and distance mapping I found that enrichment with replicated genes weakened with larger SNP to gene mapping distances and FDR values. The best enrichment was found with FDR values around 0.05 to 0.1 and a mapping distance of 20 kb. Only the CD data set showed enrichment when including only SNPs between the start and end locations of a gene (Table 3.5 and 3.6). Second, I assessed how well the SNP or gene-wide association p-values classify loci as replicated using the GWAS Catalogue entries as the benchmark. I calculated receiver operating characteristic curves and the AUC for the single SNP analysis and gene-wide association results at different p-value thresholds for six out of the seven diseases (the GWAS of HT did not have GWAS Catalogue entries with p-values $< 5\text{x}10^{-8}$). Before describing the results of this analysis I would like to remind the reader that the interpretation of a significant p-value depends among other things on the number of hypothesis tested during the course of the experiment. I have tested approximately one order of magnitude fewer genes than SNPs, so simple comparison of SNPs and gene p-values is not straightforward. In Figure 3.4 I present the AUC values with the estimated 95% confidence intervals for the SNP and gene-wide association analyses of the six different diseases. To facilitate comparison with the results presented in Tables 3.2 and 3.3, Figure 3.4 also gives the multiple testing correction thresholds used in these previous analyses, i.e. a p-value = $5\text{x}10^{-8}$ for single SNP and a p-value equivalently to a FDR of 0.1 for gene-wide association results. The results presented in Figure 3.4 show that after multiple testing correction both the single SNP analyses and gene-wide association methods find a significant proportion of true disease loci, as was already suggested by the results presented in Tables 3.2 and 3.3. The AUC profiles allow us to assess the ability of these methods to find disease loci at different significance thresholds, which was not a feature of my previous analyses. I found that single SNP p-values correctly classify many of the replicated loci at p-values above of 5 x 10-8. The amount of signal remaining above this threshold, as measured by the SNP association p-value at which the AUC ≤ 0.5, seems to vary between datasets. A second result emerging from this figure is that there is little difference between gene-wide association methods regarding their AUC profile. Finally and more importantly, these results show that in general gene-wide association methods have higher AUC values than the single SNP results. This pattern is more prominent in GWAS where single SNP analyses provided several significant hits, e.g., GWAS of RA or CD as compared to those of BD or CAD. This pattern is also observed in the results of gene-set analyses (see later in this chapter) and I discuss it further in Chapter VI of this thesis. Overall, the AUC analysis results lend additional support to the observations obtained at the fixed significance thresholds that I presented in Tables 3.2 and 3.3.

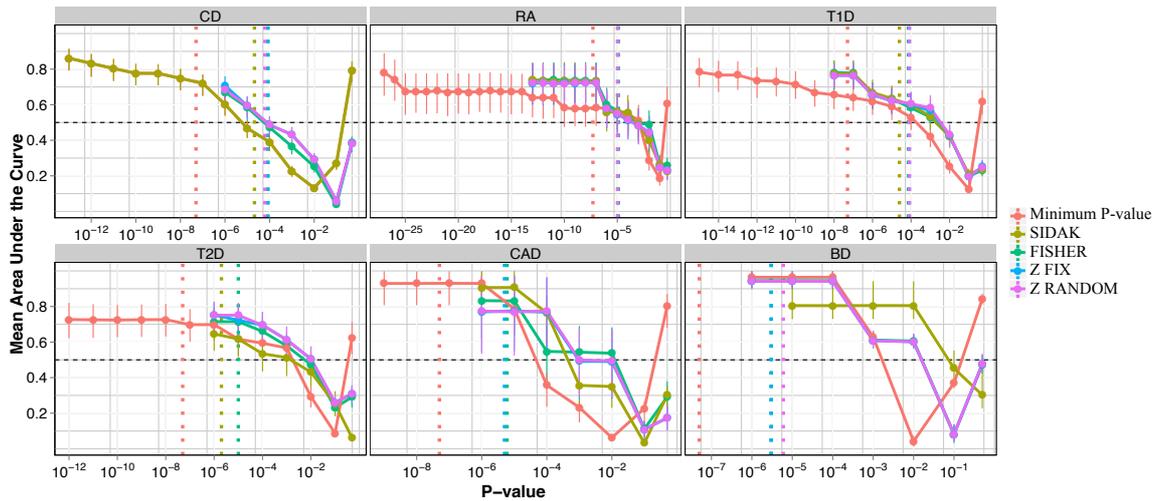

**Figure 3.4. Area under the curve for SNP and gene p-values results benchmarked against GWAS Catalogue.** Each panel presents the AUC values and their 95% confidence interval for one of six GWAS. Results from single SNP and gene-wide association methods are colour coded as indicated on the legend. Vertical dashed lines mark a p-value of 5 x 10-8 and the gene-wide p-value equivalent to a FDR = 0.1 for each method (see colour code in legend). Horizontal dashed line marks the AUC = 0.5, which marks what would be expected by chance. CD = Crohn's disease, T1D = type 1 diabetes, T2D = type 2 diabetes, RA = rheumatoid arthritis, BD = bipolar disorder and CAD = coronary artery disease. Please note for the panel of the CD data that the line of the SIDAK and minimum p-value methods overlap completely. To improve visualisation I set different scales for the x-axis on each panel.

Significant gene-wide associations were often physically clustered due to LD extending beyond gene boundaries and my allowance for SNPs to map multiple genes. I verified that clustered genes do not drive the results by calculating the overlap between gene-wide results and the replication sets in terms of recombination intervals containing significant or replicated genes instead of individual genes. I defined recombination intervals as the regions between adjacent recombination hot-spots midpoints (Frazer et al., 2007). The results remained significant (Tables 3.7 and 3.8).

I acknowledge that false positives exist among the significant regions from gene-based analyses which have not been identified in other studies. However, I hypothesise that some associations are true disease susceptibility loci that have not yet been identified by sufficiently large GWAS meta-analyses. To explore this possibility I rebuilt the replication set for CD using the NHGRI GWAS Catalogue (data taken on 9 February, 2011) (Hindorff et al., 2009), which included the latest meta-analysis analysing over 50,000 samples (Franke et al., 2010) and recalculated the overlap with the WTCCC results. The new replication set had 91 genes. Only the FISHER method found additional replication in the new set. The FISHER method deemed 84 genes to be significant for CD, with 23 overlapping the new replication set. I also further examined the results of the CAD GWAS because, unlike the CD GWAS, it did not have SNPs reaching significance in the original analysis (WTCCC, 2007). It may be possible that gene-based association methods could help prioritise targets for follow up. I performed a literature search in PubMed for the 22 genes (within 16 recombination intervals)

| Data Set | Genes found (replicated) | | | | | Significance | | | | | GWAS Catalogue |
|---|---|---|---|---|---|---|---|---|---|---|---|
| | SNP | SIDAK | FISHER | Z FIX | Z RANDOM | SNP | SIDAK | FISHER | Z FIX | Z RANDOM | |
| **CD** | 4(4) | 5(5) | 31(13) | 17(8) | 17(8) | $4\times10^{-11}$ | $9\times10^{-14}$ | $2\times10^{-26}$ | $3\times10^{-17}$ | $3\times10^{-17}$ | 21 |
| **T1D** | 5(5) | 6(6) | 35(10) | 28(9) | 33(10) | $6\times10^{-13}$ | $2\times10^{-15}$ | $7\times10^{-17}$ | $7\times10^{-16}$ | $3\times10^{-17}$ | 31 |
| **T2D** | 2(2) | 1(1) | 28(6) | 19(5) | 24(5) | $7\times10^{-6}$ | 0.003 | $1\times10^{-10}$ | $1\times10^{-9}$ | $5\times10^{-9}$ | 22 |
| **RA** | 2(2) | 2(2) | 15(4) | 14(2) | 17(2) | $4\times10^{-6}$ | $4\times10^{-6}$ | $2\times10^{-8}$ | $3\times10^{-4}$ | $5\times10^{-4}$ | 16 |
| **BD** | 0 | 0 | 15(0) | 10(0) | 14(0) | NA | NA | NA | NA | NA | 3 |
| **CAD** | 1(0) | 1(0) | 14(1) | 17(0) | 17(0) | NA | NA | 0.007 | NA | NA | 4 |
| **HT** | 0 | 0 | 21(0) | 17(0) | 21(0) | NA | NA | NA | NA | NA | 0 |
| **WTCCC Combined** | 14(13) | 15(14) | 159(34) | 122(24) | 143(25) | $7\times10^{-36}$ | $1\times10^{-38}$ | $6\times10^{-61}$ | $2\times10^{-42}$ | $7\times10^{-43}$ | 97 |

**Table 3.7. Overlap between recombination intervals with significant SNPs or genes among replicated loci.** Gene counts correspond to the number of recombination intervals, as defined by the HapMap project phase 2, with genes highlighted by single SNP analyses or gene-based methods. Within parentheses is the number of genes indexed in the NIHGRI GWAS Catalogue for that disorder (Hindorff et al., 2009). Enrichment p-values are calculated with the binomial distribution. CD = Crohn's disease, T1D = type 1 diabetes, T2D = type 2 diabetes, RA = rheumatoid arthritis, BD = bipolar disorder, CAD = coronary artery disease and HT = hypertension.

| Data Set | Genes found (replicated) | | | | Significance | | | | GWAS Catalogue |
|---|---|---|---|---|---|---|---|---|---|
| | SIDAK | FISHER | Z FIX | Z RANDOM | SIDAK | FISHER | Z FIX | Z RANDOM | |
| **CD** | 1(1) | 27(9) | 13(4) | 13(4) | 0.003 | $1\times10^{-17}$ | $3\times10^{-8}$ | $3\times10^{-8}$ | 21 |
| **T1D** | 4(4) | 33(8) | 26(7) | 31(8) | $2\times10^{-10}$ | $4\times10^{-13}$ | $5\times10^{-12}$ | $2\times10^{-13}$ | 31 |
| **T2D** | 0 | 27(5) | 18(4) | 23(4) | NA | $9\times10^{-9}$ | $1\times10^{-7}$ | $4\times10^{-7}$ | 22 |
| **RA** | 0 | 15(4) | 13(1) | 16(1) | NA | $2\times10^{-8}$ | 0.002 | 0.003 | 16 |
| **BD** | 0 | 15(0) | 10(0) | 14(0) | NA | NA | NA | NA | 3 |
| **CAD** | 0 | 13(1) | 16(0) | 16(0) | NA | 0.006 | NA | NA | 4 |
| **HT** | 0 | 21(0) | 17(0) | 21(0) | NA | NA | NA | NA | 0 |
| **WTCCC Combined** | 5(5) | 151(27) | 113(16) | 134(17) | $1\times10^{-14}$ | $2\times10^{-46}$ | $2\times10^{-26}$ | $4\times10^{-27}$ | 97 |

**Table 3.8. Overlap between recombination intervals with significant genes among replicated loci after excluding significant SNPs.** Gene counts correspond to the number of recombination intervals, as defined by the HapMap project phase 2, with genes highlighted by gene-based methods after excluding genes with SNP p-values $< 5\times10^{-8}$. Within parentheses is the number of genes indexed in the NIHGRI GWAS Catalogue for that disorder (Hindorff et al., 2009). Enrichment p-values are calculated with the binomial distribution. CD = Crohn's disease, T1D = type 1 diabetes, T2D = type 2 diabetes, RA = rheumatoid arthritis, BD = bipolar disorder, CAD = coronary artery disease and HT = hypertension.

highlighted by the Z FIX method. Five of them had known or possible links with CAD[5]:

i)      SEMA3C has mechanistic roles in cardiac neural crest cells migration during heart development (Lepore et al., 2006; Toyofuku et al., 2008; Kodo et al., 2009);

ii)     Patients with CAD had higher levels of OGDH protein compared with controls in a proteomic study (Banfi et al., 2010);

iii)    Common variants within PTGER2 have been associated with acute coronary syndrome in CAD (Szczeklik et al., 2008);

iv)     Mutations in MTMR14 are associated with autosomal dominant centronuclear myopathy (Tosch et al., 2006) and its gene product has been shown to participate in the regulation of calcium signalling (Shen et al., 2009) and muscle contraction (Dowling et al., 2010);

v)      Finally, SNPs within MIA3 have been associated with Early Onset Myocardial Infarction and CAD (Samani et al., 2007; Kathiresan et al., 2009). The WTCCC samples were included in the study of Kathiresan et al. (2009).

---

[5] It is difficult to asses if finding five genes with links to CAD out of the twenty-two deemed significant is more than it would be expected by chance. Without compelling single SNP associations reaching a p-value of $5 \times 10^{-8}$, gene-based association may be able to highlight genes and genomic regions that researchers might prioritise for additional research after a detailed review of the known functions of the associated genes plus the investigator own knowledge on the disease's molecular aetiology. I acknowledge this is not an interpretation of the data free of biases but for many diseases the molecular aetiology is not well defined and a GWAS is just the first step towards its elucidation.

### 3.3.3. Using gene-wide p-values for Gene-Set Analysis of GWAS

I performed gene-set analysis with the Z FIX gene-wide p-values results[6] and found significant gene-sets after correction for multiple testing (FDR < 0.05) for four diseases (Additional Table 3.1 at the end of this chapter). I identified in the CD data set association with PPIN sub-networks of genes involved in interleukin (IL) 23/12 (PPIN-4224, FDR = 4 x $10^{-4}$) and IL-6 signalling pathway (PPIN-4339, FDR = 1 x $10^{-3}$), closely matching the known aetiology of CD and several previous reports (reviewed by Wang et al. (2010)). For HT, there was an association with a PPIN sub-network of genes (PPIN-2605, FDR = 0.04) involved in cortisone, androstenedione and testosterone biosynthesis and metabolism. For RA, I identified several PPIN sub-networks involved in the interaction between cell and the extracellular matrix and immune response, including antigen presentation and T cell receptor (TCR) signalling. Several gene-sets reached significance in the T1D GWAS but I could only obtain bioinformatic characterisation for genes in GO0031497 (FDR = $3x10^{-3}$) and PPIN-5144 (FDR = $8x10^{-3}$), which are involved in chromatin assembly and DNA repair.

I performed hierarchical clustering of diseases and gene-sets using gene-sets with nominal significance, an uncorrected p-values < 0.05 in at least one study, and a more stringent significance level of FDR < 0.05 in at least one study (Figure 3.5). By using the nominally significant gene-sets, I obtained a cluster with T1D, RA and CD, all of which are immune system related diseases (Figure 3.5-A). Furthermore, I found a cluster with T2D and CAD, both metabolic diseases. In contrast, none of these groups were found by clustering of the more stringent significance criteria. For example, T1D and RA were placed outside the tree (Figure 3.5-B), suggesting that their sub-networks

---

[6] I choose the Z FIX method because the gene-gene correlation calculated from the gene-set analysis methods implemented on FORGE are better suited to deal with gene p-values from the Z FIX methods than from the other methods. See Chapter II section 2.5.1.1 for details.

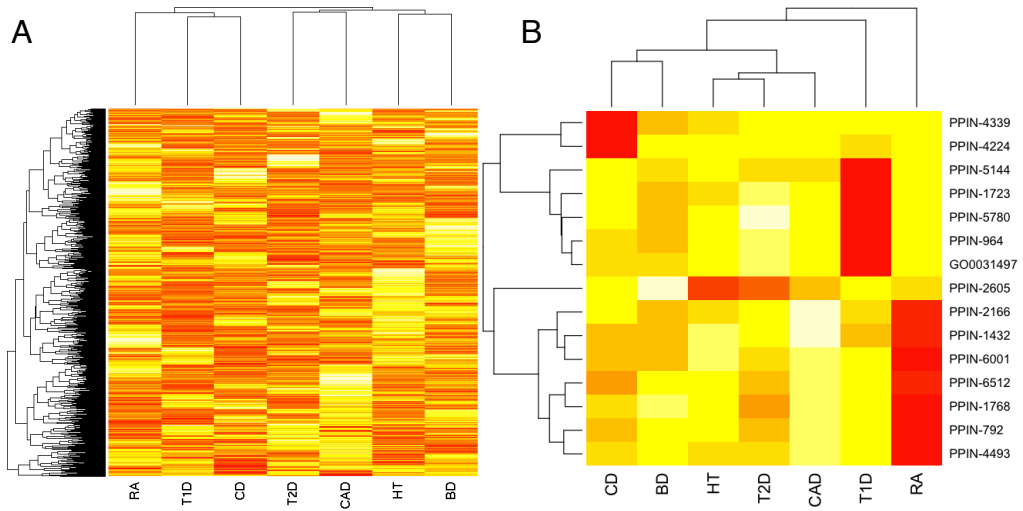

**Figure 3.5. Hierarchical clustering of diseases and gene-sets.** Clustering was performed with the complete linkage method with distances calculated using the Euclidean distance between gene-sets z-scores obtained from the association of gene-sets with diseases. Colour scale represents the significance of the association between a gene-set and a disease, red=strong associattion and light-yellow=weak association. Panel A and B present clustering results from 3183 and 15 gene-sets that were nominally significant (uncorrected $p < 0.05$) and significant at an FDR $< 0.5$ in at least one disease, respectively.

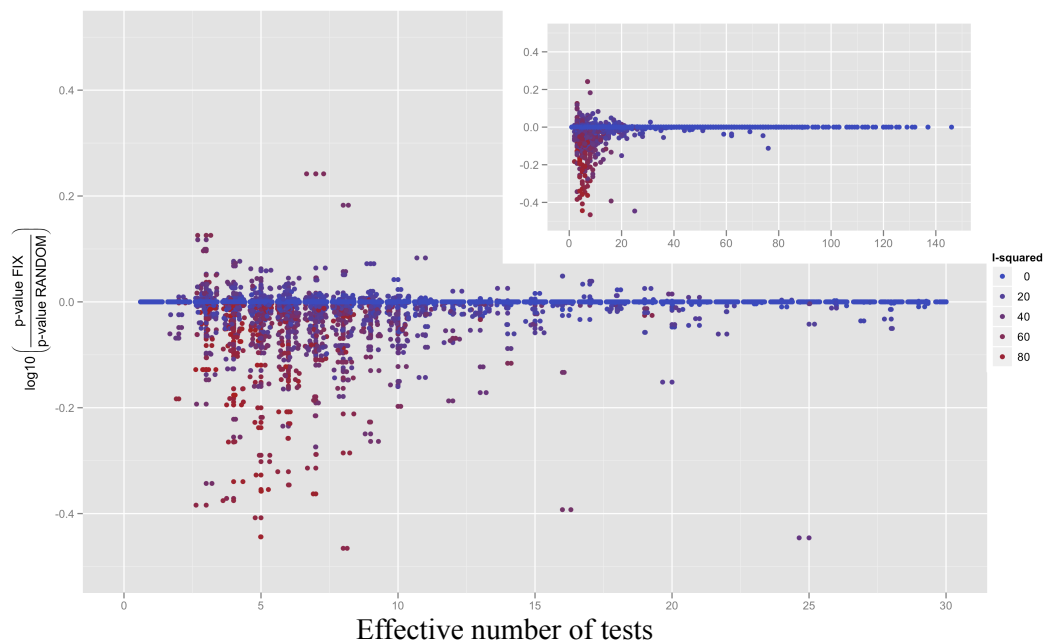

**Figure 3.6. Comparison of fixed and random-effects estimates.** Plotted for each gene (N ~20000) is the log10 of the ratio of fix and random-effects gene p-values. Insert (top right corner) presents data for the whole range of observed effective number of tests and main panel presents data on the range of effective number of tests where most differences between the two gene-wide p-value are seen. Points are coloured (see figure legend) by the gene's statistical heterogeneity measure $I^2$, which has range from 0 (no evidence of heterogeneity) to 100% (maximum evidence of heterogeneity).

may be disease specific.

### 3.3.4.    Comparison of fix and random-effects gene-wide statistics

Figure 3.6 presents a comparison of the fixed and random-effects gene-wide statistics of the BD GWAS. My results show that differences are commonly observed only in the case of (relatively small) genes with less than 30 effective SNPs or ~45 SNPs. Moreover, these differences are small, being less than one order of magnitude of the gene p-value. For larger genes, the statistical heterogeneity measure $I^2$ is rarely different from zero.

## 3.4. Discussion

Our results indicate that multivariate gene-based methods are useful for finding disease susceptibility loci. The gene-based methods found most of the replicated genes highlighted by single SNP analysis as well as additional genes that reached significance in other studies (Tables 3.2 and 3.3 and Figure 3.4). However, gene-based methods failed to perform well in disorders where SNP analysis did not perform well, e.g., BD, but succeeded in finding multiple loci that later reached the genome-wide significance level in the other disorders, such as CD (Tables 3.2 and 3.3 and Figure 3.4).

One potential confounder in my analysis is the fact that several of the studies indexed in the NHGRI GWAS catalogue include some of the WTCCC data in their analyses or as part of meta-analyses. This possibly complicates the use of this catalogue to define a replication set despite our removal of the original WTCCC reported data. While this is problematic, I do not think it is a major confounder as it means that loci found by gene-based methods might (if taken into the replication stage) reach genome-wide significance by subsequent meta-analysis of the single SNP association results.

Multivariate gene-wide association methods can potentially increase the power of GWAS to find true disease susceptibility loci by alleviating the effect of allelic heterogeneity or the incomplete tagging of causal variants (Furney et al., 2010; Liu et al., 2010; Luo et al., 2010). For example, the association highlighted between the TSPAN8 (tetraspanin 8) and T2D (Table 3.4) is supported by two SNPs, rs14953977 (p-value = 1.3 x 10$^{-6}$) and rs7961581 (p-value = 3.6 x 10$^{-6}$), in low LD, $r^2$ = 0.26 and D' = 0.85. Haplotype analysis is also expected to improve statistical power in these scenarios (Schaid, 2004). Interestingly, Browning and Browning (2008) found this region and another three to be of genome-wide significance through haplotype analysis but not single SNP analysis in the same GWAS analysed here (see Table 1 in Browning and Browning (2008)). Three of these regions were found by all multivariate gene-wide methods at FDR < 0.1. None of my methods highlighted the haplotype association at chromosome 15q26.2 in HT, which to the best of my knowledge has not been replicated yet. While both multivariate gene-wide association and haplotype analysis provides increased power, haplotype analysis is more computationally intensive and requires access to raw genotype data whereas gene-based association is more efficient, needing only the more easily obtainable summary data. Importantly, gene-wide statistics are useful for gene-set analyses, as I have shown for several diseases (Additional Table 3.1), and have the potential to be used in other strategies, including data integration for the establishment of convergent evidence towards molecular systems underpinning human diseases. Haplotype analyses results might be useful for gene-set analyses but considerations regarding the effect of gene-size bias would need to be taken into account. Additionally, I showed that gene-set results can be used to group disease in physiological meaningful categories. The hierarchical clustering results correctly grouped together immune-system diseases, CD, T1D and RA, and metabolic diseases,

T2D and CAD (Figure 3.5-A). These results suggest that disease classification may be easier using findings of association that reach only nominal significance rather than the very few that achieve genome-wide significance. It is possibly the top findings may be more relevant to understanding disease specific features whereas more general biological signatures emerge at lower significance levels. Although my analysis of the PPIN are equivalent to those of Baranzini et al. (2009), I failed to find as many associations with PPIN sub-networks and, intriguingly, with those related with the immune-system in T1D. Differences may have arisen due to the following: i) they used a different gene p-value method (Sidak's correction); ii) I performed correction for multiple testing on my results and they declared significance at a gene-set z-score > 3 (p < 0.0013) which matches a FDR > 0.2 in my analyses; and iii) I explicitly corrected my gene-set statistics using the correlation among the genes' statistics, which can be particularly problematic in regions of high LD when most genes belong to the same biological processes, such as the human leukocyte antigen (HLA) region.

My comparison of gene p-values derived with fixed and random-effects models suggest that random-effects models, or at least those implemented here, are unlikely to give very different results. When differences do occur they are in small genes (Figure 3.6). This is to be expected since only a small fraction of genetic variation is likely to be associated with any given phenotype. Therefore, in large genes (compared with small genes) a smaller proportion of the genetic variants is expected to be associated with any given phenotype[7]. For larger genes, most of the evidence will point to lack of association and the statistics will be more homogeneous (low $I^2$), indicating the relationship between statistical heterogeneity and genetic heterogeneity may not be simple. Thus statistical heterogeneity (i.e., non-uniformity of effect sizes) and random-effects may not be an effective strategy to tackle genetic heterogeneity (i.e., unlinked associated variants), contrary to what some authors have proposed (Lebrec et al., 2010). Development of a method to quantify genetic heterogeneity still remains an unsolved challenge.

_______________

[7] This argument comparing small and large genes does not include variables like LD, allele frequency distribution, haplotype diversity or functional effects of the genetic variation in a gene's product among other variables that likely differ between genes and are relevant to identify genetic association. Therefore, this argument should be interpreted as the expected pattern while keeping all these additional variables constant between genes.

I acknowledge several limitations of my study. By necessity I had to validate my results against the results from single SNP analysis rather than each method's own replication success. This could have favoured the SNP based analysis, as I cannot account for genes that would have been found and replicated only by gene-wide methods. Despite this limitation, my validation is useful in showing that gene-wide methods provide replicable findings. Furthermore, based on my benchmarking strategy, the SNP univariate analyses had higher precision (a higher proportion of significant genes replicated in later studies) than the gene-wide methods. Apart from bias in favour of replication by single SNP analysis, this can be partially explained by an increased tagging power of gene-based methods that enables the gene-based associations results to extend into overlapping genes and gene clusters in high LD across several genes (like the HLA region). I showed that accounting for overlapping genes does not change my results (Table 3.3). Likewise, multivariate methods could amplify technical artefacts such as genotyping error and be more sensitive to population stratification, thereby generating false positives. Although I did observe minor effects of this kind of confounders, I appreciate that in the future their influence will be reduced due to rapidly improving methods to deal with population stratification (Kang et al., 2010) and genotype calling (Browning and Yu, 2009).

Multivariate gene-wide association methods aid in the identification of disease loci. Furthermore, they benefit from being fast and applicable to summary statistics. As the cost of genotyping large cohorts decreases, GWAS is becoming an increasingly ubiquitous genetic research tool. Application of these methods can help by enhancing both power and interpretability of GWAS.

# 3.5. Additional Tables

| GWAS | Gene Set | Z-score | FDR | Biological Process/Description | Genes |
|---|---|---|---|---|---|
| CD | PPIN-4224 | 5.78 | $4 \times 10^{-4}$ | Immune response_IL-23 signalling pathway ($P = 2 \times 10^{-11}$), Immune response_IL-12 signalling pathway ($P = 3 \times 10^{-10}$), Inflammation Jak-STAT Pathway ($P = 8 \times 10^{-10}$) | IL12B IL12RB1 IL12RB2 IL23A IL23R JAK2 STAT3 STAT4 |
| | PPIN-4339 | 5.52 | $1 \times 10^{-3}$ | Inflammation_IL-6 signalling ($P = 2 \times 10^{-14}$), Erythropoietin pathway ($P = 2 \times 10^{-12}$), Inflammation_TREM1 signalling ($P = 2 \times 10^{-10}$) | AKT1 EPOR GAB1 GRB2 HYAL2 JAK2 MST1 PIK3R1 PLCG1 RELA SFN SHC1 SRC YES1 YWHAB YWHAE YWHAH YWHAQ YWHAZ |
| HT | PPIN-2605 | 5.08 | 0.04 | Cortisone biosynthesis and metabolism ($P = 5 \times 10^{-5}$), Androstenedione and testosterone biosynthesis and metabolism p.3 ($P = 2 \times 10^{-5}$) | CYP11A1 CYP11B1 CYP11B2 FDX1 |
| | PPIN-4493 | 7.77 | $4 \times 10^{-9}$ | Role of tetraspanins in the integrin-mediated cell adhesion ($P = 3 \times 10^{-4}$), Role of cell-cell and ECM-cell interactions in oligodendrocyte differentiation and myelination($P = 5 \times 10^{-4}$), Phagosome in antigen presentation ($P = 5 \times 10^{-4}$) | CD63 CD74 CD82 CTAG1B HLA-DMA HLA-DMB HLA-DRA HLA-DRB1 HLA-DRB3 HLA-DRB4 HLA-DRB5 MAGEC1 MBP PRPF40B |
| | PPIN-792 | 6.32 | $1 \times 10^{-5}$ | Cell-matrix interactions ($P = 1 \times 10^{-8}$), Inhibition of remyelination in multiple sclerosis: role of cell-cell and ECM-cell interactions ($P = 6 \times 10^{-7}$), Antigen presentation by MHC class II ($P = 2 \times 10^{-6}$), ERBB-family signalling ($P = 7 \times 10^{-6}$) | AP4M1 CANX CD19 CD1D CD4 CD44 CD63 CD74 CD82 CD9 CTSF CTSL1 DPM1 DPM2 DPM3 EGFR ERBB2 ERBB3 HBEGF HLA-DMA HLA-DMB HLA-DOA HLA-DOB HLA-DPA1 HLA-DPB1 HLA-DQA1 HLA-DQB1 HLA-DRA HLA-DRB1 HLA-DRB5 IGSF8 ITGA3 ITGA4 ITGA6 ITGB1 ITGB2 LRP1 MBP MIF NUP88 PI4K2A P1GA P1GC PRPF40B PTGFRN TIMP1 VANGL1 |

| GWAS | Gene Set | Z-score | FDR | Biological Process/Description | Genes |
|------|----------|---------|-----|-------------------------------|-------|
| RA | PPIN-6001 | 5.61 | $4\times10^{-4}$ | Neuropeptide signalling pathways (P = $4\times10^{-15}$) | ATP5L CTAG1B ECHDC2 GTF2F1 H2AFZ HLA-DMA HLA-DRA HLA-DRB1 HLA-DRB3 HLA-DRB4 HLA-DRB5 MAGEC1 MRPL19 POMC ROCK2 RUSC2 TRA@ TRB@ |
| | PPIN-2166 | 5.09 | $4\times10^{-3}$ | Neuropeptide signalling pathways (P = $2\times10^{-18}$), Melanocyte development and pigmentation (P = $3\times10^{-5}$) | CTAG1B FGF10 HLA-DRB1 HLA-DRB3 HLA-DRB4 HLA-DRB5 MAGEC1 MC1R MC2R MC4R MC5R MEP1A NRD1 PAM POMC VTN |
| | PPIN-1432 | 4.96 | $7\times10^{-3}$ | Immune response TCR signalling (P = $9\times10^{-8}$), Immune response Antigen presentation (P = $1\times10^{-7}$) | CANX CD1D CD247 CD3D CD3E CD3G HLA-A HLA-B HLA-DRB1 HSPA5 LEF1 TRA@ TRAT1 TRB@ TRD@ TRG@ |
| | PPIN-1768 | 4.96 | $7\times10^{-3}$ | Role of tetraspanins in the integrin-mediated cell adhesion (P = $2\times10^{-8}$), Gastrin in cell growth and proliferation (P = $1\times10^{-7}$), Immune response Antigen presentation (P = $2\times10^{-7}$) | AP4M1 APTX ATN1 ATXN1 BMP2K CANX CD19 CD1D CD4 CD44 CD63 CD74 CD82 CD9 CDK12 CLK1 CTAG1B CTDSP1 CTSF CTSL1 DPM2 DYRK1B EFHA1 EGFR ERBB2 ERBB3 HBEGF HLA-DMA HLA-DMB HLA-DOA HLA-DOB HLA-DPA1 HLA-DPB1 HLA-DQA1 HLA-DQB1 HLA-DRA HLA-DRB1 HLA-DRB5 HTT IGSF8 IRAK1 ITGA3 ITGA4 ITGA6 ITGB1 ITGB2 LRP1 MAG MAPK1 MAPK11 MAPK13 MAPK14 MAPK15 MAPK3 MAPKAPK5 MBP MELK MIF MKNK1 MMP7 NEK9 NUP88 PAK1 PI4K2A PLP1 PRKACA PRKCA PRKCD PRKCI PRKCZ PRMT1 PRMT5 PRPF40B PTGFRN RAN RPS6KA5 RTN4 SRPK1 STK16 STK39 STK4 TIMP1 TLK1 TLK2 TRA@ TRAF6 TRB@ TRPM7 VANGL1 |
| | PPIN-6512 | 4.61 | 0.03 | Immune response Antigen presentation (P = $2\times10^{-7}$), Immune response TCR signalling (P = $9\times10^{-8}$) | CANX CD1D CD247 CD3D CD3E CD3G CD63 CD74 CD82 CTAG1B HLA-A HLA-B HLA-DMA HLA-DMB HLA-DOA HLA-DRA HLA-DRB1 HLA-DRB3 HLA-DRB4 HLA-DRB5 HSPA5 LEF1 MAGEC1 MBP PRPF40B TRA@ TRAT1 TRB@ TRD@ |
| T1D | PPIN-1723 | 6.61 | $5\times10^{-4}$ | NS | HIST1H1A HIST1H1C HIST1H3G HIST1H3H HIST2H3C LUC7L2 NASP ULK2 |
| | GO0031497 | 6.09 | $3\times10^{-3}$ | Chromatin assembly | |
| | PPIN-5144 | 5.82 | $8\times10^{-3}$ | Role of Brca1 and Brca2 in DNA repair (P = $6\times10^{-4}$) | BCCIP BRCA2 CAPZB CDKN1A FAM46A HIST1H2BN ITSN2 PTN RAD51 RPL23 |
| | PPIN-964 | 5.47 | 0.03 | NS | DGCR6L SUFU ZNF165 ZNF193 ZNF232 ZNF24 ZNF263 ZNF434 ZNF446 ZNF496 ZSCAN16 |
| | PPIN-5780 | 5.41 | 0.03 | NS | DGCR6L EMG1 NOL12 SLC25A38 SUFU ZNF165 ZNF193 ZNF232 ZNF24 ZNF263 ZNF434 ZNF446 ZNF496 ZSCAN16 |

| GWAS | Gene Set | **Z-score** | **FDR** | Biological Process/Description | Genes |
|---|---|---|---|---|---|

**Additional Table 3.1. Gene-Set Analysis with gene-wide p-values in the WTCCC phase 1 study.** Gene-sets from Gene Ontology Database ([www.geneontology.org](http://www.geneontology.org)) are described with the database's description. Gene-sets derived from the protein-protein interaction (PPI) network were characterised using GeneGO MetaCore database and we report the three best biological categories and the over-representation p-value (calculated using the hypergeometric distribution). CD = Crohn's disease, HT = hypertension, RA = rheumatoid arthritis and T1D = type 1 diabetes.

# Chapter IV

# Integrative biology analyses of bipolar disorder GWAS

# 4.1. Introduction

Bipolar disorder (BD) is a chronic and episodic psychiatric condition characterised by extremes of mood ranging from mania to severe depression, usually accompanied by disturbances in cognition and behaviour. Psychotic features such as delusions and hallucinations often occur. Despite a convincing and substantial genetic contribution to the aetiology of BD (McGuffin et al., 2003), its genetic and molecular underpinnings remain largely unknown. Its diagnosis is based solely on observed clinical features. Genome-wide association studies (GWAS) and linkage studies have highlighted several genomic regions (Lewis et al., 2003; Ferreira et al., 2008) and recently replicated evidence implicating specific loci has also been reported (Ferreira et al., 2008; O'Donovan et al., 2008; Breen et al., 2011).

These and other large scale association studies of common genetic variation have reinforced the notion that many low risk genetic variants are involved in the aetiology of BD (Ferreira et al., 2008; Scott et al., 2009; Purcell et al., 2009). Analyses of GWAS of Crohn's disease and multiple sclerosis have highlighted genetic heterogeneity within single genes and biological processes, i.e. 2 or more polymorphisms within a gene being independently associated and different sets of genes within the same pathway associated in different studies (Baranzini et al., 2009; Wang et al., 2009). There is some evidence to suggest this may also be true for BD (Schulze et al., 2009). Thus, recent findings lend support to the idea that genetic heterogeneity is a major feature of the genetic architecture of common risk variants of complex traits, for example see McClellan and King (2010). Although the extent to which this is true for different phenotypes is not yet clear. Thus allelic heterogeneity (i.e. statistical association at unlinked alleles in a locus), and locus heterogeneity (i.e. different regions across the genome), are a challenge for ongoing meta-analytical strategies that aim to combine the increasing volume of GWAS data now available for psychiatric disorders (Psychiatric GWAS Consortium, 2009).

Genetic heterogeneity might have multiple sources:

i)   When the effect-size of a single genetic variant is not consistent across multiple measurements or studies. This may be quantified with statistics commonly used in meta-analyses, such as the Q statistic or $I^2$ index (Huedo-Medina et al., 2006). A possible scenario of heterogeneity at a single site occurs when an allele shows association only in one or a few studies; for example: in the extreme case in which an allele is significantly associated in a single study. This would lead to an overall heterogeneous association pattern, showing signs of allelic heterogeneity or locus heterogeneity (see below).

ii)  Allelic heterogeneity may occur when alleles at different genomic positions appear to be correlated with the phenotype because they are proxies or tag either a single same causal variant or multiple, potentially independent, causal variants. Allelic heterogeneity can occur simply due to differences in recombination patterns between populations or samples. Currently, there is not good quantitative measure for allelic heterogeneity.

iii) Locus heterogeneity refers to alleles at different loci that are correlated with the phenotype. This kind of heterogeneity is implicitly related to allelic heterogeneity since the key differences reside in how we define the physical limits of a locus. In this work the gene's coordinates define the limits.

If allelic associations are homogenous they will be found by SNP based meta-analysis, provided there is enough statistical power. However, in the case of allelic or locus heterogeneity, SNP based meta-analyses are likely to be less successful. An alternative strategy would be to aggregate or group allelic associations, e.g., within genes or biological processes. An assumption built into this strategy is that the relevant commonality of genetic risk across populations, samples and potentially individuals does not relate to the alleles themselves but their underlying functional effects. The underlying commonalities may be partially a reflection of the biological function of the genes where these association lie. This leads to the hypothesis that by grouping genetic associations into their respective biological processes one can tackle genetic heterogeneity. This strategy can potentially help under both heterogeneity models and we employ it here to enhance the information extracted from GWAS of BD.

A systems biology oriented approach to tackle heterogeneity and improve both power and interpretation of GWAS is feasible because disease susceptibility variants cluster within genes and biological processes (recently reviewed in (Wang et al., 2010)). This approach has started to reveal valuable information from BD GWAS. Analysis of the Wellcome Trust Case Control Consortium (WTCCC) BD GWAS and its meta-analysis with another GWAS provided evidence of association within biological processes involved in the modulation of transcription and cellular activity, including that of hormone action and adherens junctions (Baranzini et al., 2009; Holmans et al., 2009; O'Dushlaine et al., 2011). Also, an association between the gamma-aminobutyric acid (GABA) A receptor pathway genes and the BD schizoaffective subtype has been the only replicated system-level finding in BD to date (Breuer et al., 2010; Craddock et al., 2010). (This finding is analysed in detail in Chapter V.)

Increasing evidence suggests that gene-expression studies can be used successfully to prioritise GWAS results (see refs (Hsu et al., 2010; Zhong et al., 2010a; Zhong et al., 2010b)). For example, Zhong et al. (2010a) showed that gene-expression changes and genetic-susceptibility variants conferring risk to type 2 diabetes cluster in common biological pathways.

Here we present a systems biology oriented analysis aiming to identify biological processes associated with BD. By combining three GWAS with gene-expression studies we provide convergent and robust evidence of association with BD despite locus heterogeneity.

## 4.2. Methods

### 4.2.1. Samples and genotype data

We re-analysed GWAS of BD from the Wellcome Trust Case Control Consortium study (2007), Cichon et al. (2011) and Sklar et al. (2008) studies, which we refer to as WTCCC, Bonn and Sklar respectively. We used individual level genotype and phenotype data from the Bonn study and summary statistics from the two other studies. The quality control undertaken in the WTCCC BD GWAS was described in Chapter III and for the Sklar and Bonn studies can be found in the original reports. BD samples of all studies met DSM-IV criteria to establish BD diagnosis. Statistical association in the Bonn GWAS was performed with a logistic regression model using PLINK (Purcell et al., 2007). We also analysed summary statistics from six GWAS in common non-psychiatric disorders reported by the WTCCC. These were GWAS of Crohn's disease, type 1 diabetes, type 2 diabetes, rheumatoid arthritis, coronary artery disease and hypertension (WTCCC, 2007). Detailed description of each sample can be found in the WTCCC report. We used the summary data provided by the WTCCC and excluded SNPs as per WTCCC (WTCCC, 2007).

### 4.2.2.    Gene and network analyses

We calculated gene-wide p-values using the FORGE software including in our analyses approximately 20000 protein-coding, lincRNA (long non-coding RNA)[8] and miRNA (micro RNA) genes annotated in Ensembl v59. We mapped SNPs to genes if the SNP was within 20 kb of the annotated coordinates of the gene. We utilised the asymptotic Z FIX gene-wide association p-values calculated by FORGE and performed gene-set analyses with these gene-wide p-values and gene-sets derived from the protein-protein interaction network (PPIN), as described in Chapter III. For all analyses we used the HapMap Phase 3 CEU genotypes (Altshuler et al., 2010) to calculate SNP-SNP correlations. We interpreted the biology of the significant PPIN sub-networks using MetaCore (GeneGo, Inc.; www.genego.com). GeneGO provides biological networks constructed by literature reviews of genes and gene product interactions and bioinformatic tools to analyse genomic data sets for enrichment in these biological networks. GeneGO's tools assess the significance of the overlap between our sub-networks and their manually curated biological processes using hypergeometric distribution statistics. Please see Section 3.2.3 for additional details on the analyses performed with GeneGO's biological processes annotations. We also tested the enrichment of the gene p-values obtained from the test of differential expression of between BD and control's brains among PPIN sub-networks. This enrichment was assessed using the parametric analysis of gene set enrichment (PAGE) method (Kim and Volsky, 2005). To select significant genes and sub-networks we used the local false discovery rate (FDR) (Efron et al., 2001; Efron, 2010) as implemented in the package "fdrtools" version 1.2.6 (Strimmer, 2008) for R (R Development Core Team, 2010).

---

[8] lincRNA is a particular class of RNA molecules that do not code for protein products. Other classes of non-protein coding RNAs include the regulatory RNAs known as miRNA or the tRNA involve in protein translation. We have included lincRNA in our analyses because recent evidence suggest they have major roles in regulation of cellular processes and association with them is of potential interest. We refer the reader to (Guttman et al., 2009; Khalil et al., 2009; Marques and Ponting, 2009; Loewer et al., 2010) for additional information.

### 4.2.3.    Meta-analyses

Gene and network level statistics from different studies were combined using the fixed-effect model described in Chapter II for the gene-wide testing. We set study specific weights calculated as the square root of the sample size: WTCCC =70.71, Sklar = 59.16 and Bonn = 44.72. When combining GWAS and gene-expression studies, equal weights were used. The use of weights equal to the square root of the sample sizes is a common practice in meta-analyses when no additional information is available to quantify the uncertainty of the individual association statistic. We refer the interested reader to a recent discussion in the topic by Zaykin (2011).

### 4.2.4.    cDNA microarray analysis

The gene-expression data of dorsolateral prefrontal cortex (DLPFC) tissue from 61 subjects (30 control and 31 bipolar patients) and orbitofrontal cortex (OFC) tissue from 21 subjects (10 controls and 11 bipolar patients) reported by Ryan et al. (2006) was downloaded from the ArrayExpress database (Parkinson et al., 2009) under accession number E-GEOD-5392. Raw intensity values of Affymetrix Human Genome U133A arrays were normalized with the Robust Multi-Average (RMA) algorithm (Irizarry et al., 2003). Pre-filtering removed transcripts not detected (marked as 'absent' by using MAS5 detection call algorithm) in all samples and were not considered further. Please see Ryan et al. (2006) for details of demographic variables of cases and controls. We used as covariates variables that showed differences between cases and controls or were correlated with RNA quality measures and therefore can be considered as potential confounders. The generalised linear model (GLM) with covariates was used to assess differential expression for each probe in each brain region: for the DLPFC samples we used the GLM model with disease status (control/BD) as the main effect while controlling for brain pH and fluphenazine equivalents; and for the OFC samples, we used fluphenazine equivalents as a covariate.

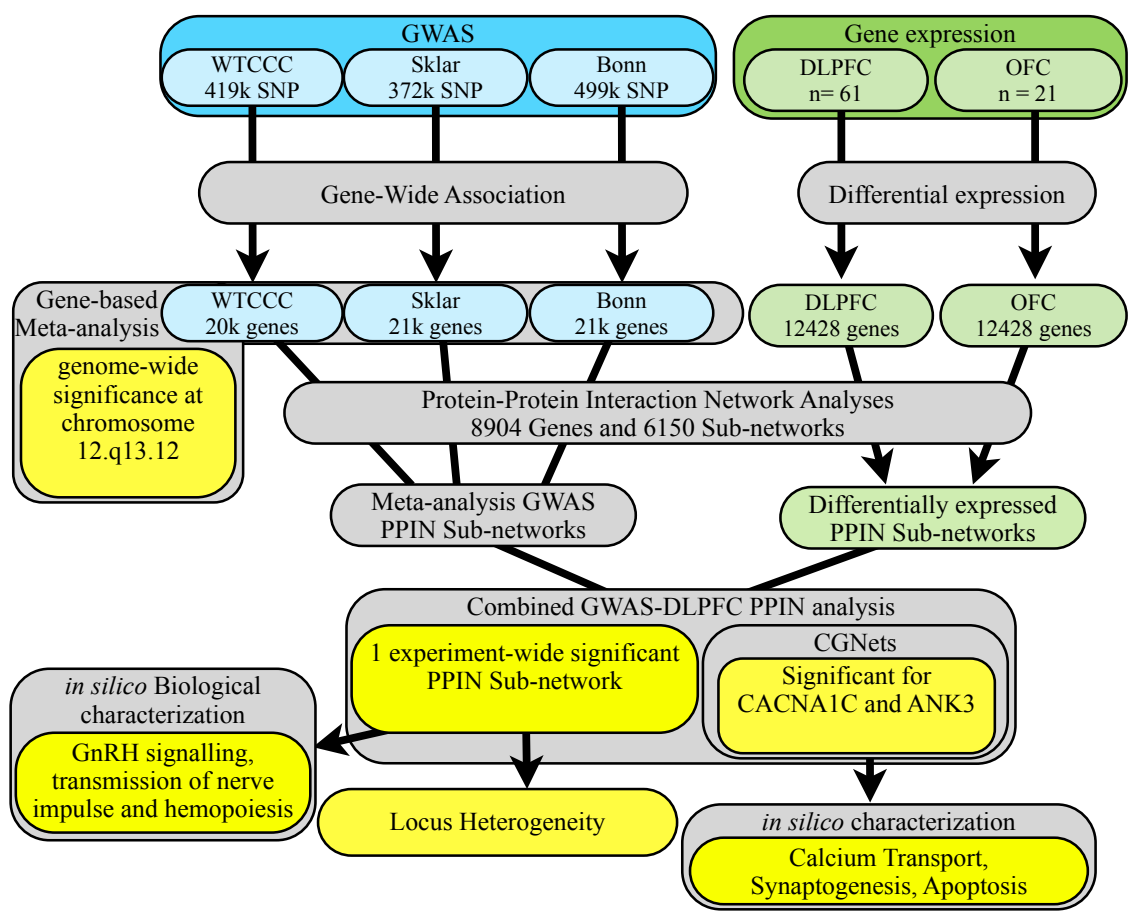

**Figure 4.1. Data analysis strategy.** Gene-wide p-values for association with BD were calculated on four GWAS and 2 gene-expression studies. Gene-wide results were mapped to the PPIN and a greedy search strategy (Ideker et al., 2002) was used to identify subnetworks of interacting proteins enriched with BD association signals. A meta-analysis of the GWAS gene-wide association p-values and PPIN subnetworks results was carried out. GWAS PPIN results were combined with gene-expression subnetwork results and significant subnetworks were characterised with regard to their biological functions and locus heterogeneity. Colour codes are: Grey = statistical/ bioinformatic analyses, yellow = final results, blue = GWAS data and green = gene-expression data. DLPFC = doroslateral prefrontal cortex and OFC = orbitofrontal cortex tissue.

## 4.3. Results

### 4.3.1. Gene-based association

We combined results from three GWAS (WTCCC, 2007; Sklar et al., 2008; Cichon et al., 2011) and two gene-expression studies of BD (Ryan et al., 2006). The combined data was analysed using gene and network analyses to identify biological processes enriched among BD associated genetic variants and gene-expression changes (Figure 4.1). Detailed descriptions of these datasets have been provided elsewhere. All studies used DSM-IV criteria to establish BD diagnosis and quality control included sample and genetic marker filters to exclude low quality data.

| Gene Symbol | GWAS | | | | | Gene expression | | GWAS + Gene expression | |
|---|---|---|---|---|---|---|---|---|---|
| | WTCCC | Sklar | Bonn | $Z_{GWAS}$ | FDR | $Z_{DLPFC}$ | $Z_{OFC}$ | $Z_{GWAS/DLPFC}$ | $Z_{GWAS/OFC}$ |
| DHH | 3.31 | 2.61 | 2.31 | 4.79 | 0.04 | - | - | - | - |
| LMBR1L | 3.09 | 2.61 | 2.31 | 4.65 | 0.049 | 1.97 | 1.06 | 4.68 | 4.04 |
| LMAN2L [b] | 2.86 | 1.79 | 3.29 | 4.44 | 0.073 | 0.36 | 1.65 | 3.39 | 4.31 |
| RHEBL1 | 3.61 | 1.69 | 1.70 | 4.21 | 0.108 | - | - | - | - |
| ENSG00000211987 | 3.57 | 0.34 | 3.28 | 4.09 | 0.129 | - | - | - | - |
| MURC | 2.90 | 1.96 | 2.03 | 4.02 | 0.143 | - | - | - | - |
| DUSP28 | 4.37 | 0.65 | 1.30 | 3.96 | 0.157 | - | - | - | - |
| ITIH3 [a] | 2.66 | 1.79 | 2.37 | 3.90 | 0.169 | -0.22 | 1.33 | 2.6 | 3.70 |
| ENSG00000246653 | 2.86 | 1.30 | 2.65 | 3.88 | 0.174 | - | - | - | - |

**Table 4.1: Gene-based meta-analysis of bipolar disorder GWAS.** Table reports statistics for genes reaching FDR < 0.2 in the gene-based GWAS meta-analysis. For each GWAS dataset we report z-scores obtained from the gene p-values using the normal distribution inverse cumulative distribution function. For the meta-analysis of GWAS results we report the pooled z-scores estimate ($Z_{GWAS}$) and the FDR value for each gene. For the gene-expression data we report z-scores obtained from the p-values calculated with the test of differential gene-expression between cases and controls for each brain region ($Z_{OFC}$ and $Z_{DLPFC}$). For the combined analysis of GWAS and gene-expression we report z-scores obtained by pooling the $Z_{GWAS}$ with each brain region z-score, which gave raise to two meta-analysis z-scores, $Z_{GWAS-DLPFC}$ and $Z_{GWAS-OFC}$. DLPFC= doroslateral prefrontal cortex and OFC = orbitofrontal cortex tissue. Loci with previously reported genetic association with bipolar disorder by [a] Scott et al. (2009) and [b] Ferreira et al. (2008).

We calculated gene-wide association p-values for the three GWAS using the FORGE software and for the gene-expression studies using a standard differential expression analysis with covariates (see Methods). Gene-based GWAS meta-analysis showed significant (FDR < 0.05) and suggestive evidence (FDR < 0.2) for 2 and 7 genes, respectively (Table 4.1). The two significant genes, DHH and LMBR1L, are within a region on chromosome 12q13.12 with high LD spanning several genes that had FDR < 0.2, including RHEBL1. The haplotype block containing ITIH3 and ENSG00000246653 has been reported to have genome-wide significance for BD (Breen et al., 2011). The loci of LMAN2L and DUSP28 had high but sub-threshold significance in the meta-analysis by Ferreira et al. (2008) (rs2314398 p-value = 3 x 10$^{-6}$) and WTCCC report (WTCCC, 2007). Finally, the MURC region has not been previously associated with BD and ENSG00000211987 is a miRNA gene predicted based on its similarity to MIR548C (hsa-mir-548c). We found no evidence of expression for ENSG00000211987 in public databases. Of the genes with FDR < 0.2 only LMBR1L had evidence of differential expression in brains of BD patients compared with controls (Table 4.1).

To identify groups of interacting gene products enriched among BD association signals, we mapped gene-level results from the GWAS and gene-expression studies to the protein-protein interaction network (PPIN) and tested the association of approximately 6100 sub-networks between 2 and 200 nodes (gene products) in size. We did not identify significant sub-networks by meta-analysis of the GWAS results but several reached suggestive levels. To strengthen the results from the genetic data alone, we combined the evidence of GWAS and gene-expression studies at the PPIN sub-network level. There were 3 and 1 sub-networks with FDR < 0.2 in the meta-analysis of GWAS and DLPFC or OFC gene-expression, respectively (Table 4.2). The best sub-network, PPIN-1576 (GWAS-DLPFC meta-analysis z-score = 4.73, p-value = 1 x 10[-6] and FDR = 0.1) had > 60% overlap with the other two sub-networks (Table 4.3), which suggests that all three sub-networks captured the same signal. We focused on PPIN-1576, depicted in Figure 4.2, for additional bioinformatic analyses to characterise its possible biological functions.

| Subnetwork ID | GWAS | | | | | Gene Expression | | | | GWAS + Gene expression | | | |
|---|---|---|---|---|---|---|---|---|---|---|---|---|---|
| | | | | | | DLPFC | | OFC | | | | | |
| | WTCCC | Sklar | Bonn | $Z_{GWAS}$ | FDR | Z | FDR | Z | FDR | $Z_{GWAS/DLPFC}$ | FDR | $Z_{GWAS/OFC}$ | FDR |
| PPIN-1576 | 1.17 | 2.88 | 1.20 | 3.22 | 1 | 3.48 | 0.003 | 2.68 | 0.06 | 4.73 | 0.10 | 4.17 | 0.14 |
| PPIN-5572 | 0.78 | 2.60 | 2.07 | 3.29 | 1 | 3.09 | 0.009 | 2.07 | 0.08 | 4.51 | 0.15 | 3.79 | 0.26 |
| PPIN-5042 | 0.62 | 2.96 | 1.20 | 2.98 | 1 | 3.29 | 0.004 | 1.85 | 0.11 | 4.43 | 0.17 | 3.41 | 0.43 |

**Table 4.2: PPIN subnetworks with FDR < 0.2.** Table reports network association statistics for GWAS, gene-expression results and their meta-analysis. For each GWAS dataset we report z-scores obtained from the network association (larger values mean more significant association). For the meta-analysis of GWAS results we report a pooled z-scores estimate ($Z_{GWAS}$) and the FDR value for each sub-network. For the gene-expression data we report z-scores obtained from the PAGE analysis for each brain region ($Z_{OFC}$ and $Z_{DLPFC}$). For the combined analysis of GWAS and gene-expression we report z-scores obtained by pooling the $Z_{GWAS}$ with each brain region z-score, which gave raise to two meta-analysis z-scores, $Z_{GWAS-DLPFC}$ and $Z_{GWAS-OFCS}$, and their corresponding FDR values. DLPFC= dorsolateral prefrontal cortex and OFC = orbitofrontal cortex tissue.

Genes of PPIN-1576 were strongly enriched (FDR < 0.001) among the GeneGO biological categories gonadotropin-releasing hormone (GnRH) signalling pathway (p-value = 2 x 10[-9]), transmission of nerve impulse (p-value = 2 x 10[-9]) and hemopoiesis and erythropoietin pathway (p-value = 3 x 10[-7]) (Table 4.4). In common with previous reports (Baranzini et al., 2009; Wang et al., 2009; O'Dushlaine et al., 2011), we found that genes driving the network associations were different in each data set (Figure 4.3 and Additional Table 4.1). Only SYNCRIP1 out of 181 PPIN-1576 genes had nominally significant gene p-values in more than one GWAS. In line with gene-set analysis results of schizophrenia GWAS (O'Dushlaine et al., 2011) there was no evidence of association with PPIN-1576 in six non-psychiatric complex diseases (Table 4.5). This may suggest some degree of specificity to psychiatric illness. However, this evidence is relatively weak because we do have less statistical power on these six GWAS than on our combined BD GWAS data set. Furthermore, we are not testing gene-expression data on these other six disorders, which in the case of bipolar disorder was important to identify the association with PPIN-1576.

| | | | |
|---|---|---|---|
| PPIN-1576 | 182 | | |
| PPIN-5572 | 0.63 | 115 | |
| PPIN-5042 | 0.63 | 0.99 | 121 |

**Table 4.3. Fraction of genes shared between PPIN subnetworks with FDR < 0.2.** Diagonal reports the total number of genes in each subnetwork.

### 4.3.2.    Candidate-Gene-Networks of bipolar disorder GWAS hits

The significant PPIN-1576 network was identified after a hypothesis free analysis of the PPIN. However, if disease susceptibility genes are known it is possible to define regions of the PPIN that are likely to harbour additional disease genes (Goh et al., 2007; Wu et al., 2008). Thus, we hypothesised that the network environment of BD susceptibility genes should be over-represented with genetic susceptibility variants and gene expression changes, potentially enabling us to strengthen the evidence for these putative susceptibility genes. We identified nine genes as candidates based on their proximity (< 20 kb) to genetic variants reported with p-values $< 5 \times 10^{-8}$ in BD GWAS: ANK3, CACNA1C, PBRM1, NT5DC2, NCAN, HALPN4 and TM6SF2 from three meta-analyses (Ferreira et al., 2008; Cichon et al., 2011) and PALB2, DCTN5 and DGKH from two individual studies (WTCCC, 2007; Baum et al., 2008). We then tested for association 45 Candidate-Gene-Networks (CGNets, PPIN sub-networks including

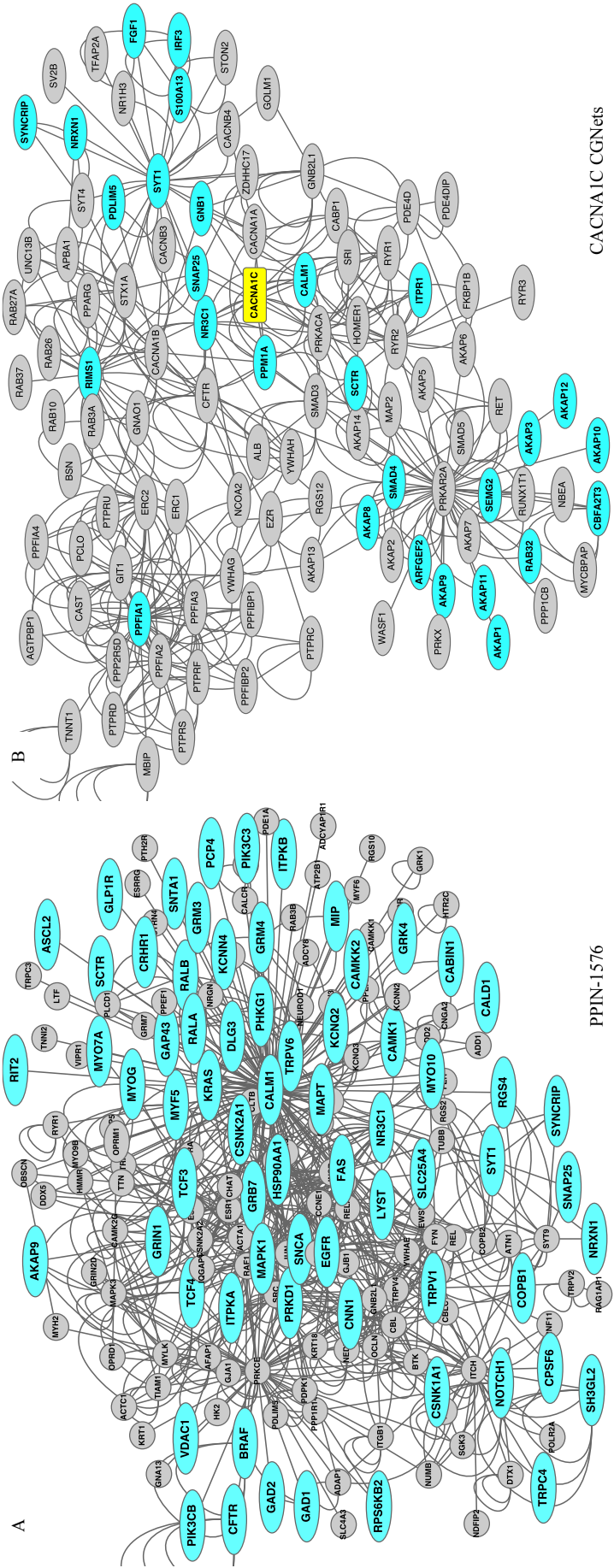

**Figure 4.2. Significant network from PPIN-wide search and CGNets analysis.** Blue nodes had a p-value < 0.05 in at least one study of the GWAS. Grey node had p-values > 0.05 in all studies. A) PPIN-1576 sub-network found by PPIN-wide search and B) CACNA1C sub-networks merged. CACNA1C is highlighted in yellow.

| Biological Category | Genes in/total | P-value |
|---|---|---|
| Reproduction: GnRH signaling pathway | 17/91 | $2 \times 10^{-9}$ |
| Neurophysiological process: Transmission of nerve impulse | 20/129 | $2.1 \times 10^{-9}$ |
| Reproduction: Gonadotropin regulation | 17/116 | $9 \times 10^{-8}$ |
| Development: Hemopoiesis, Erythropoietin pathway | 15/98 | $3 \times 10^{-7}$ |
| Development: Beta-adrenergic receptors transactivation of EGFR | 8/22 | $7.4 \times 10^{-7}$ |
| Deregulation of PSD-95-dependent signaling in Huntington's disease | 8/22 | $7.4 \times 10^{-7}$ |
| Signal transduction: NOTCH signaling | 20/185 | $1 \times 10^{-6}$ |
| Signal transduction: Neuropeptide signaling pathways | 14/97 | $1.6 \times 10^{-6}$ |
| Signal transduction: ESR1-membrane pathway | 11/61 | $2.4 \times 10^{-6}$ |
| Development: Neuromuscular junction | 14/101 | $2.6 \times 10^{-6}$ |
| Development: Gastrin in cell growth and proliferation | 10/43 | $3 \times 10^{-6}$ |
| Signal transduction: WNT signaling | 16/134 | $3.7 \times 10^{-6}$ |
| Neurophysiological process: Long-term potentiation | 9/42 | $4.6 \times 10^{-6}$ |
| Neurophysiological process: Receptor-mediated axon growth repulsion | 9/37 | $6.5 \times 10^{-6}$ |
| Neurophysiological process: Dopamine D2 receptor transactivation of PDGFR in CNS | 5/9 | $8.4 \times 10^{-6}$ |
| Development: Delta-type opioid receptor signaling via G-protein alpha-14 | 6/17 | $2.4 \times 10^{-5}$ |
| Development: G-Proteins mediated regulation MARK-ERK signaling | 7/26 | $3.6 \times 10^{-5}$ |
| Development: Angiotensin signaling via PYK2 | 7/26 | $3.6 \times 10^{-5}$ |
| Development: Angiotensin activation of ERK | 6/19 | $5 \times 10^{-5}$ |
| Development: Alpha-2 adrenergic receptor activation of ERK | 7/28 | $6.1 \times 10^{-5}$ |

**Table 4.4. Top 20 biological categories characterizing PPIN-1756 sub-network's genes.** Gene counts denote the number of PPIN-1756 genes within the biological category and the category's total size. Significance for the overlap of PPIN-1756 genes and those annotated on the biological process was calculated p-value was calculated assuming a null hypergeometric distribution (see Methods for additional details). All categories reported reached an FDR < 0.001.

the BD candidate genes) with less than 50 nodes to capture information from the genes'

adjacent network environment. We found a CGNet for the gene CACNA1C that was

significant after multiple testing correction (PPIN-6152 GWAS meta-analysis corrected

p-value = 0.01) that was supported by the gene-expression datasets (Table 4.6 and

Figure 4.2). Two other CACNA1C CGNets, PPIN-4800 and PPIN-4877, were not

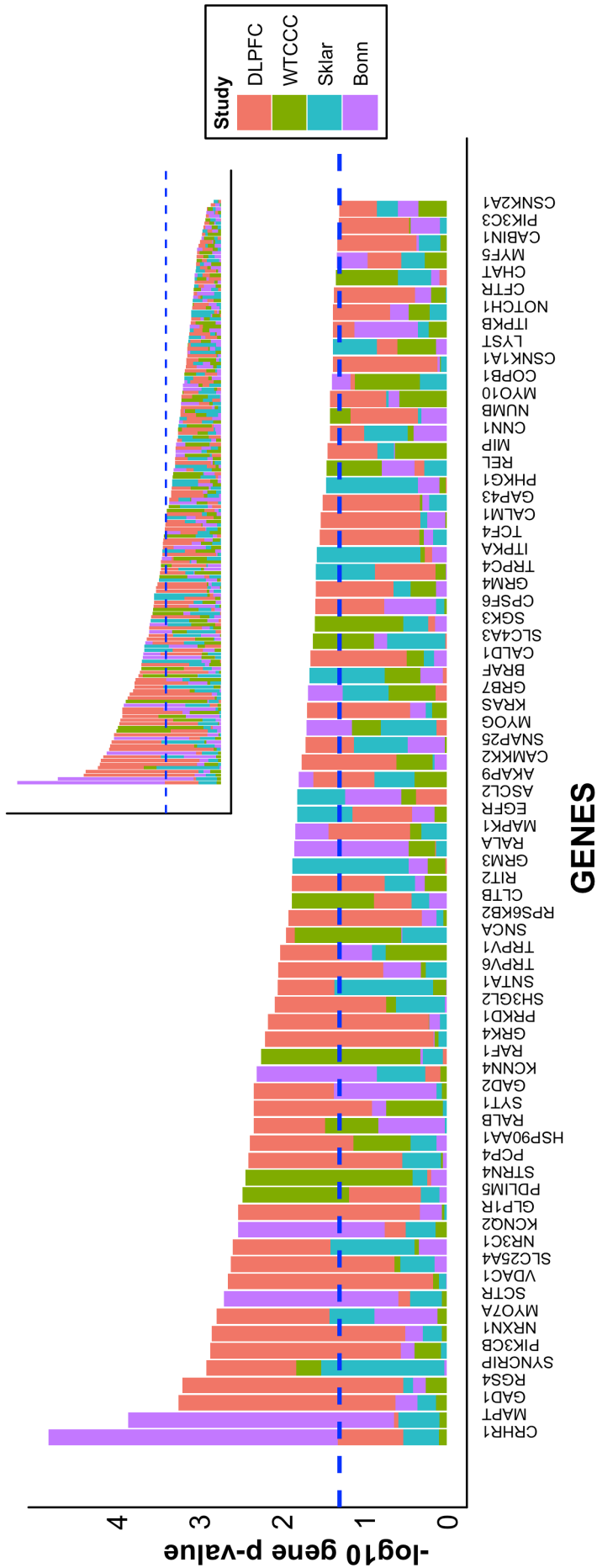

**Figure 4.3. Gene-wide p-values for PPIN-1576.** Plotted are -log10 gene-wide p-values on the y-axis for each gene within PPIN-1576. Different studies have been coloured as indicated in the legend and their bar plots were stacked to avoid over-plotting and allow comparison. The main figure presents data for genes with p-values < 0.05 in at least one study GWAS or the gene-expression DLPFC. Insert (top right) presents the same information for all genes regarding of their p-value. Dashed line marks -log10 equal to 0.05.

significantly associated in the GWAS data but had significant evidence of differential

gene-expression (Table 4.6 and Figure 4.2). Two ANK3 CGNets, PPIN-3052 and

PPIN-6701, reached significance when the GWAS and DLPFC gene-expression results

were combined (Table 4.7). In Additional Table 4.2, we report results for the 45 CGNets

tested for the 6 candidate genes present in the PPIN. ANK3 CGNets had almost

complete overlap and their genes were best characterised by biological processes related

to apoptosis signalling cascades, such as FAS (p-value $= 7 \times 10^{-12}$) or IAP-proteins

signalling (p-value $= 3 \times 10^{-8}$). Biological categories over-represented in CACANA1C

CGNets included synaptic vesicle exocytosis (p-value $= 4 \times 10^{-16}$), synaptogenesis (p-value $= 4 \times 10^{-15}$), calcium transport (p-value $= 4 \times 10^{-12}$) and GABAergic neurotransmission (p-value $= 1 \times 10^{-9}$). We repeated the CGNet analysis by removing the candidate gene and the associations remained significant, suggesting that these CGNet associations were not driven by the candidate gene statistics (data not shown).

## 4.4. Discussion

Our results show that genetic variants and gene-expression changes associated with

BD are not randomly distributed across genes, but cluster into discrete groups. These

| GWAS | P-value | Z-score |
|:---:|:---:|:---:|
| CAD | 0.26 | 0.64 |
| CD | 0.14 | 1.08 |
| HT | 0.09 | 1.34 |
| RA | 0.16 | 0.99 |
| T1D | 0.73 | -0.61 |
| T2D | 0.45 | 0.12 |

**Table 4.5. Enrichment of association signals of 6 non-psychiatric disorders in sub-network PPIN-1576.** Table reports p-values for the gene-set analysis of PPIN-1576 in the 6 non-psychiatric GWAS results of the WTCCC (2007). Null hypothesis of no enrichment was tested. I have added z-scores for easier comparison with the results presented on Table 4.2. CAD = coronary artery disease, CD = Crohn's disease, HT = hypertension, RA = rheumatoid arthritis, T1D = type 1 diabetes and T2D = type 2 diabetes.

| Candidate Gene | GWAS | | | | | Gene expression | | | | | |
|---|---|---|---|---|---|---|---|---|---|---|---|
| | WTCCC | Sklar | Bonn | $Z_{GWAS}$ | Corrected P-value | DLPFC | $Z_{GWAS\text{-}DLPFC}$ | Corrected P-value | OFC | $Z_{GWAS\text{-}OFC}$ | Corrected P-value |
| **CANCA1C** | **PPIN-6152** | | | | | | | | | | |
| | 1.58 | 0.43 | 2.35 | 3.49 | 0.01 | 1.17 | 3.29 | 0.02 | 1.98 | 3.86 | 0.002 |
| | **PPIN-4877** | | | | | | | | | | |
| | 0.83 | 1.65 | -0.59 | 2.80 | 0.11 | 2.78 | 3.95 | 0.002 | 2.59 | 3.81 | 0.003 |
| | **PPIN-4800** | | | | | | | | | | |
| | 0.72 | 0.32 | 0.37 | 2.70 | 0.14 | 1.53 | 2.99 | 0.06 | 2.92 | 3.97 | 0.002 |

**PPIN-6152**

AGTPBP1 BSN CACNA1B **CACNA1C** CAST ERC1 ERC2 GIT1 MBIP NCOA2 PCLO **PPFIA1** PPFIA2 **PPFIA3 PPFIA4 PPFIBP1** PPFIBP2 PPP2R5D PTPRC PTPRD PTPRF PTPRS PTPRU **RAB10** RAB26 RAB27A RAB37 **RAB3A** RIMS1 SNAP25 STX1A SYT1 TNNT1 **UNC13B** YWHAG YWHAH

Synaptic vesicle exocytosis (p-value = 5 x10⁻¹⁶), Synaptogenesis (p-value = 5 x10⁻¹⁵), GABAergic neurotransmission (p-value = 8 x10⁻⁶), Role of CDK5 in presynaptic signaling (p-value = 3 x10⁻⁵), Synaptic contact (p-value = 5 x10⁻⁵)

**PPIN-4877**

APBA1 **CARP1 CACNA1A** CACNA1B **CACNA1C** CACNB3 CACNB4 **CALM1 FGF1** GNAO1 **GNB1** GNB2L1 GOLM1 **IRF3** NR1H3 **NR3C1 NRXN1 PDLIM5** PPARG **PPM1A** PRKACA RGS12 **RIMS1 S100A13 SNAP25** SRI STON2 STX1A SV2B **SYNCRIP SYT1** SYT4 TFAP2A ZDHHC17

calcium transport (p-value = 4 x10⁻¹²), GABAergic neurotransmission (p-value = 1 x10⁻⁹), ACM regulation of nerve impulse (p-value = 8 x10⁻⁹), Synaptogenesis (p-value = 2 x10⁻⁸), delta-type opioid receptor in the nervous system (p-value = 5 x10⁻⁷)

**PPIN-4800**

**AKAP1 AKAP10 AKAP11 AKAP12** AKAP13 AKAP14 AKAP2 **AKAP3** AKAP5 AKAP6 AKAP7 **AKAP9** ALB **ARFGEF2** cAMP **CACNA1C CBFA2T3** CFTR EZR FKBP1B GNB2L1 HOMER1 **ITPR1** MAP2 MYCBPAP NBEA PDE4D PDE4DIP PPP1CB PRKACA PRKAR2A PRKX **RAB32** RET RUNX1T1 RYR1 RYR2 RYR3 **SCTR SEMG2 SMAD3 SMAD4** SMAD5 SRI WASF1

PKA signaling (p-value = 1 x10⁻²⁰), Beta-adrenergic receptors signaling via PKA (p-value = 2 x10⁻⁶), ACM regulation of smooth muscle contraction (p-value = 3 x10⁻⁶), Muscle contraction(p-value = 3 x10⁻⁶), Calcium transport (p-value = 6 x10⁻⁵)

**Table 4.6. Candidate-Gene-Networks of CACNA1C.** Table reports z-scores for GWAS and gene-expression studies and their meta-analysis results. Significance p-values of the meta-analysis results were corrected by the 45 networks tested using Sidak's correction. For each sub-network we present all its genes and in bold those that had a p-value < 0.05 in at least one GWAS. Significantly over-represented biological processes among each sub-network's genes are also reported (all reported had an FDR < 0.01). $Z_{GWAS}$, $Z_{GWAS\text{-}DLPFC}$ and $Z_{GWAS\text{-}OFC}$ are the meta-analysis z-scores by pooling GWAS and GWAS with each brain gene-expression results separately, respectively. DLPFC = dorolsateral prefrontal cortex and OFC = orbitofrontal cortex tissue

| Candidate Gene | GWAS | | | | | Gene expression | | | | | |
|---|---|---|---|---|---|---|---|---|---|---|---|
| | WTCCC | Sklar | Bonn | $Z_{GWAS}$ | Corrected P-value | DLPFC | $Z_{GWAS-DLPFC}$ | Corrected P-value | OFC | $Z_{GWAS-OFC}$ | Corrected P-value |
| **ANK3** | | | | | | | | | | | |
| PPIN-3052 | 1.15 | 2.08 | -0.44 | 2.07 | 0.58 | 2.52 | 3.25 | 0.03 | 1.27 | 2.36 | 0.34 |
| PPIN-6701 | 1.13 | 1.81 | -0.66 | 1.96 | 0.68 | 2.42 | 3.10 | 0.04 | 1.34 | 2.34 | 0.35 |

**PPIN-3052 genes:** ANK3 APAF1 BTK C14orf1 **C1orf103 CALM1** CASP8 CASP8AP2 **CD47 CRMP1 DAXX** EEF1A1 **EGFR** EZR FADD **FAF1 FAIM2** **FAS FASLG FBF1 FEM1B** FYN **HIPK3** ISG15 LCK **MET** **PDCD6** PRKCA **PTPN13** PTPN6 RAPIA RIPK1 **SUMO1** **TNFSF13** TRADD UBA7 UBE2I USP18

**PPIN-3052 processes:** death domain receptors & caspases in apoptosis (p-value = 1 x10^-14), FAS signaling cascades (p-value = 7 x10^-12), cytoplasmic/mitochondrial transport of proapoptotic proteins Bid, Bmf and Bim (p-value = 4 x10^-11), role of IAP-proteins in apoptosis (p-value = 3 x10^-8), apoptosis stimulation by external signals (p-value = 6 x10^-8)

**PPIN-6701 genes:** ANK3 APAF1 BTK C14orf1 **C1orf103 CALM1** CASP8 CASP8AP2 **CD47 CRMP1 DAXX** EEF1A1 **EGFR** EZR FADD **FAF1 FAIM2** **FAS FASLG FBF1 FEM1B** FYN **HIPK3** LCK MET **PDCD6** **SUMO1 TNFSF13** PRKCA PTPN13 PTPN6 RAPIA RIPK1 TRADD UBA7 UBE2I

**PPIN-6701 processes:** death domain receptors & caspases in apoptosis (p-value = 5 x10^-12), FAS signaling cascades (p-value = 2 x10^-11), cytoplasmic/mitochondrial transport of proapoptotic proteins Bid, Bmf and Bim (p-value = 2 x10^-11), role of IAP-proteins in apoptosis (p-value = 2 x10-8), apoptosis stimulation by external signals (p-value = 8 x10^-7)

**Table 4.7: Candidate-Gene-Networks of ANK3.** Table reports z-scores for GWAS and gene-expression studies and their meta-analysis results. Significance p-values of the meta-analysis results were corrected by the 45 networks tested using Sidak's correction. For each sub-network we present all its genes and in bold those that had a p-value < 0.05 in at least one GWAS. Significantly over-represented biological processes among each sub-network's genes are also reported (all reported had an FDR < 0.01). $Z_{GWAS}$, $Z_{GWAS-DLPFC}$ and $Z_{GWAS-OFC}$ are the meta-analysis z-scores by pooling GWAS and GWAS with each brain gene-expression results separately, respectively. DLPFC = dorolateral prefrontal cortex and OFC = orbitofrontal cortex tissue

can be identified with information from disease association signals and evidence of direct or indirect interaction in the context of a protein-protein interaction network (Table 4.2). By combining GWAS and gene-expression data we provided significant convergent evidence for PPIN-1576 (Figure 4.2). PPIN-1576 is a PPIN sub-network characterised by biological processes related to GnRH signalling, transmission of nerve impulse and hemopoiesis (Table 4.4). Interestingly, post-hoc analyses showed that PPIN-1576 overlaps significantly with the human post-synaptic density (hPSD) recently described by a large proteomic study: 22% of overlap compared with 12% expected by chance (p-value = 6 x10$^{-9}$). The hPSD was shown to be enriched with genes affected by mutations causing neurological, central nervous system (CNS) and cognitive phenotypes, such as mental retardation or Alzheimer's disease (Bayes et al., 2011). Therefore, not only rare mutations with large effect-size but also common variants and gene expression changes seem to affect the hPSD as a mediator of disease risk. Furthermore, data presented here and by others highlight the potential to shed light on the molecular basis of complex diseases by combining statistical signals normally buried under strict GWAS significance thresholds with knowledge of protein function, e.g., protein-protein interactions (for examples see refs (Baranzini et al., 2009; Wang et al., 2009; Ruano et al., 2010), and gene-expression changes (Hsu et al., 2010; Zhong et al., 2010a; Zhong et al., 2010b).

This strategy also strengthened previously identified GWAS significant associations with BD. Candidate-Gene-Network analyses of GWAS and it integration with gene-expression results lent additional support to CACNA1C and ANK3, the most robust results from BD GWAS to date, and suggested that their network environment is associated with apoptosis signalling pathways and cellular processes regulated by calcium, such as vesicle exocytosis (Tables 4.6 and 4.7).

Our gene-based meta-analysis highlighted regions previously known to harbour BD susceptibility variants and pointed to a possible new locus on chromosome 12q13.12. This region has high LD spanning over 200 kb and several plausible BD candidate genes. For example, DHH is involved in the WNT signalling pathway which is a candidate pathway for BD and target of BD drug treatments, such as lithium (Gould and Manji, 2002). Unfortunately, not much biological characterisation exists for the second most associated gene, LMBR1L, except that it has been shown to be a lipocain receptor (Wojnar et al., 2001) and is expressed in many tissues, including the CNS (Wu et al., 2009).

The best GWAS gene-based association within the PPIN-1576 sub-network was with the corticotropin releasing hormone receptor 1 (CRHR1, Bonn GWAS gene p-values = $1 \times 10^{-5}$), a gene with key mechanistic roles in the hypothalamic-pituitary-adrenal (HPA) axis, which has been associated with stress response and mood disorders (Gillespie et al., 2009). Other noteworthy genetic associations were with MAPT, a gene associated with neurodegenerative diseases (MIM +157140), and several membrane transporters previously associated with neuropsychiatric disorders such as KCNQ2 (Hahn and Neubauer, 2009), GRM3 (Harrison et al., 2008) and GRM7 (Walsh et al., 2008; Saus et al., 2010).

Several of the biological functions over-represented in PPIN-1576 are in line with those reported in previous GSA analyses of BD GWAS (Holmans et al., 2009; O'Dushlaine et al., 2011). For example, Holmans et al. (2009) found significant association with GO0005179 (p-value = <0.0001 in Table 4 of Holmans et al., 2009), which is a parent in the Gene Ontology classification of gonadotropin hormone-releasing activity (GO0005183), where we also found association (Table 4.4). To the best of our knowledge, the hemopoiesis and erythropoietin pathway has not been previously associated with BD. There is evidence that Lithium, the most widely used drug for BD, can modulate several aspects of hemopoiesis (Young, 1980; Focosi et al., 2009) and our findings point to changes in hemopoiesis as part of the genetic aetiology of BD. Additional studies will be needed to clarify the relationship between BD and hematopoiesis.

In line with previous reports (Baranzini et al., 2009; Wang et al., 2009; O'Dushlaine et al., 2011), we found that system-level genetic associations in complex traits are heterogeneous (Figure 4.3 and Additional Table 4.1). However, this heterogeneity is approachable and GWAS can be mined for biological processes underlying complex traits, as recently reviewed in Wang et al. (2010). Although overcoming genetic heterogeneity was a major motivation of our study, we do acknowledge the merit of increasing the sample size to improve statistical power; unmistakably this has been a successful approach for some complex traits, e.g., (Teslovich et al., 2010). However, in some situations increasing sample size is not an option, for example in the case of rare diseases. Many non-replicated signals are likely to be false positives or represent true disease-susceptibility alleles whose effect is not detected in all studies due to low statistical power. It is also possible that the effect of these alleles is not expressed due to differences in haplotype background across populations, exposure to environmental factors or ascertainment differences across studies, among others possibilities. Despite this, by clustering these weak signals with prior knowledge of biological pathways and networks, it is possible to identify molecular systems underlying complex traits.

A major limitation of our approach, and those based on predefined gene-sets, e.g., those of Wang et al. (2007), Baranzini et al. (2009), Holmans et al. (2009) and O'Dushlaine et al. (2011), was the relatively low or biased coverage of the protein interactome, which only included ~8000 genes of the ~21000 genes covered by the GWAS analysed. However, this will improve as more experiments are annotated in the public databases. We think the poor replication of the GWAS findings on the OFC gene-expression compared with DLPFC can be partially explained due to true differences in gene-expression between these two brain regions (Pearson's correlation between gene-expression fold changes of both regions = 0.35, 95% confidence interval = 0.33 - 0.37, N = 7,157) and differing degrees of involvement or sensitivity to BD pathology. Perhaps primarily, the OFC's lower power was due to the smaller sample size (1/3 of the DLPFC study size).

In summary, we found a significant and replicable association with a network involved in regulation of gonadotropin hormone-releasing activity, transmission of nerve impulse and hemopoiesis. We also found that CGNet analyses can add support to genome-wide significant associations. We suggest that these approaches are highly complementary to large meta-analytical studies based on single SNP analyses. Our results warrant replication in additional genetic and gene-expression samples. Functional studies will be needed to understand the pathological effects of interactions and the interplay between rare and common variants within the PPIN-1576 or hPSD network.

## 4.5. Additional Tables

| Gene Symbol | WTCCC | Sklar | Bonn | DLPFC | Gene Symbol | WTCCC | Sklar | Bonn | DLPFC |
|---|---|---|---|---|---|---|---|---|---|
| CRHR1 | 0.819 | 0.295 | 1.5E-05 | 0.048 | MYH2 | 0.595 | 0.333 | 0.856 | 0.747 |
| MAPT | 0.392 | 0.259 | 1.4E-04 | 0.228 | GLP2R | 0.665 | 0.378 | 0.392 | 0.712 |
| GAD1 | 0.306 | 0.439 | 0.239 | 0.001 | YWHAE | 0.594 | 0.228 | 0.076 | 0.154 |
| RGS4 | 0.329 | 0.296 | 0.391 | 0.001 | TP73 | 0.081 | 0.807 | 0.869 | 0.161 |
| SYNCRIP | 0.710 | 0.030 | 0.932 | 0.001 | GRK1 | 0.276 | 0.082 | 0.882 | 0.464 |
| PIK3CB | 0.614 | 0.849 | 0.276 | 0.001 | MAPK3 | 0.311 | NA | 0.501 | 0.082 |
| NRXN1 | 0.838 | 0.510 | 0.311 | 0.001 | LTF | 0.395 | 0.708 | 0.574 | 0.083 |
| MYO7A | 0.363 | 0.038 | 0.133 | 0.002 | ESRRG | 0.684 | 0.162 | 0.483 | 0.083 |

| Gene Symbol | WTCCC | Sklar | Bonn | DLPFC | Gene Symbol | WTCCC | Sklar | Bonn | DLPFC |
|---|---|---|---|---|---|---|---|---|---|
| SCTR | 0.502 | 0.361 | 0.002 | 0.260 | IQGAP2 | 0.439 | 0.541 | 0.084 | 0.611 |
| VDAC1 | 0.146 | 0.800 | 0.975 | 0.002 | TRPV4 | 0.408 | 0.350 | 0.086 | 0.346 |
| SLC25A4 | 0.202 | 0.274 | 0.707 | 0.002 | CCND1 | 0.214 | 0.582 | 0.351 | 0.086 |
| NR3C1 | 0.639 | 0.039 | 0.455 | 0.003 | CCNE1 | 0.711 | 0.537 | 0.086 | 0.538 |
| KCNQ2 | 0.257 | 0.318 | 0.003 | 0.176 | PLCD1 | 0.645 | 0.120 | 0.596 | 0.091 |
| GLP1R | 0.567 | 0.926 | 0.473 | 0.003 | EDF1 | 0.540 | 0.096 | 0.980 | 0.523 |
| PDLIM5 | 0.429 | 0.486 | 0.814 | 0.065 | PPEF1 | 0.420 | 0.096 | NA | 0.122 |
| STRN4 | 0.375 | 0.386 | 0.646 | 0.576 | GRM5 | 0.516 | 0.490 | 0.777 | 0.097 |
| PCP4 | 0.518 | 0.287 | 0.894 | 0.004 | TRPC3 | 0.665 | 0.683 | 0.722 | 0.120 |
| HSP90AA1 | 0.635 | 0.361 | 0.744 | 0.004 | MYLK | 0.724 | 0.100 | 0.668 | 0.610 |
| RALB | 0.725 | 0.940 | 0.148 | 0.005 | ATP2B1 | 0.550 | 0.356 | 0.102 | 0.126 |
| SYT1 | 0.441 | 0.894 | 0.123 | 0.005 | KCNQ3 | 0.235 | 0.105 | 0.620 | 0.157 |
| GAD2 | 0.646 | 0.749 | 0.043 | 0.005 | TIAM1 | 0.777 | 0.456 | 0.863 | 0.109 |
| KCNN4 | 0.356 | 0.143 | 0.005 | 0.545 | RIPK1 | 0.536 | 0.245 | 0.132 | 0.110 |
| RAF1 | 0.632 | 0.512 | 0.475 | 0.888 | KCNN2 | 0.543 | 0.860 | 0.601 | 0.113 |
| GRK4 | 0.493 | 0.788 | 0.689 | 0.006 | RELA | 0.554 | 0.645 | 0.115 | 0.305 |
| FAS | 0.006 | 0.159 | 0.135 | 0.189 | ACTC1 | 0.666 | 0.375 | 0.953 | 0.618 |
| TCF3 | 0.139 | NA | 0.804 | 0.006 | RNF11 | 0.681 | 0.494 | 0.949 | 0.117 |
| PRKD1 | 0.509 | 0.821 | 0.612 | 0.007 | PPEF2 | 0.529 | 0.296 | 0.761 | 0.128 |
| DLG3 | 0.170 | 0.026 | NA | 0.008 | VIPR1 | 0.683 | 0.349 | 0.125 | 0.820 |
| SH3GL2 | 0.316 | 0.242 | 0.941 | 0.008 | CALCR | 0.154 | 0.536 | 0.127 | 0.367 |
| SNTA1 | 0.648 | 0.044 | 0.978 | 0.009 | OPRD1 | 0.379 | 0.135 | 0.603 | 0.222 |
| TRPV6 | 0.673 | 0.554 | 0.170 | 0.009 | NRGN | 0.270 | 0.262 | 0.817 | 0.415 |
| TRPV1 | 0.015 | 0.124 | 0.051 | 0.010 | IQGAP1 | 0.833 | 0.731 | 0.812 | 0.139 |
| SNCA | 0.902 | 0.290 | 0.281 | 0.011 | POLR2A | 0.140 | 0.667 | 0.854 | 0.876 |
| RPS6KB2 | 0.191 | 0.746 | 0.498 | 0.012 | NEDD9 | 0.727 | 0.788 | 0.269 | 0.146 |
| CLTB | 0.672 | 0.372 | 0.610 | 0.131 | OBSCN | 0.253 | 0.604 | 0.151 | NA |
| RIT2 | 0.833 | 0.176 | 0.413 | 0.013 | ESR2 | 0.639 | 0.219 | 0.372 | 0.152 |
| GRM3 | 0.144 | 0.013 | 0.344 | 0.950 | ATN1 | 0.327 | 0.907 | 0.773 | 0.152 |
| RALA | 0.549 | 0.736 | 0.014 | 0.718 | MYO9B | 0.317 | 0.185 | 0.164 | 0.161 |
| MAPK1 | 0.702 | 0.491 | 0.015 | 0.037 | AKAP5 | 0.194 | 0.735 | 0.547 | 0.162 |
| EGFR | 0.262 | 0.015 | 0.378 | 0.072 | FYN | 0.199 | 0.194 | 0.762 | 0.169 |
| ASCL2 | 0.806 | 0.015 | 0.058 | 0.420 | ESR1 | 0.501 | 0.168 | 0.568 | 0.188 |
| AKAP9 | 0.543 | 0.132 | 0.016 | 0.024 | SRC | 0.484 | 0.172 | 0.337 | 0.426 |
| CAMKK2 | 0.775 | 0.670 | 0.709 | 0.018 | KCNQ5 | 0.650 | 0.175 | 0.310 | NA |
| CAMK1 | 0.019 | 0.250 | 0.861 | 0.416 | RGS2 | 0.340 | 0.554 | 0.782 | 0.176 |
| SNAP25 | 0.555 | 0.075 | 0.334 | 0.020 | GNB2L1 | 0.669 | 0.182 | 0.536 | 0.225 |
| MYOG | 0.215 | 0.158 | 0.020 | 0.752 | GNA13 | 0.816 | 0.378 | 0.185 | 0.311 |
| KRAS | 0.725 | 0.551 | 0.357 | 0.020 | BTK | 0.509 | 0.262 | NA | 0.188 |
| GRB7 | 0.421 | 0.055 | 0.021 | 0.723 | HTR2C | 0.786 | 0.821 | NA | 0.190 |
| BRAF | 0.586 | 0.022 | 0.474 | 0.892 | JUN | 0.354 | 0.249 | 0.750 | 0.192 |
| CALD1 | 0.633 | 0.525 | 0.694 | 0.022 | KRT18 | 0.767 | 0.543 | 0.328 | 0.197 |
| SLC4A3 | 0.732 | 0.188 | 0.131 | 0.957 | NEUROD1 | 0.483 | 0.199 | 0.572 | 0.962 |
| SGK3 | 0.621 | 0.296 | 0.715 | 0.590 | ADCYAP1R1 | 0.749 | 0.200 | 0.820 | 0.694 |
| CPSF6 | 0.557 | 0.736 | 0.174 | 0.026 | SYT9 | 0.200 | 0.876 | 0.535 | NA |
| GRM4 | 0.129 | 0.226 | 0.742 | 0.026 | TRPV2 | 0.413 | 0.918 | 0.592 | 0.207 |

| Gene Symbol | WTCCC | Sklar | Bonn | DLPFC | Gene Symbol | WTCCC | Sklar | Bonn | DLPFC |
|---|---|---|---|---|---|---|---|---|---|
| TRPC4 | 0.128 | 0.026 | 0.975 | 0.135 | DDX5 | 0.881 | 0.790 | 0.522 | 0.225 |
| ITPKA | 0.504 | 0.026 | 0.662 | 0.542 | HMMR | 0.625 | 0.216 | 0.362 | 0.480 |
| TCF4 | 0.633 | 0.675 | 0.525 | 0.029 | PPP1R14A | 0.320 | 0.216 | 0.385 | NA |
| CALM1 | 0.793 | 0.474 | 0.581 | 0.030 | ACTA1 | 0.297 | 0.364 | 0.218 | 0.288 |
| GAP43 | 0.415 | 0.610 | 0.506 | 0.031 | ADCY8 | 0.219 | 0.434 | 0.429 | 0.357 |
| CNN1 | 0.032 | 0.100 | 0.394 | 0.038 | MYF6 | 0.224 | 0.280 | 0.240 | 0.361 |
| PHKG1 | 0.126 | 0.034 | 0.444 | 0.966 | AFAP1 | 0.760 | 0.700 | 0.269 | 0.224 |
| REL | 0.775 | 0.530 | 0.162 | 0.407 | ITGB1 | 0.818 | 0.230 | 0.401 | 0.299 |
| MIP | 0.229 | 0.143 | 0.228 | 0.036 | RAG1AP1 | 0.879 | 0.701 | 0.836 | 0.239 |
| NUMB | 0.709 | 0.447 | 0.490 | 0.068 | TTN | 0.571 | 0.239 | 0.451 | 0.511 |
| MYO10 | 0.468 | 0.183 | 0.197 | 0.038 | DTX1 | 0.253 | 0.328 | 0.399 | NA |
| COPB1 | 0.310 | 0.468 | 0.041 | 0.068 | PDE1A | 0.628 | 0.257 | 0.583 | 0.321 |
| CSNK1A1 | 0.596 | 0.849 | 0.765 | 0.042 | CBLC | 0.664 | 0.323 | 0.393 | 0.260 |
| LYST | 0.291 | 0.042 | 0.736 | 0.142 | RAB3B | 0.263 | 0.922 | 0.723 | 0.575 |
| ITPKB | 0.056 | 0.446 | 0.076 | 0.042 | PTH2R | 0.555 | 0.464 | 0.897 | 0.358 |
| NOTCH1 | 0.137 | 0.621 | 0.206 | 0.042 | STRN3 | 0.523 | 0.498 | 0.718 | 0.267 |
| CFTR | 0.112 | 0.966 | 0.412 | 0.043 | EWSR1 | 0.413 | 0.619 | 0.626 | 0.275 |
| CHAT | 0.531 | 0.255 | 0.641 | 0.806 | GRIN2D | 0.607 | 0.833 | 0.276 | 0.363 |
| GRIN1 | 0.647 | NA | 0.889 | 0.046 | CAMK2G | 0.762 | 0.434 | 0.718 | 0.292 |
| MYF5 | 0.248 | 0.280 | 0.046 | 0.109 | FER | 0.308 | 0.364 | 0.621 | 0.483 |
| CABIN1 | 0.720 | 0.457 | 0.427 | 0.047 | PRKCE | 0.436 | 0.723 | 0.627 | 0.321 |
| PIK3C3 | 0.446 | 0.822 | 0.364 | 0.049 | GJB1 | 0.663 | 0.642 | NA | 0.326 |
| CSNK2A1 | 0.667 | 0.141 | 0.255 | 0.050 | PDPK1 | NA | NA | NA | 0.337 |
| OPRM1 | 0.539 | 0.051 | 0.152 | 0.249 | RYR1 | 0.372 | 0.685 | 0.344 | 0.829 |
| GJA1 | 0.069 | 0.361 | 0.795 | 0.052 | TUBB | NA | NA | NA | 0.354 |
| STRN | 0.108 | 0.885 | 0.704 | 0.406 | ADD1 | 0.804 | 0.827 | 0.428 | 0.389 |
| AR | 0.058 | 0.793 | NA | 0.309 | PTPRA | 0.402 | 0.941 | 0.517 | 0.436 |
| HK2 | 0.059 | 0.584 | 0.338 | 0.861 | CSNK2A2 | 0.855 | 0.744 | 0.688 | 0.405 |
| ITCH | 0.713 | 0.109 | 0.494 | 0.622 | CBL | 0.484 | 0.566 | 0.760 | 0.981 |
| INSR | 0.237 | 0.862 | 0.060 | 0.473 | ADAP1 | 0.935 | 0.428 | 0.671 | NA |
| RRAD | 0.905 | 0.098 | 0.198 | 0.060 | MYOD1 | 0.591 | 0.693 | 0.450 | 0.737 |
| ADD2 | 0.492 | 0.481 | 0.903 | 0.066 | TNNI2 | 0.452 | 0.637 | 0.748 | 0.460 |
| RGS10 | 0.470 | 0.764 | 0.373 | 0.067 | CAMKK1 | 0.480 | 0.978 | 0.642 | NA |
| GRM7 | 0.482 | 0.573 | 0.515 | 0.068 | NDFIP2 | 0.865 | 0.519 | 0.756 | NA |
| KRT1 | 0.483 | 0.070 | 0.291 | 1.000 | CNGA2 | 0.675 | 0.528 | NA | NA |
| COPB2 | 0.791 | 0.071 | 0.845 | 0.290 | OCLN | NA | NA | NA | 0.945 |

**Additional Table 4.1. Gene p-values of the PPIN-1576 sub-network genes.**
For GWAS data sets we transformed the gene z-score into p-values and for the
DLPFC we report the p-values from the differential expression analysis.

| Gene Symbol | Sub-Network | Bipolar Disorder | | | | | | | | | | |
|---|---|---|---|---|---|---|---|---|---|---|---|---|
| | | GWAS | | | | | Gene expression | | | | | |
| | | WTC CC | Sklar | Bonn | $Z_{GWAS}$ | Corrected P-value | DLPFC | $Z_{GWAS-DLPFC}$ | Corrected P-value | OFC | $Z_{GWAS-OFC}$ | Corrected P-value |
| ANK3 | PPIN-3052 | 1.15 | 2.08 | -0.44 | 2.07 | 0.58 | 2.52 | 3.25 | 0.026 | 1.27 | 2.36 | 0.337 |
| | PPIN-6701 | 1.13 | 1.81 | -0.66 | 1.96 | 0.68 | 2.42 | 3.10 | 0.042 | 1.34 | 2.34 | 0.354 |
| | PPIN-3468 | 0.37 | 0.22 | -0.56 | 1.39 | 0.98 | 1.66 | 2.16 | 0.506 | 1.86 | 2.30 | 0.384 |
| | PPIN-3067 | 0.74 | 0.57 | 0.79 | 1.45 | 0.97 | -0.58 | 0.62 | 1.000 | 0.66 | 1.49 | 0.957 |
| | PPIN-6058 | -0.33 | 1.07 | -1.03 | 0.69 | 1.00 | 1.76 | 1.73 | 0.852 | 1.20 | 1.33 | 0.987 |
| | PPIN-6899 | 1.40 | -1.10 | 0.02 | 0.61 | 1.00 | -0.87 | -0.19 | 1.000 | 0.48 | 0.77 | 1.000 |
| | PPIN-4542 | -0.25 | -0.18 | -0.28 | 0.04 | 1.00 | 1.50 | 1.09 | 0.999 | 0.57 | 0.44 | 1.000 |
| | PPIN-4533 | 1.18 | 0.26 | -0.18 | 0.71 | 1.00 | -1.79 | -0.76 | 1.000 | -0.10 | 0.43 | 1.000 |
| | PPIN-3185 | 0.45 | 0.57 | -0.10 | 0.47 | 1.00 | -1.42 | -0.68 | 1.000 | 0.09 | 0.40 | 1.000 |
| | PPIN-6244 | 0.01 | 0.79 | 0.53 | 0.39 | 1.00 | -0.72 | -0.23 | 1.000 | -0.18 | 0.15 | 1.000 |
| | PPIN-1275 | 0.00 | 0.64 | -0.11 | 0.17 | 1.00 | 0.35 | 0.37 | 1.000 | -0.09 | 0.05 | 1.000 |
| | PPIN-1312 | -0.12 | -1.27 | -0.85 | -1.43 | 1.00 | -0.08 | -1.06 | 1.000 | -0.74 | -1.53 | 1.000 |
| | PPIN-2875 | -1.10 | -1.62 | -1.56 | -2.37 | 1.00 | -0.17 | -1.79 | 1.000 | -0.70 | -2.16 | 1.000 |
| | PPIN-3499 | -1.07 | -1.14 | -1.07 | -1.35 | 1.00 | 0.07 | -0.90 | 1.000 | 0.03 | -0.93 | 1.000 |
| | PPIN-3802 | -0.01 | -1.43 | -0.17 | -0.86 | 1.00 | -0.42 | -0.90 | 1.000 | -0.20 | -0.75 | 1.000 |
| CACNA1C | PPIN-2387 | 0.30 | 0.09 | -0.18 | -0.09 | 1.00 | 0.64 | 0.39 | 1.000 | -0.21 | -0.21 | 1.000 |
| | PPIN-2906 | 1.43 | -1.21 | 0.18 | -0.41 | 1.00 | 0.34 | -0.05 | 1.000 | -0.78 | -0.84 | 1.000 |
| | PPIN-374 | 0.27 | -0.59 | 1.78 | 0.95 | 1.00 | 0.55 | 1.05 | 0.999 | 0.54 | 1.05 | 0.999 |
| | PPIN-4800 | 0.72 | 0.32 | 0.37 | 2.70 | 0.14 | 1.53 | 2.99 | 0.060 | 2.92 | 3.97 | 0.002 |
| | PPIN-4877 | 0.83 | 1.65 | -0.59 | 2.80 | 0.11 | 2.78 | 3.95 | 0.002 | 2.59 | 3.81 | 0.003 |
| | PPIN-5245 | -0.66 | 0.24 | 1.13 | 1.43 | 0.97 | 1.14 | 1.82 | 0.791 | 1.82 | 2.30 | 0.384 |
| | PPIN-5688 | 1.32 | -0.04 | 0.39 | 1.88 | 0.74 | 1.32 | 2.27 | 0.412 | 1.52 | 2.41 | 0.304 |
| | PPIN-575 | 1.15 | -1.00 | -0.80 | -1.03 | 1.00 | 2.24 | 0.85 | 1.000 | -1.34 | -1.68 | 1.000 |
| | PPIN-6152 | 1.58 | 0.43 | 2.35 | 3.49 | 0.01 | 1.17 | 3.29 | 0.022 | 1.98 | 3.86 | 0.002 |
| | PPIN-6383 | 1.64 | 0.18 | -0.04 | 1.76 | 0.84 | 1.24 | 2.12 | 0.536 | 1.09 | 2.02 | 0.631 |
| | PPIN-835 | 0.18 | -0.84 | 0.44 | -0.84 | 1.00 | -0.20 | -0.74 | 1.000 | -1.03 | -1.33 | 1.000 |
| DCTN5 | PPIN-157 | 0.57 | 0.77 | 1.68 | 2.26 | 0.41 | -0.32 | 1.37 | 0.982 | 1.25 | 2.48 | 0.255 |
| | PPIN-6020 | 0.83 | 0.75 | 1.59 | 1.93 | 0.70 | -0.75 | 0.83 | 1.000 | 0.77 | 1.91 | 0.720 |

| Gene Symbol | Sub-Network | Bipolar Disorder | | | | | | | | | | |
|---|---|---|---|---|---|---|---|---|---|---|---|---|
| | | GWAS | | | | | Gene expression | | | | | |
| | | WTCC | Sklar | Bonn | $Z_{GWAS}$ | Corrected P-value | DLPFC | $Z_{GWAS-DLPFC}$ | Corrected P-value | OFC | $Z_{GWAS-OFC}$ | Corrected P-value |
| NCAN | PPIN-1521 | -0.31 | 0.66 | 1.45 | 0.80 | 1.00 | 1.02 | 1.28 | 0.991 | 2.05 | 2.01 | 0.632 |
| | PPIN-1693 | -0.61 | -0.29 | -0.81 | -0.94 | 1.00 | 0.48 | -0.33 | 1.000 | -0.15 | -0.77 | 1.000 |
| | PPIN-2141 | -1.30 | -0.72 | 0.11 | -1.26 | 1.00 | 1.28 | 0.01 | 1.000 | -0.30 | -1.11 | 1.000 |
| | PPIN-2316 | -0.72 | 0.52 | -0.48 | -0.41 | 1.00 | 0.18 | -0.16 | 1.000 | 0.35 | -0.04 | 1.000 |
| | PPIN-2487 | -0.40 | 1.03 | -0.58 | 0.06 | 1.00 | 1.67 | 1.23 | 0.995 | 2.04 | 1.49 | 0.959 |
| | PPIN-2600 | 1.46 | -1.09 | 0.20 | 0.46 | 1.00 | 0.90 | 0.96 | 1.000 | 1.20 | 1.17 | 0.997 |
| | PPIN-3253 | -1.83 | -0.64 | -0.50 | -1.85 | 1.00 | 1.28 | -0.40 | 1.000 | -0.07 | -1.35 | 1.000 |
| | PPIN-3659 | -1.68 | 0.20 | 0.99 | -0.61 | 1.00 | 2.05 | 1.02 | 0.999 | 0.64 | 0.02 | 1.000 |
| | PPIN-4039 | -0.42 | -0.07 | 0.29 | -0.20 | 1.00 | 1.78 | 1.12 | 0.998 | 0.15 | -0.04 | 1.000 |
| | PPIN-7186 | -0.31 | 0.60 | -0.74 | -0.19 | 1.00 | 1.00 | 0.57 | 1.000 | 2.29 | 1.48 | 0.960 |
| PALB2 | PPIN-6658 | 0.33 | 0.17 | -2.20 | -0.29 | 1.00 | -0.82 | -0.78 | 1.000 | 0.28 | -0.00 | 1.000 |
| | PPIN-690 | -0.27 | -0.10 | -1.66 | -0.47 | 1.00 | -0.74 | -0.86 | 1.000 | 0.37 | -0.07 | 1.000 |
| PBRM1 | PPIN-12 | 1.24 | 1.22 | -1.90 | 0.73 | 1.00 | 1.96 | 1.90 | 0.728 | 1.84 | 1.82 | 0.795 |
| | PPIN-3070 | 1.30 | 0.35 | -0.76 | 0.76 | 1.00 | 2.16 | 2.06 | 0.588 | 1.80 | 1.82 | 0.796 |
| | PPIN-3368 | 0.87 | 0.33 | -1.62 | 0.09 | 1.00 | 1.84 | 1.37 | 0.983 | 1.65 | 1.23 | 0.995 |
| | PPIN-3719 | 1.39 | 0.47 | 2.12 | 2.16 | 0.50 | -0.39 | 1.25 | 0.994 | 0.35 | 1.77 | 0.826 |
| | PPIN-973 | 1.88 | 1.12 | -1.61 | 1.24 | 0.99 | 2.04 | 2.32 | 0.368 | 1.00 | 1.59 | 0.926 |

**Additional Table 4.2. Candidate-Gene-Networks of Bipolar Disorder.** Table reports z-scores for GWAS and gene-expression studies and their meta-analysis results. Significance p-values of the meta-analysis results were corrected by the 45 networks tested using Sidak's correction. $Z_{GWAS}$, $Z_{GWAS-DLPFC}$ and $Z_{GWAS-OFC}$ are the meta-analysis z-scores by pooling GWAS and GWAS with each brain gene-expression results separately, respectively. DLPFC = doroslateral prefrontal cortex and OFC = orbitofrontal cortex tissue

**Chapter V**

# Genome-wide association of bipolar disorder sub-phenotypes

# 5.1. Introduction

Bipolar disorder (BD) is a chronic, episodic, and pathological disturbance resulting in extreme moods ranging from mania to severe depression and is usually accompanied by disturbances in cognition and behaviour. Psychotic features such as delusions and hallucinations often occur. There is robust evidence for a genetic contribution in the aetiology of the disorder, with an estimated sibling recurrence risk 7-10 times higher than the general population risk and a heritability of 80-90% (McGuffin et al., 2003; Craddock et al., 2005). Diagnosis of BD is based solely on clinical features because, as yet, there are no validated diagnostic tests such as those available for many physical illnesses. There is some evidence suggesting BD is heterogeneous in clinical presentations, genetic aetiology and course of pharmacological response (Suppes et al., 2000; Alda, 2004). A major goal of molecular psychiatric genetics is to improve the diagnostic classification of BD. With a better understanding of the biological systems underpinning BD, this could be possible and may in turn lead to more efficacious treatments.

Several genomic regions have been implicated through linkage (McQueen et al., 2005), and genome-wide association studies (GWAS) of BD (Ferreira et al., 2008; Breen et al., 2011; Cichon et al., 2011). Increasing evidence suggests an overlap in genetic susceptibility with schizophrenia (Purcell et al., 2009), a psychotic disorder with many similarities to BD, and often seen in extended families of affected individuals with BD (see review by Craddock et al. (2005)). In particular, association findings have been reported with both disorders in DAOA (D-amino acid oxidase activator)(Prata et al., 2009), DISC1 (Disrupted in schizophrenia 1)(Chubb et al., 2008), NRG1 (neuregulin 1)(Goes et al., 2009), DTNBP1 (dystrobrevin binding protein 1)(Breen et al., 2006), CACNA1C (calcium channel, voltage-dependent, L type, alpha 1C subunit) (Ferreira et al., 2008; Green et al., 2010) and ZNF804A (zinc finger protein 804A) (O'Donovan et al., 2008).

Analysis of disease sub-phenotypes can provide increased statistical power to detect disease-susceptibility associations if the subgroup is genetically more homogeneous or the sub-phenotype has a higher heritability (Burmeister et al., 2008). Hamshere et al. (2009a) performed single marker association on BD sub-phenotypes using the WTCCC BD sample and found an association within the gene B3GALT5 on chromosome 21q22.1 with the subtype schizoaffective bipolar disorder. In addition, they showed the schizoaffective bipolar disorder subtype had the strongest genetic signal among the six sub-phenotypes analysed. They suggested that individuals with broadly defined bipolar schizoaffective features either have a particularly strong genetic contribution or, as a group, are genetically more homogeneous than the other sub-phenotypes tested. In addition, Craddock et al. (2010) and Breuer et al. (2010) have reported significant association between the gamma-aminobutyric acid (GABA)-A receptor genes and the schizoaffective bipolar disorder subtype. Overall, current evidence suggests that the use of BD subtypes can help to identify additional loci which may be informative in understanding the aetiology of BD.

As discussed previously, single marker genetic association tests in moderately powered studies are unlikely to distinguish true associations from false, when the odds ratios of susceptibility alleles are low, or disease-susceptibility variants are rare (Schaid, 2004). In the case of BD and its sub-phenotypes, either or both situations are likely to apply. Gene and gene-set association methods can increase power to detect true association by clustering susceptibility variants within genes and biological processes (Baranzini et al., 2009; Wang et al., 2010; O'Dushlaine et al., 2011). We have shown this strategy can reveal additional replicated genes that eluded single marker analyses in the WTCCC samples (see Chapter III) and an integrative biology approach is useful to reveal biological processes associated with BD susceptibility (see Chapter IV). In this study, we explored the use of disease sub-phenotypes in combination with gene and gene-set association methods to identify genes and biological processes associated with specific patient subgroups and their symptoms.

## 5.2. Material and Methods

### 5.2.1.    Samples

We used the BD sample investigated in the WTCCC study. A detailed description of the sample has been provided elsewhere (WTCCC, 2007). All individuals were from the UK and over the age of 16 years. Clinical assessment included a semi-structured interview and review of case notes. BD cases (n=1868) had experienced at least one episode of clinically significant elevated mood according to the following Research Diagnostic Criteria (RDC) (Spitzer et al., 1978): bipolar I disorder (BPI; n=1316), schizoaffective disorder bipolar type (SABP; n=279), bipolar II disorder (BPII; n=171) and manic disorder (n=102). Diagnoses following were also ascertained with the DSM-IV (American Psychiatric Association, 1994) criteria and gave the following sub-classification of bipolar disorder cases: bipolar I disorder (n=1594), schizoaffective disorder bipolar type (n=98), bipolar II disorder (n=134), and bipolar disorder not otherwise specified (n=42). Controls came from two sources: (a) the UK 1958 birth cohort longitudinal epidemiological sample (n=1458), and (b) the UK Blood Donor Service (n=1480). Controls were not screened for absence of psychiatric illness or family history. It has previously been shown that these two control samples can be combined for use as controls in genetic association studies using UK disease samples, including the current BD sample (WTCCC, 2007).

### 5.2.1.1.  bipolar disorder sub-phenotypes

We extracted sub-phenotypes from the available clinical data (OPCRIT (Craddock et al., 1996)), as described previously by Hamshere et al. (2009a) and WTCCC (2007). The sub-phenotypes, analysed here, were defined using the RDC diagnosis as:

i)  Suicidality: the sub-phenotype positive group was suicidal ideation/ behaviour (n=793).

ii)  Postnatal episode: postnatal episode of mood disorder (n=378).

iii)  Lifetime Psychosis: multiple psychotic features (n=1065).

iv)  Rapid Cycling: present during lifetime (n=222).

v)  Age of Onset: age of onset for BD was used as quantitative trait.

vi)  Schizoaffective bipolar sub-type: positive diagnosis (n=279).

A more detailed breakdown of the sub-phenotypes of this samples has been described elsewhere (Hamshere et al., 2009).

### 5.2.2.  Genotypic data and statistical analyses

The WTCCC dataset comprised 469,557 single nucleotide polymorphisms (SNPs) distributed across the genome. Individuals were excluded as described in (WTCCC, 2007). For the current analysis, SNP inclusion criteria were genotype quality over 90%, a minor allele frequency of at least 1% in the total sample, call rate >99% in the total sample, Hardy-Weinberg equilibrium p-value > $10^{-5}$ in cases, and Hardy-Weinberg equilibrium p value > $10^{-3}$ in controls.

Single marker associations using a logistic regression for case-control association tests using the sub-phenotype positive groups and population controls were undertaken in PLINK (Purcell et al., 2007). We set genome-wide significance and suggestive thresholds at a p-value = 5 x $10^{-8}$ and 7.5 x $10^{-7}$ to apply a Bonferroni correction equal to 0.05/(number of tests) and 0.5/(number of test), respectively, with the number of test genome-wide equal to ~694,000 as per (Dudbridge and Gusnanto, 2008).

### 5.2.3.    Imputation

We performed imputation using the software IMPUTE (Marchini et al., 2007) with the 1000 genomes project and HapMap Phase 3 data combined as reference panels. Software and files were downloaded from the IMPUTE website (http://mathgen.stats.ox.ac.uk/impute/impute.html).

### 5.2.4.    Gene and networks analysis

We calculated gene-based association p-values using the FORGE software on approximately 20,000 protein-coding, lincRNA and miRNA genes annotated in Ensembl v59. SNPs were mapped to genes if they were located 20 kb upstream or downstream of the annotated coordinates. We utilised the Z FIX gene-association p-values calculated by the software FORGE and performed gene-set analyses with the gene p-values and gene-sets derived from the protein-protein interaction network (PPIN), as described in Chapter II. We interpreted the biology of the significant PPIN sub-networks using MetaCore (GeneGo, Inc. www.genego.com). Please see Section 3.2.3 for details on the analyses performed with the tools and data sets provided by GeneGO. Throughout the thesis we have used a FDR threshold of 0.1 to select interesting sub-networks. However, in order to account for the multiple testing performed by analysing six phenotypes we have set our FDR threshold to 0.01 (approximately 0.1/6 or 0.16) and consider an FDR < 0.1 only as suggestive evidence.

#### 5.2.4.1.    Analysis of GABA(A) receptor genes in SABP

We also applied the gene-set association method implemented in PLINK that was used by two previous studies (Breuer et al., 2010; Craddock et al., 2010) to analyse the GWAS of the SABP sub-phenotype. Details of the method can be found on the PLINK website. Briefly, the association analysis was performed using the PLINK SNP-set association method, which works with the following procedure (obtained from PLINK's website):

i)   For each SNP in the set determine which other SNPs are in LD, above a certain threshold of $r^2$;

ii)  Perform standard single SNP analysis, in this case a case-control association a chi-square test or a logistic regression with covariates;

iii) For each SNP-set, select up to N "independent" SNPs (as defined in step 1) with p-values below a p-value threshold P. The best SNP is selected first; subsequent SNPs are selected in order of decreasing statistical significance, after removing SNPs in LD with previously selected SNPs;

iv)  From these subset of SNPs, the statistic for each SNP-set is calculated as the mean chi-square with one degree of freedom statistics of these single SNP;

v)   Permute the dataset a large number of times, keeping LD between SNPs constant (i.e. permute phenotype labels);

vi)  For each permuted dataset, repeat steps 2 to 4 above;

vii) Empirical p-value for the SNP-set is the number of times the permuted set-statistic exceeds the original one for that SNP-set.

## 5.3. Results

### 5.3.1.  Genome-wide association study of BD sub-phenotypes

After quality filtering, 370,322 SNPs were tested for association in the WTCCCC BD sample. No SNP reached genome-wide significance at p-values $< 5 \times 10^{-8}$ but six reached our suggestive significance level of p-value $< 7.5 \times 10^{-7}$ (Table 5.1). Three

| Sub-phenotype | Chromosome | Position | SNP | p-value | Genes within 20 kb |
|---|---|---|---|---|---|
| Postnatal Episode | 5 | 94263154 | rs26998 | $7 \times 10^{-7}$ | MCTP1 |
| SABP | 3 | 49865617 | rs2352974 | $4 \times 10^{-7}$ | TRAIP, CAMKV |
| | 15 | 87413740 | rs16942644 | $7 \times 10^{-7}$ | -- |
| | 21 | 39954674 | rs734413 | $3 \times 10^{-7}$ | B3GALT5 |
| | 21 | 39959593 | rs4818065 | $2 \times 10^{-7}$ | B3GALT5 |
| | 21 | 39959669 | rs4818066 | $4 \times 10^{-7}$ | B3GALT5 |

**Table 5.1. GWAS results for Bipolar Disorder sub-phenotypes.** Reported SNPs reached suggestive evidence at a p-value $< 7.5 \times 10^{-7}$. Gene annotation based on Ensembl database.

SNPs associated with SABP were located at chromosome 21q22.1 (rs734413, P = 2.7 x $10^{-7}$, OR = 1.7 (95% CI= 1.4 - 2.0); rs4818065, P = 2.4 x 10-7, OR = 1.7 (1.4 - 2.0); rs4818066, P = 4.0 x $10^{-7}$, OR = 1.7 (1.4 - 2.0)) are in a region of strong LD overlapping with the genes B3GALT5 and ENSG00000225330 (Figure 5.1). We noted a second apparently independent association signal approximately 100 kb away from the main signal and overlapping both a longer isoform of B3GALT5 and the gene C21orf88 (ENSG00000184809). We re-analysed the region covering both association signals with the HyperLasso regression method as it permits the identification of independent associations signals in a region (Hoggart et al., 2008) and found supportive evidence for the presence of a second signal at rs11088486 ($r^2$ with rs734413 = 0.02), under both a dominant (P = 7.1 x $10^{-4}$, OR = 0.6 (0.5 - 0.8)) and additive (P = 0.001, OR = 0.7 (0.6 - 0.9)) genetic model. We found no evidence for interaction between these variants (P = 0.8). Imputation of typed and untyped markers confirmed these results but did show evidence of new associations surpassing them. Although, this may seem a negative results it also provides some support for the original results because imputation can improve genotype quality at poorly genotyped SNPs [9]. Imputation showed association with an untyped non-synonymous variant (rs3746887, P = 3.0 x 10^-7, OR = 0.60 (0.49 - 0.74), imputation posterior probability = 0.996) leading to a methionine/threonine change within the protein galactosyltransferase (PF01762) domain. This amino-acidic change is not predicted to have a large effect in the protein's function (Ramensky et al., 2002).

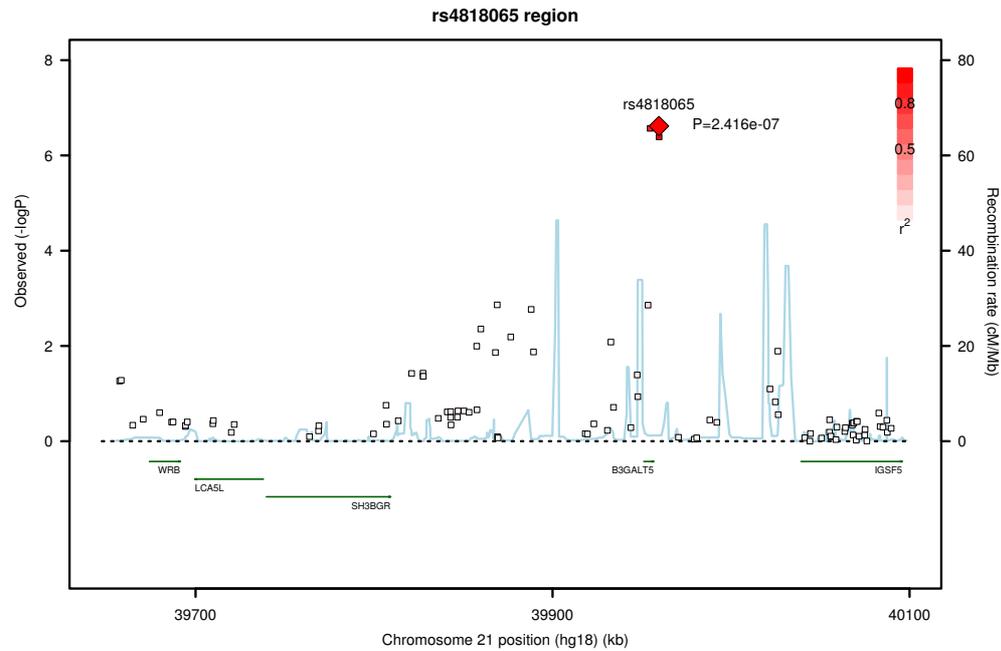

**Figure 5.1. Association at 21q22.1.** -log10 of the association p-values under an additive model are shown in the y-axis. Blue line represents the recombination rate in cM/Mb and the $r^2$ between SNPs and rs4818065 is shown as red scale. Genes are depicted as green lines and chromosomal position indicated in the x-axis.

A fourth SNP associated with SABP was located at chromosome 3p21.31 (rs2352974, P = 4.0 x 10^-7, OR = 1.6 (1.3 - 1.9)) within the first intron of TRAIP gene (Figure 5.2). Evidence of association in this region is supported by two additional SNPs located approximately 12 kb away and in strong LD (rs2271961, P = 1.5 x 10^-6, $r^2$ =

---

[9] Genotype imputation makes use of haplotype information. This can improve genotype calls at typed SNPs when their genotypes do not agree with those expected based on the haplotypic information. This strategy is implemented in many imputation algorithms (see (Marchini and Howie, 2010) for a review) and is also the basis for the BEAGLECALL method design to improve genotype calls (Browning and Yu, 2009).

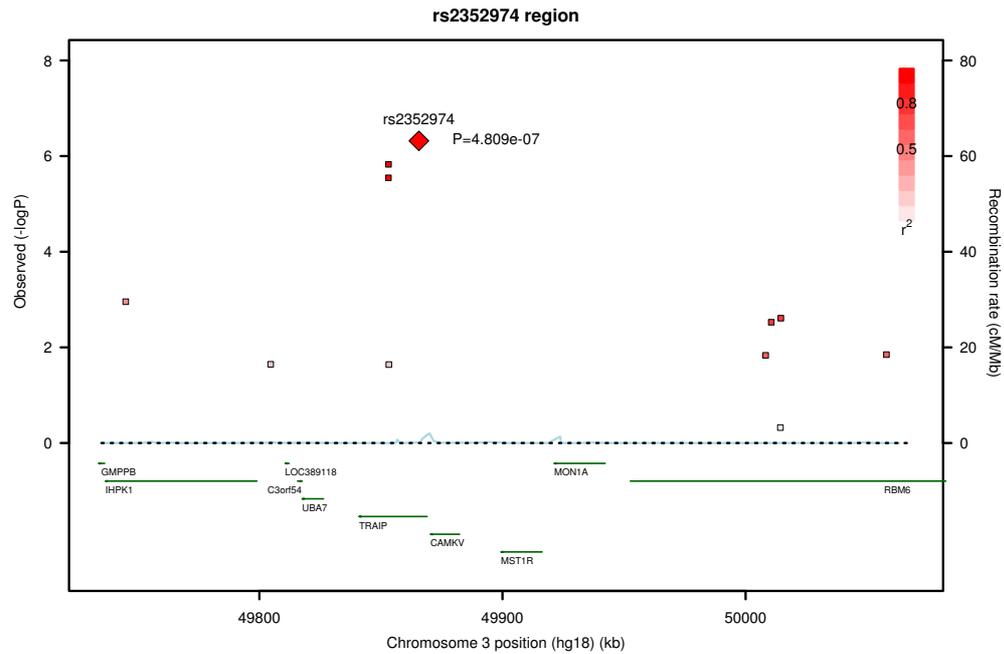

**Figure 5.2. Association at 3p21.31.** -log10 of the association p-values under an additive model are shown in the y-axis. Blue line represents the recombination rate in cM/Mb and the $r^2$ between SNPs and rs2352974 is shown as red scale. Genes are depicted as green lines and chromosomal position indicated in the x-axis.

0.97; rs2271960, P = 2.9 x $10^{-6}$, $r^2$ = 0.96) (Figure 5.2). The fifth association with SABP (rs16942644, P = 7.0 x $10^{-7}$, OR = 1.8 (1.4 - 2.2)) was at chromosome 15q26.1 in a region with a signature of transcriptional enhancer activity (data from (Ernst et al., 2011; Ernst and Kellis, 2010) as displayed at http://genome-preview.ucsc.edu). This suggests that rs16942644 may affect transcript levels of one or more genes near by. The closest gene is approximately 25 kb away and codes for an alpha/beta hydrolase domain containing protein (ABHD2).

The sixth association reaching suggestive significance was at chromosome 3p21.31 (rs26998, P = 7.0 x $10^{-7}$, OR = 1.5 (1.3 - 1.8)) within a possible transcriptional enhancer (data from (Ernst et al., 2011; Ernst and Kellis, 2010) as displayed at http://genome-preview.ucsc.edu) in an intron of the gene MCTP1 (multiple C2 domains, transmembrane 1 isoform S).

| Sub-type | Ensembl ID | Gene Symbol | p-value Z FIX | FDR |
|---|---|---|---|---|
| RDC SABP | ENSG00000180316 | PNPLA1 | $1.2 \times 10^{-6}$ | 0.005 |
| | ENSG00000164076 | CAMKV | $1.6 \times 10^{-6}$ | 0.005 |
| | ENSG00000183763 | TRAIP | $1.6 \times 10^{-6}$ | 0.005 |
| | ENSG00000173699 | SPATA3 | $2.1 \times 10^{-5}$ | 0.050 |
| SUICIDAL | ENSG00000102924 | CBLN1 | $1.8 \times 10^{-6}$ | 0.029 |

**Table 5.2. Genes with FDR < 0.05 for bipolar sub-phenotype analysis.**

### 5.3.2.    Gene-based genome-wide association of BD sub-phenotypes

We performed gene-based GWAS analysis for each of the BD sub-phenotypes. There were significant association with four genes with SABP and in one gene with Suicidal subtypes, respectively (Table 5.2). In line with the single SNP results, SABP subtype showed the greater number of significant results. Gene-based associations for SABP highlighted the 3p21.31 locus but not the B3GALT5 gene (gene p-value = 0.0014, FDR = 0.5).

Figure 5.3 shows details of the association with suicidal behaviour at the CBLN1 locus. Eight of the nine SNPs mapped to CBLN1 had p-values < 0.05 and these are localised at two association signals separated by a region of high recombination. Inspection of a wider region surrounding the genes showed that more significant associations lay beyond our 20 kb mapping limits. Two almost independent associations exist at rs12598711 and rs1816581 with p-values of $1 \times 10^{-4}$ and $3 \times 10^{-4}$, respectively, and $r^2 = 0.1$.

Figure 5.4 shows a heat-map of gene-based statistics for genes with FDR < 0.2 in the analysis of sub-phenotypes and BD diagnosis. Comparison of these gene-statistics suggests some gene associations detected in the analysis of the main BD diagnosis are strengthened in the analysis of specific sub-phenotypes, e.g., PPIL5 and CBLN1 had p-values 3 to 4 orders of magnitude smaller in the Suicidality sub-phenotype analysis, whereas others seem to be exclusively detected in sub-phenotypes, e.g., PNPLA1 had a p-value = 1 x 10$^{-6}$ and 0.24 in the SABP sub-phenotype and BD analyses, respectively.

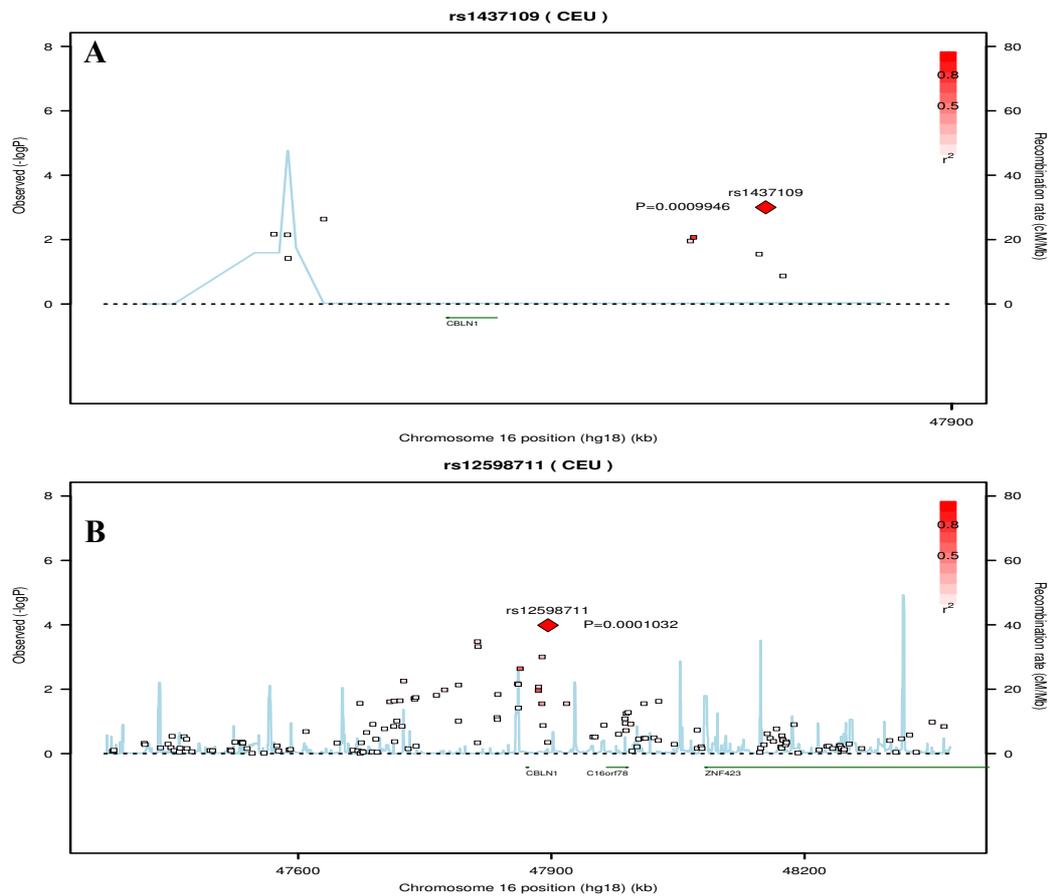

**Figure 5.3. Association at CBLN1 locus.** -log10 of the association p-values under an additive model are shown in the y-axis. Blue line represents the recombination rate in cM/Mb and the $r^2$ between SNPs and the minimum p-values in the region (panel title)is shown as red scale (see Figure legend). Genes are depicted as green lines and chromosomal position indicated in the x-axis. A) Figure including SNPs used in the gene based analysis (20 kb around the gene) and B) SNPs up to 500 kb around the gene.

### 5.3.3.    Network-based genome-wide association analyses

Next, we performed network analysis using the protein-protein interaction network (PPIN) aiming to identify groups of interacting gene's products enriched for association signals. We identified significant and suggestive sub-networks only for RDC SABP (Table 5.3). The most associated sub-network (PPIN-1156, z-score = 4.17, FDR = 0.04) contains genes previously associated with several kidney diseases in both human and the mouse (Wolf and Stahl, 2003; Hauser et al., 2009). Inspection of publicly available data provided evidence for their expression in human and mouse brain (Lein et al., 2007; Johnson et al., 2009; Uhlen et al., 2010). Interestingly, mutations in KIRREL3 have been associated with intellectual disability (Bhalla et al., 2008) and axon migration (Serizawa et al., 2006). Two other sub-networks had FDRs approaching our significance threshold. These were characterised by biological processes related to transcriptional regulation and cell proliferation, such as chromatin modification (p-value = 2 x 10$^{-15}$), negative regulation of cell proliferation (p-value = 7 x 10$^{-5}$) and anti-apoptosis mediated by external signals (p-value = 1 x 10$^{-4}$) (Table 5.3).

| | BD | PSYCHOSIS | SUICIDAL | RAPID CYCLING | PUERPERAL | RDC SABP | AGE_ONSET |
|---|---|---|---|---|---|---|---|
| CMTM8 | 5.4 | 2.5 | 4.1 | 2.8 | 1.8 | 2.2 | 0.1 |
| DUSP28 | 5.2 | 3.8 | 2.9 | 1.5 | 0.3 | 1.5 | 0.1 |
| ZNF224 | 5.1 | 3.7 | 3.1 | 1.7 | 1.3 | 2.3 | 0.1 |
| RNPEPL1 | 4.9 | 3.6 | 3.0 | 1.2 | 0.4 | 1.2 | 0.1 |
| LAMP3 | 4.7 | 3.4 | 4.4 | 2.6 | 1.9 | 3.0 | 0.4 |
| MKI67 | 4.4 | 1.8 | 4.9 | 1.0 | 3.3 | 0.2 | 0.5 |
| ZNF284 | 4.3 | 3.2 | 3.2 | 1.5 | 1.3 | 1.7 | 0.2 |
| KLHDC1 | 4.2 | 3.8 | 4.4 | 0.4 | 2.2 | 1.7 | 2.1 |
| ZNF791 | 4.2 | 3.5 | 2.3 | 1.3 | 3.2 | 0.9 | 0.3 |
| AC007346.2 | 4.1 | 3.4 | 1.8 | 1.2 | 0.7 | 1.7 | 0.2 |
| KLHDC2 | 4.1 | 3.5 | 4.1 | 0.4 | 2.1 | 1.2 | 1.8 |
| ANKMY1 | 4.1 | 2.5 | 2.9 | 2.0 | 0.3 | 0.9 | 0.0 |
| SPEF1 | 4.0 | 3.7 | 2.0 | 1.4 | 0.8 | 0.8 | 2.0 |
| CENPB | 3.9 | 3.8 | 2.0 | 1.3 | 0.8 | 0.8 | 2.1 |
| DBF4 | 3.9 | 4.5 | 2.3 | 1.1 | 1.8 | 1.5 | 0.2 |
| PPIL5 | 3.2 | 3.3 | 4.8 | 0.6 | 1.7 | 1.5 | 1.8 |
| AL139099.3 | 3.1 | 3.2 | 4.4 | 0.4 | 1.6 | 1.6 | 1.5 |
| MGAT2 | 3.1 | 3.2 | 4.4 | 0.4 | 1.6 | 1.6 | 1.5 |
| RPL36AL | 3.1 | 3.2 | 4.4 | 0.4 | 1.6 | 1.6 | 1.5 |
| SNAP47 | 3.0 | 3.7 | 1.9 | 1.4 | 0.4 | 0.5 | 0.1 |
| ZBED3 | 2.7 | 1.5 | 0.3 | 0.0 | 0.6 | 4.4 | 2.0 |
| MST1 | 2.6 | 2.2 | 0.3 | 1.2 | 0.8 | 3.8 | 0.5 |
| DPF3 | 2.6 | 3.9 | 1.4 | 0.5 | 1.8 | 0.5 | 0.3 |
| RPS29 | 2.6 | 2.6 | 4.3 | 1.2 | 1.5 | 0.8 | 2.3 |
| AC007064.28 | 2.5 | 2.1 | 1.7 | 0.4 | 2.9 | 3.9 | 0.3 |
| XKR3 | 2.5 | 2.2 | 1.5 | 0.4 | 3.3 | 4.3 | 0.2 |
| HDAC11 | 2.4 | 4.8 | 0.6 | 0.6 | 0.6 | 1.4 | 0.0 |
| ZCCHC2 | 2.4 | 2.0 | 1.1 | 0.5 | 0.1 | 4.6 | 0.5 |
| CBLN1 | 2.3 | 1.5 | 5.7 | 1.8 | 3.7 | 0.8 | 0.2 |
| TOMM40 | 2.2 | 2.1 | 0.7 | 0.0 | 1.4 | 4.3 | 0.6 |
| AC064801.1 | 2.1 | 2.1 | 1.7 | 0.6 | 0.1 | 4.3 | 0.5 |
| IRGM | 2.1 | 2.8 | 1.3 | 0.1 | 5.1 | 0.7 | 0.7 |
| AC006021.1 | 2.0 | 2.4 | 1.3 | 1.2 | 0.5 | 4.0 | 0.1 |
| UBQLN4 | 1.8 | 3.8 | 1.8 | 0.1 | 1.6 | 2.0 | 1.8 |
| SERGEF | 1.7 | 4.1 | 0.5 | 0.4 | 1.1 | 1.4 | 0.2 |
| GABRB1 | 1.7 | 1.8 | 2.0 | 0.2 | 0.3 | 3.9 | 2.0 |
| CAMKV | 1.2 | 1.7 | 0.4 | 0.6 | 0.3 | 5.8 | 0.8 |
| TRAIP | 1.2 | 1.7 | 0.4 | 0.6 | 0.3 | 5.8 | 0.8 |
| SPATA3 | 1.1 | 3.2 | 1.5 | 0.6 | 1.8 | 4.7 | 0.3 |
| TNNC2 | 1.1 | 1.3 | 0.7 | 0.7 | 0.6 | 4.2 | 0.4 |
| AL355916.1 | 1.1 | 1.4 | 1.5 | 1.3 | 4.9 | 0.6 | 1.2 |
| AC009879.1 | 1.0 | 1.3 | 1.0 | 0.2 | 5.0 | 1.6 | 0.8 |
| PNPLA1 | 0.6 | 1.3 | 0.3 | 1.4 | 0.1 | 5.9 | 0.1 |
| G0S2 | 0.3 | 1.0 | 0.4 | 0.1 | 0.1 | 4.4 | 1.3 |
| TRIM9 | 0.2 | 0.1 | 0.2 | 0.3 | 1.1 | 0.1 | 5.1 |
| HYDIN | 0.1 | 0.1 | 0.4 | 0.3 | 0.0 | 0.7 | 4.6 |
| PRUNE | 0.0 | 0.2 | 0.0 | 0.1 | 0.1 | 0.3 | 4.9 |

**Figure 5.4. Association of genes with FDR < 0.2 in different BD sub-phenotypes.** Cells are coloured by -log10 of the gene p-value with more intense tones representing larger values.

### 5.3.3.1.    Analysis of GABA(A) receptor genes in SABP

We explored the association between the 19 GABA(A) receptor genes and SABP reported by Craddock et al. (2010) and Breuer et al. (2010) with p-values = 7x10$^{-6}$ and 0.009, respectively. The authors of these reports did not explore the robustness of their

| Subnetwork | Z score | FDR | Genes | Biological Processes |
|---|---|---|---|---|
| PPIN-1156 | 4.17 | 0.04 | **CD2AP KIRREL** KIRREL3 **NPHS1** NPHS2 | no significant results |
| PPIN-12 | 4.01 | 0.06 | ACTB ACTL6A AHR ARID1A ARID1B ARID2 BRCA1 BRWD1 CBX5 CCNE1 CEBPB CHD4 CHMP5 CREB1 CTNNB1 ESR1 ETS2 **FANCA H3F3A** HSF1 IKZF1 MPHOSPH6 **MPP6** NR3C1 PAX6 PBRM1 PHB RAP1A RB1 RELB RFXAP SMARCA4 **SMARCB1** SMARCC1 SMARCE1 SS18 STAT2 STK11 TMF1 **TP53** ZMYND11 | Chromatin modification (P = $2 \times 10^{-15}$), Relaxin signaling (P = $6 \times 10^{-5}$), Negative regulation of cell proliferation (P = $7 \times 10^{-5}$), Anti-Apoptosis mediated by external signals by Estrogen (P = $1 \times 10^{-4}$) |
| PPIN-973 | 3.96 | 0.07 | ACTB ACTL6A AHR ARID1A ARID1B ARID2 BRCA1 BRWD1 CBX5 CCNE1 CEBPB CHD4 CHMP5 CIITA CREB1 CTNNB1 ESR1 ETS2 **FANCA H3F3A** HDAC2 HDAC4 HDAC5 HSF1 IKZF1 MPHOSPH6 **MPP6** NR3C1 PAX6 PBRM1 PHB RAF1 **RAP1A** RB1 RELB RFX5 RFXANK RFXAP SMARCA4 **SMARCB1** SMARCC1 SMARCE1 SS18 STAT2 STK11 TMF1 **TP53** ZMYND11 | Chromatin modification (P = $4 \times 10^{-12}$), Relaxin signaling (P = $4 \times 10^{-4}$), Negative regulation of cell proliferation (P = $2 \times 10^{-4}$) |

**Table 5.3. PPIN subnetwork with FDR < 0.1 for SABP bipolar subtype.** For each sub-network we report its z-score from the test of association with the phenotype and its FDR. Genes in bold had a p-value < 0.05. Significance for the overlap of each sub-network and GeneGO's biological process was calculated assuming a null hypergeometric distribution (see Methods for additional details). All categories reported reached an FDR < 0.001

results regarding the choice of $r^2$, p-value and maximum number of SNPs in the PLINK

SNP-set analysis thresholds. They choose thresholds that are usually regarded as

standard. However, despite using values widely accepted, it is still necessary to explore the sensitivity of the results and conclusion to these values. We repeated the analysis with the same data used by Craddock et al. (2010), with exception of their initial SNP pruning step in Haploview, and utilised different LD $r^2$ (0.1, 0.3, 0.5, 0.7 and 1), p-value (1, 0.5, 0.05, 0.01, $1\times10^{-3}$ and $1\times10^{-5}$) and a maximum number of SNPs thresholds. In practice the maximum number of SNPs was affected by the other thresholds, so the maximum number ranged from 5 to 207 out of the 383 SNPs mapped to the 19 genes. Figure 5.5 presents the results of these analyses. It seems clear that the choice of parameters influences the empirical p-value, which ranged from $5\times10^{-5}$ to $1\times10^{-2}$.

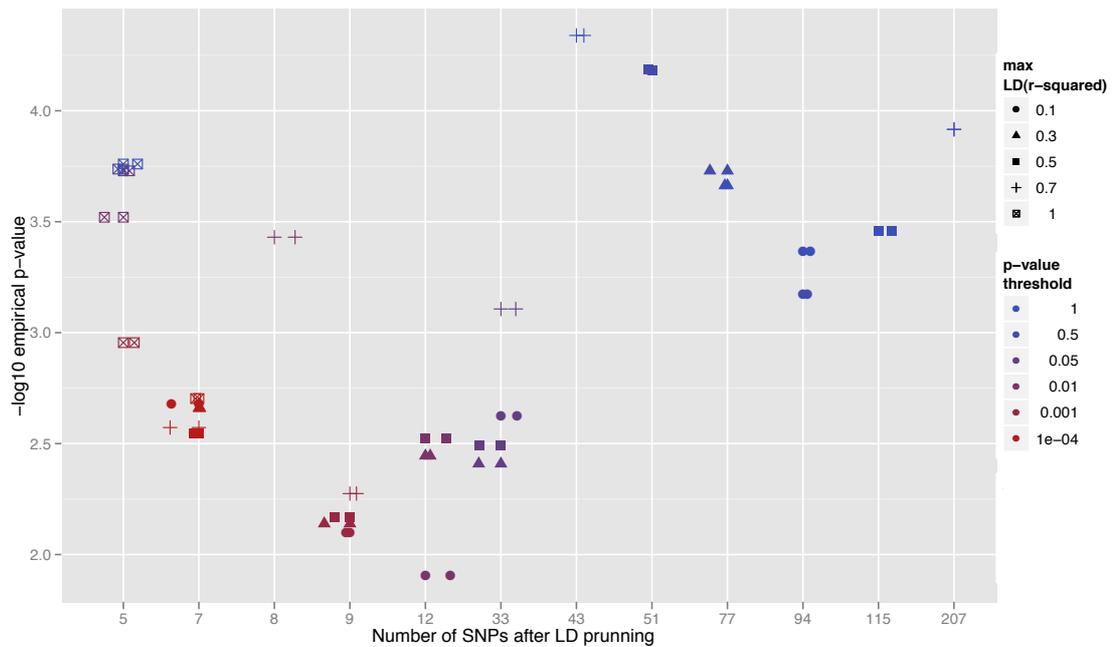

**Figure 5.5. Effect of analysis parameters on association of GABA(A) genes and SABP.** -log10 of the association p-value is shown in the y-axis and x-axis shows the maximum number of SNPs included in the association SNP-set. Note that along the x-axis we have plotted only the values used in our analysis and not a continuous range. Different combinations of p-value and r2 thresholds were used the figure legend indicates their coding.

We found only a marginal association (p-value = 0.065) when analysing the 19 GABA(A) receptor genes as a gene-set using the gene-set analysis methods implemented in FORGE (described in Chapter II).

### 5.3.4.    Further exploration of Schizoaffective bipolar disorder subtype associations

Throughout our analysis and as previously highlighted by others (Hamshere et al., 2009b), we found that the SABP sub-phenotype provided better genetic mapping results. Therefore, we focused our replication efforts on the SABP findings. When performing these analyses, we did not have access to a SABP sample for replication and decided to follow an alternative strategy to explore this association and provide additional supporting evidence. Quantitative genetic evidence has shown that while SABP has substantial heritability, genetic risk is completely shared with schizophrenia (SCZ) and bipolar disorder, suggesting that few, if any, SABP specific genes exist (Cardno et al., 2002). Therefore, we explored the association of the two SABP genetic associations in SCZ and additional BD samples. However, it is important to note that we could not exclude SABP cases from the replication samples' cases. This affects our interpretation of the association with a broader phenotype and suggests that a positive association may be explained by a non-ascertained, cryptic, SABP subgroup, whereas negative association may suggest a lack of association with a broader diagnosis. We selected for replication three SNPs present in either or both the Illumina and Affymetrix arrays used in the replication samples available, which had complete or near complete tagging (e.g., $r^2 = 1$, rs3746887 for rs734413) of the four markers. We found no evidence of association for either locus in the remaining BD samples of the WTCCC (2007), Sklar et al. (2008) or Cichon et al. (2011) (see Table 5.4 for sample details). We failed to replicate the chromosome 3p21.31 locus and the second weaker signal at 21q22.1 in the SCZ samples from the SGENE consortium (rs6802890 P = 0.80; rs11700626 as proxy for rs11088486 P = 0.68). However, we did find mild evidence for association for rs3746887 in a combined analysis of the SGENE and SGENE-plus studies (Table 5.4).

| Phenotype | Study | SNP | P-value[d] | Odd ratio (95 % CI) | N° Cases | N° Controls |
|---|---|---|---|---|---|---|
| SABP | WTCCC SABP [a] | rs734413 | $2.7 \times 10^{-7}$ | 1.67 (1.34-2.0) | 279 | 2938 |
| Bipolar Disorder | WTCCC non-SABP Bipolar [b] | rs734413 | 0.36 | 0.95 (0.85-1.06) | 1589 | 2938 |
| | WTCCC Bipolar [c] | | 0.40 | 1.04 (0.94-1.16) | 1868 | 2938 |
| | Sklar et al (2008) | | 0.46 | 1.05 (0.93-1.18) | 1461 | 2008 |
| | Bipolar Bonn | rs3746887 | 0.92 | 0.99 (0.84-1.17) | 682 | 1300 |
| Schizophrenia | SGENE | rs3746887 | $4.7 \times 10^{-3}$ | 1.13 (1.04-1.23) | 2662 | 13780 |
| | SGENE-plus | | 0.57 | 1.02 (0.95-1.1) | 3553 | 9889 |
| | Meta-analysis | | $2.5 \times 10^{-2}$ | 1.06 (1.01-1.13) | 6215 | 23665 |

**Table 5.4. Replication results of association for rs734413.** [a] WTCCC BD cases with RDC SABP diagnosis. [b] WTCCC BD cases without RDC SABP diagnosis. [c] ALL BD cases. [d]Replication p-values are one-tailed.

| | Sklar | Bonn | Meta-analysis | |
|---|---|---|---|---|
| | | | z-score | p-value |
| **Genes** | | | | |
| TRAIP | -0.23 | -0.07 | -0.21 | 0.58 |
| SPATA3 | -0.06 | -0.12 | -0.12 | 0.55 |
| PNPLA1 | -0.28 | -0.61 | -0.63 | 0.74 |
| CAMKV | -0.23 | 0.09 | -0.10 | 0.54 |
| **Subnetworks** | | | | |
| PPIN-12 | 1.22 | -1.90 | -0.48 | 0.68 |
| PPIN-1156 | -0.57 | 0.43 | -0.10 | 0.54 |
| PPIN-973 | 1.12 | -1.61 | -0.35 | 0.64 |

**Table 5.5. Association of Schizoaffective bipolar type genes and networks in other BD samples.** Sklar and Bonn refer to the GWAS reported by Sklar et al. (2008) and Cichon et al. (2011).

We found no evidence of association in two independent samples of BD for the SABP genes and networks (Table 5.5). In the future, we will attempt to replicate this association in SCZ and SABP samples.

# 5.4. Discussion

### 5.4.1.    Genome-wide association of BD sub-phenotypes

Despite its high heritability (approximately 80%) (McGuffin et al., 2003), the genetic underpinnings of BD remain largely unknown. Increasing evidence suggests substantial phenotypic heterogeneity for BD and other neuropsychiatric disorders, and analysis of sub-phenotypes has been suggested as a strategy to increase genetic homogeneity (Alda, 2004; Burmeister et al., 2008). A higher number of genetic associations are found with SABP than expected by chance compared to other BD sub-types (Hamshere et al., 2009a; Tuller et al., 2009), suggesting a molecular genetic examination of the SABP group may be fruitful.

We identified two regions that reach suggestive significance after Bonferroni correction for SABP. Lacking SABP replication samples, it was attempted to explore further this associations on other relevant phenotypes. We choose a cross-phenotype analysis in multiple BD and SCZ cohorts based on the quantitative genetic evidence available for SABP (Cardno et al., 2002). While we failed to show association for one region, 3p21.31, our results did suggest that the genetic variants identified in 21q22.2 confer susceptibility for SCZ and not BD (rs734413, P = 2.7 x $10^{-7}$, OR = 1.7 (95% CI= 1.4 - 2.0) for SABP, P = 0.02, OR = 1.06 (95% CI= 1.01 - 1.13) for SCZ, P = 0.7 for BP). Several genes in the proximity to the 21q22.1 locus are interesting candidates. B3GALT5, a member of the beta-1,3-galactosyltransferase gene family, participates in the synthesis of glycoproteins, glycolipids and other glycan structures. There has long been interest in abnormal levels of glyocolipids in SCZ (Cherayil, 1969; Tkachev et al., 2007) and there is evidence for reduced expression of genes involved in N- and O-linked glycan and glycosphingolipid biosynthesis pathways as well as oligodendrocyte marker genes in SCZ and BD patients compared with controls (Tkachev et al., 2003; Narayan et al., 2009). Glycosphingolipids are crucial for the glycoso-synapse, involved in myelin-axon communication, suggesting a possible link with the white matter deficit reported in SCZ, including decreased oligodendrocyte number, white matter volume and myelin-associated gene expression (Davis et al., 2003; Kubicki et al., 2005). The benign nature of the amino-acid change at rs3746887 (Adzhubei et al., 2010) does not directly support it as a causal variant but further functional studies would be needed for verification. In addition, other genes in the genomic area may be of potential interest. For example, two genes, SH3BGR (SH3 domain binding glutamic acid-rich protein) and WRB (tryptophan rich basic protein), map to the Down's Syndrome Congenital Heart Disease region at 21q22.2 and are widely expressed in the central nervous system. PCP4 (Purkinje cell protein 4) is a calmodulin-binding protein involved in intracellular $Ca^{+2}$ signalling (Erhardt et al., 2000; Slemmon et al., 2000; Kanazawa et al., 2008).

These results extend the findings from previous studies showing genetic variation in 21q22.2 as risk factor for SABP by providing mild evidence for its involvement in the aetiology of SCZ.

### 5.4.2.   Gene and gene-sets set associations

In addition to genes at the 3p21.31 locus, gene-based analysis provided significant result for PNPLA1 (patatin-like phospholipase domain containing 1), a gene with little biological characterisation. It is known to code for a papatin-like lipid hydrolase (Kienesberger et al., 2009) and is widely expressed in the brain, including in Purkinje neurones (Ponten et al., 2009). Our PPIN network results pointed to sub-networks involved in cell-to-cell contact as well as regulation of chromatin and cell proliferation. Common variants within genes involved in cell-to-cell adhesion have been previously associated with SCZ and BD (O'Dushlaine et al., 2011) and copy number variants affecting cell adhesion genes associated with SCZ and autism (Stefansson et al., 2008; Betancur et al., 2009; Bucan et al., 2009; Rujescu et al., 2009). This supports the involvement of cell-to-cell adhesion in the aetiology of neuropsychiatric disorders.

Our gene-based results pointed to CBLN1 (cerebellin 1 precursor) as a susceptibility loci for suicidal behaviour (Table 5.2 and Figure 5.3). The association extends over the gene with the gene-based association explained by the contribution of two signals in low LD ($r^2$ =0.1). CBLN1 has been shown to be important for the integrity of synapses and is considered a necessary member of a synaptic organiser complex also involving neurexins and the glutamate receptor delta-2 (Matsuda et al., 2010; Uemura et al., 2010). We did not find evidence of association for CBLN1 in the Sklar (Z FIX p-values = 0.1) or Bonn (Z FIX p-values = 0.78) GWAS datasets. Replication in independent samples ascertained for suicidal behaviour will be needed.

Despite only reaching a suggestive significance level, the PPIN-12 sub-network might be of special interest. Valproate, a widely used antipsychotic drug, has direct effects on chromatin remodelling through its action on histone deacetylases (Rosenberg, 2007). Furthermore, both valproate and lithium have neurogenic or neuroprotective effects (Lagace and Eisch, 2005) and valproate has been shown to facilitate cell differentiation (Huangfu et al., 2008). Our results suggest valproate may counteract the effect of risk alleles influencing cell differentiation, potentially through their effect in chromatin remodelling pathways. A potential relevance of this network analysis results is to suggests that networks associated in GWAS could be used as signatures to identify potential new pharmacological treatments, for example with the methods proposed by Lamb et al. (2006).

Detailed inspection of the method used by Craddock et al. (2010) and Breuer et al. (2010) showed it to be sensitive to the parameter values chosen ($r^2$, p-value and number of SNPs in the SNP-set). We showed that when the parameter values are varied the empirical p-value varies considerably (Figure 5.5), complicating the interpretation of the reported association between the GABA(A) receptor genes and SABP. Using our GWAS gene-set analysis methods, we found a weak evidence of association between the 19 GABA(A) receptor genes and SABP (p-value = 0.065). There are major differences between the methods used in the reports of Breuer et al. (2010) and Craddock et al. (2010) and those used throughout our studies. For example, we only need specify the reference population to calculate SNP-SNP correlations and can make use of all SNP association information (details of the method are provided in Chapter II).

Finally, we acknowledge that our gene and network association results must be considered as preliminary due to the small sample size and low statistical power of our BD sub-phenotype samples. As discussed above, replication of SABP genetic mapping results in BD and SCZ studies is encouraged by quantitative genetic research in SABP. However, from our results, it seems unlikely that an association with SABP will be replicated in BD samples (Figure 5.4) as SABP associations appear to be specific to this sub-phenotype (also observed by Hamshere et al. (2009a)). Therefore, replication of our findings in SABP samples sets would be desirable.

## 5.5. Conclusion

Overall, we demonstrate the merit of applying GWAS and system biology approaches to psychiatric sub-phenotypes. First, sub-phenotypes are often characterised by more pronounced homogenous clinical and phenotypic presentations, enabling a better interpretation of the genotype phenotype relationship. For example, from Figure 5.4, it is reasonable to hypothesise the association at CMTM8 with BD diagnosis may be more relevant in understanding the genetic basis of suicidal behaviour than BD age of onset or schizoaffective features. In addition, small sample sizes are expected for many sub-phenotypes of clinical relevance. Therefore, methods to identify biologically relevant themes by clustering mild associations could be of special interest for these samples. However, caution is needed when interpreting results from studies with small sample sizes because many false positives are expected due to low statistical power. Nevertheless, these methodological strategies may be helpful to tip the fine balance between the risk of identifying false positives and highlighting interesting biological findings that may help to guide the design of follow up studies.

In summary, our results confirm the utility of SABP for genetic mapping. We did not find compelling evidence of replication in our single SNP results in either BD or SCZ samples. However, our data suggests the effect of SABP susceptibility variants may be diluted in samples ascertained only for a BD diagnosis. We highlighted associations with genes and biological processes for SABP, which warrant further exploration and replication efforts. The analysis of BD sub-phenotypes is a useful strategy to improve the understanding of the genetic effect of BD risk variants.

# Chapter VI

# Discussion and future directions

# 6.1. Discussion

Genome-wide association studies constitute a valuable tool to study the genetic basis of human diseases. However, they are not without limitations and the design of a study will might large dictate them. For example, lack of statistical power, incomplete coverage of genetic variants, and phenotypic heterogeneity can burden genetic association and their replication success. It is likely that many of these limitations will be overcome with new technologies leading to GWAS primarily based on whole-genome sequencing and imputation (Altshuler et al., 2010; Durbin et al., 2010). Disease diagnosis will also improve with better understanding of genetic and environmental risk factors and development of new diagnostic technologies (Auffray et al., 2010). However, with these improvements increasing challenges will also arise for the interpretation and discovery of genetic association signals, particularly in highly complex diseases (Donnelly, 2011).

In the previous chapters evidence is presented to suggest that it is possible to augment the amount and interpretability of the information extracted from GWAS using gene and gene-set associations and integrative biology analyses with gene-expression and protein-protein interaction data-sets. Several lessons emerged from these results and these are discussed below.

### 6.1.1.    Gene-based association is a good complement to single SNP analyses

Asymptotic gene p-values were found to agree well with empirical estimates and are faster than p-values calculated by phenotype permutations or sampling methods, such as VEGAS (Liu et al., 2010). However, sampling-based gene p-values seem to be a good compromise between the speed of asymptotic estimates and the accuracy of phenotype permutations (the current gold standard). Gene-based association results using these methods were much closer to the expected distribution on the null hypothesis (uniform distribution of p-values). Hence, these are considered to represent an improved statistical estimate and the method of choice if the computing power is available.

In the future, it will be interesting to see if gene-based association analysis becomes a popular tool to analyse GWAS. If these results prove generalizable, many studies without significant SNPs may suggest plausible candidates for follow up studies, which in the long run may be useful in improving our biological understanding and devising new treatments.

Gene-based association results highlighted most regions found by single marker analyses and additional ones that were replicated in other studies. Even in their simplest form, i.e. correcting the minimum p-value, gene-based association complemented single SNP analyses. Hence, gene-based analyses should become an integral step of the GWAS analysis pipeline, as imputation is increasingly so, for example (Marchini and Howie, 2010).

### 6.1.2.   GWAS can be used to reveal biological processes associated with disease susceptibility

Gene-based association can be used in conjunction with gene-set analyses to highlight biological processes enriched with genetically associated variants. For example, in a GWAS of Crohn's disease (CD) we identified associations with the IL12 and IL23 signalling pathways; in the case of rheumatoid arthritis (RA), we identified enrichment in the biological categories of immune system and extracellular matrix; and with hypertension (HT), we identified a PPIN sub-network involved in cortisone, androstenedione and testosterone biosynthesis and metabolism. There is evidence for the role of immune-system biological processes in the aetiology of CD and RA and for the association between cortisone and testosterone hormone levels in HT and heart disease (Stewart et al., 1996; English et al., 1997; Smith et al., 2005; Jones, 2010; Perusquia and Stallone, 2010), supporting the validity of our results.

In bipolar disorder (BD), convergent evidence[10] emerged from GWAS for a PPIN sub-network enriched among risk alleles in genes that are also differentially expressed in brain in BD cases compared with controls. Interestingly, the genes of this network mapped to the human post-synaptic density (hPSD), a large protein complex critical for neuronal function which is altered in several behavioural phenotypes (Bayes et al., 2011). This convergent evidence warrants additional study on independent genetic and gene-expression data sets. If this association gains strength, it may support the initiation of follow up studies on several fronts. For example, the hPSD is currently estimated to comprise about ~1500 proteins organised in a modular architecture with modules differentiated by their biological functions (Armstrong et al., 2006; Pocklington et al., 2006a; Pocklington et al., 2006b; Bayes et al., 2011). Because of their relatively small number (compared with the ~20,000 annotated protein coding genes), these genes might be the target of sequencing efforts in large sample sets. A second possibility would be to use the PPIN sub-network genes as a gene signature in drug repositioning studies. This strategy has shown potential for several diseases, see (Lamb et al., 2006), and has not been explored as an application of GWAS.

My preliminary attempt to compare seven diseases at the gene-set level showed that nominally associated gene-sets (uncorrected p-values < 0.05) recovered meaningful groups: three immune-system diseases together (RA, type 1 diabetes (T1D) and CD) and a group of two comorbid conditions (coronary artery disease (CAD) and T2D (Choy et al., 2008; Huang et al., 2008; Guh et al., 2009)). Interestingly, by clustering diseases using only the most significant PPIN sub-networks, we recovered a group with HT, T2D and CAD driven by the PPIN-2605. These sub-networks are involved in cortisone, androstenedione and testosterone hormone biosynthesis and metabolism, abnormal levels of which have been associated with risk for these three diseases (English et al., 1997). Similar analyses including diseases with poor etiological characterisation may provide a new hypothesis for their genetic underpinning based on a guilt-by-similarity approach.

---

[10] Here by convergent evidence I refer to the fact of finding supportive association results from two different genomic platforms measuring different biological molecules. This evidence should not be regarded as replication since the quantities compared are not the same but it is rather convergent because two independent lines of evidence coalesced into the same conclusion.

### 6.1.3.    Combination of GWAS and gene-expression studies

Gene-expression studies constitute one of the most common genomic data-sets available and meta-analysis of multiple studies has shown that biological signatures are pervasive, even in the presence of technical confounders (Dudley et al., 2009; Zheng-Bradley et al., 2010). However, a major limitation of gene-expression studies is that mRNA levels are affected by both etiological causes and consequences of disease pathology. On the other hand, GWAS studies are not confounded in this manner but it is difficult to differentiate true biological signals because the signal-to-noise ratio is low. Therefore, by bringing together the two approaches in integrative analyses, it may be possible to identify, using readily available datasets, genes and biological processes that are more likely to be drivers of pathology. These are of interest when developing new pharmacological interventions and disease diagnostics (Schadt, 2009). Driver genes can also be identified with system genetic studies but these need both molecular phenotypes and genetic variant measures on the same sample sets (see review by Schadt (2009)), which is feasible but such data sets are not available for many diseases. For example, in the case of mental illnesses, it is difficult to access the affected tissues. Even when human brains are available, it is difficult to select the appropriate brain region for study because of the complexity and our still limited understanding of brain function. We think this makes genetic studies all the more valuable since gene-expression and measurement of other molecules will be affected by the availability of tissue, brain region of choice and environmental factors (e.g., drugs) whereas genotypes are not affected by these factors. Recently, genotype data has been released for the samples used in our gene-expression analyses (Choi et al., 2011). It would be a logical next step to perform eQTL analysis on the genes of the significant sub-network to assess if BD susceptibility risk alleles also drive gene-expression changes.

### 6.1.4.    Candidate-Gene-Networks

A prediction from network biology was used (Goh et al., 2007; Park et al., 2009) to generate a new test for genes associated via GWAS [11]. This showed that alleles conferring risk for BD and other diseases cluster within groups of genes with (often) known biological function. Our Candidate-Gene-Networks (CGNets) analysis can be seen as a way to strengthen the evidence for genes previously associated with a trait on the basis of genetic or other information by showing that their network environment is also associated with the phenotype. If an association is found, it provides *internal consistency* in the sense that it agrees with what we already know about genes associated with human diseases. At the moment the negative predictive value of the CGNet results are unknown.

---

[11]  As discussed previously through the thesis, specifically on Chapters I and IV. Network analyses of genes associated with human diseases have shown that when a gene is associated with a phenotype, its interacting partners or other genes involved in the same biological processes are more likely to be associated with the same phenotype than expected by chance (Goh et al., 2007; Ideker and Sharan, 2008; Wu et al., 2008; Park et al., 2009). This knowledge has led to bioinformatic methods to prioritise genes for diseases (e.g. (Wu et al., 2008; Yang et al., 2009)) and propose functions for uncharacterised genes (Jensen et al., 2009; Yang et al., 2009). We make use of the same concept and expect that if a gene is associated with a disease, its network partners should also be more associated than expected by chance.

Using this strategy it was possible to strengthen the evidence for CACNA1C and ANK3, two loci robustly associated with BD by meta-analysis of GWAS (Ferreira et al., 2008). This approach can be used with any trait and it is similar in principle to the text mining tool GRAIL (Raychaudhuri et al., 2009). GRAIL identifies links between loci by analysing the co-citation pattern in PubMed (www.pubmed.org). These links are used to improve evidence supporting a gene. An advantage of the CGNet strategy is that it does not require a previous publication record for two genes to be associated, which may be beneficial for poorly characterised genes with known interaction information.

### 6.1.5. Gene and gene-set association are also affected by statistical power

Re-analysis of the WTCCC phase 1 study of seven different GWAS in common disorders, showed that in general gene and gene-set analyses had the same pattern of results as single SNP analyses. For example, association was found with gene-sets in CD, T1D and RA, three autoimmune diseases which have had successful results in GWAS, but no associations with BD were found using a single study. At the moment it is not possible to distinguish whether these different results are explained by differences in the distribution of effect sizes (e.g., effect sizes for CD may be larger than for BD), in locus heterogeneity or in frequency spectrum of risk alleles. Large meta-analyses of GWAS of BD and other psychiatric diseases are underway (Psychiatric GWAS Consortium, 2009). Availability of large and well powered studies for a variety of phenotypes will help the comparison of findings at the SNP, gene and gene-set level [12].

---

[12] Statistical power is affected by several factors, e.g. whether the phenotype is discrete or quantitative. It is a common mistake to equate power to sample size, see Yang et al (2010b), meaning that comparison of findings across phenotypes will need to account for differences in power.

These may reveal differences and commonalities with valuable information in order to understand the genetic architecture of different diseases. In turn, this information may guide future studies.

### 6.1.6.    The importance of replication

In the BD study, significant association were only found by pooling the results from three independent GWAS and a gene-expression study. This suggests that, although risk alleles cluster in common biological processes, their total combined effect (at least as measured by our methods) is weak and use of several GWAS cohorts is crucial to establish compelling evidence of association. This contrasts with the scenario of immune-system related diseases (CD, T1D or RA) where with a single GWAS we could identify relevant biological process. It may also be useful to attempt replication in different genomic data sets to highlight targets for follow up studies, e.g., signalling pathways where both risk alleles and functional changes co-occur.

Gene-set associations of BD found in this study are similar in significance to the results presented for GSA of schizophrenia GWAS (O'Dushlaine et al., 2011). Although comparison of p-values may not be adequate, the fact that both studies highlight association around p-values $\sim 10^{-3}$ may suggest that replication samples and meta-analyses will be essential for GSA in psychiatric genetics. A major concern for GWAS meta-analysis is the so called winner's curse effect, also known as decline effect or regression to the mean (Yu et al., 2007; Zollner and Pritchard, 2007; Lehrer, 2010). It refers to the fact that effect size estimates tend to decrease (sometimes to zero!) as more studies are analysed. It is usually assumed to arise from biases in experimental design but mere chance and publication biases are likely to play a role (Lehrer, 2010). It will be interesting to evaluate if gene-set analyses are equally sensitive to the winner's curse effect. Gene-set analyses may be more robust to it than single SNP analyses because they do not rely on a single very significant finding but rather on several weak ones clustered using prior information of common biological processes. If so, this facet may add to the value of the GSA approach as an information mining tool for GWAS.

To date, the only replicated gene-set association with BD was the association between the GABA(A) receptor genes and the SABP subtype (Breuer et al., 2010; Craddock et al., 2010). After detailed inspection of their analyses, the methodology employed in these studies, was found to be sensitive to the choice of different parameters, the influences of which the authors did not explore. The association between the GABA(A) receptor genes and the SABP subtype may be real but its statistical significance is uncertain. Pedroso and Breen (2009) also noted other problems with reports of gene-set analyses of GWAS. For example, several authors did not correct for the number of variants in the gene or gene-set (Torkamani et al., 2008; Walsh et al., 2008; Askland et al., 2009) and others did not assess the effect of model parameters (Holmans et al., 2009; Breuer et al., 2010; Craddock et al., 2010). GSA of GWAS is maturing but it is still an exploratory technique and researchers must be cautious when applying and interpreting these reports.

### 6.1.7. Genetic heterogeneity is a double-edged sword

A major motivation of the studies presented in this thesis was to explore the ability of systems biology analyses to overcome genetic heterogeneity. In the BD associated network, we found that different genes drive the association in different studies, suggesting that these single SNP associations are relevant and worth of study even though we found no replicated evidence of association. A similar pattern of gene (or locus) heterogeneity has been reported in GWAS of CD (Wang et al., 2009) and multiple sclerosis (Baranzini et al., 2009).

What does this heterogeneity mean? Are there two kinds of genetic effects: those that consistently replicate in different studies and those that do not? Is the proportion of these two effects different between diseases (their genetic architecture)? Or do we see this heterogeneity because we failed to account for factors that alter the genetic effects, e.g. environmental exposures? Let us assume we do acknowledge that allelic heterogeneity is common (which is widely accepted for genes associated with Mendelian inheritance) and even that gene heterogeneity is also present. Will we also acknowledge pathway or biological process heterogeneity? How far should we push the heterogeneity argument to maximise the information extracted but without breaking a basic principle of the scientific practice, i.e. the replication of results?

### 6.1.8. Smaller sample size to achieve a more tightly defined phenotype may also increased power

An important challenge of psychiatric genetics is to help refine clinical diagnosis. My exploration of genetic mapping in BD sub-phenotypes suggests that this approach may be fruitful. It is possible that the genetic architecture of some disease subtypes may be simpler and lead to faster improvement of diagnostic tools, and possibly allow patient stratification by drug response which would be very important, even if only applicable to a small patient population.

### 6.1.9. GWAS as a starting point for system biology analyses of mental illness

By identifying biological processes of relevance to understanding disease aetiology, GWAS can guide researchers to new experimental models. For example, GWAS pointed to autophagy as an important pathological mechanism to explain CD (Lees et al., 2011). This cellular process is now a target of studies aiming to better understand the mechanism of this association to identify new treatments (Sirota et al., 2009; Behrends et al., 2010; Fleming et al., 2011). A recent report provided results from a large scale proteomic study of the autophagy pathway (Behrends et al., 2010). This represents a unique resource to explore the mechanistic links and biological roles of CD susceptibility alleles at the biological systems levels. Psychiatric genetics may benefit in the same way from additional studies describing the post-synaptic density (Bayes et al., 2011). Studies exploring disease aetiology using functional approaches to describe biological networks have revealed great complexity in the structure and functions of signalling cascades and how they are affected by disease associated perturbations (Bakal et al., 2008; Chen et al., 2008; Emilsson et al., 2008; Yang et al., 2009; Jorgensen and Linding, 2010; Pennell et al., 2010). Studies aiming to understand the consequences of susceptibility alleles or genes on the dynamic behaviour of these functional networks will take us away from discussing the small effect size of risk alleles into more complex representations and modelling of disease biology (for examples see (Chen et al., 2008; Emilsson et al., 2008; Schadt, 2009)). This is in no way an easy task but it is one, which, in the long term will allow thoughtful design of new treatments and diagnostic systems based on accurate knowledge of disease causation. For example, large scale system genetic analyses[13] in both human and mouse were able to identify a causal network associated with metabolic traits (Chen et al., 2008; Emilsson et al., 2008). In these studies the causal network was a co-expression network highly enriched with genes that have a causal or driver role on metabolic traits. An important lesson from these studies is that despite the fact that these metabolic traits are complex and therefore underpinned by many alleles it is possible to identify broad underpinning biological processes. By revealing a causal network the authors shifted the focus of research and discussion from the many alleles of small effect affecting the traits and expression of the network's genes to the biological function of this network, its cell type of action and developmental and physiological roles. The authors showed that the network identified in human and mouse overlap significantly and both correspond to a signalling pathway involved in macrophage biology. Because this network has a causal role it may be possible to plan intervention that would lead to a phenotypic change or disease prevention. It is important to note that these studies included more than 30,000 human subjects and 1000 mice. This represents far more subjects than most individual human and mouse studies use, suggesting than not only a technological revolution is needed to undertake these approaches but also a major change in mind set when designing biological experiments. If causality is to be disentangled, then quantitative studies on well powered samples must be used together with a systems biology an experimental design that accounts for systems biology.

_________________

[13] System genetics refers to the combination of genetic and functional biology data, e.g., gene expression measurements. The task is to identify causal association between the biological molecules and a trait of interest. The basic idea behind the methods used is that genotypes are unaffected by the phenotypes therefore is we can identify a correlation between a genotype and a biological molecule (i.e., an expression quantitative trait in the case of gene-expression studies) and a correlation between this molecule and the trait of interest there must be a causal pathway between the genotype to the phenotype mediated by the changes in gene-expression. For our discussion it is important to note that whereas an allele can have a weak correlation with a trait it can have a strong correlation with a molecular phenotype (e.g., gene-expression). Therefore, despite that the effect size of the trait correlation is small needing large samples to find it, the causal influence on the gene-expression change can be found more easily. We refer the interested reader to a selection of references in this wide topic (Schadt et al. 2005; Friedman, 2004; Chen et al., 2008; Emilsson et al., 2008; Schadt, 2009; Schadt et al. 2009; Schadt et al. 2009a; Schadt et al. 2009b; Hsu et al., 2009; Millstein et al., 2009; Yang et al., 2009; Zhu et al., 2008).

GWAS are showing good potential to help delineate biological processes of relevance in order to understand the mechanisms underpinning psychiatric illness and other human diseases. Nevertheless, many limitations still remain but current advances, as we have shown in this thesis, are promising.

## 6.2. Limitations of my study

### 6.2.1. Assumptions of the asymptotic methods

In my analyses I made several assumptions. I think the most important is that the correlation between the SNPs approximates the correlation between the SNPs' test statistics (see Chapter II). This has been shown to be the case for the chi-square statistics with one degree of freedom (Han et al., 2009). Although to my knowledge this has not been proven for other statistical tests, e.g., logistic regression with covariates, several authors have also used it in their methods (for examples see (Nyholt, 2004; Li and Ji, 2005; Galwey, 2009)). This is likely to be an important factor in explaining the non-uniformity of the asymptotic gene-wide p-values [14]. Gene p-values were calculated by phenotype permutations (i.e. statistics for the SNPs in the gene were calculated and the null distribution was derived by randomising the phenotype and calculating the statistics again) for over 900 genes and showed asymptotic estimates which approximate the empirical p-values well (Pearson correlation r > 0.98). Simulation-based p-values were also implemented, which we found produce gene p-values with a more uniform distribution than those produced by asymptotic methods. Possibilities to overcome the limitations introduced by using this assumption were evaluated and there may be some alternatives. For example, given the correlation matrix between the SNPs, it is possible to obtain a correlation matrix that explains the observed similarity between the test statistics. This would allow a better variance-covariance matrix to be obtained for the test statistics and would allow fast asymptotic methods to be used.

### 6.2.2.    Statistical power of gene and gene-set association methods

Simulated data could be used to quantify statistical power differences between gene-based association methods. However, this approach was not employed mainly due to time constraints. Instead, different methods were compared in several different real GWAS, a strategy also followed by others, e.g., (Donnelly and Marchini, 2009). Other authors have also evaluated the improvement in replication provided by gene-wide association and found results in line with the results I presented in Chapter III (Peng et al., 2010). An important difference of their study was that they compared the replication success analysing independent GWAS, a strategy free of the possible biases we may have introduced by using the GWAS Catalogue. These independent results support the conclusions of this study.

---

[14]As discussed in Chapter II, gene-wide p-values should agree with the uniform distribution only when the genetic associations are representative of the null hypothesis. Some of the data sets analysed in this thesis, e.g., GWAS of Crohn's disease, have substantial genetic signals and therefore their quantile-quantile plots deviate form the null distribution, particularly at the tail of low p-values. Nonetheless, uniformity of p-values is usually considered a criteria for a valid study, for an example see review by McCarthy et al. (2008). Some criteria like these were described in the earlier GWAS when many studies had small samples sizes and results which did not deviate from the null hypothesis. On the other hand, more recent large meta-analyses with tens of thousands of samples have quantile-quantile plots that deviate significantly from the null distribution. Criteria like uniformity of p-values are evolving but are still valuable in cases where a significant excess of signals is not expected.

### 6.2.3.   SNP versus gene and gene-set meta-analyses

Meta-analyses of GWAS of BD have provided evidence of association for the CACNA1C, ANK3, NCAN and PBRM1 loci (Ferreira et al., 2008; Breen et al., 2011; Cichon et al., 2011). None of these genes reached significance in the gene-based meta-analysis but several of the regions highlighted by the gene-based meta-analysis had high but sub-threshold significance in these studies. Results obtained by SNP and gene-based meta-analyses were not compared because access to genotype data from all BD studies was lacking.

As part of an ongoing parallel project (with unpublished results) I have analysed three GWAS that had the same set of SNPs and performed gene-based analysis on the SNP meta-analysis results and on the three individual GWAS separately. The top 20 genes from the SNP and gene-based meta-analysis overlapped significantly, suggesting that if the GWAS overlap completely in their SNP set using the SNP meta-analysis results or meta-analysing gene-based results is largely equivalent.

It is likely that SNP based meta-analysis will be a more powerful strategy, particularly when the causal variant is common and well tagged. This is likely to be the case with the increasing accuracy of imputation methods and increasingly deep coverage of reference sample sets (Altshuler et al., 2010; Durbin et al., 2010). However, in the presence of allelic heterogeneity and low frequency variants, gene-based association will also be a useful complement. In the future, simulation studies may help clarify the merit of each approach.

### 6.2.4. The choice of gene-sets

There are numerous sources from which to construct gene-sets and interaction networks. These will introduce different biological information and biases into the analyses (Lehner and Lee, 2008; Fernandes et al., 2010). In our studies, we used data from the Human Protein Reference Database (Keshava Prasad et al., 2009), a highly curated set of interactions which had been previously applied to analyses of GWAS (Baranzini et al., 2009) and the data presented here suggests that it provides better results than other gene-sets, e.g., Gene Ontology (Ashburner et al., 2000). Different interaction information is informative on different biological aspects but our analyses have been relatively basic in terms of exploiting the interactions beyond simply knowing that two proteins can interact with each other. For example, we do not exploit the direction of the interaction or its nature, i.e. phosphorylation, degradation, etc. It has been shown that combining different information sources can lead to an improved representation of cell and organism level behaviour (Lage et al., 2007; Lee et al., 2008; Zhu et al., 2008; Lee et al., 2010). By combining different interaction sets, it is possible to construct probabilistic networks, in which interactions seen multiple times (and with higher confidence) are assigned a higher probability of being true. An additional advantage of these integrated networks is the increased coverage across the interactome (Lehner and Lee, 2008). In the future, it would be appropriate to develop networks of specific relevance for psychiatric diseases, as has been done for Alzheimer's disease (Soler-Lopez et al., 2011) and cardiac pathologies (Lage et al., 2010). A starting point may be to build a good representation of the interaction between genes in the human post-synaptic density, as this is becoming increasingly important to study behaviour in both humans and mice. However, it should be noted that there is not a clear benchmark to compare the results obtained with different interaction networks, making the choice of the network and the information used to build it sensitive to biases.

## 6.3. Future development of gene and GSA of GWAS

### 6.3.1. Simultaneous analysis of multiple data-sets

At the moment, all methods consider different studies independently but it may prove advantageous to combine the evidence at the gene-set level but using information from the different studies simultaneously. This may be achieved by using Bayesian hierarchical models (Gelman et al., 2004; Ji and Liu, 2010). This strategy would allow the inclusion of prior distributions for the SNP, gene and gene-set effects. Addition of priors may also enable a better integration with different non-genetic data-sets, e.g., gene-expression or proteomics, because the difference in effect size distribution will be learnt from the data. For example, one of the issues encountered during our integrative analyses of BD GWAS (Chapter V) was that the gene-expression data found numerous significant gene-sets on its own, probably because gene-set analyses were designed for gene-expression data and that gene-expression changes are much larger than changes in allele frequencies. In practical terms, it suggests that there is a different expectation regarding the changes in magnitude found in GWAS and gene-expression studies, i.e., an odds-ratio of 2 is very interesting in a GWAS but not in a gene-expression study. Development of models considering multiple data sets and data types and allowing for different distribution of effect sizes may be very relevant as biomedical research in mental health moves into systems biology studies integrating genetic, neuroimaging, gene-expression and proteomics (for example see (Furney et al., 2010; Thambisetty et al., 2010; Simmons et al., 2011)).

### 6.3.2. Construction of interaction network for mental disorders

Publicly available protein interaction networks only contain information for a fraction of annotated human genes, e.g., HPRD used here has ~ 8000. This certainly had an impact on our analyses since GWAS will cover most of the annotated genes. For example, the gene HAPLN4 (best SNP p-values = 4 x $10^{-6}$, gene p-value = 7 x $10^{-6}$, FDR = 0.13) is in the region reported genome-wide significant for BD by Cichon et al. (2011). HAPLN4 is a structural protein involved in the brain extracellular matrix and has been associated with brain remodelling (Spicer et al., 2003). There was no interaction for HAPLN4 in our datasets, so it was left out of our network analyses. Furthermore, most interaction networks have major biases towards protein-coding genes and are depleted of interaction involving non-coding RNA, lipids, and polysaccharides. Non-coding RNA are becoming a centre of attention for biomedical research and psychiatry is not an exception. Therefore, construction of an interaction network including many more molecule types and data sets would be beneficial to our systems biology analyses. There is an increasing quantity of functional genomics datasets, which include gene-expression, epigenomics and transcription factor binding, which could be mined to construct probabilistic networks with better coverage of the interactome. Unbiased experiments, like those based on high throughput sequencing, will allow for more non-coding RNA to be included. System genetic analyses on datasets containing functional data and genetic variation may allow us to go beyond gathering correlation information and enable us to establish causal relationships. Extensive data from model organisms, like the mouse or fly, will provide across species comparisons to borrow information from other organisms. Interaction networks derived from a holistic strategy will undoubtedly lead to better systems biology analyses and in turn better interpretation of GWAS.

The methods we used are all frequentist in statistical nature. A natural extension of our analyses would be to apply Bayesian statistics. Several Bayesian methods to analyse all SNPs within a gene or genome simultaneously have been proposed and implemented, e.g., see (Hoggart et al., 2008; Omont et al., 2008), but have not yet been widely applied to GWAS. An additional attractive possibility of Bayesian statistics is to perform model comparisons between different gene association methods. For instance, one could use three models: a) the best SNP within the gene, b) combined information and c) combined information but including the interaction effects between SNPs in the same gene. Model selection would allow selection of the model that best explains the association within the gene while accounting for the complexity of each model, e.g., tests of interaction effects increase the multiple testing so that a model should be selected only if it improves the association considerably. A similar approach has been proposed to determine whether associations in a region are better explained by a single causal variant or by multiple causal variants (Donnelly and Marchini, 2009).

### 6.3.3. Analysis of large-scale sequencing projects and rare variants

During the coming years there will be an avalanche of data from sequencing projects, e.g., 1000 genomes (www.1000genomes.org) or UK10K (www.uk10k.org). These data pose new challenges for statistical genetic and bioinformatics analyses. Post-hoc analyses (i.e. after quality control), will need to cope with large amounts of data and probably with noise different from the error present in the current genotypes generated by microarray genotyping. There is increasing interest in development of new analytical methods for association of low frequency variants. Risk for several mental illnesses has been associated with rare variants identified by exome sequencing (Vissers et al., 2010).

It is hypothesised that rare variants may play a prominent role in the aetiology of psychiatric diseases (Uher, 2009), some of the most recent exome sequencing studies in schizophrenia support this hypothesis (Girard et al., 2011).

Due to the low statistical power to detect association with rare variants gene, region-based analyses constitute the main analysis strategy (Asimit and Zeggini, 2010; Bansal et al., 2010; Morris and Zeggini, 2010). There have not been reports on gene-set analyses using rare variants, but the current strategies could be extended to gene-sets. However, we may find that power to detect rare variant associations at the gene-set level will be reduced due to multiple testing. Reduction of multiple testing by focusing on variants having evident functional effects on protein coding genes may be a possibility, but will undoubtedly miss interesting risk variants associated with DNA or RNA regulation. In principle, the methods presented in this thesis could be used for a combined analysis of genetic variants through the frequency spectrum. Weighting the variants by the predicted functional relevance or frequency may help the analysis of rare variants (Li et al., 2010).

Simultaneous analysis of all individual genetic variants by incorporating information on genes and gene networks may prove useful in recovering biological processes associated with diseases. This may be achieved using hierarchical regression models as it was demonstrated by Heron et al. (2011) to include SNPs functional annotation (e.g., if the SNP is conserved or if it is a known disease quantitative trait loci). This strategy may also incorporate additional information like parent of origin effects, which can be detected in studies using a family design (Kong et al., 2009). Furthermore, it might be possible to incorporate information emerging from the increasing amount of functional genomics data available to develop increasingly complex but also more biologically rich analyses. For example, the known tissue expression level or epigenetic state of a gene could be used to dismiss or down-weight contribution of its risk variants and help detect genes and variants of relevance for different tissues.

By far the greatest of all challenges will be to deliver personalised medicine (or to convince biomedical researchers that such a thing is not possible). Personalised medicine has been the great promise of the human genome project and, without disregarding important achievements, it has not been delivered to date. Whole genome sequencing studies may bring us closer to this promise. As noted in the introduction, a key problem with this challenge is our poor understanding of causality in complex biological systems. If we do not grasp what causal relationship risk alleles (of small or large effects) have with molecular phenotypes and physiological syndromes (diseases), it will be increasingly daunting to transform linear strings of information (millions of DNA sequence positions) and a plethora of 3D data points (gene and protein expression, brain and whole body imaging over time and space) into an accurate prediction of disease risk.

### 6.3.4.    Additional integration with other omics datasets

Gene-based analyses may have an important role for bioinformaticians seeking to integrate disease signatures from different genomics platforms. Several repositories, such as dbGAP (www.ncbi.nlm.nih.gov/gap) or EGA (www.ebi.ac.uk/ega), provide summary statistics from published GWAS. These could be use to derive gene-based statistics without the need to access raw genotype data and perform cumbersome quality control procedures. Provision of gene based association statistics may also go some way towards allaying concerns about data protection, when individual SNP association data is released (Homer et al. 2008). If methods, like the ones used here, become popular, data repositories may also opt to provide gene-wide association results. This may improve the role of GWAS in data integration efforts, particularly those looking across numerous disease types. For example, availability of pre-computed post-quality control data for gene-expression microarray experiments has had a positive impact on data availability for the wider scientific community, who are not experts in data-analysis. We have taken some preliminary steps to establish such bioinformatic resources. Our first step has been to provide a web server implementing our FORGE analyses (https://compbio.brc.iop.kcl.ac.uk:8443/forgeweb). The web server was developed in collaboration with David To and Richard Dobson at the NIHR Biomedical Research Centre at IoP and SLAM (KCL, UK).

I and Dr. Gerome Breen (IoP, KCL) are currently developing approaches to mine the Connectivity Map (CMAP) (Lamb et al., 2006) database for drug-disease interactions. The CMAP project provides gene-expression values for drug treatments of cell lines. We have an initial set up to identify drugs that induce gene-expression changes in genes identified from GWAS (Inti Pedroso and Gerome Breen, unpublished). Preliminary results are promising and we have been able to recover true drug-disease pairs. Current efforts are focused on improving the bioinformatic pipeline and quantifying the false positive/false negative rate of this approach. We plan on providing this pipeline as a drug repositioning tool as part of the FORGE web server for the wider scientific community.

## 6.4. Final remarks

I started this thesis with a brief outline of GWAS and the state-of-the-art gene-set analyses of GWAS. Several major challenges facing large scale genetic studies were immediately apparent. Using these challenges as a starting point, this thesis aimed to provide a strategy that would help psychiatric genetics to use GWAS as starting point to define biological systems for detailed mechanistic studies. Hopefully this will lead to a new generation of diagnostic tools and treatments. This thesis aims and main achievements were:

i)   To develop, implement and provide the wider scientific community with software tools to perform gene-set and network analyses of GWAS. In Chapter II I described the development of the FORGE software suite to perform these analyses. Availability of the software has generated collaborations with researchers working on amyotrophic lateral sclerosis ( Prof Ammar Al-Chalabi), anorexia nervosa (Prof David Collier), autism (Dr Sarah Curran) and unipolar depression (IoP Depression Consortium, led by Prof Peter McGuffin). In collaboration with David To and Dr Richard Dobson, I have developed a web server to facilitate the use of these methods (https://compbio.brc.iop.kcl.ac.uk:8443/forgeweb).

ii)  Assess the ability of these methods to find replicated genetic associations. Additionally in Chapter III I presented evidence showing that gene-set analyses find true disease loci which eluded single marker association. Applied to BD in Chapter IV, a combination of gene and gene-set analyses allowed the discovery of convergent evidence of association of a protein interaction network. From the evidence presented in Chapter IV I was able to describe locus heterogeneity in BD and show how systems biology analyses help overcome its challenges.

iii) Evaluate the feasibility of GWAS for data integration with other genomic datasets in a system biology framework. Through Chapters II to V I successfully integrated BD GWAS results with the protein interaction network and gene-expression studies.

iv)  Provide association evidence for genes and biological processes to help the wider scientific community to define biological systems for the detailed study of risk variants. In Chapter III, I showed that systems biology analyses (i.e. network analyses) of GWAS can point to biological processes of etiological relevance for Crohn's disease, rheumatoid arthritis and type 1 diabetes. In Chapter IV, I provided convergent genomic evidence of association with BD for a group of interacting gene products. These genes were characterised by biological functions of relevance, such as transmission of nerve impulse, and a significant fraction of them were mapped to the human post-synaptic density. In Chapter V, I showed that these methods may also be useful to study BD sub-phenotypes and identify biological processes that can be related to more discrete symptoms.

Genome wide association studies in one shape or form are likely to continue to play a significant role in biomedical research for the foreseeable future. The technologies used to execute these studies have evolved tremendously during the course of this PhD, which started in the era of the SNP chip and ends in the era of next generation sequencing (NGS). Both technologies are likely to continue in parallel for the next few years, as microarray genotyping continues to be the most cost effective method for large cohorts, complemented with NGS exon sequencing. As NGS technologies become more affordable, whole-genome re-sequencing is likely to become the method of choice for GWAS, ultimately ushering in the era of the personal genome. When this era arrives, the tools described here may go some way to help uncover the genetic basis of human diseases at the level of individual genes and biological processes. It is the hope of the author that the software, methods and experience presented in this thesis will help others to mine GWAS to improve translational research and subsequently the quality of life of patients.

# Appendix A

## 8.1. Overview

This Appendix present the Perl code of the FORGE software suite. We have only included the main pieces of code and not example files and utility scripts. If you wish to check the code of the software not included here please visit the FORGE code website https://github.com/inti/FORGE.

## 8.2. Appendix Table of contents

## 8.3. Files in the distribution

A copy of the FORGE software suite programs is available in this Appendix A. The files distributed with FORGE are:

- **forge.pl**: script to perform gene-based association and gene-set association with a SNP-to-gene-set mapping strategy.

- **gsa.pl**: script to perform gene-set analyses based on gene p-values.

- **meta_analysis.pl**: script to perform meta-analyses. Its output included fixed and random-effects statistics, and statistical heterogeneity estimates. All statistical methods used in this script are described in section 2.1 under the fixed and random-effects models.

- **CovMatrix.pm**: perl module with routines to calculate correlation matrixes using a shrinkage approach. Original methods are described in (Schafer and Strimmer, 2005) and implemented in the R library 'corpcor'. Our Perl routines are implementations of these R methods and we performed tests on synthetic data to ensure no coding errors were introduced and results were identical within double precision when possible.

- **GWAS_IO.pm**: perl routines to deal with input and output of commonly use format in GWAS.

- **GWAS_STATS.pm**: perl routines to calculate some statistics, e.g. meta-analysis statistics.

- **Pareto_Distr_Fit.pm**: perl routines to estimate p-values using extreme value distribution theory as described in (Knijnenburg et al., 2009) but estimating the parameters of the generalised Pareto distribution by the method of Zhang (2010) instead of maximum likelihood.

- **affy_to_rsID.tab**: files with mapping between Affymetrix and rs-ids.

- Utilities: folder with utility scripts

  - **Calculate_add_LFDR_values.R**: R script to calculate False Discovery Rate values and add as columns to the input files.

  - **mapped_assoc_to_genes.pl**: using a file with SNP association and SNP-to-gene mappings it outputs a new file with both informations, i.e., SNP associations with chromosomal positions and genes it is mapped to.

  - **row_merge.pl**: script to merge files based on one of more columns with ids.

  - **snp_2_geneAnnot.pl**: script to generate SNP-to-gene mapping files using the Ensembl database human gene annotation.

- example: folder with example files

  - **example.assoc**: SNP association file as produced by PLINK.

  - **whole_genome_example.tab**: SNP association file with SNPs representing a whole-genome analysis.

  - **example.bed**, **example.bim** and **example.fam**: genotype files in pedigree binary format for gene ILR23 and a random dichotomous phenotype.

  - **example.prob**: same as above but is IMPUTE genotype probability format.

  - **example.map**, **example.ped**: same as above but in pedigree format.

  - **example.gene_pvals**: file with gene p-values.

  - **example.gmt**: an example gene-set definition file.

  - **example.snp_weight**, **example.snp_weight2**: example files with SNP weights to use in gene-based analyses.

  - **example.snpmap**: example SNP-to-gene mapping file.

- gene_sets: folder with gene-set definition files.

  - **biocarta.gmt**: 301 gene-sets from BioCarta (www.biocarta.com).

- **human.gol4.gmt**: 2058 gene-sets from Gene Ontology database (Ashburner et al., 2000).

- **kegg.gmt**: 212 gene-sets from KEGG data base (Kanehisa et al., 2008).

- **ppin_hprd_13April2010_min2_max500.gmt**: 7391 gene-sets derived from the Human Protein Reference Database (Keshava Prasad et al., 2009) protein-protein interaction network (PPIN).

- **ppin_hprd_22March2009_min10_max500.gmt**: 6888 gene-sets from the PPIN. Same as above but a different data base release.

• **README.txt**: some description and example commands.

## 8.4. forge.pl

Calculation of gene and gene-set association from GWAS summary statistics.

```perl
#!/usr/bin/perl -w
use strict;
use Getopt::Long;
use Pod::Usage;
use PDL;
use PDL::Matrix;
use PDL::GSL::CDF;
use PDL::Primitive;
use PDL::NiceSlice;
use PDL::Stats::Basic;
use PDL::Bad;
use Data::Dumper;
use Carp qw( confess );
$SIG{__DIE__} =  \&confess;
$SIG{__WARN__} = \&confess;

# Load local functions
use GWAS_IO;
use GWAS_STATS;
use CovMatrix;

our ( $help, $man, $out, $snpmap, $bfile, $assoc, $gene_list,
    @genes, $all_genes, $analysis_chr, $report, $spearman,
    $affy_to_rsid, @weights_file, $w_header, $v, $lambda,
    $print_cor, $pearson_genotypes,$distance, $sample_score,
    $ped, $map, $ox_gprobs,$sample_score_self, $w_maf,
    $ss_mean, $gc_correction,$g_prob_threshold,
     $bgl_gprobs, $flush, $include_gene_type, $exclude_gene_type,
$gmt,
    $gmt_min_size,$gmt_max_size, $use_ld_as_corr,$mnd_sim_target,
    $mnd_sim_max, $mnd_sim_wise_correction_methods, $mnd
);

GetOptions(
    'help|h' => \$help,
    'man' => \$man,
    'ped=s' => \$ped,
    'map=s' => \$map,
    'bfile=s'    => \$bfile,
    'out|o=s'    => \$out, #name of the output file
    'assoc|a=s' => \$assoc,
    'gene_list|g=s'    => \$gene_list,
    'genes=s' => \@genes,
    'all_genes'        => \$all_genes,
    'chr=s'=> \$analysis_chr,
    'snpmap|m=s@'        => \$snpmap,
    'report=i'    => \$report,
    'correlation|cor=s' => \$spearman,
    'affy_to_rsid=s' => \$affy_to_rsid,
    'verbose|v' => \$v,
    'lambda=f' => \$lambda,
    'gc_correction' => \$gc_correction,
    'print_cor' => \$print_cor,
    'pearson_genotypes' => \$pearson_genotypes,
    'use_ld' => \$use_ld_as_corr,
    'distance|d=i' => \$distance,
    'sample_score' => \$sample_score,
    'weights|w=s' => \@weights_file,
```

```perl
    'w_header' => \$w_header,
    'ox_gprobs=s' => \$ox_gprobs,
    'bgl_gprobs=s' => \$bgl_gprobs,
    'g_prob_threshold=f' => \$g_prob_threshold,
    'weight_by_maf|w_maf' => \$w_maf,
    'ss_mean' => \$ss_mean,
    'flush=i' => \$flush,
    'gmt=s@' =>   \$gmt,
    'gmt_min_size=i' =>\$gmt_min_size,
    'gmt_max_size=i' =>\$gmt_max_size,
    'gene_type|type=s@' => \$include_gene_type,
    'exclude_gene_type|exclude_type=s@' => \$exclude_gene_type,
    'mnd' => \$mnd,
    'mnd_target=i' => \$mnd_sim_target,
    'mnd_max=i' => \$mnd_sim_max,
    'mnd_methods=s' => \$mnd_sim_wise_correction_methods,
) or pod2usage(0);

pod2usage(0) if (defined $help);
pod2usage(-exitstatus => 2, -verbose => 1) if (defined $man);
pod2usage(0) if (not defined $assoc);

my $LOG = new IO::File;
$LOG->open(">$out.log") or print_OUT("I can not open [ $out.log ] to
write to",$LOG) and exit(1);

print_OUT("Check http://github.com/inti/FORGE/wiki for updates",$LOG);
print_OUT("LOG file will be written to [ $out.log ]",$LOG);

# define parameters for mnd simulation
defined $mnd_sim_target or $mnd_sim_target = 10;
defined $mnd_sim_max or $mnd_sim_max = 100_000;
if (defined $mnd_sim_wise_correction_methods){
    $mnd_sim_wise_correction_methods = [ split(/\,/,
$mnd_sim_wise_correction_methods) ];
} else { $mnd_sim_wise_correction_methods = [0,1,2,3]; }

if (defined $mnd){
    use PDL::LinearAlgebra qw (mchol);
    use Pareto_Distr_Fit qw (Pgpd);
    print_OUT("Will run multivariate normal distribution simulations
to estimate significance",$LOG);
    print_OUT("   '-> max number [ $mnd_sim_max ] or until statistic
is seen [ $mnd_sim_target ] times",$LOG);
    my @m = ('sidak','fisher','z_fix','z_random');
    @m = @m[@$mnd_sim_wise_correction_methods];
    print_OUT("   '-> for each gene best p-value will be selected
among the methods [ @m ] ",$LOG);
}

# defines min and max size for gene-set analyses
defined $gmt_min_size or $gmt_min_size = 2;
defined $gmt_max_size or $gmt_max_size = 999_999_999;

# flush output every $flush number of genes are ready
defined $flush or $flush = 1000;

# define distance threshold,
defined $distance or $distance = 20;
print_OUT("Max SNP-to-gene distance allowed [ $distance ] kb",$LOG);

# $report is use to specify how often report the advance when reading
input files
defined $report or $report = 50_000;
# tell if analysis is restricted to a specific chromosome
```

```perl
defined $analysis_chr and print_OUT("Restricting analysis to
chromosome [ $analysis_chr ]",$LOG);
#  defined lambda value for Genomic Control correction
defined $lambda or $lambda = 1;
# defined threshold value for genotype probabilities
defined $g_prob_threshold or $g_prob_threshold = 1.0;
# tell if user wants to print the *.correlation file
defined $print_cor and print_OUT("Defined -print_cor: I will print the
*.correlation file (it is bulky)",$LOG);
#set output file if not set already.
defined $out or $out = "gene_based_fisher_v041.OUT";
# tell if user wants to correct p-values by genomic control
if ($lambda != 1){ print_OUT("SNP p-value will be corrected with
lambda = [ $lambda ]",$LOG);}

# define option to read genotype probability files
my $geno_probs = undef;
my $geno_probs_format = undef;
# find out which format the file is with the command line options
if (defined $ox_gprobs) {
    $geno_probs = $ox_gprobs ;
    $geno_probs_format = 'OXFORD';
    print_OUT("Genotype probabilities in OXFORD format will be read
from [ $geno_probs ]",$LOG);
} elsif (defined $bgl_gprobs) {
    $geno_probs = $bgl_gprobs ;
    $geno_probs_format = 'BEAGLE';
    print_OUT("Genotype probabilities in BEAGLE format will be read
from [ $geno_probs ]",$LOG);
}

# generate an index of the genotype probability files
# this speads up the IO
my ($gprobs, $gprobs_index);
if (defined $geno_probs){
    $gprobs = IO::File->new();
    $gprobs_index = IO::File->new();
    my $index_name = "$geno_probs.idx";
    $gprobs->open("<$geno_probs") or print_OUT("I can not open
genotype probability file [ $geno_probs ]",$LOG) and exit(1);
    if (not -e "$geno_probs.idx") {
    print_OUT("   '-> Making index for genotype probabilities in
[ $index_name ] file",$LOG);
        $gprobs_index->open("+>$index_name") or print_OUT("Can't open
$index_name for read/write: $!\n",$LOG);
        build_index(*$gprobs, *$gprobs_index);
    } else {
        print_OUT("   '-> Found [ $geno_probs.idx ] file for genotype
probabilities",$LOG);
        $gprobs_index->open("<$geno_probs.idx") or print_OUT("Can't
open $index_name for read/write: $!\n",$LOG) and exit(1);
    binmode($gprobs_index);
    }
}

# print header for output file
my $OUT = new IO::File;
$OUT->open(">$out") or print_OUT("I can not open [ $out ] to write
to",$LOG) and exit(1);

# header for MND sampling analysis
if (not defined $mnd){
    print $OUT "Ensembl_ID\tHugo_id\tgene_type\tchromosome\tstart
\tend";
    print $OUT "\tmin_p\tmin_p_SIDAK\tFISHER\tFISHER_chi-square
\tFISHER_df";
```

```perl
        print $OUT "\tB_fix\tVar_fix\tB_P_fix\tB_random\tVar_random
\tB_P_random";
        print $OUT "\tI-squared\tQ\tQ_p-value\ttau_squared";
        print $OUT "\tn_effect_Galwey\tn_effect_Gao\tn_snps\n";
} else { # header for asymptotic analysis
        print $OUT "Ensembl_ID\tHugo_id\tgene_type\tchromosome\tstart
\tend";
        print $OUT "\tmin_p\tVEGAS\tSIM_SIDAK\tSIM_FISHER\tSIM_Z_FIX
\tSIM_Z_RANDOM";
        print $OUT "\tI-squared\tQ\tQ_p-value\ttau_squared";
        print $OUT "\tSIM_BEST_P\tSIM_BEST_method\tN_SIM";
        print $OUT "\tSEEN_VEGAS\tSEEN_SIDAK\tSEEN_FISHER\tSEEN_Z_FIX
\tSEEN_RANDOM";
        print $OUT "\tGPD_VEGAS\tGPD_VEGAS_LOW\tGPD_VEGAS_UP";
        print $OUT "\tGPD_SIDAK\tGPD_SIDAK_LOW\tGPD_SIDAK_UP";
        print $OUT "\tGPD_FISHER\tGPD_FISHER_LOW\tGPD_FISHER_UP";
        print $OUT "\tGPD_Z_FIX\tGPD_Z_FIX_LOW\tGPD_Z_FIX_UP";
        print $OUT "\tGPD_Z_RANDOM\tGPD_Z_RANDOM_LOW\tGPD_Z_RANDOM_UP";
        print $OUT "\tn_effect_Galwey\tn_effect_Gao\tn_snps\n";
}

# i will read the gene_list and i will load data for just this genes
to speed up.
if ( not defined $all_genes and not defined @genes and not defined
$gene_list){
    $all_genes = 1;
    print_OUT("Note: You did not provide an option for the set of genes
to be analyzed. I will analyze all genes covered by the SNP
association file. Check documentation for options -genes and -
gene_list otherwise",$LOG);
}

if ( not defined $all_genes ) { # in case user want to analyze all
genes
    if ( not defined @genes ) { # in case user gave a list of genes in
the command line
        print_OUT("Reading Gene List from [ $gene_list ]",$LOG);
        # read file with gene list and store gene names.
        open( GL, $gene_list ) or print_OUT("I can not open
[ $gene_list ]",$LOG) and exit(1);
        @genes = <GL>;
        chomp(@genes);
        close(GL);
    } else {
        print_OUT("Read Gene List command line [ @genes  ]",$LOG);
    }
} else {
        print_OUT("Going to analyze all genes on [ @$snpmap ] file.",
$LOG);
}

# Now lets going to read the affy id to rsid mapping. This is used to
keep all ids in the
# same nomenclature
my %affy_id = ();
if ( defined $affy_to_rsid ) { # if conversion file is defined
    print_OUT("Reading AFFY to rsID mapping from [ $affy_to_rsid ]");
    open( AFFY, $affy_to_rsid ) or print_OUT("I can not open
[ $affy_to_rsid ]",$LOG) and exit(1);
    while (my $affy = <AFFY>){
        chomp($affy);
        my @b = split(/\t+/,$affy);
        $affy_id{$b[0]} = $b[1];
    }
    close(AFFY);
}
```

```perl
# Read file with genetic association results.
print_OUT("Reading association file: [ $assoc ]",$LOG);
# create hash to store SNP information
my $assoc_chis = [] if (defined $gc_correction);

my %assoc_data = ();
open( ASSOC, $assoc ) or print_OUT("I can not open [ $assoc ]",$LOG)
and exit(1);
my $line = 0;
my %header = ();
while (  my $a = <ASSOC> ) {
  $a =~ s/^\s+//;
  $a =~ s/^\t+//;
  $a =~ s/\s+/\t/g;
  # here I get the header line of the file. With the name of the
columns I can use the cols
  # SNP and P to extract the information.
  my @data = split( /[\s+\t+]/, $a );
  if ( $line == 0 ){
      %header = %{get_header(\@data)};
      $line++;
      next;
  }
  # In case there is not cols with SNP and P names.
  exists $header{"SNP"} or print_OUT("Not p-value columns available,
these are the headers i found [ " . (keys %header) . " ]",$LOG) and
exit(1);
  exists $header{"P"} or print_OUT("Not p-value columns available,
these are the headers i found [ " . (keys %header) . " ]",$LOG) and
exit(1);
  # if there is a cols specifying the association test done. Only use
result from ADD tests, this is only for compatibility with PLINK
  if ( exists $header{"TEST"}){
      next if ( $data[$header{"TEST"}] ne "ADD");
  }
  #if (defined $v ) { print $data[$header{"SNP"}]," ", $data[$header
{"P"}],"\n"; }
      # if there is an affy id convert it to rsid.
  if ( defined $affy_to_rsid ) {
      if ($data[$header{"SNP"}] !~ m/^rs/){
          if (exists $affy_id{$data[$header{"SNP"}]}){ $data[$header
{"SNP"}] = $affy_id{$data[$header{"SNP"}]};}
      }
  }
  # skip if no P-value or p-value equal NA
  next if ( $data[$header{"P"}] eq "NA");
  next if ( $data[$header{"P"}] eq "");

  #generate a pseudo-hash for each snp with the association info
  my $A2 = "NA";
  my $A1 = "NA";
  my $OR = 1;
    my $SE = undef;
  my $BETA = undef;
  my $R2 = undef;
  my $STAT = undef;
  if (exists $header{"A2"}){ $A2 = $data[$header{"A2"}];}
  if (exists $header{"A1"}){ $A1 = $data[$header{"A1"}];}
  if (exists $header{"OR"}){
      $OR = $data[$header{"OR"}];
      if ((exists $header{"L95"}) and (exists $header{"U95"})){
          $SE = $header{"U95"} - $header{"L95"};
      }
  }
  if (exists $header{"BETA"}){ $BETA = $data[$header{"BETA"}]; }
```

```perl
  if (exists $header{"R2"}){ $R2 = $data[$header{"R2"}]; }
  if (exists $header{"STAT"}){ $STAT = $data[$header{"STAT"}]; }
  if ((not defined $SE) and (exists $header{"SE"})){ $SE = $data
[$header{"SE"}]; }
  my $effect = undef;
  if (exists $header{"OR"}){
    $effect = "or";
  } elsif (exists $header{"BETA"}){
      $effect = "beta";
  } elsif (exists $header{"STAT"}){
      $effect = "stat";
  }
  # do some checking
  if (defined $sample_score){
    # if there are not direction of effect defined quite analysis and
spit and error.
    if (not defined $effect){
        print_OUT("ERROR: No effect size measure or direction of
effect provided. I can not perform Sample Score analysis",$LOG);
        print_OUT("ERROR: Please provide an odd-ratio, beta or
regression coefficient value under the header OR, BETA or STAT,
respectively",$LOG);
        exit(1);
    }
  }
  $assoc_data{ $data[$header{"SNP"}] } = {
                                          'pvalue' => 1 ,
                                          'id' => $data[$header{"SNP"}],
                                          'a1' => $A1,
                                          'a2' => $A2,
                                          'or' => $OR,
                                                            'se' =>
$SE,
                                          'beta' => $BETA,
                                          'stat' => $STAT,
                                          'r2' => $R2,
                                          'effect_size_measure' =>
$effect,
                                        };
    # correct for genomic control if a lambda > 1 was specified.
    if ($lambda == 1) {
      push @{ $assoc_chis }, $data[$header{"P"}];
      $assoc_data{ $data[$header{"SNP"}] }->{'pvalue'} = $data[$header
{"P"}];
    } elsif ($lambda > 1) {# transform the p-value on a chi-square,
correct it by the inflation factor and transform it again on a p-value
      $assoc_data{ $data[$header{"SNP"}] }->{ 'pvalue' }= 1 -
gsl_cdf_chisq_P( gsl_cdf_chisq_Pinv( $data[$header{"P"}], 1 )/$lambda,
1 );
    } else {
      print_OUT("\nPlease check the lambda value is correct\n",$LOG);
      exit(1);
    }
}
close(ASSOC);

if (scalar keys %assoc_data == 0){
    print_OUT("\nNo SNPs with genetic association to used in the
analysis\n",$LOG);
    exit(1);
}
print_OUT("[ " . scalar (keys %assoc_data) . " ] SNPs with association
data",$LOG);

# Genomic Control adjustment
if (defined $gc_correction){
```

```perl
    print_OUT("Calculating lambda for genomic control correction",
$LOG);
    my $gc_lambda = get_lambda_genomic_control($assoc_chis);
    print_OUT("   '-> lambda (median) of [ $gc_lambda ]",$LOG);
    if ($gc_lambda > 1){
        print_OUT("   '-> Applying GC correction",$LOG);
        my $assoc_chis = [];
        foreach my $snp (keys %assoc_data) {
            if ( $assoc_data{ $snp }->{ 'pvalue' } == 1){
                push @{ $assoc_chis }, $assoc_data{ $snp }->
{ 'pvalue' };

                next;
            }
            my $snp_chi = gsl_cdf_chisq_Pinv ( 1 - $assoc_data
{ $snp }->{ 'pvalue' }, 1 );
            $snp_chi /=  $gc_lambda;
            $assoc_data{ $snp }->{ 'pvalue' }= 1 -
gsl_cdf_chisq_P( $snp_chi, 1 );
            push @{ $assoc_chis }, $assoc_data{ $snp }->
{ 'pvalue' };
        }
        $gc_lambda = get_lambda_genomic_control($assoc_chis);
        print_OUT("   '-> After correction the lambda is
[ $gc_lambda ]",$LOG);
    } else {
        print_OUT("   '-> GC correction not applied because lambda
is less than 1",$LOG);
    }
}

#read snp-to-gene mapping and store in a hash with key equal gene name
and value
# an array with the snps in the gene.
my @bim = ();
my @fam = ();
my $ped_map_genotypes;
if (defined $bfile) {
    # read the bim file with snp information and fam file with sample
information
    @bim = @{ read_bim("$bfile.bim",$affy_to_rsid,\%affy_id) };
    @fam = @{ read_fam("$bfile.fam") };
    print_OUT("[ " . scalar @bim .  " ] SNPs and [ " . scalar @fam .
" ] samples in genotype file",$LOG);
} elsif (defined $ped and defined $map){
    my ($fam_ref,$bim_ref);
    ($fam_ref,$ped_map_genotypes,$bim_ref) = read_map_and_ped($ped,
$map,$affy_to_rsid,\%affy_id);
    @fam = @$fam_ref;
    @bim = @$bim_ref;
print_OUT("[ " . scalar @bim .  " ] SNPs and [ " . scalar @fam .  " ]
samples in genotype file",$LOG);
} elsif (defined $geno_probs){
    print_OUT("Getting list of genotyped SNPs from [ $geno_probs ]",
$LOG);
    @bim = @{ get_snp_list_from_ox_format($gprobs, $gprobs_index) }
if ($geno_probs_format eq 'OXFORD');
    @bim = @{ get_snp_list_from_bgl_format($gprobs, $gprobs_index) }
if ($geno_probs_format eq 'BEAGLE');
    print_OUT("[ " . scalar @bim .  " ] SNPs in genotype file",$LOG);
}

my %bim_ids = ();
my $index = 0;
map {
  $bim_ids{$_->{snp_id}} = $index;
```

```perl
        $index++;
} @bim;

print_OUT("Loading SNP-2-Gene mapping");

for (my $i = 0; $i < scalar @$snpmap; $i++){
        if ($snpmap->[$i] =~ m/\#\d+\-\d+\#/) {
                print_OUT("   '-> Found [ # ] key on [ $snpmap->[$i] ]. I
will generate file names for chromosome interval.");
                my ($s,$e) = ($snpmap->[$i] =~ m/\#(\d+)\-(\d+)\#/);
                push @{$snpmap}, @{ make_file_name_array_interval($snpmap-
>[$i],$s,$e) };
                splice(@$snpmap,$i,1);
        }
        if ($snpmap->[$i] =~ m/\#/) {
                print_OUT("   '-> Found [ # ] key on [ $snpmap->[$i] ]. I
will generate file names for 26 chromosomes.");
                push @{$snpmap}, @{ make_file_name_array($snpmap->[$i]) };
                splice(@$snpmap,$i,1);
        }
}

my %gene = ();
my %snp_to_gene = ();
my %ids_map = ();
foreach my $snp_gene_mapping_file (@$snpmap){
        if (not -e $snp_gene_mapping_file){
                print_OUT("   '-> File [ $snp_gene_mapping_file ] does not
exist, moving on to next file",$LOG);
                next;
        }
        open( MAP, $snp_gene_mapping_file ) or print_OUT("Can not open
[ $snp_gene_mapping_file ] file",$LOG) and exit(1);
        print_OUT("   '-> Reading [ $snp_gene_mapping_file ]",$LOG);
        while ( my $read = <MAP> ) {
                chomp($read);
                # the line is separate in gene info and snps. the section are
separated by a tab.
                my ($chr,$start,$end,$ensembl,$hugo,$gene_status,$gene_type,
$description,@m) = split(/\t+/,$read);
                #check if gene was in the list of genes i want to analyze
                unless ( defined $all_genes ) {
                        next unless ( ( grep $_ eq $hugo, @genes ) or ( grep
$_ eq $ensembl, @genes ) );
                }
                if (defined $analysis_chr){
                        next if ($analysis_chr ne $chr);
                }

                my @first_snp_n_fields =  split(/\:/,$m[0]);
                if (4 !=  scalar @first_snp_n_fields){ $description .= splice
(@m,0,1); }

                # get all mapped snps within the distance threshold,
                my @mapped_snps = ();
                foreach my $s (@m) {
                        my ($id,$pos,$allele,$strand) = split(/\:/,$s);
                        next if (not defined $id);
                        if (( $pos >= $start) and ($pos <= $end)){ push
@mapped_snps, $id; }
                        elsif ( ( abs ($pos - $start) <= $distance*1_000 ) or
( abs ($pos - $end) <= $distance*1_000 )) { push @mapped_snps, $id; }
                }

                next if (scalar @mapped_snps == 0);
                # create a pseudo-hash with the gene info
```

```perl
        $gene{$ensembl} = {
                'hugo'          => $hugo,
                'ensembl'       => $ensembl,
                'chr'           => $chr,
                'start'         => $start,
                'end'           => $end,
                'gene_type'     => $gene_type,
                'snps'          => [],
                'minp'          => -9,
                'genotypes'     => null,
                'geno_mat_rows' => [],
                'cor' => null,
                'weights' => null,
                'pvalues' => [],
                'effect_size' => undef,
                'effect_size_se' => undef,
                'gene_status' => $gene_status,
                'desc' => $description,
        };

        # go over mapped snps and change convert affy ids to rsid.
        # and make a non-redundant set.
        my %nr_snps = ();
        foreach my $s (@mapped_snps) {
                if ( defined $affy_to_rsid ) {
                        if ($s !~ m/^rs/){
                                if (exists $affy_id{$s}){ $s = $affy_id{$s};}
                        }
                }
                # exclude snps not in the association file nor in the
bim file
                next unless ( exists $assoc_data{$s} );
                next unless ( exists $bim_ids{$s});
                $nr_snps{$s} = "";
        }
        @mapped_snps = keys %nr_snps;

        # go over the snps mapped to the gene and check if they are in
the map
        # and association files. If so, store the min p-value for the
gene.
        # if any of the snps is in the files the remove the gene from
the analysis.
        foreach my $s (@mapped_snps) {
                if (defined $v){ print_OUT("Mapping [ $s ] to
[ $ensembl ]",$LOG);}
                next if ( grep $_ eq $s, @{ $gene{$ensembl}->{snps} } );
                push @{ $snp_to_gene{$s} }, $ensembl;
                push @{ $gene{$ensembl}->{snps} }, $s;
                if ( $gene{$ensembl}->{minp} == -9) {
                        $gene{$ensembl}->{minp} = $assoc_data{$s}->
{pvalue};
                } elsif ( $assoc_data{$s}->{pvalue} < $gene{$ensembl}->
{minp} ) {
                        $gene{$ensembl}->{minp} = $assoc_data{$s}->
{pvalue};
                }
        }
        # remove gene if none of its snps is in the analysis.
        if ( scalar @{ $gene{$ensembl}->{snps} } == 0 ) {
                delete( $gene{$ensembl} );
        } else {
                if (defined $v){ print_OUT("Gene $ensembl $hugo
included in the analysis with [ ", scalar @{ $gene{$ensembl}->
{snps} }, " ] mapped SNPs",$LOG); }
                $ids_map{$hugo} = $ensembl;
```

```perl
                }

        }
        close(MAP);
}
print_OUT(" '->[ " . scalar (keys %gene) . " ] Genes read from SNP-2-
Gene Mapping files",$LOG);
print_OUT(" '->[ " . scalar (keys %snp_to_gene) . " ] SNPs mapped to
Genes and with association results will be analyzed",$LOG);

# complain if there is genes ledt for analysis
if (scalar keys %gene == 0){
        print_OUT("No genes mapped",$LOG);
        exit(1);
}

# read gene-set definition file
if (defined $gmt){
    my $total_gene_sets=0;
    print_OUT("Reading gene-set definitions",$LOG);
    foreach my $gene_set_file (@$gmt){
            print_OUT(" '-> Reading [ $gene_set_file ]",$LOG);
            open (GMT,$gene_set_file) or print_OUT("I can not open
[ $gene_set_file ] to read from.",$LOG) and die $!;
            while (my $line = <GMT>){
                chomp($line);
                my ( $p_name, $p_desc, @p_genes ) = split( /\t+/,
$line );

                my @gene_with_snp = ();
                # loop over the genes and check if the ids match
with any with SNPs
                foreach my $gn (@p_genes) {
                    if ( $gn =~ m/\// ) {
                        $gn =~ s/\s+//g;
                        my @genes = split( /\/{1,}/, $gn );
                        map {
                            if ( exists $ids_map{$_} ){
                                push @gene_with_snp,
$ids_map{$_};

                            } elsif ( exists $gene{$_} ){
                                push @gene_with_snp, $_;
                            } else {
                                next;
                            }
                        } @genes;
                    } else {
                        if ( exists $ids_map{$gn} ){
                            push @gene_with_snp, $ids_map
{$gn};

                        } elsif ( exists $gene{$gn} ){
                            push @gene_with_snp, $gn;
                        } else {
                            next;
                        }
                    }
                }
                my %tmp = ();
                map { $tmp{$_} = ""; } @gene_with_snp;
                @gene_with_snp = keys %tmp;
                # skip gene-set if does not complain with size
constrains
                next if (scalar @gene_with_snp < $gmt_min_size);
                next if (scalar @gene_with_snp > $gmt_max_size);

                %tmp = ();
```

```perl
                my ($chrs,$starts,$ends) = "";

                foreach my $g (@gene_with_snp) {
                    $chrs .= "$gene{$g}->{chr},";
                    $starts .= "$gene{$g}->{start},";
                    $ends .= "$gene{$g}->{end},";
                    foreach my $s (@{ $gene{$g}->{snps} }){
                        $tmp{$s} = "";
                    }
                }
                my @p_snps = keys %tmp;
                next if (scalar @p_snps < 2);

                map { push  @{ $snp_to_gene{ $_ } }, $p_name; }
@p_snps;

                $gene{$p_name} = {
                    'chr'       => $chrs,
                    'start'     => $starts,
                    'end'       => $ends,
                    'gene_type' => 'gene_set',
                    'snps'      => [@p_snps],
                    'minp'      => -9,
                    'genotypes' => null,
                    'geno_mat_rows' => [],
                    'cor' => null,
                    'weights' => null,
                    'pvalues' => [],
                    'effect_size' => undef,
                    'effect_size_se' => undef,
                    'gene_status' => 'KNOWN',
                    'ensembl' => $p_name,
                    'hugo' => $p_desc,
                    'desc' => $p_desc,
                    'name' => $p_name,
                    'genes' => [@gene_with_snp],
                };
                $total_gene_sets++;
            }
        }

    print_OUT("  '-> Just read [ $total_gene_sets ] gene-sets with
mapped genes with size [ $gmt_min_size ] and [ $gmt_max_size ]",$LOG);
}

# exclude genes if gene type were specified
if (defined $exclude_gene_type or defined $include_gene_type){
    print_OUT("Going to filter genes based on user defined options",
$LOG);
    if (defined $exclude_gene_type){
        print_OUT("   '-> Will exclude gene-types [ @
$exclude_gene_type ]",$LOG);
    }
    if (defined $include_gene_type){
        print_OUT("   '-> Will include gene-types [ @
$include_gene_type ]",$LOG);
    }
    foreach my $gn (keys %gene){
        # apply filter by gene type
        if (defined $exclude_gene_type){
            delete ($gene{$gn}) if ( grep $_ eq $gene{$gn}->
{gene_type} , @$exclude_gene_type );
        }
        if (defined $include_gene_type){
            if (exists $gene{$gn}) {
```

```perl
                                delete ($gene{$gn}) if (not grep $_ eq $gene
{$gn}->{gene_type}, @$include_gene_type );
                        }
                }
        }
}

print_OUT("Will analyse [ " . scalar (keys %gene) . " ] genes",$LOG);

# start a hash to store the SNP-to-SNP correlation values
my %correlation = ();

# if provided get the SNP-to-SNP correlation values from a tab
separated file with 3 cols:snp1 snp2 correlatio_value
if (defined $spearman){
    print_OUT("Reading SNP correlation from [ $spearman ]",$LOG);
    open( SPRMN, $spearman ) or print_OUT("I cannot open
[ $spearman ]",$LOG) and exit(1);
    while (my $ln = <SPRMN>){
        chomp($ln);
        my @a = split(/\s+/,$ln);
        # take the square of the correlation if they are r2 and user
wants to use r values
        $correlation{$a[0]}{$a[1]} = $a[2];
        $correlation{$a[1]}{$a[0]} = $a[2];
        # set self correlation to 1
        $correlation{$a[0]}{$a[0]} = 1;
        $correlation{$a[1]}{$a[1]} = 1;
    }
}
# start output file and print its header
print_OUT("Output file will be written to [ $out ]",$LOG);

# create a variable that will store a ref to a hash with the weights
my $weights = {};
if (defined @weights_file){
    print_OUT("Starting to read SNP weigths",$LOG);
    # create hash refs to store the name of the weight categories
    my $w_classes = {};
    my $w_counter = 0; # category counter, in case the file has not a
header.
    # loop over the files and read the weights
    foreach my $w_f (@weights_file){
        ($weights,$w_counter,$w_classes) = read_weight_file($w_f,
$w_counter,$w_classes,$weights, $w_header,\%snp_to_gene);
    }
    print_OUT("  '-> [ " . scalar (keys %{$weights}) . " ] weights
read",$LOG);
    # make a single weight vector for each snp
    foreach my $snp (keys %{$weights}){
        # get the snp weights sorted by category name
        my @tmp_all_w = map { $weights->{$snp}{$_}; } sort {$a cmp $b}
keys %{ $w_classes };
        # make piddle
        $weights->{$snp} = [@tmp_all_w];
    }
}

#start count to report advance
my $count = 0;
print_OUT("Starting to Calculate gene p-values",$LOG);

# if there are more than 100 genes change the $report variable in
order to report every ~ 10 % of genes.
unless (scalar keys %gene < 100){
```

```perl
    $report = int((scalar keys %gene)/100 + 0.5)*10;
}

# the genotype stack will store SNP genotypes if the SNP was mapped to
more than 1 gene.
# the SNP genotype will be deleted after is no longer needed.
# this reduces the IO and speeds up
my %snp_genotype_stack = ();

# if user defined a genotypes file. read genotypes and store a
genotype matrix (rows: samples, cols: genotypes)for each gene
if (defined $geno_probs) { # in case not plink binary files provided
and only a genotype prob file is given
    print_OUT("Reading genotype probabilities from [ $geno_probs ]",
$LOG);
    foreach my $gn (keys %gene){
        my $snp_list = [];
        my $lines = [];
        foreach my $mapped_snp (@{$gene{$gn}->{snps}}){
            next if (not exists $assoc_data{ $bim[$bim_ids
{$mapped_snp}]->{snp_id} } );
            if (defined $v){ print_OUT("Adding SNP [  $bim
[ $bim_ids{$mapped_snp} ]->{snp_id}  ] to genotypes of $gn",$LOG); }
            push @{$snp_list}, $mapped_snp;
            push @{$gene{$gn}->{geno_mat_rows}}, $mapped_snp;
            push @{$gene{$gn}->{pvalues}}, $assoc_data
{ $mapped_snp }->{pvalue};
            push @{$lines}, $bim_ids{$mapped_snp} + 1;
        }
        my ($p_mat,$d_mat) = extract_genotypes_for_snp_list
($snp_list,$lines,$g_prob_threshold,$geno_probs_format,$gprobs,
$gprobs_index);
        $gene{$gn}->{genotypes} = $d_mat;
        # check the range of the values. if the range is 0 then
the SNP is monomorphic and shoudl be dropped
        my ($min,$max,$min_d,$max_d)= $gene{$gn}->{genotypes}-
>minmaximum;
        my $non_zero_variance_index = which(($max - $min) != 0);
        # check if any SNPs needs to be dropped
        if ($non_zero_variance_index->isempty()){
            print_OUT(" [ $gn ] All SNPs are monomorphic, going
to next gene",$LOG) if (defined $v);
            next;
        }
        my $old_size = scalar @{ $gene{$gn}->{geno_mat_rows} };
        my $new_size = scalar list $non_zero_variance_index;
        if ( $old_size != $new_size){
            $gene{$gn}->{genotypes} = $gene{$gn}->{genotypes}->
(,$non_zero_variance_index);
            $gene{$gn}->{geno_mat_rows} = [@{$gene{$gn}->
{geno_mat_rows}}[$non_zero_variance_index->flat->list] ];
            $gene{$gn}->{pvalues} = [@{$gene{$gn}->{pvalues}}
[$non_zero_variance_index->flat->list] ];
            if (defined $v){
                my $diff = $old_size - $new_size;
                print_OUT("Dropping [ $diff ] monomorphic
SNPs",$LOG);
            }
        }

        # Calculate the genotypes correlation matrix
        my $more_corrs = "";
        ($gene{$gn}->{cor},$gene{$gn}->{cor_ld_r},$more_corrs)  =
deal_with_correlations($gene{$gn},\%correlation,$use_ld_as_corr);
        %correlation = (%correlation,%{$more_corrs});
```

```perl
        # Calculate the weights for the gene
        $gene{$gn}->{weights} = deal_with_weights(\@weights_file,
$gene{$gn},$w_maf,$weights);

        # Calculate average max genotype prob and use it as
weitghs
        $gene{$gn}->{weights} *= daverage $gene{$gn}->{genotypes};
        $gene{$gn}->{weights} /= $gene{$gn}->{weights}->sumover;

        # calculate gene p-values
        my $z_based_p = z_based_gene_pvalues($gene{$gn},$mnd);
        if (ref($z_based_p) ne 'HASH' and $z_based_p == -9){
                $z_based_p = {
                        'B_stouffer_fix' => "NA",
                        'B_stouffer_random' => "NA",
                        'B_fix' => "NA",
                        'B_random' => "NA",
                        'V_fix' => "NA",
                        'V_random' => "NA",
                        'Chi_fix' => "NA",
                        'Chi_random' => "NA",
                        'Z_P_fix' => "NA",
                        'Z_P_random' => "NA",
                        'Q' => "NA",
                        'Q_P' => "NA",
                        'I2' => "NA",
                        'tau_squared' => "NA",
                        'N' => scalar @{ $gene{$gn}->
{geno_mat_rows} },
                };
        }
        my $pvalue_based_p = gene_pvalue($gn);

        # check which analysis strategy to follow
        if (not defined $mnd){ # asymptotic
                print $OUT join "\t",($gene{$gn}->{ensembl},$gene
{$gn}->{hugo},$gene{$gn}->{gene_type},$gene{$gn}->{chr},$gene{$gn}->
{start},$gene{$gn}->{end},
                $gene{$gn}->{pvalues}->min,
                $pvalue_based_p->{sidak_min_p},
                $pvalue_based_p->{fisher},
                $pvalue_based_p->{fisher_chi},
                $pvalue_based_p->{fisher_df},
                $z_based_p->{'B_fix'},
                $z_based_p->{'V_fix'},
                $z_based_p->{'Z_P_fix'},
                $z_based_p->{'B_random'},
                $z_based_p->{'V_random'},
                $z_based_p->{'Z_P_random'},
                $z_based_p->{'I2'},
                $z_based_p->{'Q'},
                $z_based_p->{'Q_P'},
                $z_based_p->{'tau_squared'},
                $pvalue_based_p->{Meff_Galwey},
                $pvalue_based_p->{Meff_gao},
                scalar @{ $gene{$gn}->{geno_mat_rows} });
                print $OUT "\n";
        } else { # MND Sampling
                my $simulated_p = {
                        'sidak' => 'NA',
                        'fisher' => 'NA',
                        'z_fix' => 'NA',
                        'z_random' => 'NA',
                        'vegas' => 'NA',
                        'wise_p' => 'NA',
```

```perl
                        'wise_method' => 'NA',
                        'N' => 0,
                        'seen_vegas'      => 'NA',
                        'seen_sidak'      => 'NA',
                        'seen_fisher' => 'NA',
                        'seen_fix' => 'NA',
                        'seen_random' => 'NA',
                        'pareto_sidak_Phat' => 'NA',
                        'pareto_sidak_Phatci_low'  => 'NA',
                        'pareto_sidak_Phatci_up'      =>   'NA',
                        'pareto_fisher_Phat' => 'NA',
                        'pareto_fisher_Phatci_low' => 'NA',
                        'pareto_fisher_Phatci_up' =>  'NA',
                        'pareto_fix_Phat' => 'NA',
                        'pareto_fix_Phatci_low' => 'NA',
                        'pareto_fix_Phatci_up' => 'NA',
                        'pareto_random_Phat'     => 'NA',
                        'pareto_random_Phatci_low' =>  'NA',
                        'pareto_random_Phatci_up' =>  'NA',
                        'pareto_vegas_Phat'         => 'NA',
                        'pareto_vegas_Phatci_low' =>   'NA',
                        'pareto_vegas_Phatci_up'     => 'NA',
                };

                if ($z_based_p->{'Z_P_fix'} =~ m/[\d+]/){
                        $simulated_p = simulate_mnd($mnd_sim_target,
$mnd_sim_max,$pvalue_based_p->{sidak_min_p},$pvalue_based_p->
{Meff_gao},$pvalue_based_p->{fisher},$z_based_p->{'Z_P_fix'},
$z_based_p->{'Z_P_random'},$gene{$gn},
$mnd_sim_wise_correction_methods);
                }

                print $OUT join "\t",($gene{$gn}->{ensembl},$gene
{$gn}->{hugo},$gene{$gn}->{gene_type},$gene{$gn}->{chr},$gene{$gn}->
{start},$gene{$gn}->{end},
                        $gene{$gn}->pvalues->min,
                        $simulated_p->{vegas},
                        $simulated_p->{sidak},
                        $simulated_p->{fisher},
                        $simulated_p->{z_fix},
                        $simulated_p->{z_random},
                        $z_based_p->{'I2'},
                        $z_based_p->{'Q'},
                        $z_based_p->{'Q_P'},
                        $z_based_p->{'tau_squared'},
                        $simulated_p->{wise_p},
                        $simulated_p->{wise_method},
                        $simulated_p->{N},

                        $simulated_p->{'seen_vegas'}, # number of times
stats was seen
                        $simulated_p->{'seen_sidak'},
                        $simulated_p->{'seen_fisher'},
                        $simulated_p->{'seen_fix'},
                        $simulated_p->{'seen_random'},

                        $simulated_p->{'pareto_vegas_Phat'}, # VEGAS
                        $simulated_p->{'pareto_vegas_Phatci_low'},
                        $simulated_p->{'pareto_vegas_Phatci_up'},

                        $simulated_p->{'pareto_sidak_Phat'},
                        $simulated_p->{'pareto_sidak_Phatci_low'},
                        $simulated_p->{'pareto_sidak_Phatci_up'},

                        $simulated_p->{'pareto_fisher_Phat'}, # fisher
                        $simulated_p->{'pareto_fisher_Phatci_low'},
```

```perl
                    $simulated_p->{'pareto_fisher_Phatci_up'},

                    $simulated_p->{'pareto_fix_Phat'},# z fix
                    $simulated_p->{'pareto_fix_Phatci_low'},
                    $simulated_p->{'pareto_fix_Phatci_up'},

                    $simulated_p->{'pareto_random_Phat'}, # z random
                    $simulated_p->{'pareto_random_Phatci_low'},
                    $simulated_p->{'pareto_random_Phatci_up'},

                    $pvalue_based_p->{Meff_Galwey},
                    $pvalue_based_p->{Meff_gao},
                    scalar @{ $gene{$gn}->{geno_mat_rows} });
                print $OUT "\n";
            }

            # remove LD measures and genotypes from the stack that
    will not use again to free memory
            foreach my $snp (  @{ $gene{$gn}->{geno_mat_rows} } ){
                if (scalar @{ $snp_to_gene{ $snp } } == 1){
                        foreach my $pair ( keys %{$correlation{$snp}})
    {
                            delete($correlation{$snp}{$pair});
                            delete($correlation{$pair}{$snp});
                    }
                    delete($snp_genotype_stack{ $snp });
                    delete($snp_to_gene{ $snp });
                } else {
                    splice(@{ $snp_to_gene{ $snp } },0,1);
                }
            }
            # delete the gene's data to keep memory usage low
            delete($gene{$gn});
            $count++;
            &report_advance($count,$report,"Genes");

        }
} elsif (defined $bfile) {
    print_OUT("Reading genotypes from [ $bfile.bed ]",$LOG);
    # open genotype file
    my $bed = new IO::File;
    $bed->open("<$bfile.bed") or print_OUT("I can not open binary PLINK
    file [ $bfile ]",$LOG) and exit(1);
    binmode($bed); # set file type to binary
    # check if the file is a PLINK file in the proper format by checking
    the first 3 bytes
    my ($buffer,$n_bytes);
    my $plink_bfile_signature = "";
    read $bed, $plink_bfile_signature, 3;
    if (unpack("B24",$plink_bfile_signature) ne
    '0110110000011011000000001'){
        print_OUT("Binary file is not in SNP-major format, please check
    you files\n",$LOG);
        exit(1);
    } else { print_OUT("Binary file is on SNP-major format",$LOG); }
    # calculate how many bytes are needed  to encode a SNP
    # each byte has 8 bits with information for 4 genotypes
    my $N_bytes_to_encode_snp = (scalar @fam)/4; # four genotypes per
    byte
    # if not exact round it up
    if (($N_bytes_to_encode_snp - int($N_bytes_to_encode_snp)) != 0  )
    { $N_bytes_to_encode_snp = int($N_bytes_to_encode_snp) + 1;}
    # loop over all genes and extract the genotypes of the SNPs
    foreach my $gn (keys %gene){

        # this will store the genotypes
```

```perl
    my $matrix;
    # loop over the snps mapped to the gene
    foreach my $mapped_snp (@{$gene{$gn}->{snps}}){
            # skip if it does not have association information
            next if (not exists $assoc_data{ $bim[$bim_ids
{$mapped_snp}]->{snp_id} } );

            if (defined $v){ print_OUT("Adding SNP [  $bim[ $bim_ids
{$mapped_snp} ]->{snp_id}  ] to genotypes of $gn",$LOG); }
            if (exists $snp_genotype_stack{$mapped_snp}) {
            if (defined $v){
                    print_OUT("   '-> SNP [ $mapped_snp] already
read",$LOG);
            }
            push @{ $matrix }, $snp_genotype_stack{$mapped_snp};
        } else {
            # because we know the index of the SNP in the
genotype file we know on which byte its information starts
            my $snp_byte_start = $N_bytes_to_encode_snp*$bim_ids
{$mapped_snp};
            # here i extract the actual genotypes
            my @snp_genotypes = @{ extract_binary_genotypes
(scalar @fam,$N_bytes_to_encode_snp,$snp_byte_start,$bed) };
            # store the genotypes.
            # if a snp does not use the 8 bits of a byte the
rest of the bits are fill with missing values
            # here i extract the number of genotypes
corresponding to the number of samples

            my $maf = get_maf([@snp_genotypes[0..scalar @fam -
1]] ); # check the maf of the SNP
            next if ($maf == 0 or $maf ==1);  # go to next if it
is monomorphic
            push @{ $matrix }, [@snp_genotypes[0..scalar @fam -
1]];
            $snp_genotype_stack{$mapped_snp} = [@snp_genotypes
[0..scalar @fam - 1]];
        }

        # add snp id to matrix row names
        push @{ $gene{$gn}->{geno_mat_rows} }, $mapped_snp;
        # store the p-value of the snp
        push @{ $gene{$gn}->{pvalues} }, $assoc_data
{ $mapped_snp }->{pvalue};
        my $effect_measure = $assoc_data{ $mapped_snp }->
{effect_size_measure};
            if (defined $effect_measure){
            if ($effect_measure eq 'or'){
                    push @{ $gene{$gn}->{effect_size} }, log
$assoc_data{ $mapped_snp }->{$effect_measure};
            } else {
                    push @{ $gene{$gn}->{effect_size} },
$assoc_data{ $mapped_snp }->{$effect_measure};
            }
            if (defined $assoc_data{ $mapped_snp }->{se}){
                    if ($effect_measure eq 'or'){
                            #push @{ $gene{$gn}->{effect_size_se} },
abs log $assoc_data{ $bim[ $bim_ids{$mapped_snp} ]->{snp_id} }->{se};
                            push @{ $gene{$gn}->{effect_size_se} },
$assoc_data{ $mapped_snp }->{se};
                    } else {
                            push @{ $gene{$gn}->{effect_size_se} },
$assoc_data{ $mapped_snp }->{se};
                    }
            }
        }
```

```perl
    }
    # generate the genotype matrix as a PDL piddle
    $gene{$gn}->{genotypes} = pdl $matrix;

    # Calculate the genotypes correlation matrix
    my $more_corrs = "";
    ($gene{$gn}->{cor},$gene{$gn}->{cor_ld_r},$more_corrs)  =
deal_with_correlations($gene{$gn},\%correlation,$use_ld_as_corr);
    %correlation = (%correlation,%{$more_corrs});

    # Calculate the weights for the gene
    $gene{$gn}->{weights} = deal_with_weights(\@weights_file,$gene
{$gn},$w_maf,$weights);
    # calculate gene p-values
    my $z_based_p = z_based_gene_pvalues($gene{$gn},$mnd);
    if (ref($z_based_p) ne 'HASH' and $z_based_p == -9){
            $z_based_p = {
                    'B_stouffer_fix' => "NA",
                    'B_stouffer_random' => "NA",
                    'B_fix' => "NA",
                    'B_random' => "NA",
                    'V_fix' => "NA",
                    'V_random' => "NA",
                    'Chi_fix' => "NA",
                    'Chi_random' => "NA",
                    'Z_P_fix' => "NA",
                    'Z_P_random' => "NA",
                    'Q' => "NA",
                    'Q_P' => "NA",
                    'I2' => "NA",
                    'tau_squared' => "NA",
                    'N' => scalar @{ $gene{$gn}->{geno_mat_rows} },
            };
    }
    my $pvalue_based_p = gene_pvalue($gn);

    # check which analysis strategy to follow
    if (not defined $mnd){ # asymptotic
            print $OUT join "\t",($gene{$gn}->{ensembl},$gene{$gn}->
{hugo},$gene{$gn}->{gene_type},$gene{$gn}->{chr},$gene{$gn}->{start},
$gene{$gn}->{end},
                    $gene{$gn}->{pvalues}->min,
                    $pvalue_based_p->{sidak_min_p},
                    $pvalue_based_p->{fisher},
                    $pvalue_based_p->{fisher_chi},
                    $pvalue_based_p->{fisher_df},
                    $z_based_p->{'B_fix'},
                    $z_based_p->{'V_fix'},
                    $z_based_p->{'Z_P_fix'},
                    $z_based_p->{'B_random'},
                    $z_based_p->{'V_random'},
                    $z_based_p->{'Z_P_random'},
                    $z_based_p->{'I2'},
                    $z_based_p->{'Q'},
                    $z_based_p->{'Q_P'},
                    $z_based_p->{'tau_squared'},
                    $pvalue_based_p->{Meff_Galwey},
                    $pvalue_based_p->{Meff_gao},
                    scalar @{ $gene{$gn}->{geno_mat_rows} });
            print $OUT "\n";
    } else { # MND Sampling
            my $simulated_p = {
                    'sidak' => 'NA',
                    'fisher' => 'NA',
                    'z_fix' => 'NA',
                    'z_random' => 'NA',
```

```perl
                    'vegas' => 'NA',
                    'wise_p' => 'NA',
                    'wise_method' => 'NA',
                    'N' => 0,
                    'seen_vegas'    => 'NA',
                    'seen_sidak'    => 'NA',
                    'seen_fisher' => 'NA',
                    'seen_fix' => 'NA',
                    'seen_random' => 'NA',
                    'pareto_sidak_Phat' => 'NA',
                    'pareto_sidak_Phatci_low'  => 'NA',
                    'pareto_sidak_Phatci_up'   => 'NA',
                    'pareto_fisher_Phat' => 'NA',
                    'pareto_fisher_Phatci_low' => 'NA',
                    'pareto_fisher_Phatci_up' => 'NA',
                    'pareto_fix_Phat'     => 'NA',
                    'pareto_fix_Phatci_low' => 'NA',
                    'pareto_fix_Phatci_up' => 'NA',
                    'pareto_random_Phat'  => 'NA',
                    'pareto_random_Phatci_low' =>  'NA',
                    'pareto_random_Phatci_up' =>  'NA',
                    'pareto_vegas_Phat'        => 'NA',
                    'pareto_vegas_Phatci_low' =>  'NA',
                    'pareto_vegas_Phatci_up'   => 'NA',
            };

        if ($z_based_p->{'Z_P_fix'} =~ m/[\d+]/){
                $simulated_p = simulate_mnd($mnd_sim_target,
$mnd_sim_max,$pvalue_based_p->{sidak_min_p},$pvalue_based_p->
{Meff_gao},$pvalue_based_p->{fisher},$z_based_p->{'Z_P_fix'},
$z_based_p->{'Z_P_random'},$gene{$gn},
$mnd_sim_wise_correction_methods);
            }

        print $OUT join "\t",($gene{$gn}->{ensembl},$gene{$gn}->{hugo},
$gene{$gn}->{gene_type},$gene{$gn}->{chr},$gene{$gn}->{start},$gene
{$gn}->{end},
                    $gene{$gn}->{pvalues}->min,
                    $simulated_p->{vegas},
                    $simulated_p->{sidak},
                    $simulated_p->{fisher},
                    $simulated_p->{z_fix},
                    $simulated_p->{z_random},
                    $z_based_p->{'I2'},
                    $z_based_p->{'Q'},
                    $z_based_p->{'Q_P'},
                    $z_based_p->{'tau_squared'},
                    $simulated_p->{wise_p},
                    $simulated_p->{wise_method},
                    $simulated_p->{N},

                    $simulated_p->{'seen_vegas'}, # number of times stats
was seen
                    $simulated_p->{'seen_sidak'},
                    $simulated_p->{'seen_fisher'},
                    $simulated_p->{'seen_fix'},
                    $simulated_p->{'seen_random'},

                    $simulated_p->{'pareto_vegas_Phat'}, # VEGAS
                    $simulated_p->{'pareto_vegas_Phatci_low'},
                    $simulated_p->{'pareto_vegas_Phatci_up'},

                    $simulated_p->{'pareto_sidak_Phat'},
                    $simulated_p->{'pareto_sidak_Phatci_low'},
                    $simulated_p->{'pareto_sidak_Phatci_up'},
```

```perl
                    $simulated_p->{'pareto_fisher_Phat'}, # fisher
                    $simulated_p->{'pareto_fisher_Phatci_low'},
                    $simulated_p->{'pareto_fisher_Phatci_up'},

                    $simulated_p->{'pareto_fix_Phat'},# z fix
                    $simulated_p->{'pareto_fix_Phatci_low'},
                    $simulated_p->{'pareto_fix_Phatci_up'},

                    $simulated_p->{'pareto_random_Phat'}, # z random
                    $simulated_p->{'pareto_random_Phatci_low'},
                    $simulated_p->{'pareto_random_Phatci_up'},

                    $pvalue_based_p->{Meff_Galwey},
                    $pvalue_based_p->{Meff_gao},
                    scalar @{ $gene{$gn}->{geno_mat_rows} });
                    print $OUT "\n";
        }

        # remove LD measures and genotypes from the stack that will not
    use again to free memory
        foreach my $snp (  @{ $gene{$gn}->{geno_mat_rows} } ){
            if (scalar @{ $snp_to_gene{ $snp } } == 1){
                    foreach my $pair ( keys %{$correlation{$snp}}){
                        delete($correlation{$snp}{$pair});
                        delete($correlation{$pair}{$snp});
                    }
                    delete($snp_genotype_stack{ $snp });
                    delete($snp_to_gene{ $snp });
            } else {
                    splice(@{ $snp_to_gene{ $snp } },0,1);
            }
        }
        # delete the gene's data to keep memory usage low
        delete($gene{$gn});
        $count++;
        &report_advance($count,$report,"Genes");
    }
} elsif (defined $ped and defined $map){
    my $genotypes = pdl @{ $ped_map_genotypes };
    foreach (my $index_snp = 0; $index_snp <  scalar @bim; $index_snp
++){
        # store SNP genotypes only if it has association data
            if (exists $assoc_data{ ${ $bim[$index_snp] }{snp_id} } ){
                # for every gene mapped to this SNP, push inside the
    genotype matrix this SNP genotypes.
                    foreach my $gn (@{ $snp_to_gene{${ $bim
    [$index_snp] }{snp_id} } }){
                        if (defined $v){ print_OUT("Adding SNP [  $
    { $bim[$index_snp] }{snp_id}  ] to genotypes of $gn",$LOG); }
                        next if ( grep $_ eq ${ $bim[$index_snp] }
    {snp_id} , @{ $gene{$gn}->{geno_mat_rows} } );

                        my $maf = get_maf([ $genotypes(,$index_snp)-
    >list ] ); # check the maf of the SNP
                        next if ($maf == 0 or $maf ==1);  # go to next
    if it is monomorphic

                        # add the genotypes to the genotype matrix
                        $gene{$gn}->{genotypes} = $gene{$gn}->
    {genotypes}->glue(1,$genotypes(,$index_snp));
                        push @{ $gene{$gn}->{geno_mat_rows} }, ${ $bim
    [$index_snp] }{snp_id};
                        push @{ $gene{$gn}->{pvalues} }, $assoc_data
    { ${ $bim[$index_snp] }{snp_id} }->{pvalue};
```

```perl
                        my $effect_measure = $assoc_data{ ${ $bim
[$index_snp] }{snp_id}  }->{effect_size_measure};
                        if (defined $effect_measure){
                            if ($effect_measure eq 'or'){
                                push @{ $gene{$gn}->
{effect_size} }, log $assoc_data{ ${ $bim[$index_snp] }{snp_id} }->
{$effect_measure};
                            } else {
                                push @{ $gene{$gn}->
{effect_size} }, $assoc_data{ ${ $bim[$index_snp] }{snp_id} }->
{$effect_measure};
                            }
                            if (defined $assoc_data{ ${ $bim
[$index_snp] }{snp_id} }->{se}){
                                if ($effect_measure eq 'or'){
                                    #push @{ $gene{$gn}->
{effect_size_se} }, abs log $assoc_data{ $bim[ $bim_ids
{$mapped_snp} ]->{snp_id} }->{se};
                                    push @{ $gene{$gn}->
{effect_size_se} }, $assoc_data{ ${ $bim[$index_snp] }{snp_id} }->
{se};
                                } else {
                                    push @{ $gene{$gn}->
{effect_size_se} }, $assoc_data{ ${ $bim[$index_snp] }{snp_id} }->
{se};
                                }
                            }
                        }
                    if (scalar @{ $gene{$gn}->{snps} } == scalar @
{ $gene{$gn}->{geno_mat_rows} }){
                            # Calculate the genotypes correlation
matrix
                            my $more_corrs = "";
                            ($gene{$gn}->{cor},$gene{$gn}->
{cor_ld_r},$more_corrs)  = deal_with_correlations($gene{$gn},\
%correlation,$use_ld_as_corr);
                            %correlation = (%correlation,%
{$more_corrs});

                            # Calculate the weights for the gene
                            $gene{$gn}->{weights} =
deal_with_weights(\@weights_file,$gene{$gn},$w_maf,$weights);

                            # calculate gene p-values
                            my $z_based_p = z_based_gene_pvalues
($gene{$gn},$mnd);

                            if (ref($z_based_p) ne 'HASH' and
$z_based_p == -9){

                                $z_based_p = {
                                    'B_stouffer_fix' => "NA",
                                    'B_stouffer_random' => "NA",
                                    'B_fix' => "NA",
                                    'B_random' => "NA",
                                    'V_fix' => "NA",
                                    'V_random' => "NA",
                                    'Chi_fix' => "NA",
                                    'Chi_random' => "NA",
                                    'Z_P_fix' => "NA",
                                    'Z_P_random' => "NA",
                                    'Q' => "NA",
                                    'Q_P' => "NA",
                                    'I2' => "NA",
                                    'tau_squared' => "NA",
                                    'N' => scalar @{ $gene{$gn}-
>{geno_mat_rows} },
                                };
```

```perl
                                }
                                my $pvalue_based_p = gene_pvalue($gn);

                                print $OUT join "\t",(($gene{$gn}->
{ensembl},$gene{$gn}->{hugo},$gene{$gn}->{gene_type},$gene{$gn}->
{chr},$gene{$gn}->{start},$gene{$gn}->{end},
                                $gene{$gn}->{pvalues}->min,
                                $pvalue_based_p->{sidak_min_p},
                                $pvalue_based_p->{fisher},
                                $pvalue_based_p->{fisher_chi},
                                $pvalue_based_p->{fisher_df},
                                $z_based_p->{'B_fix'},
                                $z_based_p->{'V_fix'},
                                $z_based_p->{'Z_P_fix'},
                                $z_based_p->{'B_random'},
                                $z_based_p->{'V_random'},
                                $z_based_p->{'Z_P_random'},
                                $z_based_p->{'I2'},
                                $z_based_p->{'Q'},
                                $z_based_p->{'Q_P'},
                                $z_based_p->{'tau_squared'},
                                $pvalue_based_p->{Meff_Galwey},
                                $pvalue_based_p->{Meff_gao},
                                scalar @{ $gene{$gn}->
{geno_mat_rows} });
                                print $OUT "\n";
                                # remove LD measures and genotypes from
the stack that will not use again to free memory
                                foreach my $snp ( @{ $gene{$gn}->
{geno_mat_rows} } ){
                                        if (scalar @{ $snp_to_gene
{ $snp } } == 1){
                                                foreach my $pair ( keys %
{$correlation{$snp}}){
                                                        delete($correlation
{$snp}{$pair});
                                                        delete($correlation
{$pair}{$snp});
                                                }
                                                delete($snp_genotype_stack
{ $snp });
                                                delete($snp_to_gene
{ $snp });
                                        } else {
                                                splice(@{ $snp_to_gene
{ $snp } },0,1);
                                        }
                                }
                                # delete the gene's data to keep memory
usage low
                                delete($gene{$gn});
                                $count++;
                                &report_advance($count,$report,"Genes");
                        } # close if
                } # close foreach my $gn
        } # close if exists
    } # close foreach loop over snps
} else {
    print_OUT("WARNING: Gene p-values will be calculated with the
precomputed correlation only. If correlation for some SNPs pairs are
missing you may get wrong results, please check your inputs for
completeness",$LOG);
}
```

```perl
# if the user want to get the correlation values print the
*.correlation file
if (defined $print_cor){
  open (COR,">$out.correlation") or print_OUT("Cannot open
[ $out.correlation ] to write to",$LOG) and exit(1);
  foreach my $snp1 (keys %correlation) {
    foreach my $snp2 (keys %{$correlation{$snp1}}  ) {
      next if ($snp1 eq $snp2);
        printf COR ("$snp1 $snp2 %.3f\n",abs($correlation{ $snp1 }
{ $snp2 }));
      delete($correlation{ $snp1 }{ $snp2 });
    }
  }
  close(COR);
}

print_OUT("Well Done!!",$LOG);

$LOG->close();
$OUT->close();

exit(0);

sub deal_with_weights {
    my $w_files = shift;
    my $gn = shift;
    my $w_by_maf =shift;
    my $weights = shift;
    my $back = null;
    my $N = scalar @{ $gn->{geno_mat_rows} };
    if (defined @$w_files){
        $back = generate_weights_for_a_gene($gn->{geno_mat_rows},
$weights);
    } else {
        $back = ones $N;
        $back *= 1/$N;
        $back /= $back->dsum;
    }
    # if desired weigth by the 1/MAF
    if (defined $w_by_maf){
        my $MAF_w = get_maf_weights($gn->{genotypes});
        $back *= $MAF_w->transpose;
    }

    # Correct the weights by the LD in the gene.
    # the new weight will be the weigthed mean of the gene.
    # the new weight will be the weigthed mean of the gene weights.
    # the weights for the mean are the correlation between the SNP,
In that way the
    # weights reflect the correlation pattern of the SNPs
    # weights reflect the correlation pattern of the SNPs
    my $C;
    if (defined $gn->{cor_ld_r}){
        $C = $gn->{cor_ld_r};
    } else {
        $C = $gn->{cor};
    }
    my $w_matrix = $back * abs($C); # multiply the weights by the
correaltions
    my @dims = $w_matrix->dims();
    $w_matrix = pdl map { $w_matrix->(,$_)->flat->sum/$back->dsum; }
0 .. $dims[1] - 1; # sum the rows divided by sum of the weights used
    if ($w_matrix->min == 0){ $w_matrix += $w_matrix->(which
($w_matrix == 0))->min/$w_matrix->nelem; } # make sure NO weights
equal 0
```

```perl
    $w_matrix /= $w_matrix->sum; # make sure weights sum 1

    $back = $w_matrix/$w_matrix->sum;
    return($back);

}

sub simulate_mnd {
    my $target = shift; # min times stat must be seen to finish
simulations
    my $MAX = shift; # max number of simulations
    my $sidak = shift;
    my $k = shift;
    my $fisher = shift;
    my $z_fix = shift;
    my $z_random = shift;
    my $gene_data = shift; # hash ref with gene informatio
    my $compare_wise_p = shift; # ARRAY ref

    my $max_step_size = 100_000;
    my $total = 0;
    my $step= 1000;

    my ($cov,$status) = check_positive_definite($gene_data->{cor},
1e-8);
    if ($status == 1){
        print "Matrix never positive definite $gene_data->
{ensembl}\t$gene_data->{hugo}\t",scalar @{$gene_data->
{'geno_mat_rows'}},"\n";
        return(-9);
    }
    my $cholesky = mchol($cov);
    my $SEEN = zeroes 4;
    my $stats_bag = [];
    my $fake_gene = {
        'pvalues' => "",
        'effect_size' => $gene_data->{'effect_size'},
        'effect_size_se' => $gene_data->{'effect_size_se'},
        'cor' => $gene_data->{'cor'},
        'weights' => $gene_data->{'weights'},
        'geno_mat_rows' => $gene_data->{'geno_mat_rows'},
    };

    my $vegas_stat = dsum gsl_cdf_chisq_Pinv(1-$gene_data->{pvalues},
1);
    my $vegas_count = 0;
    my $fix_null_stats = [];
    my $random_null_stats = [];
    my $sidak_null_stats = [];
    my $fisher_null_stats = [];
    my $vegas_null_stats = [];

    while ($SEEN->min < $target){
        my ($sim,$c) = rmnorm($step,0,$cov,$cholesky);
        my $sim_chi_df1 = $sim**2;
        $vegas_count += dsum( $sim_chi_df1->xchg(0,1)->dsumover >=
$vegas_stat);
        push @{ $vegas_null_stats }, $sim_chi_df1->xchg(0,1)-
>dsumover->list;
        my $sim_p =  1 - gsl_cdf_chisq_P($sim_chi_df1,1);
        $sim_chi_df1 = '';
        $sim = '';
        for (my $sim_n = 0; $sim_n < $sim_p->getdim(0); $sim_n++){
            $fake_gene->{'pvalues'} = $sim_p->($sim_n,)->flat;
```

```perl
                    my $sim_n_gene_p = z_based_gene_pvalues($fake_gene,
$mnd);

                    my ($sim_fisher_chi_stat,$sim_fisher_df) =
get_makambi_chi_square_and_df($gene_data->{cor},$gene_data->{weights},
$fake_gene->{'pvalues'} );
                    my $sim_fisher_p_value = sclr double  1 -
gsl_cdf_chisq_P($sim_fisher_chi_stat, $sim_fisher_df );
                    my $sim_sidak = 1 - (1 - $fake_gene->{'pvalues'}-
>min)**$k;

                    $SEEN->(0)++ if ( $sidak >= $sim_sidak );
                    $SEEN->(1)++ if ( $fisher >= $sim_fisher_p_value );
                    $SEEN->(2)++ if ( $z_fix >= $sim_n_gene_p->
{'Z_P_fix'} );
                    $SEEN->(3)++ if ( $z_random >= $sim_n_gene_p->
{'Z_P_random'} );
                    push @{ $fix_null_stats    }, -1 *
gsl_cdf_ugaussian_Pinv( $sim_n_gene_p->{'Z_P_fix'} );
                    push @{ $random_null_stats }, -1 *
gsl_cdf_ugaussian_Pinv( $sim_n_gene_p->{'Z_P_random'} );
                    push @{ $sidak_null_stats  }, -1 *
gsl_cdf_ugaussian_Pinv( $sim_sidak );
                    push @{ $fisher_null_stats }, -1 *
gsl_cdf_ugaussian_Pinv( $sim_fisher_p_value );

                    my @sim_gene_ps = ( $sim_sidak,$sim_fisher_p_value,
$sim_n_gene_p->{'Z_P_fix'},$sim_n_gene_p->{'Z_P_random'} );
                    push @{$stats_bag}, [ @sim_gene_ps[ @
$compare_wise_p ] ];
            }
            $total += $step;
            if ($SEEN->min != 0){
                    $step = 1.1*(10*($total)/$SEEN->min);
            } elsif ($step < $max_step_size){
                    $step *=10;
            }
            if ($step > $MAX){ $step = $MAX; }

            last if ($total > $MAX);
    }
    my $back = {
            'sidak' => sclr ($SEEN->(0)+1)/($total +1),
            'fisher' => sclr ($SEEN->(1)+1)/($total +1),
            'z_fix' => sclr ($SEEN->(2)+1)/($total +1),
            'z_random' => sclr ($SEEN->(3)+1)/($total +1),
            'vegas' => ($vegas_count +1)/($total +1),
            'N' => $total,
    };

    $fix_null_stats         = pdl $fix_null_stats;
    $random_null_stats = pdl $random_null_stats;
    $sidak_null_stats  = pdl $sidak_null_stats;
    $fisher_null_stats = pdl $fisher_null_stats;
    $vegas_null_stats  = pdl $vegas_null_stats;

    my $fix_observed    = -1 * gsl_cdf_ugaussian_Pinv( $z_fix );
    my $random_observed = -1 * gsl_cdf_ugaussian_Pinv( $z_random );
    my $sidak_observed  = -1 * gsl_cdf_ugaussian_Pinv( $sidak );
    my $fisher_observed = -1 * gsl_cdf_ugaussian_Pinv( $fisher );

    my ($pareto_fix_Phat,    $pareto_fix_Phatci_low,
    $pareto_fix_Phatci_up)          = Pareto_Distr_Fit::Pgpd
( $fix_observed,  $fix_null_stats,  250,0.05);
```

```perl
    my ($pareto_random_Phat,$pareto_random_Phatci_low,
       $pareto_random_Phatci_up)= Pareto_Distr_Fit::Pgpd
( $random_observed,    $random_null_stats,    250,0.05);
    my ($pareto_sidak_Phat,  $pareto_sidak_Phatci_low,
       $pareto_sidak_Phatci_up) = Pareto_Distr_Fit::Pgpd
( $sidak_observed,      $sidak_null_stats,     250,0.05);
    my ($pareto_fisher_Phat,$pareto_fisher_Phatci_low,
       $pareto_fisher_Phatci_up)= Pareto_Distr_Fit::Pgpd
( $fisher_observed,     $fisher_null_stats,    250,0.05);
    my ($pareto_vegas_Phat,  $pareto_vegas_Phatci_low,
       $pareto_vegas_Phatci_up) = Pareto_Distr_Fit::Pgpd
( $vegas_stat,          $vegas_null_stats,     250,0.05);

    $back->{'seen_vegas'}    = $vegas_count;
    $back->{'seen_sidak'}    = sclr $SEEN->(0);
    $back->{'seen_fisher'}   = sclr $SEEN->(1);
    $back->{'seen_fix'}      = sclr $SEEN->(2);
    $back->{'seen_random'}   = sclr $SEEN->(3);
    # pareto p-values
    # sidak
    $back->{'pareto_sidak_Phat'}        = sclr $pareto_sidak_Phat;
    $back->{'pareto_sidak_Phatci_low'}  = $pareto_sidak_Phatci_low;
    $back->{'pareto_sidak_Phatci_up'}   =  $pareto_sidak_Phatci_up;

    # fisher
    $back->{'pareto_fisher_Phat'}       = sclr $pareto_fisher_Phat;
    $back->{'pareto_fisher_Phatci_low'} = $pareto_fisher_Phatci_low;
    $back->{'pareto_fisher_Phatci_up'}  =  $pareto_fisher_Phatci_up;

    # fix
    $back->{'pareto_fix_Phat'}              = sclr
$pareto_fix_Phat;
    $back->{'pareto_fix_Phatci_low'}    = $pareto_fix_Phatci_low;
    $back->{'pareto_fix_Phatci_up'}     =  $pareto_fix_Phatci_up;

    # random
    $back->{'pareto_random_Phat'}       = sclr $pareto_random_Phat;
    $back->{'pareto_random_Phatci_low'} =  $pareto_random_Phatci_low;
    $back->{'pareto_random_Phatci_up'}  =  $pareto_random_Phatci_up;

    # vegas
    $back->{'pareto_vegas_Phat'}        = sclr $pareto_vegas_Phat;
    $back->{'pareto_vegas_Phatci_low'} =  $pareto_vegas_Phatci_low;
    $back->{'pareto_vegas_Phatci_up'}   =  $pareto_vegas_Phatci_up;

    my @methods = ('sidak','fisher','z_fix','z_random');
    @methods = @methods[@$compare_wise_p];
    my $methods_p = pdl ($back->{sidak},$back->{fisher},$back->
{z_fix},$back->{z_random});
    $stats_bag = pdl $stats_bag;
    $stats_bag = gsl_cdf_ugaussian_Pinv($stats_bag);
    my $stat_cor = cov_shrink($stats_bag);
    my ($stats_Meff_gao,$stats_Meff_galwey) = number_effective_tests
(\$stat_cor->{cor});

    my $min = $methods_p->minimum_ind();
    $back->{'wise_p'} = sclr 1 - ( 1 - $methods_p->( $min ) )**
$stats_Meff_gao;
    $back->{'wise_method'} = $methods[$min];
    $back->{'stats_gao'} = $stats_Meff_gao;
    return( $back );
}

sub deal_with_correlations {
```

```perl
    my $gn = shift;
    my $correlation = shift;
    my $use_ld = shift;
    my %new_corrs = ();
    my $shrunken_matrix = cov_shrink($gn->{genotypes}->transpose);
    my $pearson_cor = $shrunken_matrix->{cor};
    my $n = scalar @{ $gn->{'geno_mat_rows'} };
    my $cor_ld_r = undef; #zeroes $n,$n;
    if (defined $use_ld){
            $cor_ld_r = stretcher(ones $n);
            for (my $i = 0; $i < $n; $i++){
                for (my $j = $i; $j < $n; $j++){
                    next if ($j == $i);
                    if (exists $correlation->{$gn->
{'geno_mat_rows'}->[$i]}{$gn->{'geno_mat_rows'}->[$j]}){
                        set $cor_ld_r, $i, $j, $correlation->
{$gn->{'geno_mat_rows'}->[$i]}{$gn->{'geno_mat_rows'}->[$j]};
                        set $cor_ld_r, $j, $i, $correlation->
{$gn->{'geno_mat_rows'}->[$i]}{$gn->{'geno_mat_rows'}->[$j]};
                    } else {
                        my $ld = calculate_LD_stats([ $gn->
{'genotypes'}->(,$i)->list ],[ $gn->{'genotypes'}->(,$j)-
>list ]);

                        set $cor_ld_r, $i, $j, $ld->{r};
                        set $cor_ld_r, $j, $i, $ld->{r};

                        $new_corrs{ $gn->{'geno_mat_rows'}->
[$i] }{ $gn->{'geno_mat_rows'}->[$j] } = $ld->{r};
                        $new_corrs{ $gn->{'geno_mat_rows'}->
[$j] }{ $gn->{'geno_mat_rows'}->[$i] } = $ld->{r};
                    }
                }
            }

    } else {
            $cor_ld_r = undef;
    }
    return($pearson_cor,$cor_ld_r,\%new_corrs);
}

sub get_maf_weights {
    my $genotypes = shift;
    my $MAF_w = $genotypes->sumover/(2*$genotypes->getdim(0));
    my $maf_to_flip = which($MAF_w >0.5);
    $MAF_w->index($maf_to_flip) .= 1 - $MAF_w->index($maf_to_flip);
    $MAF_w /= $MAF_w->sum;
    $MAF_w = $MAF_w->flat;
    return($MAF_w)
}

sub generate_weights_for_a_gene {
    my $snps = shift;
    my $weights = shift;
    # get the weights for each snp in the gene
    # if there are no weights for an SNP all will be set to 0. Meaning
    # This SNP will get the min weight in all categories present.
    my @w_mat_rows = ();
    foreach my $s (@{$snps}) {
        if (not exists $weights->{$s}) { $weights->{$s} = []; }
        push @w_mat_rows, $weights->{$s};
    }
    # make a matrix with the weigths and get its dimensions
    my $W = pdl @w_mat_rows;
    if (scalar @w_mat_rows == 1){
        $W = ones 1;
```

```perl
        } else {
            $W = abs($W);
            @w_mat_rows = ();
            my @dims = $W->dims();
            # re-scale the weights to be between 0 and 1 (columns)
            for (my $i = 0; $i < $dims[0]; $i++) {
                next if ($W->($i,)->flat->sumover == 0);
                $W->($i,) = ($W->($i,) - $W->($i,)->min) / $W->($i,)->max;
            }
            # sum the weight for each snp
            $W = pdl map { $W->(,$_)->flat->sum; } 0 .. $dims[1] - 1;
            # make the weight for the gene sum 1.
            if ($W->min == 0){
                $W += $W->(which($W > 0))->min/$dims[0];
            }
            $W /= $W->sum;
        }
        return($W);
}

sub make_file_name_array {
    my $file = shift;
    my @back = ();
    my @body = split(/\#/,$file);
    for my $chr (1..26){
        push @back, join "$chr", @body;
    }
    return([@back]);
}
sub make_file_name_array_interval {
    my $file = shift;
    my $s = shift;
    my $e = shift;
    my @back = ();
    my @body = split(/\#.*\#/,$file);
    for my $chr ($s..$e){
        push @back, join "$chr", @body;
    }
    return([@back]);
}

sub gene_pvalue {
    my $gn = shift;
    if (defined $v){ print_OUT("____ $gn ____",$LOG); }
    my $n_snps = scalar @{ $gene{$gn}->{geno_mat_rows}};
    # if the gene has just 1 SNP we make that SNP's p value the gene
p-value under all methods
    if ($n_snps < 2){
                if (defined $v){ printf (scalar localtime() . "\t$gn\t
$gene{$gn}->{hugo}\t$gene{$gn}->{gene_type}\t$gene{$gn}->{chr}\t$gene
{$gn}->{start}\t$gene{$gn}->{end}\t%0.3e\t%0.3e\tNA\tNA1\t1\n",$gene
{$gn}->{minp},$gene{$gn}->{minp},$gene{$gn}->{minp}); }
                return({
                    'fisher' => -1,
                    'fisher_df' => -1,
                    'fisher_chi' => -1,
                    'sidak_min_p' => -1,
                    'Meff_gao' => -1,
                    'Meff_Galwey' => -1,
                });
    }

    if (defined $v){ print_OUT("Calculating effective number of tests:
",$LOG); }
```

```perl
    # calculate number of effective tests by the Gao ($k) and Galwey
($Meff_galwey) method.
    my ($k,$Meff_galwey) = number_effective_tests(\$gene{$gn}->{cor});

    # get the log of the SNP p-value
    if (ref($gene{$gn}->{pvalues}) eq "ARRAY"){
        $gene{$gn}->{pvalues} = pdl @{ $gene{$gn}->{pvalues} };
    }
    # calculate the chi-square statistics for the Makambi method and
its p-value

    #
    #### WORK ON THE GENE WEIGHTS ####
    #

    if (defined $v){ print_OUT("Weigth = [ $gene{$gn}->{weights} ]",
$LOG); }

    my ($forge_chi_stat,$forge_df) = get_makambi_chi_square_and_df
($gene{$gn}->{cor},$gene{$gn}->{weights},$gene{$gn}->{pvalues});
    my $fisher_p_value = sclr double  1 - gsl_cdf_chisq_P
($forge_chi_stat, $forge_df );

    my $sidak = sclr double 1-(1- $gene{$gn}->{pvalues}->min)**$k;
    $forge_df = sclr $forge_df;
    $forge_chi_stat = sclr $forge_chi_stat;
    my $back = {
            'fisher' => $fisher_p_value,
            'fisher_df' => $forge_df,
            'fisher_chi' => $forge_chi_stat,
            'sidak_min_p' => $sidak,
            'Meff_gao' => $k,
            'Meff_Galwey' => $Meff_galwey,
    };
    # print out the results

    if (defined $v){ printf (scalar localtime() . "\t$gn\t$gene{$gn}->
{hugo}\t$gene{$gn}->{gene_type}\t$gene{$gn}->{chr}\t$gene{$gn}->
{start}\t$gene{$gn}->{end}\t%0.3e\t%0.3e\t%0.3e\t%0.5f\t%0.5f\t%0.3e\t
%0.5f\t%0.5f\t%2d\t%3d || $gene{$gn}->{weights}\t$gene{$gn}->
{pvalues}->log\t@{ $gene{$gn}->{geno_mat_rows} }\n",$gene{$gn}->
{minp},$sidak,$fisher_p_value,$forge_chi_stat,$forge_df,$Meff_galwey,
$n_snps,$k); }
    return($back);
}

# this subroutine take an array and return a hash were every element
of the line is
# a key and the value is the index in the array
sub get_header {
    my $in = shift;
    my %back = ();
    for (my $i = 0;$i< scalar @$in; $i++){
        $back{$$in[$i]} = $i;
    }
    return(\%back);
}

# this subroutine print the advance of of something
```

```perl
# it receives 3 parameter: index, rep and tag. index is the advance so
far, report how often to report and tag is name for the printing
# it will print a report when index if divisible by rep
sub report_advance {
    my ($index,$rep,$tag) = @_;
    if (( $index/$rep - int($index/$rep)) == 0) {print_OUT(" '->Done
with [ $index ] $tag",$LOG); }
}

sub print_OUT {
    my $string = shift;
    my @file_handles = @_;
    print scalar localtime(), "\t$string\n";
    unless (scalar @file_handles == 0){
        foreach my $fh (@file_handles){
            print $fh scalar localtime(), "\t$string\n";
        }
    }
}

sub read_weight_file {
    my $file = shift; # file name
    my $w_counter = shift; # weigth counter, to be use if the file has
not header
    my $class_names =shift; # header of weights
    my $weights = shift; # weight for each snp
    my $header_present = shift;
    my $snp_2_gene = shift;
    print_OUT("  '-> Reading weigths from [ $file ]",$LOG);
    open (WEIGHTS,$file) or print_OUT("I cannot open [ $file ]",$LOG)
and exit(1);
    my $count = 0;
    my $w_header = ();
    while(my $line = <WEIGHTS>){
        chomp($line);
        my ($snp_id,@w) = split(/[\s\t\,]/,$line);
        if ( defined $affy_to_rsid ) {
            if ($snp_id !~ m/^rs/){
                if (exists $affy_id{$snp_id}){ $snp_id = $affy_id
{$snp_id};}
            }
        }
        # if it is the first line check if the user declared header in
the file
        if ($count == 0){
            if (defined $header_present){
                $w_header = get_header([@w]);
                $count++;
                next;
            } else {
                map {    $w_header->{$w_counter} = $w_counter;
                        $w_counter++;
                    } @w;
            }
            $count++;
        }
        next if (not exists $snp_2_gene->{$snp_id});
        foreach my $category (sort { $w_header->{$a} <=> $w_header->
{$b} } keys %{$w_header}){
            my $val = $w[$w_header->{$category}];
            $val = 0 if ($val eq "NA");
            $val = 0 if ($val eq "");
            $weights->{$snp_id}{$category} = $val;
            $class_names->{$category} = "";
        }
```

```
    }
    close(WEIGHTS);
    return($weights,$w_counter,$class_names);
}

__END__

=head1 SYNOPSIS

script [options]

    -h, --help          print help message
    -m, --man           print complete documentation
    -report             how often to report advance
    -verbose, -v        useful for debugging

    Input Files:
    -ped                Genotype file in PLINK PED format
    -map                SNP info file in PLINK MAP format
    -bfile              Files in PLINK binary format, corresping file
with extension *.bed, *.bim and *.fam.
    -ox_gprobs      Genotype probabilities in OXFORD format.
    -bgl_gprobs     Genotype probabilities in BEAGLE format.
    -assoc, -a      SNP association file, columns header is
necessary and at leat columns with SNP and P names must be present
    -snpmap, -m         Snp-to-gene mapping file
    -affy_to_rsid       Affy id to rs id mapping file
    -gmt                Gene-set definition file

    Output Files:
    -out, -o            Name of the output file
    -print_cor          print out SNP-SNP correlations

    Analsis modifiers:
    -gene_list, -g          Only analyse genes on the file provided
    -genes              Provide some gene name in command line, only
these genes will be analyzed
    -all_genes          Analyze all genes in the snp-to-gene mapping
file
    -chr                Anlyze a specific chromosome
    -gene_type, -type   Only analyses genes of this type, e.g.
protein_coding
    -exclude_gene_type, -exclude_type    Exclude genes of this type
from the analysis, e.g. pseudogenes
    -distance, -d   Max SNP-to-gene distance allowed (in kb)
    -correlation, -cor SNP-SNP correlation file
    -pearson_genotypes Calculate SNP-SNP correlation with a pearson
correlation for categorical variables
    -use_ld                 Use Linkage Disequilibrium as measure of
SNP-SNP correlation
    -lambda             lambda value to correct SNP statistics, if
genomic control is used
    -gc_correction          Determine the lamda for the genomic
control correction from the input p-values
    -gmt_min_size     Min number of genes in gene-sets to be
analysed. default = 2
    -gmt_max_size     Max number of genes in gene-sets to be
analysed. default = 999999999
    -g_prob_threshold  Floating number. Genotype probabilites below
this threshold will be set to missing

    Multivariate Normal Distribution sampling
    -mnd                Estimate significance by sampling from a
multivatiate normal distribution
```

```
    -mnd_target         minimum number of times the statistics must be
seen to calculate the empirical p-value (default=10)
    -mnd_max            maximim number of simulation (default =
1000000)
    -mnd_methods choose best method and correct using simulation
backgroud. must specify integers: 0=sidak,1=fisher,2=z-fix,3=z-random,
4=VEGAS
                            for example to choose between the sidak
and z-random method use -mnd_methods 0,3

    Weigthing
    -weights, -w           File with SNP weights
    -w_header              Indicate if the SNP weight file has a
header.
    -weight_by_maf, -w_maf  Weight each SNP its the minor allele
frequency (MAF). The actual weight is 1/MAF.
```

=head1 OPTIONS

=over 8

=item B<-help>

Print help message

=item B<-man>

print complete documentation

=item B<-report>

How often to report advance. Provide an integer X and the program will report adnvance after X genes are analyzed.

=item B<-verbose, -v>

useful for debugging

=item B<-ped>

Genotype file in PLINK PED format. See http://pngu.mgh.harvard.edu/~purcell/plink/data.shtml for details on data format

=item B<-map>

SNP info file in PLINK MAP format

=item B<-bfile>

Files in PLINK binary format, with extension *.bed, *.bim and *.fam.

=item B<-ox_gprobs>

File with genotype probabilities in OXFORD format. See http://www.stats.ox.ac.uk/%7Emarchini/software/gwas/file_format_new.html for details in the file format. An example is
```
1   rs10489629   67400370 2 4 1 0 0 0 1 0 1 0 0
```
Columns are chromosome, SNP id, SNP position, minor alelle (A), major alelle (B), prob for AA at sample1,  prob for AB at sample1,  prob for BB at sample1, prob for AA at sample2, etc.

=item B<-bgl_gprobs>

Genotype probabilities in BEAGLE format. Please see http://faculty.washington.edu/browning/beagle/beagle.html for details on the format

=item B<-assoc, -a>

SNP association file, columns header is necessary and at leat columns with SNP and P names must be present.

=item B<-snpmap, -m>

SNP-to-gene mapping file. Format is tab separated with fields:chromosome,start,end,Ensembl id, hugo id, description, SNP1, SNP2, ..., SNPN.
Each SNP has 4 fields separated by colons: id,position,alleles and strand. This may look overcomplicated but allows to map any kind of variation and store its information without having to use additional files. An example look like:

1       67632083        67725662        ENSG00000162594 IL23R   KNOWN
protein_coding  Interleukin-23 receptor Precursor (IL-23R)
[Source:UniProtKB/Swiss-Prot;Acc:Q5VWK5]
rs7538938:67132262:T/C:1        rs72669476:67132582:C/T:1
rs72019237:67132863:C/-:1       rs61197134:67132871:C/-:1
rs11208941:67133111:T/C:1

=item B<-affy_to_rsid>

Affy id to rs id mapping file. Tab separated file, looks like:
SNP_A-8389091   rs7593668

=item B<-gmt>

Gene-set definition file. Format is (tab separated)
NAME Description  GENE1 GENE2 GENEN

=item B<-out, -o>

Name of the output file. Output file will be tab separated with the following columns: Ensembl_ID,Hugo
id,gene_type,chromosome,start,end,min p-value in the gene, Sidak corrected minimum p-value, FORGE p-value, FORGE chi-square, FORGE degrees of freedom, Eigen value ratio method gene p-value, EVR chi-square, EVR degrees freedom, number of snps in gene, number effective tests(Gao et al;PDMID:19434714)
An example would look like:ENSG00000162594       IL23R
     protein_coding  1       67632083        67725662        3.491e-02
     4.132e-01       2.128e-01       8.42305         6.04941         1.119e-01
     28.17326        10.116917       15

=item B<-print_cor>

print out SNP-SNP correlations

=item B<-gene_list, -g>

Only analyse genes on the file provided. Ids must match either Ensembl Ids or Hugo ids present in the SNP-to-gene mapping file

=item B<-genes>

Provide some gene name in command line, only these genes will be analyzed. Use like -genes IL23R

=item B<-all_genes>

Analyze all genes in the snp-to-gene mapping file

=item B<-chr>

Anlyze a specific chromosome, Chromosome code must match that of the SNP-to-gene mapping file

=item B<-gene_type, -type>

Only analyses genes of this type, e.g. protein_coding

=item B<-exclude_gene_type, -exclude_type>

Exclude genes of this type from the analysis, e.g. pseudogenes

=item B<-distance, -d>

Max SNP-to-gene distance allowed (in kb)

=item B<-correlation, -cor>

SNP-SNP correlation file. Space separated file with 3 columns, first 2 the SNP ids the the 3th the correlation between them. Like:

rs6660226 rs11209018 0.089

=item B<-pearson_genotypes>

Calculate SNP-SNP correlation with a pearson correlation for categorical variables as described on S. Wellek, A. Ziegler, Hum Hered 67, 128 (2009).

=item B<-use_ld>

Use Linkage Disequilibrium as measure of SNP-SNP correlation. Default is Pearson's correlation.

=item B<-lambda>

lambda value to correct SNP statistics, if genomic control is used

=item B<-gc_correction>

Correct the SNP pvalues by the genomic control method. It will calculate the lambda from the data itself. PLEASE MAKE SURE YOU DO NOT FILTER THE SNPS BEFORE RUNNING FORGE OR THE CORRECTION MAY BE ERRONEOUS.

=item B<-gmt_min_size>

Min number of genes in gene-sets to be analysed. default = 2

=item B<-gmt_max_size>

Max number of genes in gene-sets to be analysed. default = 999999999

=item B<-g_prob_threshold>

Floating number. Genotype probabilites below this threshold will be set to missing

=item B<-mnd>

Estimate significance by sampling from a multivatiate normal distribution

=item B<--mnd_target>

minimum number of times the statistics must be seen to calculate the empirical p-value (default=10)

=item B<--mnd_max>

maximim number of simulation (default = 1000000)

=item B<--mnd_methods>

choose best method and correct using simulation backgroud. must specify integers: 0=sidak,1=fisher,2=z-fix,3=z-random,4=VEGAS for example to choose between the sidak and z-random method use --mnd_methods 0,3
Is you used results of the method VEGAS please cite as ... "gene p-values calculated with the method VEGAS (Liu, 2010) as implemented in FORGE (Pedroso, submitted)" ...
Liu JZ, McRae AF, Nyholt DR, Medland SE, Wray NR, Brown KM, Hayward NK, Montgomery GW, Visscher PM, Martin NG, Macgregor S: A versatile gene-based test for genome-wide association studies. Am J Hum Genet 2010;87:139-145

=item B<--weights, -w>

File with SNP weights. Multiple files with weights can be used by providing the --weights for each file. the file with weigths can be tab, space or comma separated.
the first columns must be the snp id and the rest the weights. It is possible to have a header in the file, in which case you must use the option --w_header as well. SNP weights are use in the gene analysis and to calculate the sample scores. How to use different sources of information to make a simple weight for the SNP is a not trivial issue. To simplify things I: a) for each gene's SNPs I make the weight of each category sum 1,
b) then I sum these re-sacel weights for each SNP and c) finally make sure the weight within the gene sum 1.
Please note that SNPs without weights will internally get the weights set to zero.

=item B<--w_header>

Indicate if the SNP weight file has a header.

=item B<--weight_by_maf, -w_maf>

Weight each SNP its the minor allele frequency (MAF). The actual weight is 1/MAF.

=head2 DESCRIPTION

TODO

=head3 Examples

=item B<1. Basic Analysis>

To perform a basic gene-based testing with the example files run:

>perl forge.pl -bfile example/example -assoc example/example.assoc -snpmap example/example.snpmap -out test

=item B<2. Adding gene-sets to the analysis>

To perform a basic gene-based and gene-set testing with the example files run:

>perl forge.pl -bfile example/example -assoc example/example.assoc -snpmap example/example.snpmap -out test -gmt example/example.gmt

You will need to run a whole-genome analysis when using gene-sets. Check option below.

=item B<3. Using simulations to estimate significance>

To perform a basic gene-based analysis and estimate significance by simulation

>perl forge.pl -bfile example/example -assoc example/example.assoc -snpmap example/example.snpmap -out test -mnd

=item B<4. Running a whole-genome analysis>

To perform a whole-genome analysis using our SNP-to-gene annotation files. Please note we do not provide example files for this. The # symbol in the SNP-to-gene annotation file name means "analyse from chromosome 1 to 26".

>perl forge.pl -bfile whole_genome_genotypes -assoc whole_genome_genotypes.assoc -snpmap Ensemble_gene_SNP_v59/ ensemblv59_SNP_2_GENE.chr#.txt -out whole_genome.txt -mnd

This syntax can be modified a bit to analysis sub-sets of chromosomes by chaging ensemblv59_SNP_2_GENE.chr#.txt with ensemblv59_SNP_2_GENE.chr#20-22#.txt, in this case to analyse chromosomes 20 to 22.

## 8.5. gsa.pl

Calculation of gene-set association with gene p-values.

```perl
#!/usr/bin/perl -w
use strict;
use Getopt::Long;
use Pod::Usage;
use PDL;
use PDL::Matrix;
use PDL::NiceSlice;
use PDL::GSL::CDF;
use PDL::Stats::Basic;
use PDL::Ufunc;
use PDL::LinearAlgebra qw(mchol);
use Data::Dumper;
use IO::File;
use IO::Handle;
use IO::Uncompress::Gunzip qw(gunzip);

# Load local functions
use GWAS_IO;
use GWAS_STATS;
use CovMatrix;

our (     $help, $man, $gmt, $pval,
    $perm, $out, $max_size,
    $min_size, $recomb_intervals,
    $report,$gene_sets,$ref_list,
    $set_stat, $input_z, $verbose_output,
    $append, $add_file_name,$all_are_background,
    $cgnets,$complete_interval_sampling,
    $best_x_interval, $gs_coverage, $interval_merge,
    $interval_merge_by_chr, $node_similarity, $cgnets_all,
    $Neff_gene_sets, $max_processes, $snp_assoc, $bfile,
    $snpmap, $distance, $affy_to_rsid,$gene_set_list,
    $quick_gene_cor,$gene_cor_max_dist,$print_ref_list,
    $mnd,$mnd_N,$gene_p_type
);

GetOptions(
    'help' => \$help,
    'man' => \$man,
    'gmt=s@' => \$gmt,
    'file=s' => \$pval,
    'perm=i' => \$perm,
    'out|o=s' => \$out,
    'max_size=i' => \$max_size,
    'min_size=i' => \$min_size,
    'recomb_intervals=s' => \$recomb_intervals,
    'report=i' => \$report,
    'gene_sets=s@' => \$gene_sets,
    'gene_set_list=s'    => \$gene_set_list,
```

```perl
    'ref_list=s' => \$ref_list,
    'set_stat=s' => \$set_stat, # stat to calculate over the
sub-networks
    'z_score' => \$input_z,
    'verbose_output|v' => \$verbose_output,
    'append' => \$append,
    'add_file_name' => \$add_file_name,
    'backgroung_all_genes' => \$all_are_background,
    'cgnets=s' => \$cgnets,
    'best_per_interval' => \$best_x_interval,
    'complete_interval_sampling' => \
$complete_interval_sampling,
    'gs_coverage=f' => \$gs_coverage,
    'interval_merge=i' => \$interval_merge,
    'interval_merge_by_chr' => \$interval_merge_by_chr,
    'node_similarity=s'=> \$node_similarity, # NOT documented
    'cgnets_all' => \$cgnets_all,
    'Neff_gene_sets' => \$Neff_gene_sets,
    'n_runs=i' => \$max_processes,
    'snp_assoc=s@' => \$snp_assoc,
    'bfile=s'    => \$bfile,
    'snpmap|m=s@' => \$snpmap,
    'distance|d=i' => \$distance,
    'affy_to_rsid=s' => \$affy_to_rsid,
    'quick_gene_cor' => \$quick_gene_cor,
    'gene_cor_max_dist=i' => \$gene_cor_max_dist,
    'print_ref_list' => \$print_ref_list,
    'mnd' => \$mnd,
    'mnd_n=i' => \$mnd_N,
    'gene_p_type' => \$gene_p_type,
) or pod2usage(0);

pod2usage(0) if (defined $help);
pod2usage(-exitstatus => 2, -verbose => 2) if (defined $man);
pod2usage(0) if (not defined $pval);
pod2usage(0) if (not defined $out);
pod2usage(0) if (not defined $gmt);

my $LOG = new IO::File;
$LOG->open(">$out.log") or print_OUT("I can not open
[ $out.log ] to write to",$LOG) and exit(1);

print_OUT("Check http://github.com/inti/FORGE/wiki for updates",
$LOG);
print_OUT("LOG file will be written to [ $out.log ]",$LOG);

if (defined $gs_coverage){
    print_OUT("Removing pathway with less than [ " .
$gs_coverage*100 . " % ] coverage",$LOG);
}
defined $gs_coverage or $gs_coverage = 0;
defined $interval_merge_by_chr and $interval_merge = "inf";
if (defined $interval_merge and defined $best_x_interval){
    print_OUT("Using segment distance of [ $interval_merge ] to
join recombination intervals during sampling",$LOG);
}
```

```perl
if (defined $perm){
    print_OUT("Will run [ $perm ] permutations to estimate the
mean and std deviation of the statistics undel the null",$LOG);
    if (defined $best_x_interval){
            print_OUT("Permutations with bext-per-interval option
not supported at the moment.",$LOG);
            exit(0);
    }
}
if (defined $input_z and not defined $perm){ $perm = 10_000; }

defined $interval_merge or $interval_merge = 0;
defined $distance or $distance = 20;
defined $report or $report = 250;
defined $max_size or $max_size = 99_999_999;
defined $min_size or $min_size = 10;

if (defined $mnd){
    defined $mnd_N or $mnd_N = 1_000_000;
    use Pareto_Distr_Fit qw( Pgpd );
    if (defined $gene_p_type) {
            unless (grep $_ eq $gene_p_type,
('sidak','fisher','z_fix','z_random')){
                    print_OUT("I do not recognise the gene p-valyes
type entered [ $gene_p_type ]\n");
                    print_OUT("Option are sidak, fisher, z_fix and
z_random");
                    print_OUT("bye!");
                    exit(1);
            }
    } else { $gene_p_type = 'z_fix'; }
    print_OUT("Will use multivariate normal distribution
sampling to estimate gene-gene correlations. Using [ $mnd_N ]
simulations with the [ $gene_p_type] gene p-value");
}
print_OUT("Will analyses Gene-set between [ $min_size ] and
[ $max_size ] in size",$LOG);

if (defined $recomb_intervals) {
    if ((defined $best_x_interval) and (defined
$complete_interval_sampling)){
            print_OUT("Please choose one of the two sampling
scheems -best_per_interval or -complete_interval_samplig\nFor
more information type gsa.pl -man",$LOG);
            exit(1);
    }
    if (not defined $complete_interval_sampling){
            $best_x_interval = 1;
            print_OUT("Using best gene per recombination interval
for permutation sampling",$LOG);
    } elsif (defined $complete_interval_sampling) {
            print_OUT("Using all genes in recombination interval
for permutation sampling",$LOG);
    }
}
```

```perl
defined $set_stat or $set_stat = 'mean';
my $network_stat = defined_set_stat($set_stat);
print_OUT("Gene-set statistics set to [ $set_stat ]",$LOG);

my %ref_genes = ();
if (defined $ref_list){
    open (REF,$ref_list) or die $!;
    my @tmp = <REF>;
    chomp(@tmp);
    map {$ref_genes{lc($_)} = ""; } @tmp;
    print_OUT("[ " . scalar (keys %ref_genes) . " ] read from
[ $ref_list ]",$LOG);
}

my %gene_data = ();
print_OUT("Reading gene stats [ $pval ]",$LOG);
open( PVAL, $pval ) or die $!;
while (my $line = <PVAL>) {
    chomp($line);
    my ( $gn, $p ) = split( /\t/, $line );
    $gn = lc($gn);
    if (defined $ref_list){
        next if (not exists $ref_genes{$gn});
    }
    next if ( $p eq "NA" );
    unless (defined  $input_z){
        next if ( $p !~ m/\d/);
        if ( $p == 0 ) {
            print_OUT("WARNING: This gene [ " . uc($gn) .
" ] has p-value equal [ $p ], I do not know how to transformit
to Z score",$LOG);
            next;
        }
    }
    my $z = 0;
    if (defined $input_z){
        $z = abs($p);
    } else {
        if ($p == 1){ $p = 0.99999999; }
        $z = -1 * gsl_cdf_ugaussian_Pinv($p);
    }
    $z = eval($z);
    if ( exists $gene_data{$gn} ) {
        if ( $gene_data{$gn}->{stat} > $z ) {
            $gene_data{$gn}->{stat} = $z;
            $gene_data{$gn}->{pvalue} = $p;
        }

    } else {
        $gene_data{$gn} = {
                        'stat' => $z,
                        'pvalue'=> $p,
                        'recomb_int' => [],
                        'node_similarity'=> null,
                        'genotypes' => null,
                        'snps' => undef,
```

```perl
                              'chr' => undef,
                              'start' => undef,
                              'end' => undef,
                };
        }
}
close(PVAL);
print_OUT("  '-> [ " . scalar (keys %gene_data) . " ] genes
will be analysed",$LOG);

# if a CGNets analysis is performed.
# read the list of genes for which gene-sets are to be analysed
# discard the ones without gene-statistic and report how many
are left

my @candidate_genes = ();
if (defined $cgnets){
    open (LIST, $cgnets) or die $!;
    print_OUT("Reading CGNets seed genes from [ $cgnets ]",
$LOG);
    @candidate_genes = <LIST>;
    close(LIST);
    chomp(@candidate_genes);
    my %tmp = ();
    map { $tmp{lc($_)} = ""; } @candidate_genes;
    @candidate_genes = keys %tmp;
    print_OUT("  '-> [ " . scalar @candidate_genes . " ] unique
CGNets seeds read",$LOG);
    for (my $i = 0; $i < scalar @candidate_genes; $i++){
        $candidate_genes[$i] = lc($candidate_genes[$i]);
        if (not exists $gene_data{$candidate_genes[$i]})
{ splice(@candidate_genes,$i,1); }
    }
    if (scalar @candidate_genes == 0){
        print_OUT("None of the CGNets seed genes has
statistics",$LOG);
        print_OUT("Bye for now",$LOG);
        exit(1);
    }
    print_OUT("  '-> of them [ " . scalar @candidate_genes .
" ] have statistics",$LOG);
}

my %recomb_int = ();
if ( defined $recomb_intervals) {
    %recomb_int = %{ read_recombination_intervals
($recomb_intervals,\%gene_data) };
}

if (defined $gene_sets){
    print_OUT("Reading Gene-sets from command line [ " . @
{$gene_sets} . " ]",$LOG);
}
```

```perl
if ( defined $gene_set_list ) { # in case user gave a list of
genes in the command line
    print_OUT("Reading Gene-set List from [ $gene_set_list ]",
$LOG);
    # read file with gene list and store gene names.
    open( GL, $gene_set_list ) or print_OUT("I can not open
[ $gene_set_list ]",$LOG) and exit(1);
    my @gs = <GL>;
    chomp(@gs);
    push  @{$gene_sets},@gs;
    close(GL);
}

# stats for all genes in pathways
my %genes_in_paths = ();
my %gene_2_paths_map = ();
# store all pathways and it information
my @pathways        = ();
foreach my $gene_set_file (@$gmt){
    print_OUT("Reading gene-set definitions from
[ $gene_set_file ]",$LOG);
    open( GMT, $gene_set_file ) or print_OUT("Cannot open
[ $gene_set_file ]") and die $!;
    while ( my $path = <GMT> ) {
        my ( $p_name, $p_desc, @p_genes ) = split( /\t/,
$path );
        if (defined $gene_sets){
            next unless (grep $_ eq $p_name,@$gene_sets);
        }
        my @p_stats = ();
        my @gene_with_p = ();
        my $total = 0;
        foreach my $gn (@p_genes) {
            $gn = lc($gn);
            if ( $gn =~ m/\// ) {
                $gn =~ s/\s+//g;
                my @genes = split( /\/{1,}/, $gn );
                map {
                    $total++;
                    next if ( not exists $gene_data
{$_} );
                    push @p_stats, $gene_data{$_}->
{stat};
                    push @gene_with_p, $_;
                } @genes;
            } else {
                $total++;
                next if ( not exists $gene_data{$gn} );
                push @p_stats, $gene_data{$gn}->{stat};
                push @gene_with_p, $gn;
            }
        }
        next if ( scalar @p_stats < $min_size );
        next if ( scalar @p_stats > $max_size );
        next if ( $gs_coverage > (scalar @p_stats)/$total);
```

```perl
            my $p = { 'name' => $p_name,
                      'desc' => $p_desc ,
                      'stats' => [],
                      'N_all' => $total,
                      'N_in' => undef,
                      'genes' => \@gene_with_p,
                      'gene_recomb_inter_redundant' => [],
                      'intervals' => undef,
                      'node_weigths' => null,
                      'z_stat_raw' => undef,
                      'z_stat_empi' => undef,
                      'snps' => undef,
                      };

            # this section reduces the gene sets to sets of genes
of non-overlapping recombination intervals
            # it first check in which recombination intervals the
genes are
            # then it will keep those that do not share recomb
inter with other genes.
            # the cases where more than one gene belong to the
same recombination interval
            # are solve by choosing the best p-value per
recombination interval.
            $p->{N_in} = scalar @{$p->{'genes'}} if (not defined
$p->{N_in});
            if (scalar @{ $p->{'stats'} } == 0){
                @{ $p->{'stats'} } = map { $gene_data{$_}->
{stat} } @{$p->{'genes'}};
            }
            map {
                $genes_in_paths{$_} = $gene_data{$_}->{stat};
                push @{ $gene_2_paths_map{$_} }, $p->{name};
            } @{ $p->{genes} };
            map { $genes_in_paths{$_} = $gene_data{$_}->{stat}; }
@{ $p->{"gene_recomb_inter_redundant"} };
            push @pathways, $p;
    }
    close(GMT);
}

if (defined $ref_list){
    foreach my $gn (keys %ref_genes){
            if (exists $gene_data{lc($gn)}->{stat}){
                $genes_in_paths{lc($gn)} = $gene_data{lc($gn)}-
>{stat};
            }
    }
}
# if analysing only 1 pathway then reference list are all genes
in pathways
if (scalar @pathways == 1){
    foreach my $gn (keys %gene_data){
            if (exists $gene_data{lc($gn)}->{stat}){
                $genes_in_paths{lc($gn)} = $gene_data{lc($gn)}-
>{stat};
```

```perl
            }
        }
}

print_OUT("   '-> [ " . scalar @pathways . " ] gene-sets will be
analysed",$LOG);

# Now lets going to read the affy id to rsid mapping. This is
used to keep all ids in the
# same nomenclature
my %affy_id = ();
if ( defined $affy_to_rsid ) { # if conversion file is defined
    print_OUT("Reading AFFY to rsID mapping from
[ $affy_to_rsid ]",$LOG);
    open( AFFY, $affy_to_rsid ) or print_OUT("I can not open
[ $affy_to_rsid ]") and exit(1);
    while (my $affy = <AFFY>){
        chomp($affy);
        my @b = split(/\t+/,$affy);
        $affy_id{$b[0]} = $b[1];
    }
    close(AFFY);
}

my %snps_covered = ();
if (defined $snp_assoc){
    print_OUT("Reading SNPs included in analysis",$LOG);
    foreach my $file (@$snp_assoc){
        open (ASSOC,$file) or print_OUT("Cannot open
[ $file ]") and die $!;
        while(my $line = <ASSOC>){
            my @d = split(/[\t+\s+]/,$line);
            $snps_covered{$d[0]} = "";
        }
        close(ASSOC);
    }
    print_OUT("   '->[ " . scalar (keys %snps_covered) . " ]
SNP read",$LOG);
}
my $N_bytes_to_encode_snp;
my %bim_ids = ();
my @bim = ();
my @fam = ();
my $bed = new IO::File;

if (defined $bfile){
    print_OUT("Checking genotypes on [ $bfile.bed ]",$LOG);
    # open genotype file
    $bed->open("<$bfile.bed") or print_OUT("I can not open
binary PLINK file [ $bfile ]") and exit(1);
    binmode($bed); # set file type to binary
    my $plink_bfile_signature = "";
    read $bed, $plink_bfile_signature, 3;
```

```perl
    if (unpack("B24",$plink_bfile_signature) ne
'011011000001101100000001'){
        print_OUT("Binary file is not in SNP-major format,
please check you files",$LOG);
        exit(1);
    } else {
        print_OUT("Binary file is on SNP-major format",$LOG);
    }

    # read the bim file with snp information and fam file with
sample information
    @bim = @{ read_bim("$bfile.bim",$affy_to_rsid,\%affy_id) };
    @fam = @{ read_fam("$bfile.fam") };
    print_OUT("[ " . scalar @bim . " ] SNPs and [ " . scalar
@fam . " ] samples in genotype file",$LOG);
    my $index = 0;
    map {
        $bim_ids{$_->{snp_id}} = $index;
        $index++;
    } @bim;
    # calculate how many bytes are needed  to encode a SNP
    # each byte has 8 bits with information for 4 genotypes
    $N_bytes_to_encode_snp = (scalar @fam)/4; # four genotypes
per byte
    # if not exact round it up
    if (($N_bytes_to_encode_snp - int
($N_bytes_to_encode_snp)) != 0  ){ $N_bytes_to_encode_snp = int
($N_bytes_to_encode_snp) + 1;}
}

if (defined $snpmap){
    print_OUT("Loading SNP-2-Gene mapping",$LOG);
    for (my $i = 0; $i < scalar @$snpmap; $i++){
        if ($snpmap->[$i] =~ m/\#/){
            print_OUT("   '-> Found [ # ] key on [ $snpmap-
>[$i] ]. I will generate file names for 26 chromosomes",$LOG);
            push @{$snpmap}, @{ make_file_name_array
($snpmap->[$i]) };
            splice(@$snpmap,$i,1);
        }
    }
    foreach my $snp_gene_mapping_file (@$snpmap){
        if (not -e $snp_gene_mapping_file){
            print_OUT("   '-> File
[ $snp_gene_mapping_file ] does not exist, moving on to next
file",$LOG);
            next;
        }
        my $MAP = '';
        if ( $snp_gene_mapping_file =~ m/.gz$/) {
            print_OUT ("   '-> Reading
[ $snp_gene_mapping_file ]",$LOG);
            $MAP = new IO::Uncompress::Gunzip
$snp_gene_mapping_file;
        } else {
            $MAP = new IO::Handle;
```

```perl
                my $fh = new IO::File;
                print_OUT ("   '-> Reading
[ $snp_gene_mapping_file ]",$LOG);
                $fh->open("$snp_gene_mapping_file");
                $MAP->fdopen(fileno($fh),"r");
        }

        while (my $read = $MAP->getline()) {
                chomp($read);
                # the line is separate in gene info and snps.
the section are separated by a tab.
                my ($chr,$start,$end,$ensembl,$hugo,
$gene_status,$gene_type,$description,@m) = split(/\t+/,$read);
                #check if gene was in the list of genes i want
to analyze
                my $gene_id = undef;
                if ( exists $gene_data{lc($hugo)}){
                    $gene_id= lc($hugo);
                } elsif (exists $gene_data{lc($ensembl)}){
                    $gene_id= lc($ensembl);
                } else {
                    next;
                }
                # exlude genes that are not in gene-sets
                next unless (exists $genes_in_paths{lc($hugo)}
or exists $genes_in_paths{lc($ensembl)});

                my @first_snp_n_fields =  split(/\:/,$m[0]);
                if (4 !=  scalar @first_snp_n_fields)
{ $description .= splice(@m,0,1); }

                # get all mapped snps within the distance
threshold,
                my @mapped_snps = ();
                foreach my $s (@m) {
                        my ($id,$pos,$allele,$strand) = split(/
\:/,$s);

                        next if (not defined $id);
                        if (( $pos >= $start) and ($pos <= $end))
{ push @mapped_snps, $id; }
                        elsif ( ( abs ($pos - $start) <=
$distance*1_000 ) or ( abs ($pos - $end) <= $distance*1_000 ))
{ push @mapped_snps, $id; }
                }

                next if (scalar @mapped_snps == 0);

                # go over mapped snps and change convert affy
ids to rsid.
                # and make a non-redundant set.
                my %nr_snps = ();
                foreach my $s (@mapped_snps) {
                        if ( defined $affy_to_rsid ) {
                                if ($s !~ m/^rs/){
                                        if (exists $affy_id{$s}){ $s =
$affy_id{$s};}
```

```perl
                    }
                }
                # exclude snps not in the association file
nor in the bim file
                next unless ( exists $snps_covered{$s} );
                next unless ( exists $bim_ids{$s} );
                $nr_snps{$s} = "";
            }
            @mapped_snps = keys %nr_snps;
            if ( scalar @mapped_snps == 0 ){
                delete( $genes_in_paths{ $gene_id } );
                next;
            }
            $gene_data{$gene_id}->{snps} = [@mapped_snps];
            $gene_data{$gene_id}->{chr} = $chr;
            $gene_data{$gene_id}->{start} = $start;
            $gene_data{$gene_id}->{end} = $end;
        }
        $MAP->close;
    }

    print_OUT("Reading genotypes for genes",$LOG);
}

# mean of all stats across the genome
my $background_values;
if (defined $all_are_background) {
    my @values = ();
    foreach my $gn (keys %gene_data){
        push @values, $gene_data{$gn}->{stat};
    }
    $background_values = pdl @values;
} else {
    $background_values = pdl values %genes_in_paths;
}
print_OUT("  '-> [ " . scalar $background_values->list() . " ]
genes will be used to calculate the patameters of null",$LOG);
$background_values->inplace->setvaltobad( "inf" );
$background_values->inplace->setvaltobad( "-inf" );
my $mu = $background_values->davg;
# sd of all stats across the genome
my $sd = $background_values->stdv;

print_OUT("  '-> Background mean = [ $mu ] and stdv [ $sd ]",
$LOG);

if (defined $print_ref_list){
    print_OUT("Printing reference list to [ $out.refList ]",
$LOG);
    my @ids = ();
    if (defined $all_are_background) {
        @ids = keys %gene_data;
    } else {
        @ids = keys %genes_in_paths;
    }
```

```perl
    open (REFLIST,">$out.refList") or print_OUT("I cannot write
to [ $out.refList ]") and exit(1);
    print REFLIST join "\n", @ids;
    close(REFLIST);
}

my %snp_genotype_stack = ();
my %correlation_stack  = ();
my %gene_genotype_var  = ();
my $i = 0;

if (defined $bfile){
    my @new_pathways = ();
    foreach my $p (@pathways) {
            report_advance($i++,$report,"Gene-Sets Genotypes",
$LOG);
            next if ( scalar @{ $p->{genes} } < $min_size );

            my @new_genes = ();
            foreach my $gn ( @{$p->{ genes } } ){
                next if ( not exists $genes_in_paths{$gn} );
                next if ( not exists $gene_data{$gn} );
                next if ( not defined $gene_data{$gn}->
{snps} );
                next if ( $gene_data{$gn}->{genotypes}->isempty
== 0 );

                # this will store the genotypes
                my $matrix = [];
                my @gn_snps_with_genotypes = ();
                foreach my $snp ( @{ $gene_data{$gn}->
{snps} } ){
                    if (exists $snp_genotype_stack{$snp}) {
                        push @{ $matrix },
$snp_genotype_stack{$snp};
                        push @gn_snps_with_genotypes, $snp;
                    } else {
                        # because we know the index of the
SNP in the genotype file we know on which byte its information
starts
                        my $snp_byte_start =
$N_bytes_to_encode_snp*$bim_ids{$snp};
                        # here i extract the actual
genotypes
                        my @snp_genotypes = @
{ extract_binary_genotypes(scalar @fam,$N_bytes_to_encode_snp,
$snp_byte_start,$bed) };
                        # store the genotypes.
                        # if a snp does not use the 8 bits
of a byte the rest of the bits are fill with missing values
                        # here i extract the number of
genotypes corresponding to the number of samples

                        my $maf = get_maf([@snp_genotypes
[0..scalar @fam - 1]] ); # check the maf of the SNP
                        next if ($maf == 0 or $maf ==1);  #
go to next if it is monomorphic
```

```perl
                                push @{ $matrix }, [@snp_genotypes
[0..scalar @fam - 1]];
                                push @gn_snps_with_genotypes, $snp;
                                $snp_genotype_stack{$snp} =
[@snp_genotypes[0..scalar @fam - 1]];
                        }
                }
                next if (scalar @gn_snps_with_genotypes == 0);
                $gene_data{$gn}->{snps} =
[@gn_snps_with_genotypes];
                $gene_data{$gn}->{genotypes} = pdl $matrix;
                push @new_genes, $gn;
        }

        next if (scalar @new_genes < $min_size);

        $p->{genes} = [ @new_genes ];
        my @stats = map { $gene_data{$_}->{stat} } @{ $p->
{genes} };
        $p->{stats} = [ @stats ];
        $p->{N_in} = scalar @{ $p->{genes} };
        my ($more_corr,$more_var,$G_cor) = "";
        ($p->{gene_cor_mat},$more_corr,$more_var,$G_cor) =
calculate_gene_corr_mat($p->{genes},\%gene_data,\
%correlation_stack,\%gene_genotype_var,$quick_gene_cor,
$gene_cor_max_dist,$mnd,1000,$gene_p_type);
        $p->{ genotypes_corr } = ${ $G_cor };
        %correlation_stack = ( %correlation_stack, %
${$more_corr} );
        %gene_genotype_var = ( %gene_genotype_var, %
${$more_var} );

        # store the pathway information
        push @new_pathways, $p;

        foreach my $gn (  @{ $p->{genes} } ){
                if (scalar @{ $gene_2_paths_map{$gn} } == 1){
                        delete($gene_data{$gn}->{genotypes});
                } else {
                        splice( @{ $gene_2_paths_map{$gn} },0,1);
                }
        }
    }
    @pathways = @new_pathways;

    %snp_genotype_stack = ();
    %correlation_stack = ();
    %gene_genotype_var  = ();
}

print_OUT("  '->[ " . scalar (keys %gene_data) . " ] Genes read
from SNP-2-Gene Mapping files with genotpe data",$LOG);
if (scalar (keys %gene_data) == 0){
    print_OUT("No gene-sets to analyze\n");
    exit(1);
}
```

```perl
# if permutation are performed store null distribution in here
my %null_size_dist = ();

print_OUT("Writing output to [ $out ]",$LOG);
if (defined $append){
    open (OUT,">>$out") or die $!;
} else {
    open (OUT,">$out") or die $!;
    print OUT "name\traw_p\traw_z";

    if (defined $bfile and defined $snpmap and defined
$snp_assoc){
        print OUT "\tZ_fix\tV_fix\tZ_P_fix\tZ_random
\tV_random\tZ_P_random\tI2\tQ\tQ_P\ttau_squared";
        if (defined $mnd){
            print OUT "\tSIM_Z_FIX\tSIM_Z_RANDOM\tSEEN_FIX
\tSEEN_RANDOM\tN\tpareto_fix_Phat\tpareto_fix_Phatci_low
\tpareto_fix_Phatci_up\tpareto_random_Phat
\tpareto_random_Phatci_low
\tpareto_random_Phatci_up";
        }
    }
    if (defined $perm){
        print OUT ("\tempi_p:$set_stat\tempi_z:$set_stat
\tmean_set\tmean_null\tsd_null");
    }
    print OUT ("\tN_in\tN_all\tdesc");
    if (defined $add_file_name){
        print OUT "\tdataset";
    }
    print OUT "\n";
}

print_OUT("Starting to Analyse the [ " . scalar @pathways . " ]
gene-sets",$LOG);
my $c = 0;
while (my $p = shift @pathways) {
    report_advance($c,$report,"Gene-Sets");
    $c++;
    next if ( $p->{N_in} < $min_size);
    my $m = $p->{N_in};
    my $Sm = mean( $p->{stats} );
    my $z_score = ( ( $Sm - $mu ) * (sqrt($m) ) ) / $sd;
    $p->{z_stat_raw} = $z_score;

    if ( $z_score == -0 ) {
        print OUT "$p->{name}\tNA\t$z_score";
    } else {
        printf OUT ("$p->{name}\t%.3e\t%.3f", 1 -
gsl_cdf_ugaussian_P($z_score),$z_score);
    }

    if (defined $bfile and defined $snpmap and defined
$snp_assoc){
```

```perl
        my $z = pdl $p->{stats};
        my $var = dsum($p->{gene_cor_mat});
        my $se = $z->nelem * ones $z->nelem;
        my $Z_STATS = get_fix_and_radom_meta_analysis($z,
$se,undef,$p->{gene_cor_mat});
        my $num = 100;
        my $simulated_set_stats = simulate_mnd_gene_set($p,
$Z_STATS,10,$mnd_N);
        printf OUT ("\t%.3e\t%.3e\t%.3e\t%.3e\t%.3e\t%.3e\t%.
3e\t%.2f\t%.3e\t%.3e\t%.3e\t%.3e\t%.3e\t%.3e\t%.3e\t%.3e\t
%.3e\t%.3e\t%.3e\t%.3e",
                $Z_STATS->{'B_fix'},
                $Z_STATS->{'V_fix'},
                $Z_STATS->{'Z_P_fix'},
                $Z_STATS->{'B_random'},
                $Z_STATS->{'V_random'},
                $Z_STATS->{'Z_P_random'},
                $Z_STATS->{'I2'},
                $Z_STATS->{'Q'},
                $Z_STATS->{'Q_P'},
                $Z_STATS->{'tau_squared'},
                $simulated_set_stats->{'z_fix'},
                $simulated_set_stats->{'z_random'},
                $simulated_set_stats->{'seen_fix'},
                $simulated_set_stats->{'seen_random'},
                $simulated_set_stats->{'N'},
                $simulated_set_stats->{'pareto_fix_Phat'},
                $simulated_set_stats->
{'pareto_fix_Phatci_low'},
                $simulated_set_stats->{'pareto_fix_Phatci_up'},
                $simulated_set_stats->{'pareto_random_Phat'},
                $simulated_set_stats->
{'pareto_random_Phatci_low'},
                $simulated_set_stats->
{'pareto_random_Phatci_up'},
        );

    }

    if (defined $perm){
        $Sm = &$network_stat( $p->{stats} );
        if (defined $recomb_intervals){
            my $rand_stats = pdl @{ get_null_distribution
($perm,[keys %genes_in_paths],$p,\%recomb_int, \%gene_data) };
            $null_size_dist{$m} = { 'mean' => $rand_stats-
>average, 'sd' => $rand_stats->stdv };
        } else {
            if (not defined $null_size_dist{$m}){
                my $rand_stats = pdl @
{ get_null_distribution($perm,[keys %genes_in_paths],$p,\
%recomb_int, \%gene_data)};
                $null_size_dist{$m} = { 'mean' =>
$rand_stats->average, 'sd' => $rand_stats->stdv };
            }
        }
```

```perl
            my $z_score_empirical = ( ( $Sm - $null_size_dist
{$m}->{mean}) )/$null_size_dist{$m}->{sd};

            $p->{z_stat_empi} = $z_score_empirical;

            if ( $z_score == -0 ) {
                print OUT "\tNA\tNA\t";
            } else {
                printf OUT ("\t%.3e\t%.3f\t%.3f\t%.3f\t%.3f", 1
- gsl_cdf_ugaussian_P($z_score_empirical),$z_score_empirical,
$Sm,$null_size_dist{$m}->{mean},$null_size_dist{$m}->{sd});
            }
            print OUT "\t$m\t$p->{N_all}\t$p->{desc}";
            if (defined $add_file_name){
                print OUT "\t$pval";
            }
            if (defined  $verbose_output){
                map { print OUT "\t$_:",$gene_data{$_}->
{stat}; } sort { $gene_data{$b}->{stat} <=> $gene_data{$a}->
{stat} } @{$p->{genes}};
            }
            print OUT "\n";
        } else {
            print OUT "\t$m\t$p->{N_all}\t$p->{desc}";
            if (defined $add_file_name){
                print OUT "\t$pval";
            }
            if (defined  $verbose_output){
                map { print OUT "\t$_:",$gene_data{$_}->
{stat}; } sort { $gene_data{$b}->{stat} <=> $gene_data{$a}->
{stat} } @{$p->{genes}};
            }
            print OUT "\n";
        }
}

print_OUT("Well Done",$LOG);

exit;

sub simulate_mnd_gene_set {
    my $gene_set_data = shift; # HASH ref
    my $set_stats = shift; # HASH ref
    my $target = shift;
    my $MAX = shift;

    my $max_step_size = 10_000;
    my $total = 0;
    my $step=100;

    my $se = $gene_set_data->{gene_cor_mat}->getdim(0) * ones
$gene_set_data->{gene_cor_mat}->getdim(0);
    my $SEEN = zeroes 2;

    my $cholesky = mchol($gene_set_data->{gene_cor_mat});
    my $fix_stat = [];
```

```perl
    my $random_stat = [];
    while ($SEEN->min < $target){

        my ($sim,$c) = rmnorm($step,0,$gene_set_data->
{gene_cor_mat},$cholesky);
        my $sim_chi_df1 = $sim**2;
        my $sim_p =  1 - gsl_cdf_chisq_P($sim_chi_df1,1);
        for (my $sim_n = 0; $sim_n < $sim_p->getdim(0);
$sim_n++){
            my $sim_z = flat -1*gsl_cdf_ugaussian_P($sim_p-
>($sim_n,));
            my $sim_set_stats =
get_fix_and_radom_meta_analysis($sim_z,$se,undef,$gene_set_data-
>{gene_cor_mat});

            #print $set_stats->{'Z_P_fix'}," >= ",
$sim_set_stats->{'Z_P_fix'}," || ", $set_stats->
{'Z_P_random'} ," >= ", $sim_set_stats->{'Z_P_random'},"\n";
            $SEEN->(0)++ if ( $set_stats->{'Z_P_fix'} >=
$sim_set_stats->{'Z_P_fix'} );
            $SEEN->(1)++ if ( $set_stats->{'Z_P_random'} >=
$sim_set_stats->{'Z_P_random'} );
            push @{$fix_stat}, -1*gsl_cdf_ugaussian_Pinv
($sim_set_stats->{'Z_P_fix'});
            push @{$random_stat}, -1*gsl_cdf_ugaussian_Pinv
($sim_set_stats->{'Z_P_random'});
        }
        $total += $step;

        if ($SEEN->min != 0){
            $step = 1.1*(10*($total)/$SEEN->min);
        } elsif ($step < $max_step_size){
            $step *=10;
        }
        if ($step > $MAX){ $step = $MAX; }
        last if ($total > $MAX);
    }

    my $fix_null_stats = pdl $fix_stat;
    my $random_null_stats = pdl $random_stat;
    my $fix_observed = -1*gsl_cdf_ugaussian_Pinv($set_stats->
{'Z_P_fix'});
    my $random_observed = -1*gsl_cdf_ugaussian_Pinv($set_stats-
>{'Z_P_random'});

    my ($pareto_fix_Phat,$pareto_fix_Phatci_low,
$pareto_fix_Phatci_up) = Pareto_Distr_Fit::Pgpd($fix_observed,
$fix_null_stats,250,0.05);
    my ($pareto_random_Phat,$pareto_random_Phatci_low,
$pareto_random_Phatci_up) = Pareto_Distr_Fit::Pgpd
($random_observed,$random_null_stats,250,0.05);

    my $back = {
        'z_fix' => sclr ($SEEN->(0)+1)/($total +1),
        'z_random' => sclr ($SEEN->(1)+1)/($total +1),
        'N' => $total,
```

```perl
    };

    $back->{'seen_fix'} = sclr $SEEN->(0);
    $back->{'seen_random'} = sclr $SEEN->(1);
    $back->{'pareto_fix_Phat'} = sclr $pareto_fix_Phat;
    $back->{'pareto_fix_Phatci_low'} = $pareto_fix_Phatci_low;
    $back->{'pareto_fix_Phatci_up'} =  $pareto_fix_Phatci_up;
    $back->{'pareto_random_Phat'} = sclr $pareto_random_Phat;
    $back->{'pareto_random_Phatci_low'} =
$pareto_random_Phatci_low;
    $back->{'pareto_random_Phatci_up'} =
$pareto_random_Phatci_up;

    return($back);
}
sub print_OUT {
    my $string = shift;
    my @file_handles = @_;
    print scalar localtime(), "\t$string\n";
    unless (scalar @file_handles == 0){
        foreach my $fh (@file_handles){
            print $fh scalar localtime(), "\t$string\n";
        }
    }
}

sub get_genotype_matrix_var {
    my $G = shift;
    my $C = cov_shrink($G->transpose);
    my $V = $C->{cor}->dsum();
    return($V);
}
sub check_overlap {
    my $start_uno = shift;
    my $end_uno = shift;
    my $start_dos = shift;
    my $end_dos = shift;
    my $back = 0;

=h
                    start_1 ......................... end_1
      start_2 ......................... end_2

=cut
    if (($start_uno > $start_dos) and ($start_uno < $end_dos) )
{$back = 1; }

=h
      start_1 ......................... end_1
                        start_2 .........................
end_2
=cut
    if (($start_dos > $start_uno) and ($start_dos < $end_uno) )
{$back = 1; }

=h
```

```perl
    start_1 ......................... end_1
    start_2 ......................... end_2

=cut
    if (($start_uno == $start_dos) or ($end_uno == $end_dos) )
{$back = 1; }

    # else there is no overlap
    return($back);
}

sub calculate_gene_corr_mat {
    my $genes = shift; # ARRAY ref
    my $gene_data = shift; # HASH ref
    my $gene_gene_corr = shift; # HASH reaf
    my $var = shift; # HASH ref
    my $quick_cor = shift; # 0,1
    my $max_gene_dist = shift; # integer
    my $mnd= shift; # 0,1
    my $mnd_N = shift; # integer
    my $gene_p_type = shift; # 0,1

    my %new_corrs = ();
    my %new_vars = ();
    my $G_cov = ""; # will store the correlation between
genotypes.

    # define correlation matrix
    my $corr = stretcher(ones scalar @$genes);
    if (defined $mnd){
        my $G = null;
        my $genotypes_stack = [];
        my @mat_gene_idx = ();
        my $old_end = -1;
        for (my $i = 0; $i < scalar @$genes; $i++){
            push @{$genotypes_stack}, $gene_data->{ $genes-
>[$i] }->{genotypes};
            $mat_gene_idx[ $i ]= {
                'id' => $genes->[$i],
                'start' => $old_end + 1,
                'end' =>  $old_end + 1 - 1 + $gene_data->
{ $genes->[$i] }->{genotypes}->getdim(1),
            };
            $old_end = $mat_gene_idx[ $i ]->{end};
        }
        $G = $G->glue(1,@$genotypes_stack);
        my $G_corr = cov_shrink($G->transpose);
        my $status = undef;
        ($G_cov,$status) = check_positive_definite($G_corr->
{cor},1e-8);
        if ($status == 1){
            print "Matrix never positive definite";
            exit(1);
        }
        my $cholesky = mchol($G_cov);
```

```perl
            my ($sim,$c) = rmnorm($mnd_N,0,$G_cov,$cholesky);

            my $sim_chi_df1 = $sim**2;
            my $sim_p =  1 - gsl_cdf_chisq_P($sim_chi_df1,1);
            my @null_stats = ();
            my %effective_tests = ();
            my $stats_bag = [];
            for (my $stat_sim = 0; $stat_sim < $sim_p->getdim(0);
$stat_sim++){
                my $r_p = $sim_p->($stat_sim,)->flat;
                my  @null_stats = ();
                foreach my $g (@mat_gene_idx){
                    my $fake_gene = {
                        'pvalues' => $r_p->($g->{start}:$g->
{end}),
                        'effect_size' => undef,
                        'effect_size_se' => undef,
                        'cor' => $G_cov->($g->{start}:$g->
{end},$g->{start}:$g->{end}),
                        'weights' => ((ones $r_p->($g->
{start}:$g->{end})->nelem)/$r_p->($g->{start}:$g->{end})-
>nelem),
                        'geno_mat_rows' => [ $r_p->($g->
{start}:$g->{end})->list ],
                    };
                    my $n_stat = undef;
                    if ($gene_p_type eq 'sidak' or $r_p->($g->
{start}:$g->{end})->nelem == 1){

                        if (not defined $effective_tests
{ $g->{id} }){
                            $effective_tests{ $g->{id} } =
number_effective_tests(\$G_cov->($g->{start}:$g->{end},$g->
{start}:$g->{end}));
                        }
                        $n_stat = 1 - (1 - $fake_gene->
{'pvalues'}->min)**$effective_tests{ $g->{id} };
                        $n_stat = gsl_cdf_ugaussian_Pinv
($n_stat );

                    } elsif ($gene_p_type eq 'fisher'){
                        my ($sim_fisher_chi_stat,
$sim_fisher_df) = get_makambi_chi_square_and_df($fake_gene->
{cor},$fake_gene->{weights},$fake_gene->{'pvalues'} );
                        my $sim_fisher_p_value = sclr double
1 - gsl_cdf_chisq_P($sim_fisher_chi_stat, $sim_fisher_df );
                        $n_stat = gsl_cdf_ugaussian_Pinv
($sim_fisher_p_value);
                    } elsif (($gene_p_type eq 'z_fix') or
($gene_p_type eq 'z_random')){
                        my $sim_z_gene =
z_based_gene_pvalues($fake_gene,$mnd);
                        $n_stat = gsl_cdf_ugaussian_Pinv
($sim_z_gene->{'Z_P_fix'}) if ($gene_p_type eq 'z_fix');
                        $n_stat = gsl_cdf_ugaussian_Pinv
($sim_z_gene->{'Z_P_random'}) if ($gene_p_type eq 'z_random');
```

```perl
                }
                push @null_stats, $n_stat;
            }
            push @{$stats_bag}, [@null_stats];
        }
        $stats_bag = pdl $stats_bag;
        my $gene_stats_cor = cov_shrink($stats_bag);
        $corr = $gene_stats_cor->{cor};
    } else {
        for (my $i = 0; $i < scalar @$genes; $i++){
            # get name of gene i
            my $gn_i = $genes->[$i];

            # get indexes for its SNPs in the snp
correlation matrix
            my $idx_i = sequence $gene_data->{ $gn_i }->
{genotypes}->getdim(1);
            # get its variance if it has not been
calculated already
            if (not exists $var->{ $gn_i }){
                $var->{ $gn_i } =  get_genotype_matrix_var
($gene_data->{ $gn_i }->{genotypes});
                $new_vars{ $gn_i } = $var->{ $gn_i };
            }
            for (my $j = $i; $j < scalar @$genes; $j++){
                next if ($j == $i);
                # get name of gene j
                my $gn_j = $genes->[$j];

                # if defined a quick gene-gene correlation
the correlation will only be calculate between genes in
                # the same chromosome
                if (not exists $gene_gene_corr->{ $gn_i }
{ $gn_j } and defined $quick_cor){
                    # if chromosomes are different set
the correlation to 0
                    if ($gene_data->{ $gn_i }->{chr} ne
$gene_data->{ $gn_j }->{chr} ) {
                        $gene_gene_corr->{ $gn_i }
{ $gn_j } = 0;

                        $new_corrs{$gn_j}{$gn_i} = 0;
                    } else { # if are in the same
chromosome
                        # if user provided a maximum
distance to evaluate correlations
                        if (defined $max_gene_dist){
                            # if they overlap we will
need to calculate the correlation
                            # return 0 from
check_overlap mean there is no overlap
                            if (check_overlap
($gene_data->{ $gn_i }->{start},$gene_data->{ $gn_i }->
{end},check_overlap($gene_data->{ $gn_j }->{start},$gene_data->
{ $gn_j }->{end}) == 0 )){
```

```perl
                                        # check the distance
between the gene coordinates
                                        if ( $gene_data->
{ $gn_i }->{start} > $gene_data->{ $gn_j }->{end}   ){
                                                #
start_i.....end_i
                                                #
    start_j.....end_j
                                                if
( ($gene_data->{ $gn_i }->{start} – $gene_data->{ $gn_j }->
{end}) > $max_gene_dist * 1000 ){

    $gene_gene_corr->{ $gn_i }{ $gn_j } = 0;

    $new_corrs{$gn_j}{$gn_i} = 0;
                                                        }
                                        } elsif
( $gene_data->{ $gn_j }->{start} > $gene_data->{ $gn_i }->
{end}   ){
                                                #
    start_i.....end_i

    #                                               start_j.....end_j
                                                if
( ($gene_data->{ $gn_j }->{start} – $gene_data->{ $gn_i }->
{end}) > $max_gene_dist * 1000 ){

    $gene_gene_corr->{ $gn_i }{ $gn_j } = 0;

    $new_corrs{$gn_j}{$gn_i} = 0;
                                                        }
                                                }
                                        }
                                    }
                                }

                        # next if this correlation was already
calculated
                        if (exists $gene_gene_corr->{ $gn_i }
{ $gn_j }){
                                set $corr, $i ,$j, $gene_gene_corr->
{ $gn_i }{ $gn_j };
                                set $corr, $j ,$i, $gene_gene_corr->
{ $gn_i }{ $gn_j };
                                next;
                        }
                        # get indexes for its SNPs in the snp
correlation matrix
                        my $idx_j = $gene_data->{ $gn_i }->
{genotypes}->getdim(1) + sequence $gene_data->{ $gn_j }->
{genotypes}->getdim(1);
                        # get its variance if it has not been
calculated already
```

```perl
                        if (not exists $var->{ $gn_j }){
                            $var->{ $gn_j } =
get_genotype_matrix_var($gene_data->{ $gn_j }->{genotypes});
                            $new_vars{ $gn_j } = $var->
{ $gn_j };
                        }

                        # combine the genotype data information
                        my $r_i_j = zeroes $gene_data->{ $gn_i }->
{genotypes}->getdim(1), $gene_data->{ $gn_j }-
>getdim(1);
                        for ( my $i_snp = 0; $i_snp < $gene_data->
{ $gn_i }->{genotypes}->getdim(1); $i_snp++ ){
                            my $genotype_i = $gene_data->
{ $gn_i }->{genotypes}->(,$i_snp);

                            for ( my $j_snp = 0; $j_snp <
$gene_data->{ $gn_j }->{genotypes}->getdim(1); $j_snp++ ){

                                my $genotype_j = $gene_data->
{ $gn_j }->{genotypes}->(,$j_snp);
                                my $c = corr( $genotype_j,
$genotype_i );
                                set $r_i_j, $i_snp, $j_snp, $c-
>sclr;

                            }
                        }
                        my $c_i_j = double $r_i_j->dsum()/sqrt
( $var->{ $gn_i } * $var->{ $gn_j }  );
                        set $corr, $i ,$j, $c_i_j;
                        set $corr, $j ,$i, $c_i_j;

                        $new_corrs{$gn_j}{$gn_i} = $new_corrs
{$gn_i}{$gn_j} = $c_i_j;
                    }
                }
        }
        return($corr,\%new_corrs,\%new_vars,\$G_cov);
}

sub get_makambi_df {
  my $cor = shift; # a pdl matrix with the varoable correlations
  my $w = shift; # a pdl vector with the weights;
  # make sure weiths sum one
  $w = $w*abs($cor); # multiply the weights by the correaltions
  my @dims = $w->dims();
  $w = pdl map { $w->(,$_)->flat->sum/$w->sum; } 0 .. $dims[1] -
1; # sum the rows divided by sum of the weights used
  if ($w->min == 0){ $w += $w->(which($w == 0))->min/$w-
>length; } # make sure NO weights equal 0
  $w /= $w->sum; # make sure weights sum 1
```

```perl
   # calculate the correlation matrix before the applying the
weights
   # I have change this calculation following the results of the
paper of Kost et al. this should improve the approximation of
the test statistics
   # Kost, J. T. & McDermott, M. P. Combining dependent p-values.
Statistics & Probability Letters 2002; 60: 183-190.
   # my $COR_MAT = (3.25*abs($cor) + 0.75*(abs($cor)**2));
   my $COR_MAT = (3.263*abs($cor) + 0.710*(abs($cor)**2) + 0.027*
(abs($cor)**3));
   my $second = $COR_MAT*$w*($w->transpose); # apply the weights
   ($second->diagonal(0,1)) .= 0; # set the diagonal to 0
   my $varMf_m = 4*sumover($w**2) + $second->flat->sumover; #
calculate the variance of the test statistics
   my $df = 8/$varMf_m; # the degrees of freedom of the test
statistic
   return ($df);
}

sub check_if_exist {
    my $bait = shift;
    my $array = shift;
    return( grep $_ eq $bait, @$array);
}
sub check_if_overlap {
    my $array1 = shift;
    my $array2 = shift;
    my $match = 0;

    if (scalar @$array1 < scalar @$array2){
         foreach my $g (@$array1){ $match += check_if_exist
($g, $array2); }
    } else {
         foreach my $g (@$array2){ $match += check_if_exist
($g, $array1); }
    }
    return($match);
}
sub std_dev {
    my $ar       = shift;
    my $elements = scalar @$ar;
    my $sum      = 0;
    my $sumsq    = 0;
    foreach (@$ar) {
         $sum   += $_;
         $sumsq += ( $_**2 );
    }
    return sqrt( $sumsq / $elements - ( ( $sum / $elements )
**2 ) );
}

sub mean {
    my $ar = pdl @_;
```

```perl
    $ar->inplace->setvaltobad( "inf" );
    $ar->inplace->setvaltobad( "-inf" );
    return $ar->average;
}

sub get_null_distribution {
        my $N = shift; # number of statisitics to be generated
    my $genes = shift;
    my $path_info = shift;
        my $sub_net_genes = $path_info->{genes}; # genes in the
subnetwork
        my $size = scalar @$sub_net_genes; # size of the
subnetwork
        my $comb_int = shift; # recombination interval
information
    my $gene_stats = shift;
        my @rand_stat = (); # array to store the statistics from
each permutation
        for (my $i = 0; $i < $N; $i++){
            if (defined $verbose_output){
                &report_advance($i,100,"Permutations");
            }
            # if sampling conditional to recombination intervals,
first check is any group of genes
            # from the subnetwotk are in the same recombination
interval. If not then sample normally.
            my %intervals = ();

            my @normal_sampling = ();
            my @interval_sampling = ();
            if (defined $recomb_intervals){
                if ( defined $path_info->{intervals}){
                    %intervals = %{$path_info->{intervals}};
                } else {
                    foreach my $g (@$sub_net_genes){
                        foreach my $int (@{$gene_stats->
{$g}->{recomb_int}}){
                            push @{ $intervals{$int} }, $g;
                        }
                    }
                }
                while (my ($int,$int_genes) = each %intervals){
                    if (scalar @$int_genes > 1){ push
@interval_sampling,$int_genes; } # genes that are in the same
recombination interval
                    else { push @normal_sampling, @
$int_genes;} # all other genes
                }
            } else {
                @normal_sampling = @$sub_net_genes; # if not
recombination interval information has been provided sampling is
normal
            }
            my @sampled_genes_normal = (); # to store genes from
the sampling
```

```perl
            my @sampled_genes_interval = (); # to store genes
from the sampling with recomb intervals
            # get genes from separated recombination intervals by
sampling unconditional to the recombination intervals
            if (scalar @normal_sampling > 0){
                # if sampling is not conditional on the node
degree
                my @index = @{ get_rand_index(scalar @$genes,
scalar @normal_sampling)};
                @sampled_genes_normal = @$genes
[@index];
            }
            # if some genes need to be sampled conditional on
recombination intervals

            if (scalar @interval_sampling > 0){
                # loop over all group of genes
                for (my $interval = 0; $interval < scalar
@interval_sampling; $interval++){
                    # get the genes ids
                    my @interval_genes = @{ $interval_sampling
[$interval] };
                    my @seed_intervals = map { @{ $gene_stats-
>{ $_ }->{recomb_int} } } @interval_genes;
                    my %tmp = ();
                    foreach my $int (@seed_intervals) {
                        map { $tmp{$_} = "" }  @{ $comb_int-
>{$int}->{genes} };
                    }
                    my @interval_seed = keys %tmp;
                    my @all_intervals = keys %{ $comb_int };
                    # sample one gene and then sample the rest
from its interval
                    my $idx = int(rand(scalar
@all_intervals));
                    my @interval_sampling_universe = @
{ $comb_int->{$all_intervals[$idx]}->{genes} };
                    # skip the interval if has less genes than
need to be sampled
                    if (scalar @interval_genes >
@interval_sampling_universe){
                        $interval--;
                        next;
                    }
                    my $ratio = scalar @interval_seed/scalar
@interval_sampling_universe;
                    $ratio = 1/$ratio if ($ratio >1);
                    my $rand = rand();
                    if ($ratio < $rand){
                        $interval--;
                        next;
                    }
                    my @index = @{ get_rand_index(scalar
@interval_sampling_universe, scalar @interval_genes)};
```

```perl
                    my @selected_genes = sort { $gene_stats->
{ $b }->{stat} <=> $gene_stats->{ $a }->{stat}}
@interval_sampling_universe[@index];
                    if (defined $best_x_interval){
                        my $top_per_rand_invertal = shift
@selected_genes;
                        push @sampled_genes_interval ,
{ 'id'=>$top_per_rand_invertal, 'n' => scalar
@interval_sampling_universe };
                    } else {
                        foreach my $s (@selected_genes){
                            push @sampled_genes_interval,
{ 'id'=>$s, 'n' => scalar @interval_sampling_universe };
                        }
                    }
                }
            }
            my @sampled_gene_stats = map { $gene_stats->{$_}->
{stat}; } @sampled_genes_normal;
            foreach my $sampled_gene ( @sampled_genes_interval ){
                if (defined $best_x_interval){
                    my $sidak_p = 1 - ( 1 - $gene_data
{ $sampled_gene->{id} }->{pvalue} )**$sampled_gene->{n};
                    $sidak_p -= 1e-15 if ($sidak_p ==1);
                    $sidak_p =  $gene_data{$sampled_gene->
{id}}->{pvalue} * $sampled_gene->{n} if ($sidak_p ==0);
                    push @sampled_gene_stats, -1 *
gsl_cdf_ugaussian_Pinv($sidak_p);
                } else {
                    push @sampled_gene_stats, $gene_stats->
{ $sampled_gene->{id} }->{stat};
                }
            }
            my $s = &$network_stat(\@sampled_gene_stats);
            push @rand_stat, $s;
        }
        return(\@rand_stat);
}
sub defined_set_stat {
        my $stat = shift;
        if ($stat eq 'stouffer_z_score') {
            return(\&stat_set_stouffer_z_score);
    } elsif ($stat eq 'mean'){
            return(\&stat_set_mean_z_score);
    } else { die("The statistics [ $stat ] you want to apply to
the sub-networks is not recognized\n\n"); }

}
sub stat_set_mean_z_score {
    my $stat = shift;
    my $back = 0;
    map { $back += $_ } @$stat;
    $back /= scalar @$stat;
    return($back);
}
sub stat_set_stouffer_z_score {
```

```perl
    my $stat = shift;
    $stat = pdl @$stat;
    my $back = $stat->sum/sqrt(length $stat->list);
    return($back);
}

sub get_rand_index {
    my $max = shift;
    my $N = shift;
        my @universe = (0..$max-1);
        my @index = ();
    for (1 .. $N){
              my $i = int(rand(scalar @universe));
          push @index, splice (@universe,$i,1);
    }
    return(\@index);
}

sub report_advance {
    my ($index,$rep,$tag) = @_;
    if (( $index/$rep - int($index/$rep)) == 0) {
          print_OUT("   '->Done with [ $index ] $tag",$LOG);
    }
}

sub make_file_name_array {
    my $file = shift;
    my @back = ();
    my @body = split(/\#/,$file);
    for my $chr (1..26){
          push @back, join "$chr", @body;
    }
    return([@back]);
}

__END__

=head1 NAME

 Perl implementation of PAGE: parametric analysis of gene set
enrichment. As bonus includes strategies to correct for gene
clusters.

=head1 DESCRIPTION

B<This program> will read a file with gene symbols and
statistics and performed the PAGE gene-set analysis. It reads a
at least two files: a file with gene id's and p-values and a
second file with the gene-set definitions. Please check the
original paper Kim SY, Volsky DJ: PAGE: parametric analysis of
gene set enrichment. BMC Bioinformatics 2005, 6:144. for details
of the method. In addition it implements correction for Linkage
Disequilibrium which is useful when analysing results from
```

genome-wide association studies. Please check http://github.com/
inti/ for updates and documentation.

=head1 SYNOPSIS

script [options]

```
     General options
     -h, --help        print help message
     -m, --man         print complete documentation
     -report               how often to report advance
     -gmt              gene-set definitions on GMT format
     -file             gene p-value file
     -out, -o          output file
     -max_size         max gene-set size
     -min_size         min gene-set size
     -ref_list         set reference list for the analysis

     Analysis modifiers
     -set_stat         statistics to calculate over the sub-
networks
     -z_score          input values are z_scores (the absolute
values will be used)
     -cgnets               File with gene ids. Only gene-sets
that contain them will be analyzed.
     -cgnets_all       Set all genes in gene-sets as CGNets seed.
it basically runs the CGNets analysis for all genes.
     -Neff_gene_sets       Calculate the number effective gene-
sets being analysed.
     -gs_coverage      number [0,1]. Fraction of the gene-set
that must be covered by the
                       experiment for the gene set to be
considered in the analysis
     -recomb_intervals correct for genes that are in the same
recombination interval
     -interval_merge       Integer. Recombination intervals
closer than this number will be merge. It is helpfull to
                       assess the effect of residual long range
LD. It can be very strict if you use a large number.
     -interval_merge_by_chr Same as before but will define a
whole chromosome as the interval. It is the extreme of the
                       previous option and overrides it.
     -best_per_interval    Used by default if recomb_intervals
is used. If two or more genes are in the
                       same recombination interval, the one with
the best statistic will be selected. This behavior
                       will be mantained on the permutation
sampling to calculate the null. No permutation will be run
                       with this option, I may fix this in the
feature.

     Output modifiers:
     -append               Append results to output file rather
than overwrite it
     -add_file_name        add the input file name to the
result
```

```
    Permutations:
    –perm             number of permutations
    –complete_interval_sampling Implements a method to correct
for recombination intervals but using all
                         statistics in the recombination
interval.
```

=head1 OPTIONS

=over 8

=item B<–help>

Print help message

=item B<–man>

print complete documentation

=item B<–report>

how often to report advance. Provide an integer X and the
program will report adnvance after X networks are analyzed.

=item B<–gmt>

ggene–set definitions on GMT format

=item B<–file>

gene p–value file. Tab separated file with at least two columns:
gene_id and p–value. please make sure that not p–values equal to
0 are included, those genes will be excluded from the analysis.

=item B<–perm>

number of permutations

=item B<–out>

output file: ithe output file looks like
GO0007156  9.032e–06   4.288 4.474e–06   4.441
GO0016339  1.656e–04   3.590 2.306e–04   3.502

columns are:
1) gene–set id
2) PAGE asymtotic p–value, z_score,
3) p–value calculated with the null distribution calculated via
sampling
4) z score calculated with a null distribution calculated via
sampling.

the last 2 columsn will only be printed if permutatins are run.

=item B<–max_size>

max gene-set size

=item B<-min_size>

min gene-set size

=item B<-recomb_intervals>

correct for genes that are in the same recombination interval.
On GWAS analysis one would often derive one p-value per genes
and these will be correlated if the genetic variants on
different genes are in Linkage Desequilibrium, as for example
when genes lay on the same recombination interval (pice of
genome between two recombination hot-spots). The proble is that
the statistics calculated across the sub-network asssume that
the gene p-values are independent. With this option is possible
to provide a set of genomic interval that group genes, e.g.
recombination interval. This information will be use during the
MC sampling. For example, if we have a network with 10 genes, 5
of which lay on the same recombination interval. With the -
recomb_intervals option set on the montecarlo sampling 5 of the
genes will be obtain from a single recombination interval
(elsewhere in the genome), thus assuring the genetic structure
in the subnetwotk is somehow preserved.

=item B<-interval_merge>

Integer. Recombination intervals closer than this number will be
merge. It is helpfull to assess the effect of residual long
range LD. It can be very strict if you use a large number.

=item B<-interval_merge_by_chr>

Same as before but will define a whole chromosome as the
interval. It is the extreme of the previous option and overrides
it.

=item B<-z_score>

Input values are z_score intead of p-values, for calculations
the absolute value of the z-score will be taken. Permutations
must be performed together with this option.

=item B<-append>

Append results to output file rather than overwrite it

=item B<-add_file_name>

Add the input file name to the result. Usefull is analysing
different data sets that wil be concatenated in on single file

=item B<-ref_list>

set reference list for the analysis

```
=item B<-set_stat>
```

statistics to calculate over the sub-networks. options are
stouffer_z_score and mean. mean if the default

```
=item B<-best_per_interval>
```

Used by default if recomb_intervals is used. If two or more
genes are in the same recombination interval, the one with the
best statistic will be selected. This behavior will be
mantainedon the permutation sampling to calculate the null.

```
=item B<-complete_interval_sampling>
```

Implements a method to correct for recombination intervals but
using all statistics in the recombination interval.

```
=item B<-cgnets>
```

Implements the CGNet analysis. It receives a list of genes and
will restric the analysis to gene-sets that contain them.

```
=item B<-cgnets_all>
```

Set all genes in gene-sets as CGNets seed. it basically runs the
CGNets analysis for all genes.

```
=item B<-Neff_gene_sets>
```

Calculate the number effective gene-sets being analysed. It can
be quite slow is many pathways are under analysis but it will
finish in a reasinable time. It is quite useful when performing
the CGNets analysis.

```
=back
```

```
=cut
```

# 8.6. meta_analysis.pl

Script to perform meta-analysis using fix and random-effects statistics. It also computes statistical heterogeneity statistics.

```perl
#!/usr/bin/perl -w
use strict;
use Getopt::Long;
use Pod::Usage;
use PDL;
use PDL::GSL::CDF;
use Data::Dumper;

use GWAS_STATS qw( get_lambda_genomic_control );

our (      $help,$man, $out, $w, $in_files,
           $stouffer,$w_col,$random,$stat_or,
           $stat_col,$id_col, $stat_pvalue,$header,
           $min_studies, $gc_correction,$lambda,$sdtze,
           $gc_correction_all
);

GetOptions(
       'help|h'                => \$help,
       'man'                   => \$man,
       'out|o=s'               => \$out, #name of the output file
       'file|f=s@'             => \$in_files,
       'weights|w=f@'          => \$w,
       'w_col=i@'              => \$w_col,
       'id_col=i@'             => \$id_col,
       'stat_col=i@'           => \$stat_col,
       'stat_pvalue'           => \$stat_pvalue,
       'stat_or'               => \$stat_or,
       'header'                => \$header,
       'min_studies|min_st=i'  => \$min_studies,
       'gc_correction=i@'          => \$gc_correction,
       'gc_correction_all'         => \$gc_correction_all,
       'lambda=i@'             => \$lambda,
       'sdtze=i@'              => \$sdtze,
) or pod2usage(0);

pod2usage(0) if (defined $help);
pod2usage(-exitstatus => 2, -verbose => 1) if (defined $man);
pod2usage(0) if (not defined $in_files);

defined $min_studies or $min_studies = -1;

if (not defined $stat_pvalue and not defined $stat_or){
    print_OUT("Please state wheter is a p-value [ -
stat_pvalue ] or odd-ratio [ -stat_or ] based analysis");
```

```perl
        exit(1);
}

if (scalar @$in_files < 2){
        print_OUT("Please provide at least two input files");
        exit(1);
}

my %data     = ();
my @studies  = ();
my %total_seen = (); # records how many times and ids has been
seen
# Defined weights
# weights can be either variable specific is provided with the
option -w_col
# or study specific if provided with the option -w
# if the option -w_col is defined then for each file the
variable weight will be extracted
my @study_weights = list ones scalar @$in_files;
if (defined $w){
        @study_weights = @$w;
}

# defined gc_correction for all studies
if (defined $gc_correction_all){
        $gc_correction = [ list ones scalar @$in_files ];
}

open(OUT,">$out") or die $!;
print_OUT("Writting output to [ $out ]");

foreach my $file (@$in_files) {
        # get name of input file
        my $st = $file;
        ($st) = ( $file =~ m/[\/\w+]{1,}\/(.*)/ ) if ($file =~ m/
\///);
        # get the study weights. if not provided will be 1
        my $st_w = shift @study_weights;
        # get the column number of variable specific weight
        my $st_w_col = shift @$w_col if (defined $w_col);

        # get the stat_col
        my $st_stat_col = 2;
        $st_stat_col = shift @$stat_col if (defined $stat_col);

        # get the stat_col
        my $st_id_col = 1;
        $st_id_col = shift @$id_col if (defined $id_col);

        push @studies, $st; # store the name of this study for the
print out.
        # print out some infor about the information read in the
file
        print_OUT("Reading [ $file ]");
```

```perl
    print_OUT("   '-> Variables in col [ $st_w_col ]") if
(defined $w_col);
    print_OUT("   '-> Study weight [ $st_w ]") if (defined $w);
    print_OUT("   '-> Study stat in col [ $st_stat_col ]") if
(defined $stat_col);

    open( IN, $file ) or print_OUT("I cannot open file
[ $file ]\n") and die $!;
    my $counter = 0;
    while ( my $line = <IN> ) {
        if (defined $header and $counter == 0){
            $counter++;
            next;
        }
        chomp($line);
        # split line
        my @d = split( /[\s+\t+]/, $line );
        # set the variable weight to the var of the study
        my $var_w = double $st_w;
        # modify the vartiable weight if a columns with its
variance is given
        $var_w /= $d[ $st_w_col -1 ] if ( defined $w_col );
        $var_w = sclr $var_w; # make a perl scalar to simply
calculations later
        # get the variable stat
        my $var_stat = $d[ $st_stat_col - 1 ];
        if (defined $stat_pvalue){
            if ( $var_stat == 1 ){
                $var_stat = double 1-2.2e-16;
            }
            $var_stat = -1 * gsl_cdf_ugaussian_Pinv
( $var_stat );
        } elsif (defined $stat_or){
            $var_stat = log $var_stat;
        }
        # store the data as a pseudohash
        $data{ $d[ $st_id_col -1 ] }{$st} = { 'w' => $var_w,
'stat' => $var_stat };
        $total_seen{ $d[ $st_id_col -1 ] }++;
    }
    close(IN);
}

if (defined $gc_correction){
    print_OUT("Calculating lambda for genomic control
correction");

    my %studies_p = ();
    foreach my $var (keys %data){
        my $f_counter = 0;
        foreach my $file ( @$in_files ){
            my $st = $file;
            ($st) = ( $file =~ m/[\/\w+]{1,}\/(.*)/ ) if
($file =~ m/\//);
            if (not exists $data{$var}{$st}){
                $f_counter++;
```

```perl
                        next;
                }
                next unless ($gc_correction->[$f_counter++] ==
1);
                push @{ $studies_p{$st} }, 1 -
gsl_cdf_ugaussian_P($data{$var}{$st}->{stat});
        }
    }
    foreach my $st ( keys %studies_p ){
            print_OUT(" '-> [ $st ]");
            my $gc_lambda = get_lambda_genomic_control($studies_p
{$st});
            print_OUT("   '-> lambda (median) of
[ $gc_lambda ]");
            if ($gc_lambda > 1){
                print_OUT("   '-> Applying GC correction in
[ $st ]");
                my $assoc_p = [];
                foreach my $var (keys %data) {
                    next unless (exists $data{$var}{$st});
                    my $p = 1 - gsl_cdf_ugaussian_P($data
{$var}{$st}->{stat});
                    if ( $p == 1){
                        push @{ $assoc_p }, $p;
                        next;
                    }
                    my $var_chi = gsl_cdf_chisq_Pinv ( 1 -
$p , 1 );
                    $var_chi /=  $gc_lambda;
                    $p = gsl_cdf_chisq_P( $var_chi, 1 );
                    $data{$var}{$st}->{stat} =
gsl_cdf_ugaussian_Pinv( $p );
                    push @{ $assoc_p }, 1 - $p;
                }
                $gc_lambda = get_lambda_genomic_control
($assoc_p);
                print_OUT("   '-> After correction the lambda
is [ $gc_lambda ]");
            } else {
                print_OUT("   '-> GC correction not applied
because lambda is less than 1");
            }
        }
}

print_OUT("Calculating Stats");
print OUT join "\t",
("id","stouffer_fix","B_fix","Var_fix","Chi-
square_df1_fix","B_fix_P","stouffer_fix","B_random","Var_random"
,"Chi-
square_df1_random","B_random_P","Q","Q_P","tau_squared","I2","N"
,"binomial_rank_pvalue",@studies);
print OUT "\n";
foreach my $gn ( keys %data ) {
    my $count = scalar keys %{$data{ $gn } };
```

```perl
    next if ($count < 2); # next if only one study looked at
this gene
    next if ($count < $min_studies);
    next if ($total_seen{ $gn } > scalar @studies);

    my $all_w = [];
    my $all_b = [];
    # collect stat and weights for each study
    foreach my $st ( keys %{ $data{ $gn } } ) {
            push @{$all_w}, $data{ $gn }{$st}->{'w'};
            push @{$all_b}, $data{ $gn }{$st}->{'stat'};
    }
    $all_w = pdl $all_w;
    if ($all_w->minimum() == 1 and $all_w->maximum() ==1){
            $all_w *= $all_w->nelem;
    } else {
            $all_w = 1/$all_w;
    }
    my $metaResults = get_fix_and_radom_meta_analysis($all_b,
$all_w);

    my $max_b = maximum pdl $all_b;
    my $min_p = gsl_cdf_ugaussian_P( -1 * $max_b );
    my $binomial_rank_p = 1 - gsl_cdf_binomial_P(1,$min_p,
$metaResults->{'N'});

    printf OUT ("$gn\t%.6f\t%.6f\t%.6f\t%.6f\t%.6e\t%.6f\t%.6f
\t%.6f\t%.6f\t%.6e\t%.6f\t%.6e\t%.6f\t%.6f\t$metaResults->{'N'}
\t%.6e",
                    $metaResults->{'B_stouffer_fix'},
                    $metaResults->{'B_fix'},
                    $metaResults->{'V_fix'},
                    $metaResults->{'Chi_fix'},
                    $metaResults->{'Z_P_fix'},
                    $metaResults->{'B_stouffer_random'},
                    $metaResults->{'B_random'},
                    $metaResults->{'V_random'},
                    $metaResults->{'Chi_random'},
                    $metaResults->{'Z_P_random'},
                    $metaResults->{'Q'},
                    $metaResults->{'Q_P'},
                    $metaResults->{'tau_squared'},
                    $metaResults->{'I2'},
                    $binomial_rank_p,
                    );
    map {
            if (defined $w_col){
                    if   ( exists $data{ $gn }{$_} ) { printf OUT
("\t%.6f(%.3f)",$data{ $gn }{$_}->{'stat'},1/$data{ $gn }{$_}->
{'w'}); }
                    else { print OUT "\tNA"; }
            } else {
                    if   ( exists $data{ $gn }{$_} ) { printf OUT
("\t%.6f",$data{ $gn }{$_}->{'stat'});
                    #print Dumper($data{ $gn }),"\n";
                    }
```

```perl
                    else { print OUT "\tNA"; }
            }
    } @studies;
    print OUT "\n";
    delete($data{ $gn });
}

print_OUT("Done baby");

exit;
sub get_fix_and_radom_meta_analysis {
    my $B = shift;
    my $SE = shift;
    my $external_w = shift;
    my $VarCov = shift;
    if (ref($B) eq 'ARRAY') { $B = pdl $B; }
    if (ref($SE) eq 'ARRAY') { $SE = pdl $SE; }
    if (ref($external_w) eq 'ARRAY') { $external_w = pdl
$external_w; }
    my $N = $B->nelem;
    if (not defined $VarCov) {
            $VarCov = stretcher(ones $N);
    }
    my $W = null;
    if ( (ref($external_w) eq 'PDL') and (nelem($external_w >
0) > 0) and (dsum( ($external_w - $external_w->davg)**2 ) > 0 ))
{
            $W = 1/($SE + 1/$external_w);
    } else {
            $W = 1/$SE;
    }

    # calculate fix effect estimate
    my $norm_w_fix = $W/$W->dsum;
    #my $B_fix = dsum($B*$W)/$W->dsum;
    #my $V_fix = dsum($W*$W->transpose*$VarCov);
    my $B_fix = dsum($B*$norm_w_fix)/$norm_w_fix->dsum;
    my $V_fix = dsum($norm_w_fix*$norm_w_fix->transpose*
$VarCov);
    my $fix_chi_square_df1 = ($B_fix**2)/$V_fix;

    my $stouffer_w_fix = $W/dsum($W**2);
    my $B_stouffer_fix = dsum($B*$stouffer_w_fix)/sqrt(dsum
($stouffer_w_fix*$stouffer_w_fix->transpose*$VarCov));

    # calculate heteroogeneity parameter Q
    my $Q = 0.0;
    {
            my $cor = $VarCov->copy();
            $cor->diagonal(0,1) .=0;
            $cor = (3.263*abs($cor) + 0.710*(abs($cor)**2) +
0.027*(abs($cor)**3));
            my $df = 8/(dsum($cor*$W*$W->transpose) + 4*dsum
($W**2));
            my $Q_naive = dsum ($W * (($B_fix - $B)**2));
```

```perl
            $Q = sclr gsl_cdf_chisq_Pinv( gsl_cdf_chisq_P
(  $df*0.5*$Q_naive, $df), $N - 1);
    }
    # calculate heteroogeneity parameter I-squared
    my $I_squared = 0.0;
    if ($Q > ($N - 1)){
            eval { $I_squared = 100*($Q - ($N - 1))/$Q; };
            if ($@){
                $I_squared = 0.0;
            }
    }
    # calculate tau-squared
    my $tau_squared = 0.0;
    if ($Q > ($N - 1)){
            eval { $tau_squared = ($Q - ($N - 1))/(dsum($W) -
dsum($W**2)/$W->dsum); };
            if ($@){
                # tau-squared equal 0 if was < 0
                $tau_squared = 0.0;
            }
    }
    # calculate the random effect estimate
    my $w_star = null;
    if ( (ref($external_w) eq 'PDL') and (nelem($external_w >
0) > 0) and (dsum( ($external_w - $external_w->davg)**2 ) > 0 ))
{
            $w_star = 1/( $tau_squared + $SE + 1/$external_w);
    } else {
            $w_star = 1/( $tau_squared + $SE);
    }
    my $norm_w_random = $w_star/$w_star->dsum;
    my $B_random = dsum( $B*$norm_w_random)/$norm_w_random-
>dsum;
    #my $B_random = dsum( $B*$w_star)/$w_star->dsum;
    my $V_random = dsum($norm_w_random*$norm_w_random-
>transpose*$VarCov);
    #      my $V_random = dsum($w_star*$w_star->transpose*
$VarCov);
    my $random_chi_square_df1 = ($B_random**2)/$V_random;

    my $stouffer_w_random = $w_star/dsum($w_star**2);
    my $B_stouffer_random = dsum($B*$stouffer_w_random)/sqrt
(dsum($stouffer_w_random*$stouffer_w_random->transpose*
$VarCov));

    my $back =  {
            'B_stouffer_fix' => $B_stouffer_fix,
            'B_stouffer_random' => $B_stouffer_random,
            'B_fix' => $B_fix,
            'B_random' => $B_random,
            'V_fix' => $V_fix,
            'V_random' => $V_random,
            'Chi_fix' => $fix_chi_square_df1,
            'Chi_random' => $random_chi_square_df1,
            'Q' => $Q,
            'I2' => $I_squared,
```

```perl
            'tau_squared' => $tau_squared,
            'N' => $N,
            'Z_P_fix' => gsl_cdf_gaussian_P( -1 * $B_fix ,sqrt
($V_fix )),
            'Z_P_random' => gsl_cdf_gaussian_P( -1 *
$B_random,sqrt($V_random) ),
            'Q_P' => 1- gsl_cdf_chisq_P($Q ,$N -1),
    };
    return $back;
}

sub print_OUT {
    my $string = shift;
    print scalar localtime(), "\t$string\n";
    #     print LOG scalar localtime(), "\t$string\n";
}

__END__

=head1 NAME

 Running network analysis by greedy search

 =head1 SYNOPSIS

 script [options]

 -h, --help          print help message
 -m, --man       print complete documentation
 -report             how often to report advance
 -verbose, -v        useful for debugging

 Input Files:

 =head1 OPTIONS

 =over 8

 =item B<-help>

 Print help message

 =item B<-man>

 print complete documentation

 =back

 =head1 DESCRIPTION

 TODO

 =cut
```

## 8.7. CovMatrix.pm

Perl module with routines to calculate correlation matrixes using a shrinkage approach. Original methods are described on Schafer and Strimmer (2005).

```perl
package CovMatrix;
use strict;
use warnings;
use Carp;
use Exporter qw (import);

# define verbose = undef;
my $v = undef;

# check the modules needed

eval {
    use PDL;
    use PDL::Matrix;
    use PDL::Primitive;
    use PDL::NiceSlice;
    use PDL::LinearAlgebra;
    use PDL::LinearAlgebra::Real;
};
if ($@) {
    print "Some libraries does not seem to be in you system.
quitting\n";
    exit(1);
}

our (@EXPORT, @EXPORT_OK, %EXPORT_TAGS);

@EXPORT = qw( check_positive_definite cov_shrink
make_positive_definite is_positive_definite);    # symbols to
export by default
@EXPORT_OK = qw( check_positive_definite cov_shrink
make_positive_definite is_positive_definite); # symbols to
export on request

sub wt_moments{
    my ($x, $w) = @_;
    $w = pvt_check_w($w, $x->getdim(1));
    my $h1 = 1/(1 - dsum($w * $w));
    my $m = dsumover($w * $x->transpose);
    my $v = $h1 * (dsumover($w * transpose($x**2)) - dsumover
($w * $x->transpose)**2);
    return({ 'mean' => $m, 'var' => $v});
}
sub wt_scale{
    my $x = shift;
```

```perl
    my $w = shift;
    my $center = shift;
    my $scale = shift;
    $w = pvt_check_w($w,$x->getdim(1));
    my $wm = { 'mean' => $x->xchg(0,1)->daverage, 'var' => $x-
>xchg(0,1)->stdv**2 };
    if ($center eq 'TRUE'){
        $x = $x - $wm->{'mean'};
    }
    if ($scale eq 'TRUE') {
        my $sc = sqrt($wm->{'var'});
        $x = $x/$sc;
    }
    return($x);
}
sub pvt_check_w {
    my ($w, $n) = @_;
    if (not defined $w) {
        $w = ones $n;
            $w /= $n;
            $w /= $w->dsum;
    } else {
        if ($w->getdim(0) != $n) {
            warn("Weight vector has incompatible length. Is
[ " . $w->getdim(0) . " ] should be [ $n ]\n");
        }
        $w = ones $n;
            $w /= $n;
            $w /= $w->dsum;
        my $sw = sum($w);
        if ($sw != 1){ $w = $w/$sw; }
    }
    return($w);
}

sub make_positive_definite {

    my ($m, $tol) = @_;
    my $d = $m->getdim(1);
    if ($m->getdim(1) != $m->getdim(0)) { die("Input matrix is
not square!"); }
    my ($es,$esv) = eigens $m;
    if (not defined $tol){ $tol = $d * max(abs($esv)) *
2e-16; }
    my $delta = 2 * $tol;
    my $tau = $delta - $esv;
    $tau->( which($tau < 0) ) .= 0;
    my $dm = $es x stretcher($tau) x transpose($es);
    return($m + $dm)
}

sub is_positive_definite {
    my $m = shift;
    my $tol = shift;
    my ($es,$esv) = eigens $m;
```

```perl
    if (not defined $tol){ $tol = $m->getdim(0) * max(abs
($esv)) * 2e-8; }
    if ( sum($esv > $tol) == scalar( $esv->list) ) {
        return(0);
    } else {
        return(1);
    }
}
=head
 sub is_positive_definite
 function (m, tol, method = c("Eigen", "chol"))
 {
 method = match.arg(method)
 if (!is.matrix(m))
 m = as.matrix(m)
 if (method == "Eigen") {
 eval = Eigen(m, only.values = TRUE)$values
 if (is.complex(eval)) {
 warning("Input matrix has complex eigenvalues!")
 return(FALSE)
 }
 if (missing(tol))
 tol = max(dim(m)) * max(abs(eval)) * .Machine$double.eps
 if (sum(eval > tol) == length(eval))
 return(TRUE)
 else return(FALSE)
 }
 if (method == "chol") {
 val = try(chol(m), silent = TRUE)
 if (class(val) == "try-error")
 return(FALSE)
 else return(TRUE)
 }
 }
=cut

sub cov_shrink {
    my ($x, $lambda, $lambda_var, $w, $collapse, $verbose) =
@_;
    my $n = $x->getdim(1);
    if (not defined $lambda) { $lambda = -1; }
    if (not defined $lambda_var) { $lambda_var = -1; }
    if (not defined $w) {
        $w = ones $n;
        $w /= $n;
        $w /= $w->sum;
    }
    my $sc_data = pvt_svar($x, $lambda_var, $w, $verbose);
    my $sc = $sc_data->{'vs'}->sqrt;
    my $c = pvt_powscor($x, 1, $lambda, $w, $collapse,
$verbose);
    my $cor = $c->{powr};
    # I have commented this if because I do not know when is
useful.
    # it seems to be some kind of error checking that i do not
understand.
```

```perl
    #if ( $c->isempty){
    #       $c->{powr} = $c->{powr} * $sc * $sc;
    #} else{
    $c->{powr} = transpose($c->{powr}*$sc)* $sc;
    #}
    my $back = {
            'lambda_var' => $sc_data->{"lambda_var"},
            'lambda_var_estimated' => $sc_data->
{"lambda_var_estimated"},
            'lambda_cor' => $c->{"lambda"},
            'lambda_cor_estimated' => $c->{"lambda_estimated"},
            'cov' => $c->{powr},
            'cor' => $cor,
    };
    return($back);
}

sub pvt_get_lambda {
    my ($x, $lambda, $w, $verbose, $type,$target) = @_;
    my ($kind,$func) = undef;
    my $lambda_estimated = undef;

    if ($lambda < 0) {
            if ($type eq "correlation") {
                $kind = "lambda (correlation matrix):";
                $lambda = pvt_corlambda($x, $w, $target);
            }
            if ($type eq "variance") {
                $kind = "lambda.var (variance vector):";
                $lambda = pvt_varlambda($x, $w, $target);
            }
        if ($verbose) {
            print  ("Estimating optimal shrinkage intensity
[ $kind ]");
            }
            $lambda_estimated = "TRUE";
            if ($verbose) {
             print  ("lambda [ $lambda ]");
            }
    } elsif ($lambda > 1){
            $lambda = 1;
            $lambda_estimated = "FALSE";
            if ($verbose) {
                print  ("Specified shrinkage intensity $kind
[ $lambda ]");
            }
    }
    if ($type eq "correlation") {
        return( { 'lambda' => $lambda, 'lambda_estimated' =>
$lambda_estimated } );
    }
    if ($type eq "variance") {
        return( { 'lambda_var' => $lambda,
'lambda_var_estimated' => $lambda_estimated } );
```

```perl
        }
}

sub pvt_powscor {
    my ($x, $alpha, $lambda, $w, $collapse, $verbose) = @_;
    my $xs = wt_scale($x, $w, 'TRUE', 'TRUE');
    my $h1 = 1/(1 - dsum($w * $w));
    my $z = pvt_get_lambda($xs, $lambda, $w, $verbose,
"correlation", 0);
    my $p = $xs->getdim(0);
    my $powr = null;
    # for debug
    #       $alpha = 0.5;
    if ($z->{'lambda'} == 1 or $alpha == 0) {
        if ($collapse) {
            $powr = ones $p;
        } else {
            $powr = zeroes $p,$p;
                $powr->diagonal(0,1) .= ones $p;
        }
    } elsif ($alpha == 1) {
            my $cross = $xs*$w->transpose->sqrt;
        my $r0 = $h1 * mcrossprod( $xs*$w->transpose->sqrt);
        $powr = (1 - $z->{'lambda'}) * $r0;
        $powr->diagonal(0,1) .= ones $powr->getdim(0);
    } else {
            # These methods have not been implemented
            # program will throw an error.
        my $zeros = $xs;
        my $svdxs = fast_svd($xs);
        my $m = length($svdxs->{'d'});
        my $UTWU = transpose($svdxs->{'u'}) x ( $svdxs->{'u'}*
$w);
        my $C = $UTWU*$svdxs->{'d'}*$svdxs->{'u'}->transpose;
        $C = (1 - $z->{'lambda'}) * $h1 * $C;
        $C = ($C + transpose($C))/2;
        if ($lambda == 0) {
            if ($m < $p - dsum($zeros)){
                warning("Estimated correlation matrix doesn't
have full rank - pseudoinverse used for inversion.\n");
                    $powr = $svdxs->{'v'} x ((mpower($C,
$alpha) x transpose($svdxs->{'v'})) );
                }
            } else {
                my $F = $m->diagonal(0,1) - mpower($C/$z->{'lambda'}
+ $m->diagonal(0,1), $alpha);
            $powr = ( $p->diagonal(0,1) - $svdxs->{'v'} x ($F x
transpose($svdxs->{'v'}) )) * ($z->{'lambda'})**$alpha;
        }
        $powr->diagonal(0,1) = 1;
        #rownames(powr) = colnames(xs)
        #colnames(powr) = colnames(xs)
    }
    $powr = {
            'powr' => $powr,
            'lambda' => $z->{lambda},
```

```perl
                'lambda_estimated' => $z->{lambda_estimated},
                'class' => "shrinkage",
        };
        return($powr);
}

sub fast_svd {
        my ($m, $tol) = @_;
        my $n = $m->getdim(1);
        my $p = $m->getdim(0);
        my $EDGE_RATIO = 2;
        if ($n > $EDGE_RATIO * $p) {
                return(psmall_svd($m, $tol));
        } elsif ($EDGE_RATIO * $n < $p) {
                return(nsmall_svd($m, $tol));
        } else {
                return(positive_svd($m, $tol));
        }
}
=head
 sub psmall.svd {
 my ($m, $tol) = @_;
 my $B = mcrossprod($m);
 my $s = svd($B, 0);
 if (not defined ($tol)) {
 my $tol = $B->getdim(1) * maximum($s->{'d'}) * 2.220446e-16;
 my $Positive = which($s->{'d'} > $tol);
 my $d = sqrt($s->{'d'}->($Positive));
 my $v = $s->{'v'}->(, $Positive);
 my $u = m %*% v %*% diag(1/d, nrow = length(d))
 return(list(d = d, u = u, v = v))
 }
=cut
sub pvt_svar {
        my ($x, $lambda_var, $w, $verbose) = @_;
        $w = pvt_check_w($w,$x->getdim(1));
        my $xs = wt_scale( $x,$w,'TRUE', 'FALSE');
        my $wt_mom = wt_moments($xs,$w);
        my $v = $wt_mom->{var};
        my $tgt = median($v);
        my $z = pvt_get_lambda($xs, $lambda_var, $w, 0, "variance",
$tgt);
        my $vs = $z->{'lambda_var'} * $tgt + (1 - $z->
{'lambda_var'}) * $v;
        my $back = {
                'vs' => $vs,
                "lambda_var" => $z->{'lambda_var'},
                "lambda_var_estimated" => $z->
{'lambda_var_estimated'},
                "class" => "shrinkage",
        };
        return($back);
}

sub pvt_varlambda {
        my ($xc,$w, $target) = @_;
```

```perl
    my $w2 = dsum($w * $w);
    my $h1 = 1/(1 - $w2);
    my $h1w2 = $w2/(1 - $w2);
    my $zz = $xc**2;
    my $q1 = dsumover(  $zz->transpose*$w  );
    my $q2 = dsumover( ($zz->transpose**2)*$w ) - $q1**2;
    my $numerator = dsum($q2);
    my $lambda = undef;
    my $denominator = dsum(($q1 - $target/$h1)**2);
    if ($denominator == 0) {
            $lambda = 1;
    } else {
            $lambda = ($numerator/$denominator) * $h1w2;
            if ($lambda > 1) { $lambda =1; }
    }
    return($lambda);
}

sub pvt_corlambda {
    my ($xs, $w, $target) = @_;
    my $w2 = dsum($w * $w);
    my $h1w2 = $w2/(1 - $w2);
    my $sw = sqrt($w);
    my $Q1_squared = (mcrossprod($xs*$sw->transpose))**2;
    my $Q2 = mcrossprod(($xs**2)*$sw->transpose)  - $Q1_squared;
    my $denominator = dsum($Q1_squared) - dsum( $Q1_squared-
>diagonal(0,1));
    my $numerator = dsum($Q2) - dsum( $Q2->diagonal(0,1));
    my $lambda = undef;
    if ($denominator == 0) {
        $lambda = 1;
    } else {
            $lambda = ($numerator/$denominator) * $h1w2;
            if ($lambda > 1) { $lambda = 1; }
    }
    return($lambda);
}

sub check_positive_definite {
    my $cov_in = shift;
    my $tol = shift;

    defined $tol or $tol = 1e-8;
    my $status = 0;
    my $cov = $cov_in;
    eval { mchol ($cov) };
    my $i = 0;
    while($@ ne ""){
            $@ = "";
            $i++;
            if ($i > 25){
                    last;
            }
            my ($es,$esv) = eigens $cov;
            $cov = make_positive_definite($cov,$tol);
            ($es,$esv) = eigens $cov;
```

```perl
        eval { mchol ($cov) };
    }

    if (is_positive_definite($cov,$tol) == 1){
        $cov = make_positive_definite($cov,$tol);
    }

    if(is_positive_definite($cov,$tol) == 1){
        $cov = $cov_in;
        $cov->diagonal(0,1) .= 1.0001;
    }
    if(is_positive_definite($cov,$tol) == 1){
        $cov->diagonal(0,1) .= 1.001;
    }
    if(is_positive_definite($cov,$tol) == 1){
        $cov->diagonal(0,1) .= 1.01;
    }

    if (is_positive_definite($cov,$tol) == 1){
        print "Matrix never positive definite\n";
        $status = 1;
    }
    return($cov,$status);
}

1;
```

# 8.8. GWAS_IO.pm

Perl routines to deal with input and out of commonly use format on GWAS.

```perl
package GWAS_IO;
use strict;
use warnings;
use Carp;
use Exporter qw (import);

# define verbose = undef;
my $v = undef;

# check the modules needed

eval {
    use PDL;
    use PDL::Matrix;
    use PDL::NiceSlice;
    #use PDL::LinearAlgebra; # commented until re-implement the
use of simulation to calculate p-values
    use IO::File;
    use IO::Seekable;
    use Fcntl;
};
if ($@) {
    print "Some libraries does not seem to be in you system.
quitting\n";
    exit(1);
}

our (@EXPORT, @EXPORT_OK, %EXPORT_TAGS);

@EXPORT = qw( build_index line_with_index
extract_binary_genotypes extract_genotypes_for_snp_list
get_snp_list_from_bgl_format get_snp_list_from_ox_format
read_bim read_fam read_map_and_ped );                # symbols
to export by default
@EXPORT_OK = qw( build_index line_with_index
extract_binary_genotypes extract_genotypes_for_snp_list
get_snp_list_from_bgl_format get_snp_list_from_ox_format
read_bim read_fam read_map_and_ped);                 # symbols to
export on request

# usage: build_index(*DATA_HANDLE, *INDEX_HANDLE)
sub build_index {
    my $data_file  = shift;
    my $index_file = shift;
    my $offset     = 0;
```

```perl
    while (<$data_file>) {
        print $index_file pack("N", $offset);
        $offset = tell($data_file);
    }
}

# usage: line_with_index(*DATA_HANDLE, *INDEX_HANDLE,
$LINE_NUMBER)
# returns line or undef if LINE_NUMBER was out of range
sub line_with_index {
    my $data_file   = shift;
    my $index_file  = shift;
    my $line_number = shift;

    my $size;               # size of an index entry
    my $i_offset;           # offset into the index of the entry
    my $entry;              # index entry
    my $d_offset;           # offset into the data file

    $size = length(pack("N", 0));
    $i_offset = $size * ($line_number - 1);

    seek($index_file, $i_offset, 0) or return;
    read($index_file, $entry, $size);
    $d_offset = unpack("N", $entry);
    if (not defined $d_offset){
        return('1');
    } else {
        seek($data_file, $d_offset, 0);
        return scalar(<$data_file>);
    }
}

# this subroutine read the fam file and stores the information
in an array
# the elements of the array are pseudo hashes with all the
sample's information
sub read_fam {
    my $fam = shift;
    print "Reading samples info from [ $fam ]\n";
    open( FAM, $fam ) or print "Cannot open [ $fam ] file\n"
and exit(1);
    my @back = ();
    while ( my $s = <FAM> ) {
        my @data = split( /\s+/, $s );
        push @back,
        {
                'iid'   => $data[0],
                'fid'   => $data[1],
                'mid'   => $data[2],
                'pid'   => $data[3],
                'sex'   => $data[4],
                'pheno' => $data[5],
        };
    }
    #print_OUT("[ " . scalar @back . " ] samples read");
```

```perl
      return ( \@back );
}

# this subroutine read the bim file and store information about
the SNPs
# each element of the array returned is a pseudo hash with the
SNP information
sub read_bim {
    my $bim = shift;
    my $affy_to_rsid = shift;
    my $affy_id = shift;
    print "Reading SNPs info from [ $bim ]\n";
    open( BIM, $bim ) or print "Cannot open [ $bim ] file\n"
and exit(1);
    my @back = ();
    while ( my $snp = <BIM> ) {
            chomp($snp);
            my @data = split( /\t+/, $snp );
            # if an affy to rsid mapping was provided change the
ids
            if ( defined $affy_to_rsid ) {
                    if ($data[1] !~ m/^rs/){
                            if (exists $affy_id->{$data[1]}){ $data[1]
= $affy_id->{$data[1]};}
                    }
            }
            push @back,
        {
                    'snp_id' => $data[1],
                    'chr'    => $data[0],
                    'cm'     => $data[2],
                    'pos'    => $data[3],
                    'a2'     => $data[4],
                    'a1'     => $data[5],
        };
    }
    #print_OUT("[ " .  scalar @back . " ] SNPs on BED file");
    return ( \@back );
}

sub read_map {
    my $map = shift;
    my $affy_to_rsid = shift;
    my $affy_id = shift;
    #print_OUT("Reading SNPs info from [ $map ]");
    open( MAP, $map ) or #print_OUT("Cannot open [ $map ]
file") and exit(1);
    my @back = ();
    while ( my $snp = <MAP> ) {
            chomp($snp);
            my @data = split( /\t+/, $snp );
            # if an affy to rsid mapping was provided change the
ids
            if ( defined $affy_to_rsid ) {
                    if ($data[1] !~ m/^rs/){
```

```perl
                            if (exists $affy_id->{$data[1]}){ $data[1]
= $affy_id->{$data[1]};}
                        }
                    }

            push @back, {
                    'snp_id' => $data[1],
                    'chr'    => $data[0],
                    'cm'     => $data[2],
                    'pos'    => $data[3],
                    'a2'     => 0,
                    'a1'     => 0,
        };
    }
    #print_OUT("  '->[ " . scalar @back . " ] SNPs on PED
file");
    return ( \@back );
}

sub read_map_and_ped {
    my $file = shift;
    my $map = shift;
    my $affy_to_rsid = shift;
    my $affy_id = shift;

    my @bim = @{ read_map($map,$affy_to_rsid,$affy_id) };
    #print_OUT("Reading Genotypes from [ $file ]");
    open(PED,$file) or #print_OUT("Cannot open [ $file ] file")
or exit(1);
    my @back_fam = ();
    my @geno_matrix= ();
    while (my $sample = <PED>){
            chomp($sample);
            my ($iid,$fid,$mid,$pid,$sex,$pheno,@genotypes) =
split(/\s+/,$sample);
            my $snp_counter = 0;
            for (my $g = 0; $g < scalar @genotypes; $g+=2){
                $geno_matrix[$snp_counter]->{alleles}->
{$genotypes[$g]}++;
                $geno_matrix[$snp_counter]->{alleles}->
{$genotypes[$g+1]}++;
                    push @{ $geno_matrix[$snp_counter]->
{genotypes} }, "$genotypes[$g]$genotypes[$g+1]";
                $snp_counter++;
            }
            push @back_fam,{
                    'iid'   => $iid,
                    'fid'   => $fid,
                    'mid'   => $mid,
                    'pid'   => $pid,
                    'sex'   => $sex,
                    'pheno' => $pheno,
        };
    }
    my @back_genotypes = ();
    my $snp_counter = 0;
```

```perl
    foreach my $snp (@geno_matrix){
        my @alleles = sort { $snp->{alleles}->{$b} <=> $snp->
{alleles}->{$a} } keys %{ $snp->{alleles} };
        ($snp->{alleles}->{major},$snp->{alleles}->{minor},
$snp->{alleles}->{missing}) = 0;
        for (my $i = 0; $i < scalar @alleles; $i++) {
            if ($alleles[$i] == 0) {
                $snp->{alleles}->{missing} =$alleles[$i];
            } else {
                if ( $snp->{alleles}->{major} == 0){
                    $snp->{alleles}->{major}= $alleles
[$i];
                } else {
                    $snp->{alleles}->{minor}= $alleles
[$i];
                }
            }
        }

        $bim[$snp_counter]->{a1} = $snp->{alleles}->{minor};
        $bim[$snp_counter]->{a2} = $snp->{alleles}->{major};
        foreach my $g (@{ $snp->{genotypes} }){
            my $major_homo = "$snp->{alleles}->{major}$snp-
>{alleles}->{major}";
            my $minor_homo = "$snp->{alleles}->{minor}$snp-
>{alleles}->{minor}";
            my $hetero1 = "$snp->{alleles}->{major}$snp->
{alleles}->{minor}";
            my $hetero2 = "$snp->{alleles}->{minor}$snp->
{alleles}->{major}";
            my $missing = "$snp->{alleles}->{missing}$snp->
{alleles}->{missing}";
            my $recoded = 0;
            # homozygous major allele
            if ($g == $major_homo) {
                $recoded = 3;
            } elsif ($g == $minor_homo){ # homozygous minor
allele
                $recoded = 1;
            } elsif (($g == $hetero1) or ($g == $hetero2))
{ # heterozygous
                $recoded = 2
            } elsif ($g == $missing) { # missing
                $recoded = 0;
            } else { #print_OUT(" COULD NOT RECOGNIZE THIS
GENOTYPE >$g<\n" . Dumper($snp->{alleles}) . "");
                exit(1);
            }
            push @{$back_genotypes[$snp_counter]},
$recoded;
        }
        $snp_counter++;
    }
    #print_OUT("  '-> [ " . scalar @back_fam . " ] samples");
```

```perl
        #print_OUT("  '-> [ " . scalar @back_genotypes . " ]
SNPs");
        return(\@back_fam, \@back_genotypes,\@bim);
}

sub extract_genotypes_for_snp_list{
    my $snp_list = shift;
    my $line_index = shift;
    my $g_prob_threshold = shift;
    my $geno_probs_format = shift;
    my $gprobs = shift;
    my $gprobs_index = shift;
    my @geno_probs = ();
    my @geno_hard_coded = ();
    # loop over the snps mapped to the gene
    for (my $i = 0; $i < scalar @$snp_list; $i++){
            my $line = line_with_index($gprobs, $gprobs_index,
$line_index->[$i]);
            my @genos = split(/[\t+\s+]/,$line);
            # now loop over all samples for this snps
            my $sample_counter = 0;
            # counter start from 5 because the first columns are
chromosome, SNP id, position, minor allele and major allele
            # counter increases by three because each sample has
3 genotype probabilities for the AA, AB and BB, with A the minor
allele
            my $start_index = 0;
            $start_index = 5 if ($geno_probs_format eq 'OXFORD');
            $start_index = 3 if ($geno_probs_format eq 'BEAGLE');
            for (my $g = $start_index; $g < scalar @genos; $g
+=3){
                my $snp_prob = pdl @genos[$g..$g+2];
                my $max_index = maximum_ind($snp_prob);
                my $value = undef;

                if ($snp_prob->dsum == 0){
                    $value = 0;
                } else {
                    $value = $snp_prob->($max_index);
                }
                if ($value < $g_prob_threshold) {$value = 0;}
                push @{ $geno_probs[$sample_counter] } , sclr
$value;

                my $dossage = 0*$snp_prob->(0) + 1*$snp_prob->
(1) + 2*$snp_prob->(2);
                push @{ $geno_hard_coded[$sample_counter] },
sclr $dossage;
                $sample_counter++;
            }
    }
    my $coded_mat = transpose double pdl @geno_hard_coded;
    my $prob_mat = transpose double pdl @geno_probs;
    return($prob_mat,$coded_mat);
}
sub get_snp_list_from_bgl_format {
```

```perl
    my $geno_probs = shift;
    my $geno_probs_index = shift;
    my $affy_to_rsid = shift;
    my $affy_id = shift;

    my $index = 0;
    my @back = ();
    my $desired_line = 1;
    my $eof = 0;
    while () {
            my $line = line_with_index(*$geno_probs, *
$geno_probs_index, $desired_line);
            last if ($line eq '1');
            my ($snp,$a1,$a2) = split(/\s+/,$line);
            if ( defined $affy_to_rsid ) {
                    if ($snp !~ m/^rs/){
                            if (exists $affy_id->{$snp}){ $snp =
$affy_id->{$snp};}
                    }
            }
            push @back,{
                    'snp_id' => $snp,
                    'chr'    => 0,
                    'cm'     => 0,
                    'pos'    => 0,
                    'a2'     => $a1,
                    'a1'     => $a2,
        };
            $desired_line++;
    }
    #print_OUT("[ " .  scalar @back . " ] SNPs on BEAGLE format
genotype probability file");
    return ( \@back );
}

sub get_snp_list_from_ox_format {
    my $geno_probs = shift;
    my $geno_probs_index = shift;
    my $affy_to_rsid = shift;
    my $affy_id = shift;

    my $index = 0;
    my @back = ();
    my $desired_line = 1;
    my $eof = 0;
    while () {
            my $line = line_with_index(*$geno_probs, *
$geno_probs_index, $desired_line);
            last if ($line eq '1');
            my ($chr,$snp,$pos,$a1,$a2) = split(/\s+/,$line);
            if ( defined $affy_to_rsid ) {
                    if ($snp !~ m/^rs/){
                            if (exists $affy_id->{$snp}){ $snp =
$affy_id->{$snp};}
                    }
            }
```

```perl
        push @back,{
                'snp_id' => $snp,
                'chr'    => $chr,
                'cm'     => 0,
                'pos'    => $pos,
                'a2'     => $a1,
                'a1'     => $a2,
        };
            $desired_line++;
    }
    #print_OUT("[ " .  scalar @back . " ] SNPs on OXFORD format
genotype probability file");
    return ( \@back );
}

sub extract_binary_genotypes {
    my $n_genotypes = shift; # number of genotypes per SNP
    my $bytes_per_snp = shift; # number of bytes needed to code
the SNP
    my $byte_position = shift; # starting byte position for
this SNP
    my $FH = shift; # file handle for the genotype file
    $FH->seek(3 + $byte_position,SEEK_SET); # re-set the file-
handle to position start position of the SNP of interest
    my $buffer = ""; # this will store the information read
    my $n_bytes = read $FH, $buffer, $bytes_per_snp; # read the
genotypes
    my $data_size = $bytes_per_snp*8; # the amount of data to
extract is 8 bits per byte
    my $bin_data = unpack("B$data_size",$buffer);
    my @bits = ( $bin_data =~ m/\d{8}/g );
    my @genotypes = ();
    foreach my $b (@bits){
            $b = reverse($b); # for some odd reason PLINK stores
the genotypes in reverse order
            push @genotypes, @{ get_genotypes($b)};# transform
each byte on genotypes
    }
    return(\@genotypes);
}

sub get_genotypes {
    my $b = shift; # a byte
    my @back = ();
    my @genotypes = ( $b =~ m/\d{2}/g ); # extract a pair of
number = a genotype
    foreach my $geno (@genotypes){
            # 10 indicates missing genotype, otherwise 0 and 1
point to allele 1 (minor) or allele 2 (mayor) in the BIM file,
respectively
            if    ( $geno eq '00' ) {  # homozygous 1/1
                push @back, '1';
            } elsif ( $geno eq '11' ) { # -- other homozygous 2/2
                push @back, '3';
            } elsif ( $geno eq '01' ) { # -- heterozygous 1/2
                push @back, '2';
```

```perl
            } elsif ( $geno eq '10' ) { # -- missing genotype 0/0
                push @back, '0';
            } else {
                #print_OUT("This genotype is not recognize
[ $geno ]");
            }    # genotype not recognize

    }
    return(\@back);
}
```

## 8.9. GWAS_STATS.pm

Perl routines to calculate some statistics, e.g. meta-analysis statistics.

```perl
package GWAS_STATS;
use strict;
use warnings;
use Carp;
use Exporter qw (import);

# define verbose = undef;
my $v = undef;

# check the modules needed

eval {
    use PDL;
    use PDL::Matrix;
    use PDL::GSL::CDF;
    use PDL::Primitive;
    use PDL::NiceSlice;
    use PDL::Stats::Basic;
    #use PDL::LinearAlgebra; # commented until re-implement the
use of simulation to calculate p-values
    use PDL::Bad;
};
if ($@) {
    print "Some libraries does not seem to be in you system.
quitting\n";
    exit(1);
}

our (@EXPORT, @EXPORT_OK, %EXPORT_TAGS);

@EXPORT = qw(  rmnorm get_maf get_fix_and_radom_meta_analysis
get_makambi_chi_square_and_df calculate_LD_stats
get_lambda_genomic_control number_effective_tests
z_based_gene_pvalues);      # symbols to export by default
@EXPORT_OK = qw(  rmnorm get_maf get_fix_and_radom_meta_analysis
get_makambi_chi_square_and_df calculate_LD_stats
get_lambda_genomic_control number_effective_tests
z_based_gene_pvalues); # symbols to export on request

sub rmnorm {
    my $n = shift;
    my $mean = shift;
    my $c = shift;
    my $chol = shift;
    my @back = ();
```

```perl
    my $d = $c->getdim(0);
    my $vector = mpdl grandom $d,$n;
    if (not defined $chol) { $chol = mchol($c); }
    my $z = $vector x $chol->transpose;
    my $y = transpose( $mean + transpose($z));
    return($y,$chol);
}

# this subroutine applies the makambi method to combine p-values
from correlated test
# the method receives a correlation matrix, a set of weights for
each statistic and a set of probabilies to combine.
# it returns the chi-square and the degrees of freedom of the
test

sub get_makambi_chi_square_and_df {
    my $cor = shift; # a pdl matrix with the varoable
correlations
    my $w = shift; # a pdl vector with the weights;
    # make sure weiths sum one
    $w /= $w->sumover;
    my $pvalues = shift; # a pdl vector with the p-values to be
combined
    # calculate the correlation matrix before the applying the
weights
    # I have change this calculation following the results of
the paper of Kost et al. this should improve the approximation
of the test statistics
    # Kost, J. T. & McDermott, M. P. Combining dependent p-
values. Statistics & Probability Letters 2002; 60: 183-190.
    # my $COR_MAT = (3.25*abs($cor) + 0.75*(abs($cor)**2)); #
OLD
    my $COR_MAT = (3.263*abs($cor) + 0.710*(abs($cor)**2 +
0.027*(abs($cor)**3)); # NEW
    my $second = $COR_MAT*$w*($w->transpose); # apply the
weights
    ($second->diagonal(0,1)) .= 0; # set the diagonal to 0
    my $varMf_m = 4*sumover($w**2) + $second->flat->sumover; #
calculate the variance of the test statistics
    my $df = 8/$varMf_m; # the degrees of freedom of the test
statistic
    my $chi_stat = dsum(-2 * $pvalues->log * $w); # and the
chi-square for the combine probability
    $chi_stat = ( $chi_stat/2 ) * $df;
    my $sum = dsum(-2 * $pvalues->log * $w);
    if (defined $v){
        print_OUT("df: $df; var: $varMf_m; w: $w; chi-stat:
$chi_stat; SUM: $sum");
        print_OUT("P-values: " . $pvalues . "");
    }
    return ($chi_stat,$df);
}

sub z_based_gene_pvalues {
```

```perl
    my $gene = shift;
    my $mnd = shift;
    if ( ref($gene->{'pvalues'}) eq 'ARRAY'){ $gene->
{'pvalues'} = pdl @{ $gene->{'pvalues'} }; }
    if ( ref($gene->{'effect_size'}) eq 'ARRAY'){ $gene->
{'effect_size'} = pdl @{ $gene->{'effect_size'} }; }
    if ( ref($gene->{'effect_size_se'}) eq 'ARRAY'){ $gene->
{'effect_size_se'} = pdl @{ $gene->{'effect_size_se'} }; }

    if (scalar @{$gene->{'geno_mat_rows'}} < 2){
        return(-9);
    }
    my $cov;
    if (defined $gene->{cor_ld_r}){
        $cov = $gene->{cor_ld_r};
    } else {
        $cov = $gene->{cor};
    }

    my $pvals = $gene->{'pvalues'};
    $pvals->index( which($pvals == 1) ) .= double 1-2.2e-16;
    $pvals->index( which($pvals == 0) ) .= double 2.2e-16;
    my $B = -1*gsl_cdf_ugaussian_Pinv($pvals);
    my $observed_stat = undef;
    if (defined $gene->{'effect_size_se'}){
        $observed_stat = get_fix_and_radom_meta_analysis($B,
$gene->{'effect_size_se'},undef,$cov);
    } else {
        my $se = 1/$gene->{'weights'};
        $observed_stat = get_fix_and_radom_meta_analysis($B,
$se,undef,$cov);
    }
    return($observed_stat);
}

sub get_lambda_genomic_control {
    my $p = shift;
    $p = double 1 - pdl $p;
    my $chi = gsl_cdf_chisq_Pinv($p,1);
    return $chi->median/0.456;
}

sub get_fix_and_radom_meta_analysis {
    my $B = shift;
    my $SE = shift;
    my $external_w = shift;
    my $VarCov = shift;
    if (ref($B) eq 'ARRAY') { $B = pdl $B; }
    if (ref($SE) eq 'ARRAY') { $SE = pdl $SE; }
    if (ref($external_w) eq 'ARRAY') { $external_w = pdl
$external_w; }
    my $N = $B->nelem;
    if (not defined $VarCov) {
```

```perl
                $VarCov = stretcher(ones $N);
        }
        my $W = null;
        if ( ( ref($external_w) eq 'PDL') and (nelem($external_w >
0) > 0) and (dsum( ($external_w - $external_w->davg)**2 ) > 0 ))
{
                $W = 1/($SE + 1/$external_w);
        } else {
                $W = 1/$SE;
        }
        # calculate fix effect estimate
        my $norm_w_fix = $W/$W->dsum;
        #my $B_fix = dsum($B*$W)/$W->dsum;
        #my $V_fix = dsum($W*$W->transpose*$VarCov);
        my $B_fix = dsum($B*$norm_w_fix)/$norm_w_fix->dsum;
        my $V_fix = dsum($norm_w_fix*$norm_w_fix->transpose*
$VarCov);
        my $fix_chi_square_df1 = ($B_fix**2)/$V_fix;

        my $stouffer_w_fix = $W/dsum($W**2);
        my $B_stouffer_fix = dsum($B*$norm_w_fix)/sqrt(dsum
($stouffer_w_fix*$stouffer_w_fix->transpose*$VarCov));

        # calculate heterogeneity parameter Q
        my $Q = 0.0;
        {
                my $cor = $VarCov->copy();
                $cor->diagonal(0,1) .=0;
                $cor = (3.263*abs($cor) + 0.710*(abs($cor)**2) +
0.027*(abs($cor)**3));
                my $df = 8/(dsum($cor*$W*$W->transpose) + 4*dsum
($W**2));
                my $Q_naive = dsum ($W * (($B_fix - $B)**2));
                $Q = sclr gsl_cdf_chisq_Pinv( gsl_cdf_chisq_P
(  $df*0.5*$Q_naive, $df), $N - 1);
        }
        # calculate heterogeneity parameter I-squared
        my $I_squared = 0.0;
        if ($Q > ($N - 1)){
                eval { $I_squared = 100*($Q - ($N - 1))/$Q; };
                if ($@){
                        $I_squared = 0.0;
                }
        }
        # calculate tau-squared
        my $tau_squared = 0.0;
        if ($Q > ($N - 1)){
                eval { $tau_squared = ($Q - ($N - 1))/(dsum($W) -
dsum($W**2)/$W->dsum); };
                if ($@){
                        # tau-squared equal 0 if was < 0
                        $tau_squared = 0.0;
                }
        }
        # calculate the random effect estimate
        my $w_star = null;
```

```perl
    if ( (ref($external_w) eq 'PDL') and (nelem($external_w >
0) > 0) and (dsum( ($external_w - $external_w->davg)**2 ) > 0 ))
{
            $w_star = 1/( $tau_squared + $SE + 1/$external_w);
    } else {
            $w_star = 1/( $tau_squared + $SE);
    }
    my $norm_w_random = $w_star/$w_star->dsum;
    my $B_random = dsum( $B*$norm_w_random)/$norm_w_random-
>dsum;
    #my $B_random = dsum( $B*$w_star)/$w_star->dsum;
    my $V_random = dsum($norm_w_random*$norm_w_random-
>transpose*$VarCov);
    #    my $V_random = dsum($w_star*$w_star->transpose*
$VarCov);
    my $random_chi_square_df1 = ($B_random**2)/$V_random;

    my $stouffer_w_random = $w_star/dsum($w_star**2);
    my $B_stouffer_random = dsum($B*$norm_w_random)/sqrt(dsum
($stouffer_w_random*$stouffer_w_random->transpose*$VarCov));

    my $back =  {
            'B_stouffer_fix' => $B_stouffer_fix,
            'B_stouffer_random' => $B_stouffer_random,
            'B_fix' => $B_fix,
            'B_random' => $B_random,
            'V_fix' => $V_fix,
            'V_random' => $V_random,
            'Chi_fix' => $fix_chi_square_df1,
            'Chi_random' => $random_chi_square_df1,
            'Q' => $Q,
            'I2' => $I_squared,
            'tau_squared' => $tau_squared,
            'N' => $N,
            'Z_P_fix' => gsl_cdf_gaussian_P( -1 * $B_fix ,sqrt
($V_fix )),
            'Z_P_random' => gsl_cdf_gaussian_P( -1 *
$B_random,sqrt($V_random) ),
            'Q_P' => 1- gsl_cdf_chisq_P($Q ,$N -1),
    };
    return $back;
}

sub calculate_LD_stats {
=head1 Docs

    From Lon Cardon

    The EM algorithm is used to estimate the recombination
fraction (parameter theta)
    between two genetic markers under the assumption of Hardy-
Weinberg equilibrium.

    For SNPs, consider the 3 x 3 table of marker 1 with
alleles 'A' and 'a'
```

and marker 2 with alleles 'B' and 'b'.  Number the cells as follows:

```
AA  |  Aa  |  aa
───────────────────
BB  |  0  |   1  |   2
Bb  |  3  |   4  |   5
bb  |  6  |   7  |   8
```

Note that all cells arise from unique haplotypes (e.g., cell 0 can only comprise
two 'AB' haplotypes; cell 1 has one 'AB' and one 'aB', etc), _except_ the double
heterozygote cell 4, which can have either AB/ab or Ab/aB. We need to estimate the
probability of these two events.  We assume that pair AB/ab arises when no
recombination occurs (1 - theta) whereas pair Ab/aB arises in the presence of recombination
(probability theta).  Thus, for the four haplotype possibilities at two markers
(AB,ab,Ab,aB), we have:

```
prAB=prAb=praB=prab=.25;

nAB=(float)(2*cells[0]+cells[1]+cells[3]);
nab=(float)(2*cells[8]+cells[7]+cells[5]);
nAb=(float)(2*cells[6]+cells[7]+cells[3]);
naB=(float)(2*cells[2]+cells[1]+cells[5]);

N = nAB + nab + nAb + naB + 2*cell4;

while(theta-thetaprev > CONVERGENCE_CRITERIA)
{
thetaprev=theta;
prAB=(nAB + (1-theta)*cells[4])/N;
prab=(nab + (1-theta)*cells[4])/N;
prAb=(nAb + theta*cells[4])/N;
praB=(naB + theta*cells[4])/N;
theta=(prAb*praB)/(prAB*prab + prAb*praB);
}
```

i.e., D = prAB - frq(A)*frq(B), r^2 = D^2/(frq(A)*frq(B)*frq(a)*frq(b)),
Dprime = D/Dmax, etc.

```
=cut
    my $first  = shift;
    my $second = shift;

    my %genotypes = ( 'AABB' => 0, # cell 0
    'AaBB' => 0, # cell 1
    'aaBB' => 0, # cell 2

    'AABb' => 0, # cell 3
```

```perl
    'AaBb' => 0, # cell 4
    'aaBb' => 0, # cell 5

    'AAbb' => 0, # cell 6
    'Aabb' => 0, # cell 7
    'aabb' => 0  # cell 8
    );

    for (my $i = 0; $i < scalar @$first; $i++){
        $genotypes{'AABB'}++ if (( $first->[$i] == 1) and
($second->[$i] == 1));
        $genotypes{'AaBB'}++ if (( $first->[$i] == 2) and
($second->[$i] == 1));
        $genotypes{'aaBB'}++ if (( $first->[$i] == 3) and
($second->[$i] == 1));

        $genotypes{'AABb'}++ if (( $first->[$i] == 1) and
($second->[$i] == 2));
        $genotypes{'AaBb'}++ if (( $first->[$i] == 2) and
($second->[$i] == 2));
        $genotypes{'aaBb'}++ if (( $first->[$i] == 3) and
($second->[$i] == 2));

        $genotypes{'AAbb'}++ if (( $first->[$i] == 1) and
($second->[$i] == 3));
        $genotypes{'Aabb'}++ if (( $first->[$i] == 2) and
($second->[$i] == 3));
        $genotypes{'aabb'}++ if (( $first->[$i] == 3) and
($second->[$i] == 3));
    }

    my ($nAB,$nAb,$naB,$nab) = double 0.0;
    $nAB = 2*$genotypes{'AABB'} + $genotypes{'AaBB'} +
$genotypes{'AABb'};
    $nab = 2*$genotypes{'aabb'} + $genotypes{'Aabb'} +
$genotypes{'aaBb'};
    $nAb = 2*$genotypes{'AAbb'} + $genotypes{'Aabb'} +
$genotypes{'AABb'};
    $naB = 2*$genotypes{'aaBB'} + $genotypes{'AaBB'} +
$genotypes{'aaBb'};
    my $AaBb = double $genotypes{'AaBb'};

    if (defined $v){
        print "Haplotype freqs\n";
        print "Table: $genotypes{'AABB'}  $genotypes{'AaBB'}
$genotypes{'aaBB'}\n";
        print "Table: $genotypes{'AABb'}  $genotypes{'AaBb'}
$genotypes{'aaBb'}\n";
        print "Table: $genotypes{'AAbb'}  $genotypes{'Aabb'}
$genotypes{'aabb'}\n";
        print "nAB [ $nAB ] nab [ $nab ] nAb [ $nAb ] naB
[ $naB ] nAaBb [ $AaBb ]\n";
    }

    my $N = double $nAB + $nab + $nAb + $naB + 2*$AaBb;
```

```perl
    my $theta = double 0.5;
    my $thetaprev = 2;
    my $iterations = 0;
    while( abs($theta-$thetaprev) > 0.0001 ) {
            $thetaprev = $theta;
            eval {
                $theta = (($nAb + $theta*$AaBb)*($naB + $theta*
$AaBb))/
                ((nAB + (1-$theta)*$AaBb)*($nab + (1-$theta)*
$AaBb) + ($nAb + $theta*$AaBb)*($naB + $theta*$AaBb));
            };
            if (defined $v){
                print "iter: $iterations; theta [ $theta ]\n";
            }
            if ($@){ $theta = 0.5; } #included to avoid division
by 0
            if (isbad(setnantobad($theta))) { $theta = 0.5; }
            $iterations++;
    }

    my ( $f_A ) = get_maf($first);
    my ( $f_B ) = get_maf($second);

    if ($f_B == 0 or $f_A == 0 or $f_B == 1 or $f_A == 1){
            return({ 'D'=> 0,
                'r2' => 0,
                'r' => 0,
                'theta' => -1,
                'N' =>  0,
                'd_prime' =>  0,
            });
    }

    my $D;
    my $r2;
    my $r;
    eval{
            $D  = ($nAB+(1-$theta)*$AaBb)/$N - ($f_A*$f_B);
            $r2 = $D*$D/($f_A*$f_B*(1-$f_A)*(1-$f_B));
            $r = $D/sqrt($f_A*$f_B*(1-$f_A)*(1-$f_B));
    };

    if ($@){
            $D = 0;
            $r2 = 0; #for some cases is not possible to calculate
the r2 due to a 0 in the divisor
            $r = 0;
    }

    my $Dmax = 0;
    my $d_prime;

    if ($D < 0){
            $Dmax = $f_A*$f_B if ($f_A*$f_B < ((1-$f_A)*(1-
$f_B)));
```

```perl
                $Dmax = (1-$f_A)*(1-$f_B) if ($f_A*$f_B >= ((1-$f_A)*
(1-$f_B)));
        }
        if ($D > 0){
                $Dmax = $f_A*(1-$f_B) if ($f_A*(1-$f_B) < (1-$f_A)*
$f_B);
                $Dmax = (1-$f_A)*$f_B if ($f_A*(1-$f_B) >= (1-$f_A)*
$f_B);
        }
        eval{
                $d_prime = $D/$Dmax;
        };
        if ($@){
                $d_prime = 0;
        }

        $D = sclr $D;
        $r2 = sclr $r2;
        if (abs($r2) > 1){ $r2 = $r2/abs($r2); }
        $r = sclr $r;
        if (abs($r) > 1){ $r = $r/abs($r); }
        $N = sclr $N;
        $d_prime = sclr $d_prime;
        $theta = sclr $theta;

        my $o = { 'D'=> $D,
                'r2' => $r2,
                'r' => $r,
                'theta' => $theta,
                'N' =>  $N,
                'd_prime' =>  $d_prime,
        };
        return $o;

}

sub get_maf {
    my $a = shift;
    if (ref($a) eq 'ARRAY'){
            $a = pdl $a;
    }
    my $maf = (2*dsum( $a ==1 ) + 1*dsum( $a ==2 ) )/(2*dsum
($a != 0));
    if ($maf > 0.5){ $maf = 1 - $maf;}
    return($maf);
}

sub pearson_corr_genotypes {
    # implemented as in S. Wellek, A. Ziegler, Hum Hered 67,
128 (2009).
    # genotypes must be coded as 1,2 and 3,any other coding
will be use. missing genotypes can be set to anything different
of 1, 2 or 3.
    # this method will fail is more than 2 out of the four
homozygoues haplotypes have counts 0. In such case i recommend
to use the default option of the spearman correlation.
```

```perl
    my $snp1 = shift;
    my $snp2 = shift;
    # initialize to 0 all values for the haplotype cells
    my @cells = (0,0,0,0,0,0,0,0,0);
    # count how many observation of every haplotype are
    for (my $i = 0; $i < scalar @$snp1; $i++){
            $cells[8]++ if (($$snp1}[$i] == 1) and ($$snp2}[$i]
== 1));
            $cells[7]++ if (($$snp1}[$i] == 2) and ($$snp2}[$i]
== 1));
            $cells[6]++ if (($$snp1}[$i] == 3) and ($$snp2}[$i]
== 1));
            $cells[5]++ if (($$snp1}[$i] == 1) and ($$snp2}[$i]
== 2));
            $cells[4]++ if (($$snp1}[$i] == 2) and ($$snp2}[$i]
== 2));
            $cells[3]++ if (($$snp1}[$i] == 3) and ($$snp2}[$i]
== 2));
            $cells[2]++ if (($$snp1}[$i] == 1) and ($$snp2}[$i]
== 3));
            $cells[1]++ if (($$snp1}[$i] == 2) and ($$snp2}[$i]
== 3));
            $cells[0]++ if (($$snp1}[$i] == 3) and ($$snp2}[$i]
== 3));
    }
    if (defined $v) {
            #print_OUT("  AA Aa aa");
            #print_OUT("BB $cells[0]  $cells[1]  $cells[2]");
            #print_OUT("Bb $cells[3]  $cells[4]  $cells[5]");
            #print_OUT("bb $cells[6]  $cells[7]  $cells[8]");
    }
    my $h1 = sqrt($cells[0]);
    my $h2 = sqrt($cells[2]);
    my $h3 = sqrt($cells[6]);
    my $h4 = sqrt($cells[8]);
    my $corr = 2*($h1 * $h4 - $h2 * $h3)/sqrt(4*($h1 + $h2)*
($h3 + $h4)*($h1 + $h3)*($h2 + $h4));
    return($corr);
}

# this subroutine calculate the number of effective test by the
Galwey and Gao method.
sub number_effective_tests {
    my $mat = shift;
    # calculate the Eigen value of the correlation matrix
    my $eigens = eigens ${$mat};
    # normalize the values
    my $eigens_norm =  pdl sort { $b <=> $a } list ($eigens/
(dsum $eigens) );
    # calculate number of effective test per Gao et al Gao, X.,
Becker, L. C., Becker, D. M., Starmer, J. D. & Province, M. A.
Avoiding the high Bonferroni penalty in genome-wide association
studies. Genet. Epidemiol. (2009).
    # the methos simply count how many Eigen values are needed
to explain 99.5 % of the variance
```

```perl
    my $simpleM = 0;
    for( $simpleM = 0; $simpleM < scalar list $eigens_norm;
$simpleM++){
        my $sum = sumover $eigens_norm->slice("0:$simpleM");
        if ($sum >= 0.995){
            $simpleM++;
            last;
        }
    }

    # calculate number of effective test by Galwey, N. W. A new
measure of the effective number of tests, a practical tool for
comparing families of non-independent significance tests. Genet.
Epidemiol. (2009).
    # this method calculate the square of sum of the square-
root of the Eigen values divided by the sum.
    my $numerator = 0;
    my $denominator = 0;
    foreach my $e (list $eigens) {
        if ($e > 0){
            $numerator += sqrt($e);
            $denominator += $e;
        }
    }
    my $Meff_galwey = ($numerator**2)/$denominator;
    if (defined $v){
        #print_OUT(" simpleM_Gao = $simpleM; Meff_galwey=
$Meff_galwey $numerator $denominator");
    }
    return($simpleM,$Meff_galwey);
}

1;
```

# 8.10.Pareto_Distr_Fit.pm

Perl routines to estimate p-values using extreme value distribution theory as described on (Knijnenburg et al., 2009) but estimating the parameters of the generalised Pareto distribution by the method of Zhang (2010) instead of maximum likelihood.

```perl
package Pareto_Distr_Fit;
use strict;
use diagnostics;
use PDL;
use PDL::Matrix;
use PDL::NiceSlice;
use PDL::GSL::CDF;
use PDL::Stats::Basic;
use PDL::Stats::Distr;
use PDL::LinearAlgebra qw(mchol);
use PDL::GSL::INTERP;

our (@EXPORT, @EXPORT_OK, %EXPORT_TAGS);

@EXPORT = qw( Pgpd );                  # symbols to export by
default
@EXPORT_OK = qw( Pgpd );              # symbols to export on
request

sub Pgpd {
=h
    Computing permutation test P-value of the GPD
approximation

            Input:    x0        original statistic
                      y          permutation values
                      N          number of permutation values
                      Nexc       number of permutations used
to approximate the tail
                      alpha      confidence level
            Output:  Phat     estimated P-value
                      Phatci    confidence intervals of the
estimated P-value
     Original code in Matlab by Theo Knijnenburg, Institute for
Systems Biology (Dec 11 2008)
=cut

    my $x0 = shift;
    my $y =  shift;
    my $Nexc = shift;
    my $alpha = shift;
    my $N = $y->nelem;
    my $Padth = 0.05;
```

```perl
    if ($Nexc >= $N-1) {
            $Nexc = int(0.1*$N);
    }

    #print "$N $Nexc $alpha\n";

    my ($PW2_less_005,$PA2_less_005,$W2,$A2,$cov,$a,$k,$t,
$frac) = undef;
    my $fitted = 0;
    do {
            $y = pdl reverse list qsort $y;
            # Defining the tail
            my $z = $y->(1:$Nexc);
            $t = davg($y($Nexc:$Nexc+1));
            $z = $z-$t;

            $frac = $Nexc/$N;
            # Fitting the tail and computing the Pvalue
            ($a,$k,$cov) = Jin_ZHANG_pareto_fit($z);
            #print "$Nexc $a $k $cov\n";
            my $p = 1- gsl_cdf_pareto_P($z,$a,$k);
            ($PW2_less_005,$PA2_less_005,$W2,$A2) = gpdgoft( $p ,
$k); # goodness-of-fit test
            #print "$PW2_less_005,$PA2_less_005,$W2,$A2\n";

            if ( $A2 eq 'nan'){
                    $fitted = $PA2_less_005;
            } else {
                    $fitted = $PW2_less_005;
            }
            $Nexc -= 10;
    } until ( ( $fitted == 1 ) or ($Nexc < 11) );
    if ($fitted == 1){
            my ($Phat,$Phatci_low,$Phatci_up) = gpdPval($x0-$t,
$a,$k,$cov,$alpha);
            $Phat = $frac*$Phat;
            $Phatci_up = $frac*$Phatci_up;
            $Phatci_low =$frac*$Phatci_low;
            return($Phat,$Phatci_low,$Phatci_up);
    } else {
            return(-1,-1,-1);
    }
}

sub gpdPval {
    my $x0 = shift;
    my $a = shift;
    my $k = shift;
    my $cov = shift;
    my $alpha = shift;

    my $Phat = 1 - gsl_cdf_pareto_P($x0,$a,$k);

    my $gaussian_random_mat = grandom 2, 10000;
```

```perl
    my $reapmat= ones 2, 10000;
    my $par = pdl $a,$k;
    $reapmat = $reapmat * $par;

    my $chol = mchol($cov);
    my $Q = ($gaussian_random_mat  x $chol->transpose ) +
$reapmat;
    my $sample =  1 - gsl_cdf_pareto_P($x0,$Q->(0,),$Q->(1,));
    $sample->inplace->setnantobad;
    $sample =  $sample->where(  $sample->isgood() );

    my $Phatci_low = pct($sample,$alpha/2);
    my $Phatci_up =  pct($sample,1-$alpha/2);

    return($Phat,$Phatci_low,$Phatci_up);
}

sub Jin_ZHANG_pareto_fit {
    my $x = shift; # x is the sample data from the GPD
    my $n = $x->nelem;
    $x = qsort $x;
    my $p = pdl (3 .. 9);
    $p /= 10;
    my $xp = $x->(rint($n*(1-$p)-0.5));
    my $m = 20 + rint( sqrt($n) );
    my $xq = $x->(rint($n*(1-$p*$p)+0.5));
    my $k = log($xq/$xp-1)/log($p);
    $k->inplace->setnantobad;
    $k->inplace->setbadtoval(0);
    my $a = $k*$xp/(1-$p**$k);
    $a->inplace->setnantobad;
    $a->inplace->setbadtoval(0);
    $a->where( $k == 0 ) .= (-$xp->where( $k == 0 )/log($p-
>where( $k == 0 )) ) ;
    $k = -1;
    my $tmp1 = pdl 1 .. $m;
    my $j_minus05_overm = ($tmp1 -0.5)/$m;
    my $leading = (($n-1)/($n+1))*(1/$x->($n-1));
    my $b =  $leading - ((1- $j_minus05_overm**$k)/$k) * (1/
median($a))/2;
    my @tmp = ();
    for (my $i = 0; $i < $m; $i++) {
         push @tmp, $n * lx($b->($i),$x);
    }
    my $L = flat pdl @tmp;
    @tmp = ();
    for (my $i = 0; $i < $m; $i++) {
         push @tmp, 1/dsum( exp( $L - $L->( $i ) ) );
    }
    my $w = flat pdl @tmp;
    $b = dsum($b*$w);
    $k = -davg(log(1-$b*$x));
    my $sigma = $k/$b;
```

```perl
    my $cov = mones 2,2;
    $cov(0,0) .= 2*($sigma**2)*(1-$k);
    $cov(1,0) .= $sigma*(1 - $k);
    $cov(0,1) .= $sigma*(1 - $k);
    $cov(1,1) .= (1 - $k)**2;
    return($sigma, $k,$cov);
}

sub lx {
    my $b= shift;
    my $x = shift;

    my $k = - davg( log( 1-$b*$x ) );
    if ($b == 0) {
            return( $k-1-log(davg($x)) );
    } else {
            return( $k-1+log($b/$k) );
    }
}

sub gpdgoft{
    my $p = shift;
    my $k = shift;

    $p = qsort $p;
    $p->where( (1 - $p) < 2.2e-16 ) .= 2.2e-16;

    my ($W2,$A2) =  fit_stats($p);

    my $W2_less_005 = 0; # 0 no and 1 yes
    my $A2_less_005 = 0; # 0 no and 1 yes

    $k = 0.5 if ($k > 0.5);

    my $k_vals = pdl
(-0.9,-0.5,-0.2,-0.1,0,0.1,0.2,0.3,0.4,0.5);
    my $W2_vals  = pdl
(0.115,0.124,0.137,0.114,0.153,0.16,0.171,0.184,0.201,0.222);
    my $A2_vals = pdl
(0.771,0.83,0.903,0.935,0.974,1.02,1.074,1.14,1.221,1.321);

    my $trend_A2 =  PDL::GSL::INTERP->init('linear',$k_vals,
$A2_vals);
    my $trend_W2 =  PDL::GSL::INTERP->init('linear',$k_vals,
$W2_vals);

    my $critic_w2 = undef;
    my $critic_a2 = undef;

    if ($k < -0.9 or $k > 0.5) {
            $critic_w2 = $trend_W2->eval($k,{ Extrapolate =>
1 } );
```

```perl
        $critic_a2 = $trend_A2->eval($k,{ Extrapolate =>
1 } );
    } else {
        $critic_w2 = $trend_W2->eval($k,{ Extrapolate =>
0 } );
        $critic_a2 = $trend_A2->eval($k,{ Extrapolate =>
0 } );
    }

    $A2_less_005 = 1 if ( $A2 >= $critic_a2 );
    $W2_less_005 = 1 if ( $W2 >= $critic_w2 );

    return($A2_less_005,$W2_less_005,$W2,$A2);
}

sub fit_stats {
    my $p = shift;
    my $n = $p->nelem;
    my $i = 1 + sequence $n;

    # Cramer-von Mises statistic
    my $W2 = dsum( ($p - ( (2*$i-1)/(2*$n) ))^2 ) + 1/(12*$n);
    # Anderson Darling statistic
    my $A2 = -$n - (1/$n) * dsum( ((2*$i-1) * (log($p)) + log
(double 1 - double $p->($n+1-$i - 1)) ));
    return($W2,$A2);
}

1;
```